\pdfoutput=1

\documentclass[11pt,twoside,a4paper,openany,headings=optiontotocandhead,bibliography=totoc]{scrbook}
\usepackage{scrhack}

\usepackage{graphicx}

\usepackage[hidelinks,
colorlinks=true,
linkcolor = .,
citecolor = .,
urlcolor = {myblue},
bookmarksnumbered=true
]{hyperref}

\usepackage{scrartcl_style_malcha}

\usepackage{subfiles}

\begin{document}
\frontmatter

\thispagestyle{empty}
\begin{tikzpicture}[overlay, remember picture]
\node[anchor=north east,					
      	xshift= -1  cm, 			  				
    	yshift= -1 cm] 
     	at (current page.north east) 	
     {\includegraphics[scale=1]{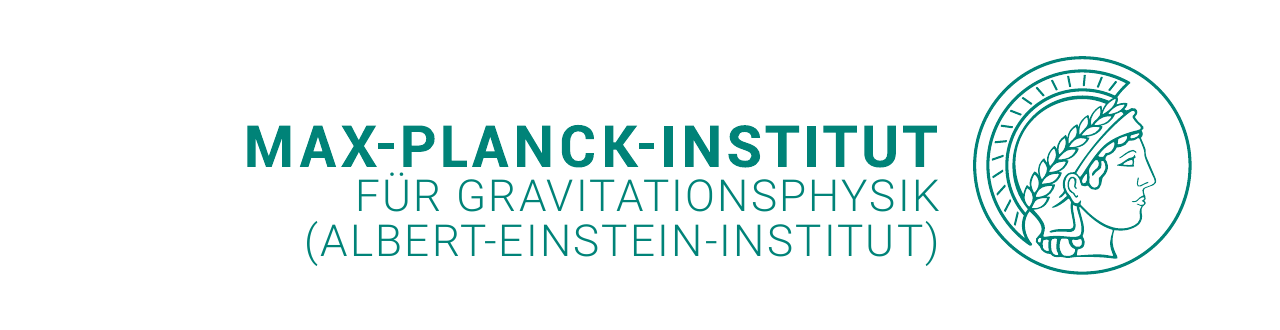}}; 
\end{tikzpicture}
\vspace*{2cm}
\begin{center}
\textsf{\textbf{\Huge The Nicolai Map and its Application}}\\[4ex]
\textsf{\textbf{\Huge in Supersymmetric Field Theories}}\\
\vspace{3.5cm}
\textsf{\LARGE Hannes Malcha} \\
\vspace*{\fill}
\textsf{2023\\[4ex]
Dissertation\\
Eingereicht an der Humboldt-Universität zu Berlin\\[4ex]
Max-Planck-Institut für Gravitationsphysik (Albert-Einstein-Institut)\\
Am Mühlenberg 1, 14476 Potsdam, Germany}
\end{center}

\newpage 		

\chapter*{Abstract}
Supersymmetric field theories can be characterized by the existence of a non-linear and non-local transformation of the bosonic fields, called the Nicolai map. It maps the interacting functional measure to that of a free theory such that the Jacobian determinant of the transformation equals the product of the fermionic determinants. In this thesis, we study the Nicolai maps of the 2-dimensional Wess-Zumino model, $\mathcal{N}=1$ super Yang-Mills and $\mathcal{N}=4$ super Yang-Mills. 

We give a constructive proof for the existence of the Nicolai map in these theories. The proof includes the derivation of the infinitesimal generator of the inverse Nicolai map, called the $\mathcal{R}_g$-operator. We use this operator to compute the Nicolai map of the 2-dimensional Wess-Zumino model up to the fifth order in the coupling. In $\mathcal{N}=1$ super Yang-Mills, we introduce the notion of `on- and off-shell' Nicolai maps, corresponding to the `on-shell' respectively `off-shell' supersymmetry in the different versions of the theory. The `on-shell' Nicolai map of $\mathcal{N}=1$ super Yang-Mills exists in $d=3$, 4, 6 and 10 dimensions but is constrained to the Landau gauge. We compute this map up to the fourth order in the coupling. The `off-shell' Nicolai map exists only in $d=4$ dimensions but for general gauges. We compute it in the axial gauge up to the second order in the coupling. In $\mathcal{N}=4$ super Yang-Mills, we give the $\mathcal{R}_g$-operator and use it to show that the Nicolai map of $\mathcal{N}=4$ super Yang-Mills can be obtained from the Nicolai map of 10-dimensional $\mathcal{N}=1$ super Yang-Mills by dimensional reduction.

Inverse Nicolai maps have a remarkable property. They map quantum correlation functions of bosonic observables to free correlation functions. Hence, Nicolai maps allow for a fermion (and ghost) free quantization of supersymmetric (gauge) theories.  We apply this property to the 10-dimensional $\mathcal{N}=1$ super Yang-Mills Nicolai map and compute the vacuum expectation value of the infinite straight line Maldacena-Wilson loop in $\mathcal{N}=4$ super Yang-Mills to order $g^6$. Thus extending the previous perturbative result by one order.

In the second part of this thesis, we derive the explicit field content of the $\frac{1}{2}$-BPS stress tensor multiplet in $\mathcal{N}=4$ super Yang-Mills, which in particular contains the $R$-symmetry current and the energy-momentum tensor.

\newpage 		

\chapter*{Acknowledgments}
First and foremost, I want to thank my supervisor Hermann Nicolai. Your continuous support and patient explanations have been invaluable to the success of my PhD. Over the past three years, I have greatly benefited from your vast knowledge and infallible intuition. In particular, I enjoyed our many discussions and always felt encouraged to pursue my own ideas. I am very much looking forward to continuing working with you at the AEI. 

I am grateful to my collaborators
Sudarshan Ananth,
Olaf Lechtenfeld,
Chetan Pandey,
and
Saurabh Pant
for their stellar contributions to our joint publications. Furthermore, I would like to thank
Olaf Lechtenfeld, 
Sudarshan Ananth
and
Jan Plefka
for many fruitful discussions. 

I am also grateful to everyone at the AEI in Potsdam for making my PhD such an enjoyable time and creating a superb working environment. During my PhD, I was supported by the IMPRS for Mathematical and Physical Aspects of Gravitation, Cosmology and Quantum Field Theory. I sincerely thank everyone involved and, in particular, Axel Kleinschmidt for the organization of lecture days and skill seminars. Moreover, I especially want to thank my fellow PhD students from the Quantum Gravity and Unified Theories division 
Matteo Broccoli,
Hugo Camargo,
Lorenzo Casarin,
Mehregan Doroudiani,
Jan Gerken,
Serena Giardino,
Caroline Jonas,
Johannes Knaute,
Benedikt König,
Lars Kreutzer,
and
Tung Tran
for the many delightful lunch and coffee breaks with stimulating discussions about physics and quite literally everything else. I want to thank Matthias Blittersdorf for the organization of weekly football matches, providing a very welcome change from research. 

Furthermore, I want to thank Lorenzo Casarin and Hermann Nicolai for helpful comments on the manuscript. 

I am grateful to my family for their continuous support and encouragement. Finally, I would like to thank Maja Leusch for the time of my life during the past three years. I am always looking forward to our amazing weekends together in Bremen, Hamburg and Potsdam. Thank you for your unconditional love and support.

\newpage 		

\chapter*{Publications by the Author}
The work presented here is mostly based on:
\begin{itemize}
\item [\cite{Ananth:2020lup}] S.~Ananth, O.~Lechtenfeld, H.~Malcha, H.~Nicolai, C.~Pandey and S.~Pant, 
\textit{Perturbative linearization of supersymmetric Yang-Mills theory},
\href{https://dx.doi.org/10.1007/JHEP10(2020)199}{JHEP \textbf{10} (2020), 199}, \\
\href{https://arxiv.org/abs/2005.12324}{\texttt{arXiv:2005.12324 [hep-th]}}.

\item [\cite{Ananth:2020jdr}] S.~Ananth, H.~Malcha, C.~Pandey and S.~Pant,
\textit{Supersymmetric Yang-Mills theory in $D=6$ without anticommuting variables},
\href{https://dx.doi.org/10.1103/PhysRevD.103.025010}{Phys. Rev. D \textbf{103} (2021) no.2, 025010}, \\
\href{https://arxiv.org/abs/2006.02457}{\texttt{arXiv:2006.02457 [hep-th]}}.

\item [\cite{Malcha:2021ess}] H.~Malcha and H.~Nicolai,
\textit{Perturbative linearization of super-Yang-Mills theories in general gauges},
\href{https://dx.doi.org/10.1007/JHEP06(2021)001}{JHEP \textbf{06} (2021), 001},
\href{https://arxiv.org/abs/2104.06017}{\texttt{arXiv:2104.06017 [hep-th]}}.

\item [\cite{Malcha:2022fuc}] H.~Malcha,
\textit{Two Loop Ghost free Quantisation of Wilson Loops in $\mathcal{N}=4$ supersymmetric Yang-Mills},
\href{https://dx.doi.org/10.1016/j.physletb.2022.137377}{Phys. Lett. B \textbf{833} (2022), 137377}, 
\href{https://arxiv.org/abs/2206.02919}{\texttt{arXiv:2206.02919 [hep-th]}}.
\end{itemize}

\tableofcontents

\mainmatter
\hypersetup{linkcolor=myblue,citecolor=mygreen}

\chapter{Introduction}
Quantum field theory is our most comprehensive theoretical framework to describe subatomic particles and forces. It was largely developed in the 20th century and unifies the principles of quantum mechanics and special relativity. In particular, the perturbative predictions of the Standard Model have an unprecedented accuracy. However, in general quantum field theory is hard and to a large extent results beyond perturbation theory are unattainable. Thus, a considerable part of modern-day research in quantum field theory revolves around possible simplifications one can impose to make it more tractable. One such simplification is supersymmetry. Supersymmetry is a spacetime symmetry between bosons and fermions. The constraints it imposes on quantum field theory are strong enough to yield some additional non-perturbative results and considerable simplifications while at the same time not prohibiting all the interesting dynamics of regular quantum field theory. Mathematically, supersymmetry is a $\mathbbm{Z}_2$-graded extension of the Poincaré algebra, where the new anti-commuting generators relate bosonic states to fermionic states and vice versa. We call this algebra the supersymmetry algebra. While every quantum field theory is invariant under the action of the Poincaré algebra, every supersymmetric field theory is invariant under the action of the supersymmetry algebra.

Supersymmetry was initially discovered in string theory in the early 1970s. The first supersymmetric quantum field theory was the 4-dimensional Wess-Zumino model \cite{Wess:1974tw}, introduced by Wess and Zumino in 1974. It describes two scalar fields, a four-component Majorana spinor and two auxiliary fields. Shortly after the first supersymmetric gauge theories were introduced, first as a supersymmetric extension of quantum electrodynamics \cite{Wess:1974jb} and later as supersymmetric extensions of Yang-Mills theories \cite{Ferrara:1974pu,Salam:1974ig,Brink:1976bc}. Today supersymmetric Yang-Mills theories and, in particular, the maximally extended $\mathcal{N}=4$ super Yang-Mills theory \cite{Gliozzi:1976qd,Brink:1976bc}, are among the best-studied examples of quantum field theories. 

A key consequence of supersymmetry is the dramatic improvement it produces in the ultraviolet behavior of quantum field theories. In theories with linearly realized supersymmetry, such as the 4-dimensional Wess-Zumino model, there is a very powerful non-renormalization theorem, stating that superpotentials of chiral superfields do not get renormalized \cite{Grisaru:1979wc,Grisaru:1982zh}. Moreover, the mass and coupling constant do not receive any renormalization besides the wave-function renormalization. Thus, the mass term is at most logarithmically divergent. In supersymmetric gauge theories, the situation is a little more complicated as the renormalization constants become gauge-dependent. However, there is a notable exception. It has been shown that the beta function of $\mathcal{N}=4$ super Yang-Mills vanishes to all orders in perturbation theory \cite{Mandelstam:1982cb,Brink:1982wv} (see also \cite{Sohnius:1981sn,Grisaru:1980nk,Howe:1982tm,Brink:1982pd,Howe:1983sr,Baulieu:2006ru}). This implies that the coupling constant does not get renormalized in any gauge and allows the theory's superconformal symmetry to extend to the quantum level. Furthermore, it was argued that, in the light-cone gauge, the $\mathcal{N}=4$ theory is completely ultraviolet finite \cite{Mandelstam:1982cb,Grisaru:1980nk,Stelle:1981gi,Howe:1982tm}.

Another universal consequence of supersymmetry is the exact vanishing of the vacuum energy \cite{Zumino:1974bg}. Fermions and bosons contribute with opposite signs to the vacuum energy and since there are just as many fermions as bosons in supersymmetric theories, their respective contributions cancel to all orders in perturbation theory. 

Among all the supersymmetric field theories one, in particular, stands out, namely $\mathcal{N}=4$ super Yang-Mills. Beyond its finiteness properties and exact quantum conformal invariance, it is ubiquitous in non-perturbative formulations of string theory (M-theory), either via the AdS/CFT correspondence \cite{Maldacena:1997re} or, in its dimensionally reduced form, via the maximally supersymmetric $d=1$ matrix model with gauge group $SU(\infty)$ \cite{deWit:1988wri,Banks:1996vh}. In the large $N$ limit $\mathcal{N}=4$ super Yang-Mills is even integrable (see \cite{Beisert:2010jr} for a review). Moreover, being a 4-dimensional, non-abelian, minimally coupled gauge theory, $\mathcal{N}=4$ super Yang-Mills is similar enough to more realistic theories such as QCD that we can hope to deduce some approximations for these theories from exact results in $\mathcal{N}=4$ super Yang-Mills. Thus, a sustained effort to study supersymmetric Yang-Mills theories from all possible perspectives is more than justified. 

Yet, despite the vast literature on supersymmetric gauge theories, and especially the $\mathcal{N}=4$ theory, important questions remain. For example, does the $\mathcal{N}=4$ theory exist beyond perturbation theory as a non-trivial quantum theory or is it simply a free theory in disguise?  Because of the conformal invariance of the theory, even at the quantum level, it does not have a mass gap. Thus, there are no asymptotic one-particle states and consequentially no $S$-matrix (at least not in any conventional sense) whose non-triviality would affirm the non-triviality of the theory. A more appropriate framework to establish the non-perturbative existence is provided by the conformal bootstrap program (see \emph{e.g.} \cite{Rychkov:2016iqz}). Here the challenge is to compute exact $n$-point (for $n \ge 4$) correlation functions and the associated conformal cross ratios. Considerable progress in this direction has been made using integrability \cite{Beisert:2010jr}, amplitude calculations \cite{Dixon:2011xs,Henn:2014yza}, and holographic duality \cite{Alday:2021odx}. 

Other issues revolve around the finiteness of $\mathcal{N}=4$ super Yang-Mills. Thus far, it has only been established in the light-cone gauge and only for the transversal degrees of freedom \cite{Mandelstam:1982cb,Leibbrandt:1984yb,Bassetto:1987uh}. In other gauges, the usual quantum field theoretical infinities appear and a wave function renormalization has to be implemented \cite{Velizhanin:2008rw}. Thus, a non-perturbative construction of the $\mathcal{N}=4$ theory would require a non-perturbative regularization both in the IR and the UV. However, non-perturbative regularizations break supersymmetry at least partially; thus, it is unclear how the supersymmetry can be utilized in a non-perturbative formulation of the theory. 

This thesis is part of an ongoing effort to develop an alternative perspective on supersymmetric quantum field theories in order to eventually address some of the above questions in a different way. The importance of anti-commuting variables in quantizing supersymmetric field theories is well known. Gauge theories, in particular, require the introduction of additional anti-commuting virtual particles, the ghost fields, to allow for a consistent definition of the path integral. However, anti-commuting variables, both real and virtual, are difficult to work with, especially in perturbation theory. Fortunately, supersymmetric field theories are overdetermined in the sense that there exist Ward identities relating bosonic and fermionic correlation functions. This observation can be formalized, giving rise to a fermion (and ghost) free quantization of supersymmetric field theories.

According to a theorem by Hermann Nicolai \cite{Nicolai:1984jg}, supersymmetric field theories can be characterized by the existence of a non-linear and non-local transformation of the bosonic fields, called the Nicolai map $\mathcal{T}_g$. It was first proposed by Nicolai in \cite{Nicolai:1979nr,Nicolai:1980hh,Nicolai:1980jc,Nicolai:1980js,Nicolai:1984jg} and further developed by Dietz, Flume and Lechtenfeld in \cite{Flume:1983sx,Dietz:1984hf,Lechtenfeld:1984me,Dietz:1985hga,Lechtenfeld:1986gd}. The Nicolai map $\mathcal{T}_g$ maps the interacting functional measure to that of a free theory such that the Jacobian determinant of $\mathcal{T}_g$ equals the product of the fermionic determinants. The Nicolai map thus allows for a more economical, purely bosonic formulation of the quantum theory of supersymmetric field theories. 

\section{A First Example and Previous Results}\label{sec:FirstExample}
The Nicolai map is best understood from examples and the by far simplest example of a Nicolai map is found in supersymmetric quantum mechanics. The theory describes a real scalar $q(t)$ and a pair of Grassman coordinates $\psi(t)$, $\bar{\psi}(t)$. Its Euclidean action is given by \cite{Witten:1981nf,Nicolai:1976xp}
\begin{align}\label{eq:QMAction}
S = \int \mathrm{d}t \ \left[ \frac{1}{2} \dot{q}^2 + \dot{q} V(q) +  \frac{1}{2} V(q)^2 + \bar{\psi} \left( \frac{\mathrm{d}}{\mathrm{d}t} + V^\prime(q) \right) \psi \right] \, , 
\end{align}
where $V(q)$ is some superpotential and the topological invariant $\dot{q}V(q)$ has been added for later convenience. The Nicolai map of \eqref{eq:QMAction} is given by 
\begin{align}\label{eq:QMNicolaiMap}
(\mathcal{T}_g \,  q)(t) = q(t) + \int \mathrm{d}t^\prime \ \theta(t - t^\prime) V(q(t^\prime)) \, ,
\end{align}
where $\theta(t)$ is the step function with $\frac{\mathrm{d}}{\mathrm{d}t} \theta(t-t^\prime) = \delta(t-t^\prime)$. It can easily be checked that $\mathcal{T}_g$ indeed maps the bosonic action to the free action
\begin{align}\label{eq:QMFreeAction}
\int \mathrm{d}t \ \left[ \frac{1}{2} \left( \textstyle{\frac{\mathrm{d}}{\mathrm{d}t}} (\mathcal{T}_g \,  q) \right)^2 \right] =  \int \mathrm{d}t \ \left[ \frac{1}{2} \dot{q}^2 + \dot{q}V(q) +  \frac{1}{2} V(q)^2 \right]
\end{align}
and that its Jacobian determinant equals the fermionic determinant
\begin{align}\label{eq:QMDet}
\det(\mathcal{J} (\mathcal{T}_g \, q)) = \det \left( \delta(t - t^\prime) \pm \theta(t - t^\prime) V^\prime(q(t^\prime)) \right) = \Delta_\mathrm{MSS}[q] \, .
\end{align}
The fermionic (or Matthews-Salam-Seiler) determinant \cite{Matthews:1955zi,Berezin:1966nc} is defined by
\begin{align}\label{eq:QMMSS}
\int \mathcal{D}\bar{\psi} \ \mathcal{D}\psi \ e^{ - \int \mathrm{d}t \,  \bar{\psi} \left( \frac{\mathrm{d}}{\mathrm{d}t} + V^\prime(q) \right) \psi} = \det \left[ \left( \textstyle{\frac{\mathrm{d}}{\mathrm{d}t}} + V^\prime(q(t)) \right) \delta(t - t^\prime) \right] \eqqcolon \det\left[ \textstyle{\frac{\mathrm{d}}{\mathrm{d}t}}  \delta(t - t^\prime) \right] \Delta_\mathrm{MSS}[q] \, .
\end{align}
The Nicolai map of supersymmetric quantum mechanics is special in several ways. There is no need for a formal derivation of the map \eqref{eq:QMNicolaiMap} as it can be inferred from the action \eqref{eq:QMAction} and the fermionic determinant \eqref{eq:QMMSS}. Moreover, it terminates after the first order. In general, Nicolai maps require tedious derivations and are non-finite perturbative series without even a closed-form formulation. Most of this work is dedicated to setting up formalisms to derive Nicolai maps in various supersymmetric (gauge) theories.

However, before outlining the scope of this thesis, let us recollect some historical results. In fact, a lot of work on the Nicolai map was done in the early 1980s until the research came to a sudden hold with the beginning of the first superstring revolution. Many problems and incomplete results were left behind and have not been solved until the active research on the Nicolai map continued very recently in 2020. A comprehensive overview of the state of research in the mid-1980s can be found in the lecture notes of Nicolai \cite{Nicolai:1984jg} and the doctoral thesis of Lechtenfeld \cite{Lechtenfeld:1984me}. By 1984 Nicolai had formulated and proven his theorem capturing the properties of the Nicolai map for the 2-dimensional Wess-Zumino model and 4-dimensional $\mathcal{N}=1$ super Yang-Mills in the Landau gauge. A crucial addition to the proof in $\mathcal{N}=1$ super Yang-Mills was made by Flume and Lechtenfeld \cite{Flume:1983sx} ensuring the distributivity of the infinitesimal generator of the inverse Nicolai map. Moreover, Nicolai found explicit expressions for the Nicolai maps of supersymmetric quantum mechanics, the 2-dimensional Wess-Zumino model and 4-dimensional $\mathcal{N}=1$ super Yang-Mills in Landau gauge. However, neither of these Nicolai maps was systematically derived, but rather they were the result of guesswork. Hence, the Nicolai maps for the 2-dimensional Wess-Zumino model and $\mathcal{N}=1$ super Yang-Mills were limited to the lowest orders of perturbation theory.

In \cite{Lechtenfeld:1984me,Dietz:1985hga} Dietz and Lechtenfeld discovered that the inverse Nicolai map $\mathcal{T}_g^{-1}$ maps interacting correlation functions of bosonic operators $\mathcal{O}_i$ to free correlation functions
\begin{align}\label{eq:NicolaiMapCorrFct}
\big< \!\! \big< \mathcal{O}_1(x_1) \ldots \mathcal{O}_n(x_n) \big> \!\! \big>_g = \big< (\mathcal{T}_g^{-1} \mathcal{O}_1)(x_1) \ldots (\mathcal{T}_g^{-1}\mathcal{O}_n)(x_n) \big>_0 \, .
\end{align}
This statement is quite remarkable. While it, of course, does not render the computation of interacting correlation functions trivial as the complexity is now hidden in the transformations $(\mathcal{T}_g^{-1} \mathcal{O}_i)$, it allows us to obtain bosonic correlation functions in supersymmetric field theories without the use of anti-commuting variables. Thus, one does not have to worry about the fermion (and ghost) loops arising on the left-hand side of \eqref{eq:NicolaiMapCorrFct} in each order of perturbation theory. But instead, simply compute the bosonic Wick contractions on the right-hand side.

Furthermore, there were several attempts to formulate a Nicolai map for $\mathcal{N}=1$ super Yang-Mills in gauges other than the Landau gauge. Most notably, Dietz and Lechtenfeld came very close to finding the Nicolai map in general gauges \cite{Lechtenfeld:1984me,Dietz:1985hga}. As we have now learned almost 40 years later, they have found the correct infinitesimal generator of the inverse map but used it in the wrong way (for more details, see chapter \ref{ch:YM1}). Also de Alfaro, Fubini, Furlan and Veneziano were investigating a variation of the Nicolai map \cite{deAlfaro:1984gw, deAlfaro:1984hb,deAlfaro:1986mti}. Their work displayed hints of a polynomial form of the map for the $\mathcal{N}=1$ and $\mathcal{N}=2$ super Yang-Mills theories in the light-cone gauge, and in terms of the light-cone components of the field strength. Unfortunately, an inspection of the relevant formulas revealed that their Nicolai map does not apply to the `real' super Yang-Mills theory. Instead, one must simultaneously invoke the light-cone gauge (which exists only for Lorentzian signature) and introduce a complexification of the basic fields, which for the fermions would only be appropriate for Euclidean spinors. On the other hand, employing a time-like axial gauge with Euclidean signature, a direct construction fails \cite{Nicolai:1982ye}. In fact, the map in \cite{Nicolai:1982ye} is an expansion in powers of the covariant derivative rather than the coupling constant. Thus, it is not applicable to \eqref{eq:NicolaiMapCorrFct}. Moreover, it only works up to the quadratic order in the covariant derivative. 

\section{Results of this Work}
In this thesis, we continue the research of the 1980s and present a comprehensive study of the Nicolai map in various supersymmetric (gauge) theories of increasing complexity. We address the problems and open questions mentioned above and demonstrate a neat application of the Nicolai map. The work presented here is mostly based on the author's publications \cite{Ananth:2020lup,Ananth:2020jdr,Malcha:2021ess,Malcha:2022fuc}. The main results of this work are the systematic calculations of the Nicolai map for the 2-dimensional Wess-Zumino model up the fifth order in the coupling, the Nicolai map of $\mathcal{N}=1$ super Yang-Mills for general gauges and the second order in the coupling and the Nicolai map of $\mathcal{N}=1$ super Yang-Mills in the Landau gauge up to the fourth order in the coupling. Furthermore, we use this last result to compute the vacuum expectation value of the infinite straight line Maldacena-Wilson loop up to order $g^6$. In a little more detail, the results of this thesis are as follows.

First, we study the Nicolai map for the 2-dimensional Wess-Zumino model. We give a pedagogical proof of Nicolai's main theorem and compute the Nicolai map to the fifth order in perturbation theory, thus extending the previous result \cite{Nicolai:1984jg} by three orders. The Nicolai map is obtained in two simple steps. First, the inverse Nicolai map is constructed from its infinitesimal generator, the $\mathcal{R}_g$-operator. Then the actual Nicolai map is obtained by formal power series inversion. The explicit expression for the map is subjected to two tests similar to \eqref{eq:QMFreeAction} and \eqref{eq:QMDet}. While the 2-dimensional Wess-Zumino model by itself is not a particularly interesting theory, it provides an excellent non-trivial example to understand the Nicolai map and its derivation. Since the Wess-Zumino model is not a gauge theory, its supersymmetry is realized linearly. This makes for a straightforward construction of the $\mathcal{R}_g$-operator and gives us a benchmark of an ideal non-trivial Nicolai map. These results have not been published before.

Then we introduce the Nicolai map and the $\mathcal{R}_g$-operator for $\mathcal{N}=1$ super Yang-Mills. In particular, we introduce the notion of an `on-shell' respectively `off-shell' Nicolai map corresponding to the `on-shell' respectively `off-shell' supersymmetry in the different formulations of the theory. We state and prove according versions of the main theorem. In the presence of `off-shell' supersymmetry, we are able to construct a Nicolai map in general gauges, satisfying the scaling relation $\mathcal{G}^a[A] = g \, \mathcal{G}^a[g^{-1} A]$. The construction is universal to all gauge theories. We compute the Nicolai map for 4-dimensional $\mathcal{N}=1$ super Yang-Mills in axial gauge. We find that the axial gauge Nicolai map is considerably more complicated than the previously known Landau gauge Nicolai map. This is in accord with the mixed success story of the axial gauge in quantum field theory \cite{Konetschny:1975he}. Moreover, `off-shell' supersymmetry (with finitely many auxiliary fields) exists for $\mathcal{N}=1$ super Yang-Mills only in 4 dimensions. Thus we subsequently resort to `on-shell' supersymmetry. Here we are confined to the Landau gauge, but we can construct the Nicolai map in $d=3$, 4, 6 and 10 dimensions. The map is much simpler than in axial gauge and we can compute and test it up to the fourth order in the coupling constant, thus extending the previously known result \cite{Ananth:2020gkt} by two orders. Besides restricting the gauge parameter to $\xi =0$ in the $R_\xi$ type gauges, \emph{i.e.} forcing us into the Landau gauge, the `on-shell-ness' in the Nicolai map is much less of a restriction than usual, where for example the equations of motion are required to close the supersymmetry algebra. We have first published our work on the `on- and off-shell' $\mathcal{N}=1$ super Yang-Mills Nicolai maps in \cite{Ananth:2020lup} and \cite{Malcha:2021ess}.

Already very early in the research on the Nicolai map, it was discovered that finite perturbative expansions of the map are not unique. This is because (in Landau gauge) Nicolai's theorem only makes statements about the derivative of the Nicolai map. In \cite{Ananth:2020jdr} we have found an alternative formulation of the $\mathcal{N}=1$ super Yang-Mills Nicolai map specifically in 6 dimensions up to the third order in perturbation theory. We briefly discuss the result and validate it.  

The third and last supersymmetric field theory we will discuss is $\mathcal{N}=4$ super Yang-Mills and its cousin 4-dimensional maximally extended $\mathcal{N}=1$ super Yang-Mills. For the latter, we can construct an `off-shell' Nicolai map. For the former, however, it seems to be out of reach due to the lack of an `off-shell' formulation. Thus, as before, we are constrained to the Landau gauge. Unfortunately, the formal construction of the Nicolai map via the $\mathcal{R}_g$-operator in the maximally extended $\mathcal{N}=1$ theory and the $\mathcal{N}=4$ theory is very sophisticated due to the large number of (auxiliary) fields. Luckily, however, we are able to prove that the $\mathcal{N}=4$ Nicolai map can be obtained from the 10-dimensional $\mathcal{N}=1$ super Yang-Mills Nicolai map by dimensional reduction. Thus, we automatically get access to the fourth-order Nicolai map we computed before. The $\mathcal{N}=4$ Nicolai map was first studied by Rupprecht in \cite{Rupprecht:2021wdj}.

We close the discussion of the Nicolai map in this thesis by giving a neat example of its application. We use \eqref{eq:NicolaiMapCorrFct} to compute the vacuum expectation value of the infinite-straight line Maldacena-Wilson loop in $\mathcal{N}=4$ super Yang-Mills to order $g^6$. The results are two-fold. Contrary to popular belief, the perturbative cancellations of the different contributions to the vacuum expectation value of the Maldacena-Wilson loop are by no means trivial and seem to resemble those of the circular Maldacena-Wilson loop at order $g^4$. Furthermore, we argue that our approach to computing quantum correlation functions with the Nicolai map is competitive with more standard diagrammatic techniques. This result was first published in \cite{Malcha:2022fuc} by the author of this thesis. 

\section{The Stress Tensor Multiplet}
If an operator is a fixed point of the inverse Nicolai map, \emph{i.e.} if $(\mathcal{T}_g^{-1} \mathcal{O}) = \mathcal{O}$, then by \eqref{eq:NicolaiMapCorrFct} its $n$-point correlation functions do not receive quantum corrections. The only known example of such an operator is the (anti) self-dual field strength tensor in 4-dimensional $\mathcal{N}=1$ super Yang-Mills with a topological term. However, in $\mathcal{N}=4$ super Yang-Mills, there are many operators whose 2- and 3-point functions are entirely determined by their classical expressions. They form several so-called short supermultiplets of the superconformal algebra. A long-term goal is to study these short supermultiplets through the Nicolai map. However, this requires us to know their explicit field content. In the second part of this thesis, we thus determine the entire field content of one of these short multiplets, namely the stress tensor multiplet in $\mathcal{N}=4$ super Yang-Mills. 

The symmetry algebra of $\mathcal{N}=4$ super Yang-Mills is the superconformal algebra $\mathfrak{psu}(2,2|4)$. A set of fields forming a representation of the superconformal algebra is called a supermultiplet. Understanding these supermultiplets is crucial for the study of many aspects of supersymmetric field theories. For example, the construction of the $\mathcal{R}_g$-operator, \emph{i.e.} the infinitesimal generator of the inverse Nicolai map, hinges on understanding the supermultiplet of the Lagrangian. Given the primary field of a supermultiplet, its descendants are obtained by repeatedly acting with the anti-commuting supersymmetry generators $Q_\alpha^A$ and $\bar{Q}_{\dot{\alpha}A}$. In $\mathcal{N}=4$ super Yang-Mills, there are 16 such generators. If the primary field is annihilated by the action of some of the supersymmetry generators, then it lives in a so-called short or semi-short multiplet. In \cite{Dolan:2002zh}, Dolan and Osborn have studied all short and-semi short representations of $\mathcal{N}=2$ and $\mathcal{N}=4$ superconformal symmetry. The most interesting short multiplet is the current (or stress tensor) multiplet $\mathcal{B}_{[0,2,0](0,0)}^{\frac{1}{2},\frac{1}{2}}$. It contains the $R$-symmetry current as well as the energy-momentum tensor. It was first discovered in perturbation theory  \cite{Arutyunov:2000ku, Arutyunov:2000im, Ferrara:1999ed, Bianchi:2001cm, Penati:2001sv} and later shown non-perturbatively \cite{Baggio:2012rr} that the elements of the stress tensor multiplet are subjected to a non-renormalization theorem and do not receive any anomalous dimension. Thus, their 2- and 3-point functions are entirely determined by their classical expressions. 

The first correlation functions to receive quantum corrections are the 4-point functions. They have been studied in a variety of ways. In particular, it has been shown that all the 4-point functions of the stress tensor multiplet are related by the superconformal algebra \cite{Eden:2000bk}. Moreover, the 4-point correlation functions can be expressed in terms of a single function of the two conformal invariants \cite{Dolan:2001tt}. However, this function is not yet fully known. 

In \cite{Dolan:2001tt} Dolan and Osborn have given a list of all states in the $\mathcal{N}=4$ stress tensor multiplet $\mathcal{B}_{[0,2,0](0,0)}^{\frac{1}{2},\frac{1}{2}}$. Moreover, they have found the field constraints as well as supersymmetry transformations of all fields in the stress tensor multiplet. However, the explicit expression for the fields corresponding to the states in the multiplet were not provided. When we want to study the multiplet via the Nicolai map, we need to know its explicit field content. Thus, starting from the chiral primary field of the stress tensor multiplet, we compute all descendant fields by repeated action of the supersymmetry transformations. With our analysis we set the foundation for future investigations of the stress tensor multiplet via the Nicolai map. This result has not been published before.

\section{Outline}
This thesis is divided into two parts. In the first part, we discuss the Nicolai map and in the second part, we discuss the stress tensor multiplet. The first part is organized as follows. 

In chapter \ref{ch:SFT}, we introduce the various supersymmetric quantum field theories studied in the subsequent chapters. These are the 2-dimensional Wess-Zumino model, $\mathcal{N}=1$ super Yang-Mills in $d=3$, 4, 6 and 10 spacetime dimensions and $\mathcal{N}=4$ as well as maximally extended 4-dimensional $\mathcal{N}=1$ super Yang-Mills. In particular, we discuss their supersymmetry transformations, quantization and, in the case of the Yang-Mills theories, gauge fixing procedure. 

Chapters \ref{ch:WZ} - \ref{ch:N4YM} each present the Nicolai map for one particular supersymmetric theory. We begin with an extensive study of the Nicolai map for the 2-dimensional Wess-Zumino model in chapter \ref{ch:WZ}. We formulate and prove the main theorem. In particular, we derive the $\mathcal{R}_g$-operator and do a step-by-step calculation of the first two orders of the Nicolai map. Moreover, we briefly introduce an alternative approach to computing the Nicolai map developed by Lechtenfeld and Rupprecht. Section \ref{sec:WZResult} contains the main result, \emph{i.e.} the Nicolai map for the Wess-Zumino model up to the fifth order in the coupling constant. We briefly comment on the radius of convergence of the Nicolai map. In the last part of this chapter, we verify our result by conducting two tests on it similar to \eqref{eq:QMFreeAction} and \eqref{eq:QMDet}. They correspond to the two statements of the main theorem. 

In chapter \ref{ch:YM1}, we discuss the Nicolai map for `off-shell' $\mathcal{N}=1$ super Yang-Mills in four dimensions. Since super Yang-Mills is a gauge theory, we adjust the main theorem from before by demanding that the Nicolai map is gauge invariant. The new theorem is then briefly proven. The `off-shell' supersymmetry necessitates the introduction of rescaled fields to derive the $\mathcal{R}_g$-operator in general gauges. The main result of this chapter is the second order `off-shell' $\mathcal{N}=1$ super Yang-Mills Nicolai map in axial gauge in section \ref{eq:YM1Result}. Similar to the previous chapter, we verify the result. In section \ref{sec:YM1OnShell}, we discuss a potential simplification of the `off-shell' Nicolai map. 

In chapter \ref{ch:YM2}, we repeat the discussions of the previous chapter but for $\mathcal{N}=1$ super Yang-Mills with `on-shell' supersymmetry. The `on-shell' supersymmetry will restrict the Nicolai map to the Landau gauge but allow for a construction in $d=3$, 4, 6 and 10 dimensions. Moreover, there is an extra step in constructing the $\mathcal{R}_g$-operator where we have to prove that it acts distributively. But we are no longer required a detour via rescaled fields to compute the Nicolai map. The main result of this chapter is the fourth-order Nicolai map. The first three orders are given in section \ref{sec:YM2Result} and the entire expression is provided in appendix \ref{app:FourthOrder}. Again the result is tested. In section \ref{sec:YM2Renormalization}, we discuss the renormalization properties of the $\mathcal{R}_g$-operator and the Nicolai map in different supersymmetric theories. In section \ref{sec:YM2Ambiguity}, we point out an ambiguity in the Nicolai map. We show that a different map up to the third order exists specifically for $d=6$ dimensions. In the last section, we test this additional result. 

In chapter \ref{ch:N4YM}, we discuss the $\mathcal{R}_g$-operators and Nicolai maps for maximally extended $\mathcal{N}=1$ super Yang-Mills and $\mathcal{N}=4$ super Yang-Mills. Both $\mathcal{R}_g$-operators are rather cumbersome compared to the $\mathcal{N}=1$ super Yang-Mills $\mathcal{R}_g$-operator from the previous chapters. Thus, we will not compute any Nicolai maps. Instead, we prove that the `on-shell' $\mathcal{N}=4$ super Yang-Mills Nicolai map can be obtained from the 10-dimensional `on-shell' $\mathcal{N}=1$ super Yang-Mills Nicolai map by the means of dimensional reduction. Furthermore, we comment on the calculation of correlation functions, BPS operators and the large $N$ limit in the context of the Nicolai map.

In chapter \ref{ch:WilsonLoop}, we demonstrate a perturbative application of the Nicolai map. We use the previously obtained fourth-order Nicolai map to compute the vacuum expectation value of the infinite straight line Maldacena-Wilson loop in $\mathcal{N}=4$ super Yang-Mills to order $g^6$ (for all $N$). This chapter marks the end of our discussion of the Nicolai map in this thesis. 

The second part of this thesis is much shorter. It consists only of the two chapters \ref{ch:SCA} and \ref{ch:BPS}. In chapter \ref{ch:SCA} we introduce the superconformal algebra $\mathfrak{psu}(2,2|4)$ and its unitary representations. Moreover, we discuss multiplet shortening, anomalous dimensions and conformal correlation functions. In chapter \ref{ch:BPS} we present the explicit field content of the $\mathcal{N}=4$ stress tensor multiplet. Some general properties of the multiplet are discussed and we demonstrate how to obtain the descendant fields in the multiplet from supersymmetry transformations of the primary field. Furthermore, we derive a generalization of \eqref{eq:NicolaiMapCorrFct} for correlation functions including spinor fields.

In the final chapter \ref{ch:Outlook}, we briefly review our most important results and give an outlook into the future of the Nicolai map. 

There are three appendices. Appendix \ref{app:Spinors} contains our notation and conventions for spinors, gamma and sigma matrices in various dimensions. Also, we collect some important formulae. In appendix \ref{app:FD} we review the calculation of bosonic and fermionic functional determinants. Finally, in appendix \ref{app:FourthOrder}, we present the $\mathcal{N}=1$ super Yang-Mills Nicolai map in Landau gauge and $d=3$, 4, 6 and 10 dimensions up to the fourth order in the coupling.

\chapter{Aspects of Supersymmetric Field Theories}\label{ch:SFT}
In this chapter we establish the technical foundation for the first part of this thesis. We begin with a brief overview of the notation and conventions. Then we discuss the super Poincaré algebra and its representations. Finally, we introduce three supersymmetric field theories. Namely the 2-dimensional Wess-Zumino model, $\mathcal{N}=1$ super Yang-Mills and $\mathcal{N}=4$ super Yang-Mills. Good references for the topics presented in this chapter are \cite{Freedman:2012zz, Peskin:1995ev, Ramond:1981pw, Sohnius:1985qm} and \cite{Wess:1992cp}. 

\section{Notation and Conventions}\label{sec:Notation}
In the following, we consider various supersymmetric field theories. Generally, we work in Minkowski space and only for the 2-dimensional Wess-Zumino model in Euclidean space. In Minkowski space we use the mostly minus metric $\eta^{\mu\nu}$ with $\mu,\nu = 0, \ldots, d-1$ and signature $(+,-, \ldots, -)$. The Clifford algebra is
\begin{align}\label{eq:CliffordAlgebra}
\{ \gamma^\mu, \gamma^\nu\} = 2 \eta^{\mu\nu} \, .
\end{align}
In the case of the 2-dimensional Euclidean space $\eta^{\mu\nu}$ is replaced by $\delta^{\mu\nu}$ and the metric is only plus, \emph{i.e.} $(+,+)$. This change of the signature will have consequences on the subsequent definitions. We will highlight them when discussing the Wess-Zumino model and stick with Minkowski space for now.

In the first part of this thesis, we predominantly work with Majorana spinors. They carry one spinor index $\alpha, \beta = 1, \ldots, r$, where $r$ counts the number of `off-shell' fermionic degrees of freedom. Majorana spinors are real spinors with $\bar{\lambda} = \lambda^T \mathcal{C}$, where $\mathcal{C}$ is the charge conjugation matrix and $\bar{\lambda}= \lambda^\dagger \gamma_0$. In general we will suppress spinor indices by writing \emph{e.g.} $(\bar{\lambda}\lambda) \equiv \bar{\lambda}_\alpha \lambda_\alpha $ or $(\bar{\lambda} \gamma^\mu \lambda) \equiv \bar{\lambda}_\alpha \gamma_{\alpha\beta}^\mu \lambda_\beta$. All other spinor conventions and some useful formulae are summarized in appendix \ref{app:Spinors}.

Traces over the spinor indices are denoted by $\tr$ with
\begin{align}\label{eq:UnityTrace}
\tr \, \mathbf{1} = r \, .
\end{align}
The trace over a single gamma matrix vanishes. Traces over products of gamma matrices are obtained recursively using the cyclicity of the trace
\begin{align}
\tr( \gamma^{\mu_1} \ldots \gamma^{\mu_n}) = \sum_{i=2}^n (-1)^i \ \eta^{\mu_1 \mu_i} \ \tr (\gamma^{\mu_2} \ldots \hat{\gamma}^{\mu_i} \ldots \gamma^{\mu_n})\, , 
\end{align}
where the hat indicates that $\hat{\gamma}^{\mu_i}$ is excluded from the product.

In a non-abelian gauge theory, the fields carry an additional index $a,b = 1, \ldots , |G|$, with $G$ the gauge Lie group and $|G|$ the dimension of $G$. The generators of the Lie algebra $\mathfrak{g}$ associated to $G$ are denoted by $t^a$. $f^{abc}$ are the real totally anti-symmetric structure constants with
\begin{align}
[t^a, t^b] = i f^{abc} t^c \, .
\end{align}
The structure constants satisfy the Jacobi identity
\begin{align}\label{eq:JacId}
f^{abc} f^{ade} + f^{abd} f^{aec} + f^{abe} f^{acd} = 0 \,.
\end{align}
The gauge group of super Yang-Mills field theories is usually $U(N)$ or $SU(N)$, depending on the application. The dimension of $U(N)$ is $N^2$ and the dimension of $SU(N)$ is $N^2-1$. In either case, the structure constants of the associated Lie algebras obey
\begin{align}
f^{abc} f^{abd} = N \delta^{cd} \, .
\end{align}
In the fundamental representation of $\mathfrak{u}(N)$ (respectively $\mathfrak{su}(N)$), the generators $t^a$ are hermitian $N \times N$ matrices with 
\begin{align}
t^a t^a = C_F \mathbf{1}
\end{align}
and the quadratic Casimir $C_F = \frac{N}{2}$ for $\mathfrak{u}(N)$ (respectively $C_F = \frac{N^2-1}{2N}$ for $\mathfrak{su}(N)$). Furthermore, in the case of $\mathfrak{su}(N)$ the $t^a$ are traceless. The trace over the representation space is denoted by $\tr_c$ with
\begin{align}\label{eq:ColorTrace}
\tr_c \, \mathbf{1} = N \quad \text{and} \quad \tr_c(t^a t^b) = \frac{1}{2} \delta^{ab} \, .
\end{align}
The remaining pieces of notation are specific to the supersymmetric field theories and will be introduced alongside the latter. 

\section{The super Poincaré Algebra}\label{sec:Poincare}
Any relativistic field theory is invariant under the Poincaré algebra 
\begin{align}\label{eq:PoincareAlgebra}
\begin{aligned}
&[P_\mu, P_\nu] = 0 \, , \\
&[M_{\mu\nu}, M_{\rho\lambda}] = i (\eta_{\mu\rho} M_{\nu\lambda} - \eta_{\mu\lambda} M_{\nu\rho} - \eta_{\nu\rho} M_{\mu\lambda} + \eta_{\nu\lambda} M_{\mu\rho} ) \, , \\
&[M_{\mu\nu}, P_\lambda] = i(\eta_{\mu\lambda} P_\nu - \eta_{\lambda\nu} P_\mu) \, ,
\end{aligned}
\end{align}
where $P_\mu$ is the generator of translations and $M_{\mu\nu}$ is the generator of Lorenz transformations. The supersymmetric Poincaré algebra is a graded extension of the Poincaré algebra. It introduces $\mathcal{N}$ generators of supersymmetric transformations $Q_\alpha^A$ (with $A = 1, \ldots, \mathcal{N}$), which transform bosonic states into fermionic states and vice versa, \emph{i.e.}
\begin{align}
Q \ket{\mathrm{boson}} = \ket{\mathrm{fermion}} \quad \text{and} \quad Q \ket{\mathrm{fermion}} = \ket{\mathrm{boson}} .
\end{align}
The new relations to supplement \eqref{eq:PoincareAlgebra} are
\begin{align}
\begin{aligned}\label{eq:susyAlgebra1}
&\{ Q_\alpha^A, \bar{Q}_\beta^B \} = \delta^{AB} (\gamma^\mu)_{\alpha\beta} P_\mu \, , \\
&\{ Q_\alpha^A, Q_\beta^B \} = \{ \bar{Q}_\alpha^A, \bar{Q}_\beta^B \} = 0 \, , \\
&[P_\mu, Q_\alpha^A] = [P_\mu, \bar{Q}_\alpha^A] = 0 \, , \\
&[M_{\mu\nu}, Q_\alpha^A] = - \frac{i}{2} (\gamma_{\mu\nu} \, Q^A)_\alpha \, , \\
&[M_{\mu\nu}, \bar{Q}_\alpha^A] = - \frac{i}{2} (\bar{Q}^A \, \gamma_{\mu\nu})_\alpha \, .
\end{aligned}
\end{align}
Supersymmetric field theories are constructed from representations of the super Poincaré algebra. Each of these representations contains an equal number of bosonic and fermionic states. Furthermore, all states in one representation have the same mass. The set of states forming a representation of the supersymmetry algebra is called a multiplet. In four dimensions, the maximum amount of supersymmetry of a multiplet with spin $\le 1$ is $\mathcal{N}=4$. If we allow for spin $\le 2$ in our multiplet, we can go as high as $\mathcal{N}=8$ supersymmetry. 

To formulate a supersymmetric field theory, we must represent the supersymmetry algebra on fields. To this end, we introduce the Majorana spinors $\varepsilon_\alpha^A$ with
\begin{align}
\{\varepsilon_\alpha^A, \varepsilon_\beta^B \} = \{ \varepsilon_\alpha^A, Q_\beta^B \} = \ldots = [P_\mu, \varepsilon_\alpha^A] = 0 
\end{align}
such that the superalgebra \eqref{eq:susyAlgebra1} can be expressed only in terms of commutators, \emph{i.e.}
\begin{align}
\begin{aligned}\label{eq:susyAlgebra2}
&[ (\bar{\varepsilon}Q), (\bar{Q}\varepsilon) ] = (\bar{\varepsilon} \gamma^\mu \varepsilon) P_\mu \, , \\
&[(\bar{\varepsilon}Q),(\bar{\varepsilon}Q)] = [(\bar{Q}\varepsilon) , (\bar{Q}\varepsilon) ] = 0 \, , \\
&[P_\mu, (\bar{\varepsilon}Q)] = [P_\mu, (\bar{Q}\varepsilon)] = 0 \, , \\
&[M_{\mu\nu}, (\bar{\varepsilon}Q)] = - \frac{i}{2} (\bar{\varepsilon} \gamma_{\mu\nu} Q) \, , \\
&[M_{\mu\nu}, (\bar{Q}\varepsilon)] = - \frac{i}{2} (\bar{Q} \gamma_{\mu\nu} \varepsilon) \, .
\end{aligned}
\end{align}
We used $(\bar{\varepsilon} Q) \equiv \bar{\varepsilon}_\alpha^A Q_\alpha^A$. A (component) multiplet $(A, \psi, \ldots)$ is a set of fields on which we define the infinitesimal supersymmetry transformation $\delta$
\begin{align}
\delta A \coloneqq (\bar{\varepsilon} Q + \bar{Q}\varepsilon) \times A \, , \quad \delta \psi \coloneqq (\bar{\varepsilon} Q + \bar{Q}\varepsilon) \times \psi \, , \quad \ldots
\end{align}
The supersymmetry generators $Q_\alpha^A$ have mass dimension $\frac{1}{2}$. Thus a field of mass dimension $l$ transforms into fields of mass dimension $l + \frac{1}{2}$ and derivatives of fields of lower mass dimension. The transformations are such that the supersymmetry algebra \eqref{eq:susyAlgebra2} closes on the multiplet
\begin{align}
[\delta_1, \delta_2] A &= [(\bar{\varepsilon}_1 \gamma^\mu \varepsilon_2) - (\bar{\varepsilon}_2 \gamma^\mu \varepsilon_1) ] P_\mu A = - i [(\bar{\varepsilon}_1 \gamma^\mu \varepsilon_2) - (\bar{\varepsilon}_2 \gamma^\mu \varepsilon_1) ] \partial_\mu A \, , \quad \ldots
\end{align}
Here we have used that $P_\mu = - i \partial_\mu$. We define $b^\mu \coloneqq [(\bar{\varepsilon}_2 \gamma^\mu \varepsilon_1) - (\bar{\varepsilon}_1 \gamma^\mu \varepsilon_2)]$ such that
\begin{align}\label{eq:susyAlgebra3}
[\delta_1, \delta_2] A = i b^\mu \partial_\mu A \, , \quad \ldots
\end{align}
If the multiplet in question is a gauge multiplet, the supersymmetry algebra closes up to a gauge transformation
\begin{align}\label{eq:susyAlgebra4}
[\delta_1, \delta_2] A^a = i b^\mu \partial_\mu A^a + \delta_\mathrm{gauge} A^a \, , \quad \ldots
\end{align}
For some supersymmetric field theories, the supersymmetry algebra \eqref{eq:susyAlgebra3} will close only upon evoking the field equations of motion. We will call such theories `on-shell' supersymmetric field theories. If the algebra closes without the equations of motion, we call the field theory `off-shell' supersymmetric. For some theories such as $\mathcal{N}=1$ super Yang-Mills in 3, 6 and 10 dimensions or $\mathcal{N}=4$ super Yang-Mills in 4 dimensions, there are no `off-shell' formulations with finitely many auxiliary fields\footnote{Despite some serious attempts (see for example \cite{Nicolai:1982gi}) this has never been proven.}.

\section{The Wess-Zumino Model}\label{sec:WZ}
In this section, we discuss various aspects of the 2-dimensional Wess-Zumino model. 
\subsection{The Action}
The Wess-Zumino model was first introduced in 1974 \cite{Wess:1974tw} as a 4-dimensional supersymmetric field theory describing a scalar field $A(x)$, a pseudoscalar field $B(x)$, a four-component Majorana spinor $\psi_\alpha(x)$ and two auxiliary fields $F(x)$ and $G(x)$. In its 2-dimensional version, the model describes a scalar field $A(x)$, a two-component Majorana spinor $\psi_\alpha(x)$ and one auxiliary field $F(x)$. Since Majorana spinors are real, the 2-dimensional Wess-Zumino model has two fermionic and two bosonic degrees of freedom. The action is given by \cite{Zumino:1974beb,Ferrara:1975nf,Browne:1975xf}
\begin{align}\label{eq:WZ1}
S_\mathrm{wz} &= \frac{1}{2} \int \mathrm{d}^2x \, \left[ (\partial_\mu A)( \partial^\mu A) + 2 i p(A) F + F^2 + (\bar{\psi} \gamma^\mu \partial_\mu \psi) + p^\prime(A) (\bar{\psi} \psi) \right] \, , 
\end{align}
where $p(A)$ is some polynomial in $A$. The spinor indices have been suppressed by using the shorthand notation $\bar{\psi}_\alpha \psi_\alpha \equiv (\bar{\psi}\psi)$. For the Wess-Zumino model, we work in Euclidean spacetime where the gamma matrices satisfy the Clifford algebra relation $\{ \gamma^\mu, \gamma^\nu\} = 2 \delta^{\mu\nu}$ with signature $(+,+)$. A priori, one would expect a factor of $i$ in front of the fermionic terms in the action. However, with the change in the signature, this factor gets absorbed in the definition of the gamma matrices. Moreover, the Euclidean signature modifies the supersymmetry algebra \eqref{eq:susyAlgebra1} such that there is an additional factor of $i$ on the right-hand side. In particular, the first relation becomes
\begin{align}\label{eq:susyAlgebra5}
\{ Q_\alpha, \bar{Q}_\beta \} = i (\gamma^\mu)_{\alpha\beta} P_\mu \, .
\end{align}
Finally, we must adjust our notion of Majorana spinors because, in Euclidean spacetime, we cannot have $\bar{\lambda} =  \lambda^\dagger \gamma_0$. There are several approaches to this problem by Zumino \cite{Zumino:1977yh}, Schwinger \cite{Schwinger:1959zz,Schwinger:1960} and Fubini, Hanson and Jackiw \cite{Fubini:1972mf}. However, here we will follow the work of Nicolai \cite{Nicolai:1978vc,Nicolai:1984jg} (see also \cite{vanNieuwenhuizen:1996tv} for a summary). Nicolai proposed to drop the usual hermiticity condition of the Euclidean action introduced by Zumino in favor of Osterwalder-Schrader positivity \cite{Osterwalder:1973dx}. This leads to $\bar{\lambda} \coloneqq \lambda^T \mathcal{C}$, with the 2-dimensional charge conjugation matrix $\mathcal{C}$, as a definition of $\bar{\lambda}$. The advantage of this approach is that it preserves all the usual Majorana spinor relations (see appendix \ref{app:Spinors}) as well as the supersymmetric Ward identities between fermions and bosons. So, in particular, it is possible to translate between correlation functions in Euclidean spacetime and Minkowski spacetime. 

Thus we have discussed all peculiarities of Euclidean space and can now study the properties of the action \eqref{eq:WZ1}. First and foremost, it is invariant under the following supersymmetry transformations
\begin{align}\label{eq:WZsusy}
\delta A = (\bar{\psi} \varepsilon)\, , \quad \delta \psi_\alpha = (\gamma^\mu \varepsilon)_\alpha \partial_\mu A - i \varepsilon_\alpha F \, , \quad \delta F = - i ( \bar{\varepsilon} \gamma^\mu \partial_\mu \psi) \, ,
\end{align}
where $\varepsilon_\alpha$ is a constant two-component Euclidean Majorana spinor. These variations satisfy the supersymmetry algebra \eqref{eq:susyAlgebra5}
\begin{align}\label{eq:WZsusy1}
[\delta_1, \delta_2] A = - b^\mu \partial_\mu A \, , \quad [\delta_1, \delta_2] \psi_\alpha = - b^\mu \partial_\mu \psi_\alpha \, , \quad [\delta_1, \delta_2] F = - b^\mu \partial_\mu F
\end{align}
with $b^\mu \coloneqq [(\bar{\varepsilon}_2 \gamma^\mu \varepsilon_1) - (\bar{\varepsilon}_1 \gamma^\mu \varepsilon_2)]$ as above. To show this, one needs to use the Fierz identity for 2-dimensional gamma matrices (see appendix \ref{app:Spinors}). The relations \eqref{eq:WZsusy1} differ from \eqref{eq:susyAlgebra3} by a factor of $i$.

The action \eqref{eq:WZ1} may be simplified by integrating out the auxiliary field $F$, \emph{i.e.} we replace all instances of $F$ by their corresponding (algebraic) equation of motion
\begin{align}
F = - i p(A) \, .
\end{align}
Furthermore, we choose the simplest non-trivial example $p(A) = mA + \lambda A^3$ for the polynomial $p(A)$, where $m$ is the mass and $\lambda$ is the coupling constant. Both $m$ and $\lambda$ are assumed to be positive such that $p^\prime(A) > 0$ for all $A$ (see \cite{Nicolai:1979nr}). Hence the action \eqref{eq:WZ1} becomes
\begin{align}\label{eq:WZ2}
S_\mathrm{wz} &= \frac{1}{2} \int \mathrm{d}^2x \, \left[ (\partial_\mu A )( \partial^\mu A) + ( m A + \lambda A^3 )^2 + (\bar{\psi} \gamma^\mu \partial_\mu \psi) + ( m + 3 \lambda A^2) (\bar{\psi} \psi) \right] \, .
\end{align}
We distinguish between the `off-shell' and `on-shell' action by writing $S_\mathrm{wz}[\lambda;A,F,\psi]$ for the former and $S_\mathrm{wz}[\lambda;A,\psi]$ for the latter. Thus the absence of the auxiliary field is immediately evident. The `on-shell' supersymmetry variations are
\begin{align}\label{eq:WZsusy2}
\delta A = (\bar{\psi} \varepsilon) \, , \quad \delta \psi_\alpha = (\gamma^\mu \varepsilon)_\alpha \partial_\mu A - \varepsilon_\alpha (mA + \lambda A^3) \, .
\end{align}
With these variations, the supersymmetry algebra \eqref{eq:susyAlgebra5} closes only upon evoking the equations of motion for $\psi_\alpha$. The `on-shell' action \eqref{eq:WZ2} has one fermionic and one bosonic degree of freedom. Compared to \eqref{eq:WZ1}, this is only half the fermionic and bosonic degrees of freedom. The bosonic degrees of freedom are reduced by the elimination of the auxiliary field. The fermionic degrees of freedom, on the other hand, are halved because $\psi_\alpha$ must obey its equation of motion to close the supersymmetry algebra. Moreover, we obtain the expressions for the bosonic and fermionic propagators from the action \eqref{eq:WZ2} by collecting the terms quadratic in $A$ respectively $\psi_\alpha$. The free, \emph{i.e.} $\lambda=0$, bosonic propagator $C(x-y)$ is defined via
\begin{align}
(-\Box + m^2) C(x-y) = \delta(x-y) 
\end{align}
with the Laplacian $\Box = \partial_\mu \partial^\mu$ and
\begin{align}\label{eq:WZFreeProp}
C(x) \coloneqq \int \frac{\mathrm{d}^2k}{(2\pi)^2} \frac{e^{ikx}}{k^2+m^2} \, .
\end{align}
The fermion propagator $\bcontraction{}{\psi}{(x)}{\bar{\psi}} \psi(x) \bar{\psi}(y) \equiv S(x,y;A)$ is defined via the Dirac equation
\begin{align}\label{eq:WZFermionProp}
\big[ \slashed{\partial} + m + 3 \lambda A^2(x) \big]_{\alpha\gamma} \bcontraction{}{\psi}{_\gamma(x) }{\bar{\psi}} \psi_\gamma(x) \bar{\psi}_\beta(y) = \delta_{\alpha\beta} \delta(x-y) \, .
\end{align}
In the limit $\lambda = 0$, we obtain the free fermionic propagator
\begin{align}\label{eq:WZFreeFermion}
S_0(x-y) = (- \slashed{\partial} + m) C(x-y) \, .
\end{align}
The bosonic propagator is symmetric under the exchange of $x$ and $y$, \emph{i.e.} $C(x-y) = C(y-x)$. Derivatives act on the first argument unless indicated differently. In particular, we have 
\begin{align}
\partial_\mu C(x-y) = \frac{\partial}{\partial x^\mu} C(x-y) = - \frac{\partial}{\partial y^\mu} C(x-y) = - \frac{\partial}{\partial y^\mu} C(y-x) = - \partial_\mu C(y-x) \, .
\end{align}

\subsection{Correlation Functions}
The correlation function of a set of operators $\mathcal{O}_1(x_1) \ldots \mathcal{O}_n(x_n)$ in the `on-shell' Wess-Zumino model is given by
\begin{align}\label{eq:WZCorr1}
\left<\!\!\left< \mathcal{O}_1(x_1) \ldots \mathcal{O}_n(x_n) \right>\!\!\right>_\lambda \coloneqq \int \mathcal{D}A \ \mathcal{D}\psi \ e^{-S_\mathrm{wz}[\lambda;A,\psi]} \ \mathcal{O}_1(x_1) \ldots \mathcal{O}_n(x_n) \, .
\end{align}
The correlation function is automatically normalized to $\left<\!\!\left< 1 \right>\!\!\right>_\lambda = 1$. It was first shown by Zumino in 1974 that this is a general feature of any supersymmetric field theory \cite{Zumino:1974bg}. 

From now on, let us assume that the operators $\mathcal{O}_1(x_1) \ldots \mathcal{O}_n(x_n)$ are purely bosonic, \emph{i.e.} they do not depend on the spinor field $\psi_\alpha$. This motivates the introduction of the free bosonic correlation function
\begin{align}
\left< \mathcal{O}_1(x_1) \ldots \mathcal{O}_n(x_n) \right>_0 = \int \mathcal{D}_0A \ e^{-S_\mathrm{wz}[0;A]} \ \mathcal{O}_1(x_1) \ldots \mathcal{O}_n(x_n) \, ,
\end{align}
where $\mathcal{D}_0A = \mathcal{D}A \ \det(\slashed{\partial} + m)^{1/2}$ is the free measure and $S_\mathrm{wz}[0;A]$ denotes the bosonic part of the action \eqref{eq:WZ2} at $\lambda =0$. The normalization constant $\det(\slashed{\partial} + m)^{1/2}$ in the free measure is chosen such that $\left<1 \right>_0 =1$, \emph{i.e.}
\begin{align}\label{eq:vev1}
\left< 1 \right>_0 = \int \mathcal{D}_0A \ e^{-S_\mathrm{B}[0;A]} = \int \mathcal{D}A \ \det(\slashed{\partial} + m)^{1/2} \ e^{ - \frac{1}{2} \int \mathrm{d}^2x \, A(-\Box + m^2) A } = 1 \, .
\end{align}
In the last step, we computed the functional determinant
\begin{align}
\int \mathcal{D}A \ \ e^{ - \frac{1}{2} \int \mathrm{d}^2x \, A(-\Box + m^2) A } = \det(- \Box + m^2)^{-1/2} 
\end{align}
and used $\det(- \Box + m^2)^{-1/2} = \det(\slashed{\partial} + m)^{-1/2}$. See appendix \ref{app:FD} for an introduction to computing functional determinants. Finally, if $\mathcal{O}_1(x_1) \ldots \mathcal{O}_n(x_n)$ is not only bosonic but also independent of the coupling $\lambda$ we obtain
\begin{align}
\left<\!\!\left< \mathcal{O}_1(x_1) \ldots \mathcal{O}_n(x_n) \right>\!\!\right>_\lambda \Big\vert_{\lambda=0} = \left< \mathcal{O}_1(x_1) \ldots \mathcal{O}_n(x_n) \right>_0 \, .
\end{align}
Free bosonic correlation functions are computed using Wick's theorem
\begin{align}
\left< A(x) A(y) \right>_0 = \bcontraction{}{A}{(x) }{A} A(x) A(y) = C(x-y) \, ,
\end{align}
where $C(x)$ is the free bosonic propagator \eqref{eq:WZFreeProp}. The free bosonic correlation function of more than two fields is the sum of all possible Wick contractions, where terms containing uncontracted fields vanish. Because the Majorana spinor $\psi_\alpha$ appears only quadratically in the action \eqref{eq:WZ2} we can integrate out the fermionic degrees of freedom in any correlation function. If the correlation function in question is purely bosonic, we obtain
\begin{align}
\begin{gathered}
\int \mathcal{D}\psi \, \exp \left[ - \frac{1}{2} \int \mathrm{d}^2x \ \bar{\psi} ( \slashed{\partial} + m + 3 \lambda A^2) \psi \right] \\
= \det(\slashed{\partial} + m + 3 \lambda A^2)^{1/2} = \det( \slashed{\partial} + m )^{1/2} \det( 1 + 3 \lambda S_0 \ast A^2)^{1/2} \, ,
\end{gathered}
\end{align}
where $S_0$ is the free fermion propagator from \eqref{eq:WZFreeFermion} and $\ast$ denotes the convolution. The first term is the normalization factor from above and the second term is the Matthews-Salam-Seiler (MSS) determinant $\Delta_\mathrm{MSS}$ \cite{Matthews:1955zi,Berezin:1966nc}. This determinant will be key to the main part of this work. In particular, we shall be interested in the perturbative expansion of its logarithm in powers of the coupling constant $\lambda$. Using the well-known relation $\det \mathcal{M} = \exp( \Tr( \log \mathcal{M}))$ we find that 
\begin{align}\label{eq:WZMSS}
\log(\Delta_\mathrm{MSS}[\lambda;A] ) = \frac{1}{2} \Tr \log \left[ 1 + 3 \lambda S_0 \ast A^2 \right] \, .
\end{align}
The capitalized trace $\Tr$ is over both the spinor indices and the convolution, \emph{i.e.} in the perturbative expansion of the logarithm it identifies the last variable with the first and integrates over the remaining free variable. Expanding the right-hand side in powers of $\lambda$ yields
\begin{align}
\begin{aligned}
\frac{1}{2} \Tr \log \left[ 1 + 3 \lambda S_0 \ast A^2 \right] &= \frac{3\lambda}{2} \Tr \left( S_0 \ast A^2 \right) - \frac{9\lambda^2}{4} \Tr \left( S_0 \ast A^2 \ast S_0 \ast A^2 \right) \\
&\quad + \frac{9\lambda^3}{2} \Tr \left( S_0 \ast A^2 \ast S_0 \ast A^2 \ast S_0 \ast A^2 \right) + \ldots 
\end{aligned}
\end{align}
For the leading term, we obtain
\begin{align}
\frac{3\lambda}{2} \Tr \left( S_0 \ast A^2 \right) &= \frac{3 \lambda}{2} \int \mathrm{d}^2 x \ \tr \left[- \slashed \partial + m \right] C(0) A^2(x) = 3 \, m \lambda \int \mathrm{d}^2 x \ C(0) A^2(x) \, .
\end{align}
The propagator $C(0)$ is formally divergent but can be regulated. For the second term, we find
\begin{align}
\begin{aligned}
&\quad-\frac{9 \lambda^2}{4} \Tr \left( S_0 \ast A^2 \ast S_0 \ast A^2 \right) \\
& = - \frac{9 \lambda^2}{4} \int \mathrm{d}^2 x \ \mathrm{d}^2y \ \tr \left[ \left( - \slashed \partial + m \right) C(x-y) A^2(y) \left( - \slashed \partial + m \right) C(y-x) A^2(x) \right] \\
&= - \frac{9\lambda^2}{2} \int \mathrm{d}^2x \ \mathrm{d}^2y \ \partial^\mu C(x-y) A^2(y) \partial_\mu C(y-x) A^2(x) \\
&\quad  - \frac{9 \, m^2 \lambda^2}{2} \int \mathrm{d}^2x \ \mathrm{d}^2y \ C(x-y) A^2(y) C(y-x) A^2(x) \, .
\end{aligned}
\end{align}
The higher orders are computed accordingly.

A correlation function where the fermionic degrees of freedom have been integrated out is denoted by a single bracket
\begin{align}
\left< \mathcal{O}_1(x_1) \ldots \mathcal{O}_n(x_n) \right>_\lambda = \int \mathcal{D}_\lambda A \ e^{-S_\mathrm{wz}[\lambda;A]} \ \mathcal{O}_1(x_1) \ldots \mathcal{O}_n(x_n) \, ,
\end{align}
with $\mathcal{D}_\lambda A = \mathcal{D}A \ \det(\slashed{\partial} + m)^{1/2} \Delta_\mathrm{MSS}[\lambda;A]$. This definition is such that $\left<\mathcal{O}_1 \ldots \mathcal{O}_n \right>_\lambda \big\vert_{\lambda=0} = \left< \mathcal{O}_1 \ldots \mathcal{O}_n \right>_0$ for a $\lambda$ independent set of bosonic operators. 

Finally, we may also integrate out the fermionic degrees of freedom in a correlation function containing spinor fields. Consider the example
\begin{align}
\big<\!\!\big< A(x) \psi(y) \bar{\psi}(z) \big>\!\!\big>_\lambda = \big< A(x) \bcontraction{}{\psi}{(y) }{\bar{\psi}} \psi(y) \bar{\psi}(z) \big>_\lambda = \big< A(x) S(y,z;A) \big>_\lambda \, .
\end{align}
Again this is Wick's theorem in action; however, when applying it to interacting, \emph{i.e.} $\lambda$ dependent, correlation functions with fermions, the contractions refer to the interacting fermion propagator $S(x,y;A)$. 

\subsection{Ward Identities}\label{sec:WZWardId}
Ward identities can be thought of as the quantum version of Noether's theorem. The invariance of the Wess-Zumino action under the supersymmetry transformations leads to the conservation of its correlation functions. Let $X[A,\psi]$ be an arbitrary string of operators depending on the scalar field $A$ and the Majorana spinor $\psi_\alpha$. We introduce the anti-commuting supersymmetry variation $\delta_\alpha$ via $\delta \equiv \varepsilon_\alpha \delta_\alpha$. The Ward identity is
\begin{align}\label{eq:WZWardId}
\left<\!\!\left< \delta_\alpha X[A,\psi] \right>\!\!\right>_\lambda = 0 \, .
\end{align}
To prove the Ward identity, we first show that
\begin{align}
\int \mathcal{D}A \ \mathcal{D}\psi \ \delta_\alpha X[A,\psi] = 0 \, .
\end{align}
Using \eqref{eq:WZsusy2}, we write the supersymmetry variation as the sum of a bosonic and a fermionic derivative
\begin{align}
\delta_\alpha \equiv - \bar{\psi}_\alpha \frac{\delta}{\delta A} +\left[ \gamma_{\alpha\beta}^\mu (\partial_\mu A) - \delta_{\alpha\beta} (mA + \lambda A^3) \right] \frac{\delta}{\delta \psi_\beta} \, .
\end{align}
Thus the integral becomes
\begin{align}\label{eq:WZWardIdIntegral}
\begin{aligned}
\int \mathcal{D}A \ \mathcal{D}\psi \ \delta_\alpha X[A,\psi] &= - \int \mathcal{D}\psi \ \bar{\psi}_\alpha \int \mathcal{D}A \ \frac{\delta X[A,\psi]}{\delta A} \\
&\quad + \int \mathcal{D}A \ \left[ \gamma_{\alpha\beta}^\mu (\partial_\mu A) - \delta_{\alpha\beta} (mA + \lambda A^3) \right] \int \mathcal{D}\psi \ \frac{\delta X[A,\psi]}{\delta \psi_\beta} \, .
\end{aligned}
\end{align}
First, we consider the second term. Let $\theta$ be a Grasssmann variable. Recall that the Taylor expansion of any function $f(\theta)$ terminates at the second order, \emph{i.e.} $f(\theta) = a + b \theta$. Moreover, recall the Berezin integral (see appendix \ref{app:FD})
\begin{align}
\int \mathrm{d}\theta \ \theta = 1 \, , \quad \int \mathrm{d}\theta = 0 \, .
\end{align}
Thus it follows
\begin{align}
\int \mathrm{d}\theta \ \frac{\mathrm{d}}{\mathrm{d}\theta} f(\theta) = \int \mathrm{d}\theta \ b = 0 \, .
\end{align}
This statement extends to functional integrals and thus, the second term in \eqref{eq:WZWardIdIntegral} vanishes because the integral over $\psi$ is zero. Then we consider the first term. Here we have a total derivative. Thus we expect to obtain a boundary term. However, we assume that all fields vanish at the boundary. Hence we conclude
\begin{align}\label{eq:WZTD}
\int \mathcal{D}A \ \mathcal{D}\psi \ \delta_\alpha X[A,\psi] = 0 \, .
\end{align}
In order to turn the left-hand side of \eqref{eq:WZTD} into a correlation function, we have to insert the exponential of the action in the integral. Since $\delta_\alpha S_\mathrm{wz}[\lambda;A,\psi] = 0$ we get
\begin{align}
e^{- S_\mathrm{wz}[\lambda;A,\psi] } \ (\delta_\alpha X[A,\psi])  = \delta_\alpha \left( e^{- S_\mathrm{wz}[\lambda;A,\psi] } \ X[A,\psi] \right) \, .
\end{align}
Together with \eqref{eq:WZTD} this implies \eqref{eq:WZWardId} and concludes the proof.

\section[\texorpdfstring{$\mathcal{N}=1$}{N=1} super Yang-Mills]{$\boldsymbol{\mathcal{N}}$\:=\;1 super Yang-Mills}
This section discusses various aspects of $\mathcal{N}=1$ super Yang-Mills. 
\subsection{The Action}
Four-dimensional $\mathcal{N}=1$ super Yang-Mills is a massless non-abelian gauge theory describing a gauge field $A_\mu^a$, a four-component Majorana spinor $\lambda_\alpha^a$ and a real auxiliary field $D^a$ \cite{Ferrara:1974pu,Salam:1974ig}. All the fields are in the adjoint representation of the gauge group $U(N)$ or $SU(N)$. The gauge invariant action is given by
\begin{align}\label{eq:YM1}
S_\mathrm{inv}^1 = \int \mathrm{d}^4x \ \left[ -\frac{1}{4} F_{\mu\nu}^a F^{a \, \mu\nu} - \frac{i}{2} 
\bar{\lambda}^a \gamma^\mu (D_\mu \lambda)^a + \frac{1}{2} D^a D^a \right] \, , 
\end{align}
with the standard definitions
\begin{align}\label{eq:FandD}
&F_{\mu\nu}^a \coloneqq \partial_\mu A_\nu - \partial_\nu A_\mu + g f^{abc} A_\mu^b A_\nu^c \,, \\
&(D_\mu \lambda_\alpha)^a \coloneqq \partial_\mu \lambda_\alpha^a + g f^{abc} A_\mu^b \lambda_\alpha^c 
\end{align}
for the field strength tensor and covariant derivative. The coupling constant is denoted by $g$. The superscript $1$ in $S_\mathrm{inv}^1$ indicates the one supersymmetry (opposed to the four supersymmetries in $\mathcal{N}=4$ super Yang-Mills). In its `off-shell' version, $\mathcal{N}=1$ super Yang-Mills has four fermionic degrees of freedom from the Majorana spinor and $3+1$ bosonic degrees of freedom from the gauge field and the auxiliary field. Notice that the gauge field only has three degrees of freedom and not four because of the gauge condition. The action \eqref{eq:YM1} is invariant under the supersymmetry variations
\begin{align}\label{eq:YMsusy1}
&\delta A_\mu^a = - i (\bar{\lambda}^a \gamma_\mu \varepsilon ) \, , 
&&\delta \lambda_\alpha^a = -\frac{1}{2} (\gamma^{\mu\nu} \varepsilon)_\alpha F_{\mu \nu}^a + i (\gamma^5 \varepsilon)_\alpha D^a \, , 
&&\delta D^a = - ( \bar{\varepsilon} \gamma^5 \gamma^\mu (D_\mu \lambda^a) ) \, .
\end{align}
These variations satisfy the supersymmetry algebra
\begin{align}\label{eq:YMsusyAlgebra1}
[\delta_1, \delta_2] A_\mu^a = i b^\nu F_{\nu\mu}^a \, , \quad [\delta_1, \delta_2] \lambda_\alpha^a = i b^\mu (D_\mu \lambda)_\alpha^a \, , \quad [\delta_1, \delta_2] D^a = i b^\mu (D_\mu D)^a \, .
\end{align}
Notice that $b^\nu F_{\nu\mu}^a = b^\nu \partial_\nu A_\mu^a - b^\nu (D_\mu A_\nu)^a$ in agreement with \eqref{eq:susyAlgebra3}. Both the invariance of the action under the supersymmetry transformations and the closing of the supersymmetry algebra requires the 4-dimensional Fierz identities for Majorana spinors (see appendix \ref{app:Spinors}). 

Similar to the Wess-Zumino model, it is possible to integrate out the auxiliary field $D^a$ in the action \eqref{eq:YM1}. The relevant equation of motion is $D^a=0$ and we obtain
\begin{align}\label{eq:YM2}
S_\mathrm{inv}^1 = \int \mathrm{d}^dx \ \left[ -\frac{1}{4} F_{\mu\nu}^a F^{a \, \mu\nu} - \frac{i}{2} 
\bar{\lambda}^a \gamma^\mu (D_\mu \lambda)^a \right] \, .
\end{align}
One particular feature of the `on-shell' $\mathcal{N}=1$ super Yang-Mills theory is that it also exists in other spacetime dimensions than four \cite{Brink:1976bc}. To this end, let us assume for a moment that $\lambda_\alpha^a$ is a Dirac spinor. In even spacetime dimension $d$ a Dirac spinor has $2 \cdot 2^{d/2}$ real degrees of freedom, \emph{i.e.} twice as many as a Majorana spinor. In odd spacetime dimensions, a Dirac spinor has $2 \cdot 2^{(d-1)/2}$ real degrees of freedom.  Since we are in the `on-shell' formulation $\lambda_\alpha^a$ must obey the Dirac equation to close the supersymmetry algebra. This halves its degrees of freedom and the spinor now has $\frac{1}{2} \cdot 2 \cdot 2^{d/2} = 2^{d/2}$ (respectively $2^{(d-1)/2}$ for odd $d$) real degrees of freedom. The `on-shell' gauge field, on the other hand, has $d-2$ degrees of freedom. We see that these numbers do not match for any $d$. Hence we must implement further constraints on the Dirac spinor to reduce its degrees of freedom. 

If the spacetime dimension is $d \equiv 1,2,3,4 \mod 8$ we may impose the Majorana condition
\begin{align}
\bar{\lambda}^a = (\lambda^{a \, T} \mathcal{C}) \, ,
\end{align}
rendering the spinor real and reducing its degrees of freedom by a factor of 2. In any even number of spacetime dimensions, we may impose the Weyl condition
\begin{align}
\lambda^a = \frac{1}{2} ( \mathbbm{1} - \gamma^{d+1}) \lambda^a \, ,
\end{align}
which also halves the number of degrees of freedom. Finally, we may impose both conditions simultaneously if the spacetime dimension is $d \equiv 2 \mod 8$.  This reduces the degrees of freedom by a factor of 4. In table \ref{tab:Spinors}, we have summarized the spacetime dimensions for which we can match the fermionic and bosonic degrees of freedom and which spinor constraints we have to impose to do so.
\begin{table}
\renewcommand{\arraystretch}{1.0}
\begin{tabular}{|c|l|c|}
\hline
\textbf{Spacetime Dimension} & \multicolumn{1}{c|}{\textbf{Spinor Type}}  & \begin{tabular}{@{}c@{}}\textbf{Real `on-shell'} \\ \textbf{fermionic DoF}\end{tabular}    \\
\hline
3 & Majorana & 1  \\
4 & Majorana (or Weyl) & 2   \\
6 & Weyl & 4 \\
10 & Majorana-Weyl & 8 \\
\hline
\end{tabular}
\caption{Possible spacetime dimensions and their spinor types for `on-shell' $\mathcal{N}=1$ super Yang-Mills.}
\label{tab:Spinors}
\end{table}
There are no solutions for $d > 10$ since the number of bosonic degrees of freedom grows linearly while the number of fermionic degrees of freedom grows exponentially and there are simply no other conditions to implement on the spinor. We conclude that `on-shell' $\mathcal{N}=1$ super Yang-Mills exists in $d= 3$, 4, 6 and 10 spacetime dimensions. Moreover, the dimension of the corresponding Clifford algebra representation is related to the number of spacetime dimensions by
\begin{align}\label{eq:rd}
r = 2(d-2) \, .
\end{align}
We will later re-derive this equation in the context of the Nicolai map. In the following, we want to continue working with Majorana spinors only. Thus the case of $d=6$ is formally excluded from our derivations. However, let us emphasize that our results do not depend on the choice of spinors and thus are also valid for $d=6$. In particular, we may repeat all calculations using Weyl spinors instead of Majorana spinors. 

Upon using the appropriate Fierz identities, the `on-shell' action \eqref{eq:YM2} is invariant under the following supersymmetry transformations
\begin{align}\label{eq:YMsusy2}
\delta A_\mu^a = - i (\bar{\lambda}^a \gamma_\mu \varepsilon ) \, , 
\quad \delta \lambda_\alpha^a = -\frac{1}{2} (\gamma^{\mu\nu} \varepsilon)_\alpha F_{\mu \nu}^a \, .
\end{align}
In the `on-shell' formulation, the supersymmetry algebra
\begin{align}\label{eq:YMsusyAlgebra2}
[\delta_1, \delta_2] A_\mu^a = i b^\nu F_{\nu\mu}^a \, , \quad [\delta_1, \delta_2] \lambda_\alpha^a = i b^\mu (D_\mu \lambda)_\alpha^a 
\end{align}
closes up to terms proportional to the equations of motion
\begin{align}
(D^\nu F_{\mu\nu})^a =  \frac{ig}{2} f^{abc} (\bar{\lambda}^b \gamma_\mu \lambda^c) \, , \quad  \gamma^\mu (D_\mu \lambda)^a = 0 \, .
\end{align}
The $\mathcal{N}=1$ super Yang-Mills fermion propagator $i \bcontraction{}{\lambda}{^a(x) }{\bar{\lambda}} \lambda^a(x) \bar{\lambda}^b(y) \equiv S^{ab}(x,y;A)$ is defined via the Dirac equation
\begin{align}\label{eq:YMFermProp}
(\gamma^\mu D_\mu \bcontraction{}{\lambda}{_\alpha)^a(x) }{\bar{\lambda}} \lambda_\alpha)^a(x) \bar{\lambda}_\beta^b(y) = - i \delta^{ab} \delta_{\alpha\beta} \delta(x-y) \, .
\end{align}
The limit $g=0$ gives us the free massless fermionic propagator $S_0(x-y)$, which obeys
\begin{align}\label{eq:YMFreeFermProp}
\gamma^\mu \partial_\mu S_0(x-y) =  \delta(x-y) \, .
\end{align}
This implies $S_0(x-y) = - \gamma^\mu \partial_\mu C(x-y)$, where $C(x)$ is the free (massless) scalar propagator\footnote{We used the same symbols as for the massive propagators in the Wess-Zumino model, but it should always be clear which one we refer to from the context.}
\begin{align}
C(x) = \int \frac{\mathrm{d}^dk}{(2\pi)^d} \ \frac{e^{ikx}}{k^2}  \, . 
\end{align}
The massless scalar propagator obeys $-\Box C(x) = \delta(x)$. The free bosonic (or Feynman) propagator $C_{\mu\nu}(x)$ is obtained from the first term in the action \eqref{eq:YM2} at $g=0$
\begin{align}
-\frac{1}{4} \int \mathrm{d}^dx \  F_{\mu\nu}^a F^{a \, \mu\nu}  \Big\vert_{g=0} = - \frac{1}{2} \int \mathrm{d}^dx \ A_\mu^a \left[-\eta^{\mu\nu} \Box + \partial^\mu \partial^\nu \right] A_\nu^a \, .
\end{align}
However, the $d \times d$ matrix $\left[-\eta^{\mu\nu} \Box + \partial^\mu \partial^\nu \right]$ is singular and cannot be inverted. The singularity stems from the gauge invariance of the action. A general gauge transformation is of the form
\begin{align}\label{eq:GaugeTransformation}
A_\mu^a(x) \to A_\mu^a(x) +  (D_\mu \alpha)^a(x) \, .
\end{align}
For all $A_\mu^a(x) = (D_\mu \alpha)^a(x)$, the field strength tensor term of the action vanishes, thus making the inverse Feynman propagator singular. Furthermore, also the path integral
\begin{align}\label{eq:PI}
\int \mathcal{D}A \ e^{-iS_\mathrm{inv}[g;A,\lambda]}
\end{align}
is not well-defined because we are redundantly integrating over field configurations related by gauge transformations \eqref{eq:GaugeTransformation}. These two issues can be resolved by a gauge fixing procedure. 

\subsection{The Faddeev-Popov Procedure}
In the following, we describe the Faddeev-Popov procedure \cite{Faddeev:1967fc}. It will fix the problem of overcounting physically equivalent field configurations and subsequently make the Feynman propagator and path integral well-defined. 

Let $\mathcal{G}^a(A)$ be an arbitrary gauge function. Physically equivalent field configurations are removed from the path integral by demanding that $\mathcal{G}^a(A)=0$. We insert a 1 in \eqref{eq:PI} in the following way
\begin{align}
1 = \int \mathcal{D}\alpha \ \delta(\mathcal{G}^a(A^\alpha)) \det\left( \frac{\delta \mathcal{G}^a(A^\alpha)}{\delta \alpha} \right) \, ,
\end{align}
where $A^\alpha$ is the gauge transformed field
\begin{align}
(A^\alpha)_\mu^a t^a = e^{i g \alpha^a t^a} \left[ A_\mu^b t^b + \frac{i}{g} \partial_\mu \right] e^{-i g \alpha^c t^c} \, .
\end{align}
The infinitesimal form of the transformation is \eqref{eq:GaugeTransformation}
\begin{align}
(A^\alpha)_\mu^a = A_\mu^a +  (D_\mu \alpha)^a \, .
\end{align}
Since the super Yang-Mills action is gauge invariant, we can replace $A$ by $A^\alpha$ in \eqref{eq:PI}. Furthermore, also the path integral measure is invariant under this transformation, \emph{i.e.} $\mathcal{D}A = \mathcal{D}A^\alpha$. Thus we have
\begin{align}\label{eq:GaugeFixingInsertion}
\int \mathcal{D}A \ e^{-iS_\mathrm{inv}[g;A,\lambda]} = \int \mathcal{D}\alpha \int \mathcal{D}A^\alpha \ e^{-iS_\mathrm{inv}[g;A^\alpha,\lambda]} \, \delta(\mathcal{G}^a(A^\alpha)) \det\left( \frac{\delta \mathcal{G}^a(A^\alpha)}{\delta \alpha} \right) \, .
\end{align}
Subsequently, we can rename $A^\alpha$ to $A$. For linear gauges the Faddeev-Popov determinant $\det\left(\frac{\delta \mathcal{G}^a(A^\alpha)}{\delta \alpha}\right)$ does not depend on $\alpha$ and hence the $\alpha$ integration factors out. Thus $\int \mathcal{D}\alpha$ is just a constant. To continue, we specify the gauge function a bit further. Let $\mathcal{G}^a(A) = \mathcal{G}^\mu A_\mu^a(x) - \omega^a(x)$ with some scalar function $\omega^a(x)$. Then the Faddeev-Popov determinant becomes
\begin{align}
\det \left( \frac{\delta \mathcal{G}^a(A^\alpha)}{\delta \alpha} \right) = \det \left(  \mathcal{G}^\mu D_\mu \right) \, .
\end{align}
Similar to \eqref{eq:vev1}, we can express the determinant as a path integral. To this end, we introduce the anti-commuting Faddeev-Popov ghost fields $C$ and $\bar{C}$ and write
\begin{align}
\det \left( \mathcal{G}^\mu D_\mu \right) = \int \mathcal{D}\bar{C} \ \mathcal{D}C \ \exp\left[ - \frac{i}{2} \int \mathrm{d}^dx \ \bar{C}^a \mathcal{G}^\mu (D_\mu C)^a \right] \, .
\end{align}
See also appendix \ref{app:FD} for a derivation of this equation. Now \eqref{eq:GaugeFixingInsertion} reads
\begin{align}
\begin{aligned}
\int \mathcal{D}A \ e^{-iS_\mathrm{inv}[g;A,\lambda]} = \left[ \int \mathcal{D}\alpha \right] \int &\mathcal{D}A \ \mathcal{D}\bar{C} \ \mathcal{D}C \ \delta(\mathcal{G}^\mu A_\mu^a - \omega^a) \\
&\quad \times e^{-iS_\mathrm{inv}[g;A,\lambda] - \frac{i}{2} \int \mathrm{d}^dx \ \bar{C}^a \mathcal{G}^\mu (D_\mu C)^a } \, .
\end{aligned}
\end{align}
This equation is true for any $\omega$. Hence we can integrate over all $\omega^a(x)$, with a Gaussian weighting function centered on $\omega^a = 0$, provided we introduce a normalization constant $N(\xi)$. Thus we obtain
\begin{align}
\begin{aligned}
\int \mathcal{D}A \ e^{-iS_\mathrm{inv}[g;A,\lambda]} &= N(\xi) \left[ \int \mathcal{D}\alpha \right] \int \mathcal{D}\omega \ e^{- \frac{i}{2\xi} \int \mathrm{d}^dx \ \omega^a \omega^a} \int \mathcal{D}A \ \mathcal{D}\bar{C} \ \mathcal{D}C \ \delta(\mathcal{G}^\mu A_\mu^a - \omega^a) \\
&\quad \quad \quad \times e^{-iS_\mathrm{inv}[g;A,\lambda] - \frac{i}{2} \int \mathrm{d}^dx \ \bar{C}^a \mathcal{G}^\mu (D_\mu C)^a } \\
&= N(\xi) \left[ \int \mathcal{D}\alpha \right] \int \mathcal{D}A \ \mathcal{D}\bar{C} \ \mathcal{D}C \\
&\quad \quad \quad \times e^{-iS_\mathrm{inv}[g;A,\lambda] - \frac{i}{2\xi} \int \mathrm{d}^dx \ (\mathcal{G}^\mu A_\mu^a) (\mathcal{G}^\nu A_\nu^a) - \frac{i}{2} \int \mathrm{d}^dx \ \bar{C}^a \mathcal{G}^\mu (D_\mu C)^a } \, ,
\end{aligned}
\end{align}
where $\xi$ is an arbitrary constant. In the second step, we used the delta function to integrate over $\omega^a$. For convenience, we will remove the unimportant normalization constant $N(\xi) \left[ \int \mathcal{D}\alpha \right]$ by simply redefining the path integral measure. Furthermore, we introduce the gauge-fixing action
\begin{align}\label{eq:YMgf}
S_\mathrm{gf} = \int \mathrm{d}^dx \ \left[ \frac{1}{2\xi} (\mathcal{G}^\mu A_\mu^a) (\mathcal{G}^\nu A_\nu^a) + \frac{1}{2} \bar{C}^a \mathcal{G}^\mu (D_\mu C)^a \right] \, .
\end{align}
When computing correlation functions, we must add this term to the gauge-invariant action \eqref{eq:YM2} to obtain a finite result. Thus the `on-shell' correlation function of some operators $\mathcal{O}_1(x_1) \ldots \mathcal{O}_n(x_n)$ is given by
\begin{align}\label{eq:YMCorr}
\left<\!\!\left< \mathcal{O}_1(x_1) \ldots \mathcal{O}_n(x_n) \right>\!\!\right>_g \coloneqq \int \mathcal{D}A \ \mathcal{D}\lambda \ \mathcal{D}\bar{C} \ \mathcal{D}C \ e^{-i S_\mathrm{inv}^1[g;A,\lambda]- i S_\mathrm{gf}[g;A,C,\bar{C}]} \ \mathcal{O}_1(x_1) \ldots \mathcal{O}_n(x_n) \, .
\end{align}
The `off-shell' version of this definition has an additional path integral over the auxiliary field $D^a$ and the corresponding `off-shell' action in the exponent. Integrating out the auxiliary field simply yields
\begin{align}
\int \mathcal{D}D \ e^{- \frac{i}{2} \int \mathrm{d}^4x \ D^a D^a} = 1 \, .
\end{align}
 Furthermore, the gauge fixing action introduces the ghost propagator $\bcontraction{}{C}{^a(x) }{\bar{C}} C^a(x) \bar{C}^b(y) \equiv G^{ab}(x,y;A)$. It is defined via
\begin{align}\label{eq:YMGhostProp}
\mathcal{G}^\mu (D_\mu \bcontraction{}{C}{)^a(x) }{\bar{C}} C)^a(x) \bar{C}^b(y) = \delta^{ab} \delta(x-y) \, .
\end{align}
In the limit $g = 0$, we obtain the free ghost propagator $G_0(x-y)$. The explicit form of the ghost propagator depends on the choice of gauge. 

\subsection{Choosing a Gauge}
We review several common gauge choices and their implications. Popular gauge choices where the Faddeev-Popov determinant does not depend on $\alpha$ are the $R_\xi$ type gauges $\mathcal{G}^a(A) = \partial^\mu A_\mu^a$ and the axial gauge $\mathcal{G}^a(A) = n^\mu A_\mu^a$ with $n_\mu n^\mu = 1$. If $n_\mu n^\mu = 0$, the axial gauge is called light-cone gauge. In the following, we simplify the notation by writing $\mathcal{G}^a(A) \equiv \mathcal{G}^\mu A_\mu^a$ with $\mathcal{G}^\mu = \partial^\mu$ or $\mathcal{G}^\mu = n^\mu$ depending on the type of gauge in question.  

At first, consider the $R_\xi$ type gauges. When adding the gauge fixing term \eqref{eq:YMgf} to the action \eqref{eq:YM2} the defining equation for the Feynman propagator becomes
\begin{align}\label{eq:FeynmanPropagator2}
\left[-\eta^{\mu\nu} \Box + \left(1- \frac{1}{\xi}\right) \partial^\mu \partial^\nu \right] C_{\nu\lambda}(x-y) = \delta_{\ \lambda}^\mu \delta(x-y) \, .
\end{align}
This equation is  well-defined and solved by
\begin{align}
C_{\mu\nu}(x-y) = \left[ \eta_{\mu\nu} - (1-\xi) \frac{\partial_\mu \partial_\nu}{\Box} \right]C(x-y) \, .
\end{align}
So far, we have kept the gauge parameter $\xi$ arbitrary. The Feynman propagator becomes particularly simple in the Feynman gauge $\xi = 1$. When computing correlation functions, this is usually the preferred choice. 

Another common choice is the Landau gauge $\xi = 0$. In the limit $\xi \to 0$ the functional
\begin{align}
e^{- \frac{i}{2\xi} \int \mathrm{d}^dx \ (\partial^\mu A_\mu^a)^2}
\end{align}
oscillates very rapidly, except near $\partial^\mu A_\mu^a = 0$. Thus in the `on-shell' formulation (\emph{i.e.} when minimizing the action), the functional acts like a delta function imposing the gauge condition.

Now consider the axial type gauges $\mathcal{G}^\mu = n^\mu$. In this type of gauge, the Feynman propagator becomes
\begin{align}\label{eq:FeynmanPropagator3}
C_{\mu\nu}(x-y) = \left[ \eta_{\mu\nu} - \frac{n_\mu \partial_\nu + n_\nu \partial_\mu}{(\partial \cdot n)} + \frac{n^2 - \xi \Box}{(\partial \cdot n)^2} \partial_\mu \partial_\nu \right] C(x-y) \, .
\end{align}
Compared to the Feynman propagator in the $R_\xi$ type gauges, this is a very complicated expression. However, there are also advantages to using axial type gauges. For example, the ghost fields decouple from the gauge field and can thus always be integrated out in the path integral. This is because for $n^\mu A_\mu^a = 0$
\begin{align}
\frac{1}{2} \int \mathrm{d}^d x \ \bar{C}^a n^\mu (D_\mu C)^a = \frac{1}{2} \int \mathrm{d}^d x \ \bar{C}^a n^\mu (\partial_\mu C)^a \, .
\end{align}
So there is no dependence on the gauge field. Furthermore, the light-cone gauge ($n^2 = 0$) has been used to prove the ultraviolet finiteness of $\mathcal{N}=4$ super Yang-Mills \cite{Mandelstam:1982cb, Brink:1982wv}.

\subsection{Correlation Functions}\label{sec:YMCorrFct}
Recall the `on-shell' correlation function of some operators $\mathcal{O}_1(x_1) \ldots \mathcal{O}_n(x_n)$
\begin{align}\label{eq:YMCorr1}
\left<\!\!\left< \mathcal{O}_1(x_1) \ldots \mathcal{O}_n(x_n) \right>\!\!\right>_g = \int \mathcal{D}A \ \mathcal{D}\lambda \ \mathcal{D}C \ \mathcal{D}\bar{C} \ e^{-i S_\mathrm{inv}^1[g;A,\lambda]- i S_\mathrm{gf}[g;A,C,\bar{C}]} \ \mathcal{O}_1(x_1) \ldots \mathcal{O}_n(x_n) \, .
\end{align}
Similar to the Wess-Zumino model also the super Yang-Mills correlation function is automatically normalized to $\left<\!\!\left< 1 \right>\!\!\right>_g = 1$. This remains true even for `on-shell' supersymmetry. 

For a purely bosonic string of operators $\mathcal{O}_1(x_1) \ldots \mathcal{O}_n(x_n)$ we introduce
\begin{align}
\left< \mathcal{O}_1(x_1) \ldots \mathcal{O}_n(x_n) \right>_0 = \int \mathcal{D}_0A \ e^{-i S_\mathrm{B}^1[0;A]} \ \mathcal{O}_1(x_1) \ldots \mathcal{O}_n(x_n) \, ,
\end{align}
with the free measure $\mathcal{D}_0A = \mathcal{D}A \ \det(\Box)^{r/4} \det(\mathcal{G}^\mu \partial_\mu)$ and the free bosonic action
\begin{align}
S_\mathrm{inv}^1[0;A] =  \int \mathrm{d}^dx \ \left[ - \frac{1}{4} (\partial_\mu A_\nu^a - \partial_\nu A_\mu^a) (\partial^\mu A^{a\, \nu} - \partial^\nu A^{a \, \mu}) + \frac{1}{2\xi} (\mathcal{G}^\mu A_\mu^a)(\mathcal{G}^\nu A_\nu^a) \right] \, .
\end{align}
The free measure is chosen such that $\left<1 \right>_0 = 1$. If, furthermore, the set of operators does not depend on the coupling $g$, we have
\begin{align}
\left<\!\!\left< \mathcal{O}_1(x_1) \ldots \mathcal{O}_n(x_n) \right>\!\!\right>_g \Big\vert_{g=0} = \left< \mathcal{O}_1(x_1) \ldots \mathcal{O}_n(x_n) \right>_0 \, .
\end{align}
Free bosonic correlation functions are computed using Wick's theorem. For example, we have
\begin{align}
\begin{aligned}\label{eq:YMFreePropagator}
&\big< A_\mu^a(x) A_\nu^b(y) \big>_0 = \bcontraction{}{A}{_\mu^a(x) }{A} A_\mu^a(x) A_\nu^b(y) = \delta^{ab} C_{\mu\nu}(x-y) \, ,
\end{aligned}
\end{align}
where $C_{\mu\nu}(x-y)$ is the gauge dependent Feynman propagator. In the Feynman gauge, the free correlator is particularly simple. Integrating out the fermion and ghost fields in a purely bosonic correlation function is straightforward. For the fermions, we obtain the Matthews-Salam-Seiler determinant $\Delta_\mathrm{MSS}$
\begin{align}
\begin{gathered}\label{eq:YMFermDet}
\int \mathcal{D}\lambda \, \exp \left[ - \frac{1}{2} \int \mathrm{d}^dx \ \bar{\lambda}^a \gamma^\mu (D_\mu \lambda)^a \right] = \det(\slashed{D})^{1/2} = \det(\slashed{\partial})^{1/2} \det( 1 - \mathbf{Y})^{1/2} \, ,
\end{gathered}
\end{align}
with the integration kernel
\begin{align}
\mathbf{Y}_{\alpha\beta}^{ab}(x,y;A) = g f^{abc} (\gamma^\mu \gamma^\nu )_{\alpha\beta} \, \partial_\mu C(x-y) A_\nu^c(y) \, .
\end{align}
$\mathbf{Y}$ is obtained by observing that $\slashed{\partial}$ is the inverse of the free fermion propagator $S_0(x-y)$ from \eqref{eq:YMFreeFermProp}. Again we are interested in the perturbative expansion of the logarithm of the Matthews-Salam-Seiler determinant. We use the well-known relation $\det \mathcal{M} = \exp( \Tr( \log \mathcal{M}))$ and obtain
\begin{align}
\log(\Delta_\mathrm{MSS}[g;A] ) = \frac{1}{2} \Tr \log \left[ 1 - \mathbf{Y} \right] \, .
\end{align}
The factor $\frac{1}{2}$ comes from the square root in \eqref{eq:YMFermDet}. The trace $\Tr$ is over all indices and variables of $\mathbf{Y}_{\alpha\beta}^{ab}(x,y;A)$. In particular, it integrates over the remaining free variable. Subsequently, we find
\begin{align}
\begin{aligned}\label{eq:YMMSS}
\log(\Delta_\mathrm{MSS}[g;A] )
&= - \frac{g}{2} f^{aab} \, \tr(\gamma^\mu \gamma^\nu ) \int \mathrm{d}^dx \ \partial_\mu C(x-x) A_\nu^b(x) \\
&\quad - \frac{g^2}{4} f^{abc} f^{bad} \, \tr(\gamma^\mu \gamma^\nu \gamma^\rho \gamma^\lambda ) \int \mathrm{d}^dx \ \mathrm{d}^dy \\
&\quad \quad \times \partial_\mu C(x-y) A_\nu^c(y) \partial_\rho C(y-x) A_\lambda^d(x) \\
&\quad - \frac{g^3}{6} f^{abc} f^{bde} f^{dam} \, \tr(\gamma^\mu \gamma^\nu \gamma^\rho \gamma^\lambda \gamma^\sigma \gamma^\tau ) \int \mathrm{d}^dx \ \mathrm{d}^dy \ \mathrm{d}^dz \\
&\quad \quad \times \partial_\mu C(x-y) A_\nu^c(y) \partial_\rho C(y-z) A_\lambda^e(z) \partial_\sigma C(z-x) A_\tau^m(x) \\
&\quad + \mathcal{O}(g^4) \, .
\end{aligned}
\end{align}
The first term cancels since $f^{aab} = 0$. For the second term we use $f^{abc} f^{bae} = - N \delta^{ce}$. In the third order, the structure constants do not simplify. Computing the traces over the gamma matrices yields the final result
\begin{align}
\begin{aligned}\label{eq:YMLogMSS}
\log(\Delta_\mathrm{MSS}[g;A]) &= \frac{r g^2 N}{4} \int \mathrm{d}^d x \ \mathrm{d}^d y \ \Big\{ \\ 
&\quad \quad + 2\, \partial^\mu C(x-y) A_\mu^a(y) \partial^\rho C(y-x) A_\rho^a(x) \\
&\quad \quad - \partial^\mu C(x-y) A^{a \, \rho}(y) \partial_\mu C(y-x) A_\rho^a(x) \Big\} \\
&\quad + \frac{r g^3}{6} f^{a d m} f^{b e m} f^{c d e} \ \int \mathrm{d}^d x \ \mathrm{d}^d y \ \mathrm{d}^d z \ \Big\{ \\
&\quad \quad - 6 \, \partial^\mu C(x-y) A_\mu^b(y) \partial^\rho C(y-z) A^{c \, \lambda}(z) \partial_\rho C(z-x) A_\lambda^a(x) \\ 
&\quad \quad + 2 \, \partial^\mu C(x-y) A_\rho^b(y) \partial^\rho C(y-z) A^{c \, \lambda}(z) \partial_\lambda C(z-x) A_\mu^a(x) \\ 
&\quad \quad + 3 \, \partial^\mu C(x-y) A_\rho^b(y) \partial^\rho C(y-z) A_\mu^c(z) \partial^\lambda C(z-x) A_\lambda^a(x) \\ 
&\quad \quad - \partial^\mu C(x-y) A_\rho^b(y) \partial^\lambda C(y-z) A_\mu^c(z) \partial^\rho C(z-x) A_\lambda^a(x) \\ 
&\quad \quad + 3 \, \partial^\mu C(x-y) A^{b\, \rho}(y) \partial^\lambda C(y-z) A_\mu^c(z) \partial_\lambda C(z-x) A_\rho^a(x) \Big\} \\
&\quad + \mathcal{O}(g^4) \, .
\end{aligned}
\end{align}
The MSS determinant is invariant under the choice of gauge. The ghost (or Faddeev-Popov) determinant, on the other hand, is gauge-dependent. However, its calculation is much simpler. In the $R_\xi$ type gauges, we find
\begin{align}
\int \mathcal{D}\bar{C} \ \mathcal{D}C \, \exp \left[ - \frac{1}{2} \int \mathrm{d}^dx \ \bar{C}^a \partial^\mu (D_\mu C)^a \right] 
= \det\left(\partial^\mu D_\mu\right) = \det(\Box) \det( 1 - \mathbf{X} )
\end{align}
with
\begin{align}
\mathbf{X}^{ab}(x,y;A) = g f^{abc} \, C(x-y) A_\mu^c(y) \partial_y^\mu \, .
\end{align}
Here the relevant observation is that in the $R_\xi$ type gauges the free ghost propagator is given by $G_0(x-y) = -C(x-y)$. The logarithm of the Faddeev-Popov determinant is given by
\begin{align}
\log( \Delta_\mathrm{FP}[g;A]) = \Tr \log[ 1 - \mathbf{X}] \, .
\end{align}
Expanding the logarithm and computing the trace yields
\begin{align}
\begin{aligned}\label{eq:YMLogFPLandau}
\log( \Delta_\mathrm{FP}[g;A]) &= - g f^{aab} \int \mathrm{d}^dx \ \partial^\mu C(x-x) A_\mu^b(x)\\
&\quad + \frac{g^2 N}{2} \int \mathrm{d}^dx \ \mathrm{d}^dy \ \partial^\mu C(x-y) A_\rho^a(y) \partial^\rho C(y-x) A_\mu^a(x) \\
&\quad - \frac{g^3}{3} f^{abc} f^{bde} f^{dam} \int \mathrm{d}^dx \ \mathrm{d}^dy \ \mathrm{d}^dz  \\
&\quad \quad \times  \partial^\rho C(x-y) A_\mu^c(y) \partial^\mu C(y-z) A_\nu^e(z) \partial^\nu C(z-x) A_\rho^m(x) \\
&\quad + \mathcal{O}(g^4)\, .
\end{aligned}
\end{align}
Again the first term vanishes since $f^{aab} = 0$. In the axial type gauges, the integration kernel for the Faddeev-Popov determinant is given by
\begin{align}
\mathbf{X}^{ab}(x,y;A) = g f^{abc} G_0(x-y) n \cdot A^c(y) \, , 
\end{align}
where $G_0(x-y)$ is the free axial ghost propagator. It solves the equation
\begin{align}
n^\mu \partial_\mu G_0(x-y) = \delta(x-y) \, .
\end{align}
In four dimensions and for $n_\mu n^\mu = 1$ it is given by
\begin{align}\label{eq:YMaxialGhostProp}
G_0(x) = \varepsilon(n \cdot x) \delta^{(3)}(x^\perp) = - G_0(-x) \, ,
\end{align}
where $\varepsilon(x)$ is the anti-symmetric step function $\varepsilon(x) \coloneqq \Theta(x) - \frac{1}{2}$ and $x_\mu^\perp$ is the transverse coordinate $x_\mu^\perp \coloneqq n_\mu (n \cdot x)$. Expanding the logarithm of the Faddeev-Popov determinant in the axial gauge, we obtain
\begin{align}
\begin{aligned}\label{eq:YMLogFPAxial}
\log( \Delta_\mathrm{FP}[g;A]) &= - g f^{aab} \int \mathrm{d}^dx \ G_0(x-x) \,  n \cdot A^b(x)\\
&\quad + \frac{g^2 N}{2}  \int \mathrm{d}^dx \ \mathrm{d}^dy \ G_0(x-y) \, n \cdot A^a(y) G_0(y-x) n \cdot A^a(x) \\
&\quad + \mathcal{O}(g^3) \, .
\end{aligned}
\end{align}
Once more, the first term vanishes since $f^{aab} = 0$. Restricting ourselves to the gauge surface, \emph{i.e.} $\mathcal{G}^a(A) = 0$, the Faddeev-Popov determinant in axial gauge becomes trivial. In the $R_\xi$ type gauges, it remains unchanged. 

As before, a correlation function where we have integrated out the anti-commuting degrees of freedom is denoted by a single bracket
\begin{align}
\left< \mathcal{O}_1(x_1) \ldots \mathcal{O}_n(x_n) \right>_g = \int \mathcal{D}_g A \ e^{-iS_\mathrm{inv}^1[g;A]-iS_\mathrm{gf}[g;A]} \ \mathcal{O}_1(x_1) \ldots \mathcal{O}_n(x_n) \, ,
\end{align}
with $\mathcal{D}_g A = \mathcal{D}A \ \det(\Box)^{r/4} \det(\mathcal{G}^\mu \partial_\mu) \Delta_\mathrm{MSS}[g;A] \Delta_\mathrm{FP}[g;A]$. This definition is such that in the limit $\left< \mathcal{O}_1 \ldots \mathcal{O}_n \right>_g \big\vert_{g=0} = \left<\mathcal{O}_1 \ldots \mathcal{O}_n \right>_0$ for a $g$ independent set of bosonic operators. 

Furthermore, we may also integrate out the anti-commuting degrees of freedom in any correlation function containing spinor or ghost fields. To this end, consider the examples
\begin{align}\label{eq:SpinorCorr}
\big<\!\!\big< A_\mu^a(x) \lambda^b(y) \bar{\lambda}^c(z) \big>\!\!\big>_g = i \big< A_\mu^a(x) \bcontraction{}{\lambda}{^b(y) }{\bar{\lambda}} \lambda^b(y) \bar{\lambda}^c(z) \big>_g =   \big< A_\mu^a(x)  S^{bc}(y,z;A) \big>_g 
\end{align}
and
\begin{align}
\big<\!\!\big< A_\mu^a(x) C^b(y) \bar{C}^c(z) \big>\!\!\big>_g = - i \big< A_\mu^a(x)  \bcontraction{}{C}{^b(y) }{\bar{C}} C^b(y) \bar{C}^c(z) \big>_g = -  i\big< A_\mu^a(x)  G^{bc}(y,z;A) \big>_g \, .
\end{align}
Notice that each contraction of two spinor fields produces a factor of $i$ and each contraction of two ghost fields produces a factor of $(-i)$. This is due to different normalizations of the spinor and ghost path integrals (see appendix \ref{app:FD} for details). Using Wick's theorem, the generalization to multiple spinor and ghost fields is immediate.

\subsection{BRST Variations and Ward Identities}
We want to construct a Ward identity for $\mathcal{N}=1$ super Yang-Mills similar to \eqref{eq:WZWardId} for the Wess-Zumino model. However, the complete $\mathcal{N}=1$ super Yang-Mills action
\begin{align}\label{eq:YM3}
S^1 = S_\mathrm{inv}^1 + S_\mathrm{gf} \, ,
\end{align}
which is used for computing correlation functions, is not invariant under the supersymmetry transformations \eqref{eq:YMsusy1} (respectively \eqref{eq:YMsusy2}). In particular $\delta_\alpha S_\mathrm{gf} \not=0$. Thus the expression \eqref{eq:WZWardId} must be modified for the super Yang-Mills field theory. 

Combining the `off-shell' supersymmetry variations \eqref{eq:YMsusy1} into a single equation, we obtain
\begin{align}
\delta_\alpha \equiv i (\bar{\lambda}^a \gamma_\mu)_\alpha \frac{\delta}{\delta A_\mu^a } - \left( \frac{1}{2} (\gamma^{\mu\nu})_{\beta\alpha} F_{\mu \nu}^a - i (\gamma^5)_{\beta\alpha} D^a \right) \frac{\delta}{\delta \lambda_\beta^a} - ((D_\mu \bar{\lambda})^a \gamma^\mu \gamma^5)_\alpha \frac{\delta}{\delta D^a} \, .
\end{align}
There is no susy partner for the ghost field. Hence it transforms trivially under the supersymmetry variation. Now let $X[A,D,\lambda,C, \bar{C}]$ be an arbitrary string of operators. Repeating the argument about integrals over fermionic and bosonic derivatives from subsection \ref{sec:WZWardId}, we obtain
\begin{align}
\int \mathcal{D}A \ \mathcal{D}D \ \mathcal{D}\lambda \ \mathcal{D}\bar{C} \ \mathcal{D}C \ \delta_\alpha X[A,D,\lambda,\bar{C},C] = 0 \, .
\end{align}
Completing the integral to a correlation function, we obtain
\begin{align}
\begin{aligned}
\big<\!\!\big< \delta_\alpha X[A,D,\bar{C},C] \big>\!\!\big>_g &= \int \mathcal{D}A \ \mathcal{D}D \ \mathcal{D}\lambda \ \mathcal{D}\bar{C} \ \mathcal{D}C \ e^{-iS_\mathrm{inv}^1 - iS_\mathrm{gf}} \, \delta_\alpha X[A,D,\lambda,\bar{C},C] \\
&= \int \mathcal{D}A \ \mathcal{D}D \ \mathcal{D}\lambda \ \mathcal{D}\bar{C} \ \mathcal{D}C \ \delta_\alpha \left( e^{-iS_\mathrm{inv}^1 - iS_\mathrm{gf}} \, X[A,D,\lambda,\bar{C},C] \right) \\
&\quad + i \int \mathcal{D}A \ \mathcal{D}D \ \mathcal{D}\lambda \ \mathcal{D}\bar{C} \ \mathcal{D}C \ e^{-iS_\mathrm{inv}^1 - iS_\mathrm{gf}} \, \left( \delta_\alpha S_\mathrm{gf} \right) X[A,D,\lambda,\bar{C},C] \\
&= i \, \big<\!\!\big< (\delta_\alpha S_\mathrm{gf}) \, X[A,D,\lambda,\bar{C},C] \big>\!\!\big>_g \, .
\end{aligned}
\end{align}
Hence the super Yang-Mills Ward identity reads
\begin{align}\label{eq:YMWardId1}
\big<\!\!\big< \delta_\alpha X[A,D,\lambda,\bar{C},C] \big>\!\!\big>_g = i \, \big<\!\!\big< (\delta_\alpha S_\mathrm{gf}) \, X[A,D,\lambda,\bar{C},C] \big>\!\!\big>_g \, .
\end{align}
In an attempt to develop a mathematically rigorous quantization of gauge theories, Becchi, Rouet, Stora and Tyutin have introduced the so-called BRST (or Slavnov) variations \cite{Becchi:1975nq, Tyutin:1975qk, Iofa:1976je} which leave the entire super Yang-Mills action \eqref{eq:YM3} invariant. For `off-shell' $\mathcal{N}=1$ super Yang-Mills, the BRST variations are
\begin{align}
\begin{aligned}\label{eq:YMBRST}
&s A_\mu^a = (D_\mu C)^a \, , &&s D^a = g f^{abc} D^b C^c \, , &&s \lambda^a = - g f^{abc} \lambda^b C^c \, , \\
&s \, \bar{C}^a = - \frac{1}{\xi} \mathcal{G}^a(A) \, , &&s\, C^a = - \frac{g}{2} f^{abc} C^b C^c \, .
\end{aligned}
\end{align}
The Slavnov operator $s$ is fermionic, \emph{i.e.} it anti-commutes with other fermionic quantities, and it is nilpotent, \emph{i.e.} $s^2 = 0$. Since $s(S^1) = 0$ the Ward identity for the Slavnov variations reads
\begin{align}\label{eq:YMWardId2}
\big<\!\!\big< s \, X[A,D,\lambda,\bar{C},C] \big>\!\!\big>_g = 0 \, .
\end{align}
Moreover, we observe that
\begin{align}
\delta_\alpha S_\mathrm{gf} = - s \int \mathrm{d}^dx \ \bar{C}^a \, (\delta_\alpha \mathcal{G}^a(A) ) \, .
\end{align}
So we can further modify \eqref{eq:YMWardId1} and obtain
\begin{align}\label{eq:YMWardId3}
\big<\!\!\big< \delta_\alpha X[A,D,\bar{C},C] \big>\!\!\big>_g =  i \, \Big<\!\!\!\Big< \left(  \int \mathrm{d}^dx \ \bar{C}^a \, (\delta_\alpha \mathcal{G}^a(A) ) \right) \, s(X[A,D,\lambda,\bar{C},C]) \Big>\!\!\!\Big>_g \, .
\end{align}

\section[\texorpdfstring{$\mathcal{N}=4$}{N=4} super Yang-Mills]{$\boldsymbol{\mathcal{N}}$\:=\;4 super Yang-Mills}
In this section, we discuss various aspects of $\mathcal{N}=4$ super Yang-Mills relevant to the first part of this thesis. Further properties of the theory specific to the second part of the thesis are given in chapter \ref{ch:SCA} and \ref{ch:BPS}.

\subsection{The Action}
The 4-dimensional $\mathcal{N}=4$ super Yang-Mills theory was originally introduced as the dimensional reduction of 10-dimensional $\mathcal{N}=1$ super Yang-Mills \cite{Gliozzi:1976qd,Brink:1976bc}. It describes a gauge field $A_\mu^a$, three scalar fields $A_i^a$ ($i,j= 1,2,3$), three pseudoscalar fields $B_i^a$  and four Majorana spinors $\lambda_{\alpha A}^a$ ($A,B = 1, \ldots, 4$). The gauge invariant action is given by
\begin{align}
\begin{aligned}\label{eq:YM4}
S_\mathrm{inv}^4 = \int \mathrm{d}^4x \ \bigg[ &-\frac{1}{4} F_{\mu\nu}^a F^{a\, \mu\nu} - \frac{1}{2} (D_\mu A_i)^a (D^\mu A^i)^a - \frac{1}{2} (D_\mu B_i)^a (D^\mu B^i)^a \\
& - \frac{i}{2} \bar{\lambda}_A^a \gamma^\mu (D_\mu \lambda_A)^a + \frac{g}{2} f^{abc} \bar{\lambda}_A^a (\alpha_{AB}^i A_i^b + i \gamma_5 \beta_{AB}^i B_i^b) \lambda_B^c \\
&- \frac{g^2}{4} f^{abc} f^{ade} \left(A_i^b A_j^c A^{d \, i} A^{e \, j} + B_i^b B_j^c B^{d \, i} B^{e \, j} + 2 A_i^b B_j^c A^{d \, i} B^{e \, j} \right) \bigg] \, ,
\end{aligned}
\end{align}
where the field strength tensor $F_{\mu\nu}^a$ and the covariant derivative are defined as before. The real anti-symmetric $4 \times 4$ matrices $\alpha^i$ and $\beta^i$ satisfy
\begin{align}
\begin{aligned}\label{eq:AlphaBetaRelations}
&\{ \alpha^i, \alpha^j \} = - 2 \delta^{ij} \, , \quad [\alpha^i , \alpha^j ] = 2 \varepsilon^{ijk} \alpha^k \, , \\
&\{ \beta^i, \beta^j \} = - 2 \delta^{ij} \, , \quad [\beta^i , \beta^j ] = - 2 \varepsilon^{ijk} \beta^k \, , \quad [\alpha^i, \beta^j] = 0  \, .
\end{aligned}
\end{align}
While working with Majorana spinors, we do not distinguish between subscript and superscript spinor indices $A, B = 1, \ldots, 4$. The action \eqref{eq:YM4} is invariant  under the supersymmetry transformations
\begin{align}
\begin{aligned}\label{eq:susyAlgebraYM3}
&\begin{aligned}
&\delta A_\mu^a = i (\bar{\varepsilon}_A \gamma_\mu \lambda_A^a) \, , &&\delta A_i^a = - (\bar{\varepsilon}_A \alpha_i^{AB} \lambda_B^a) \, , &&\delta B_i^a = - i (\bar{\varepsilon}_A \gamma_5 \beta_i^{AB} \lambda_B^a) \, , \\
\end{aligned} \\
&\delta \lambda_{\alpha A}^a = - \frac{1}{2} (\gamma^{\mu\nu} \varepsilon_A)_\alpha F_{\mu\nu}^a - i (\gamma^\mu \alpha_{AB}^i \varepsilon_B)_\alpha (D_\mu A_i)^a - (\gamma_5 \gamma^\mu \beta_{AB}^i \varepsilon_B)_\alpha (D_\mu B_i)^a \\
&\quad \quad \quad - \frac{g}{2} f^{abc} \left( \left(\alpha_{AB}^i A_i^b - i \gamma_5 \beta_{AB}^i B_i^b \right) \left(\alpha_{BC}^j A_j^c + i \gamma_5 \beta_{BC}^j B_j^c \right) \varepsilon_C \right)_\alpha \, .
\end{aligned}
\end{align}
Using the 4-dimensional Fierz identity for Majorana spinors and the definition
\begin{align}
c^a \coloneqq \left[ \bar{\varepsilon}_A^2 (\alpha_{AB}^i A_i^a + i \gamma_5 \beta_{AB}^i B_i^a) \varepsilon_B^1 - \bar{\varepsilon}_A^1 (\alpha_{AB}^i A_i^a + i \gamma_5 \beta_{AB}^i B_i^a) \varepsilon_B^2 \right]
\end{align}
we can show that the supersymmetry algebra \eqref{eq:susyAlgebraYM3} closes `on-shell'. Hence up to terms proportional to the equations of motion, we have
\begin{align}
\begin{aligned}
&[\delta_1, \delta_2] A_\mu^a = i b^\nu F_{\nu\mu}^a - (D_\mu c)^a \, , &&[\delta_1, \delta_2] \lambda_{A\alpha}^a = i b^\mu (D_\mu \lambda_{A\alpha})^a + g f^{abc} c^b \lambda_{A\alpha}^c \, , \\
&[\delta_1, \delta_2] A_i^a = i b^\mu (D_\mu A_i)^a + g f^{abc} c^b A_i^c \, , &&[\delta_1, \delta_2] B_i^a = i b^\mu (D_\mu B_i)^a + g f^{abc} c^b B_i^c \, .
\end{aligned}
\end{align}
Since the supersymmetry is only `on-shell', the theory \eqref{eq:YM4} has $2+3+3$ bosonic and $4 \cdot 2$ fermionic degrees of freedom. 

The fermion propagator of the four Majorana spinors in the action \eqref{eq:YM4} is
\begin{align}\label{eq:N4FermProp}
i \bcontraction{}{\lambda}{_A^a(x) }{\bar{\lambda}} \lambda_A^a(x) \bar{\lambda}_B^b(y) \equiv S_{AB}^{ab}(x,y;\mathscr{A} ) \, ,
\end{align}
where $\mathscr{A} = (A_\mu^a, A_i^a, B_i^a)$ is the set of bosonic fields in \eqref{eq:YM4}. The Dirac equation then reads
\begin{align}\label{eq:N4DiracEq}
\left[ \delta_{AC} \delta^{ac} \gamma^\mu D_\mu + i g f^{aec} \alpha_{AC}^i A_i^e(x) - g f^{aec} \gamma_5 \beta_{AC}^i B_i^e(x)  \right] S_{BC}^{cb}(x,y;\mathscr{A} ) = \delta_{AB} \delta^{ab} \delta(x-y) \, .
\end{align}
The corresponding free fermion propagator is the same as for the $\mathcal{N}=1$ theory.

\subsection{Dimensional Reduction}\label{sec:SFTDimRed}
We describe how to obtain the action \eqref{eq:YM4} via dimensional reduction of the 10-dimensional $\mathcal{N}=1$ super Yang-Mills action \eqref{eq:YM2}. Recall
\begin{align}\label{eq:YM10D}
S_\mathrm{inv}^1 &= \int \mathrm{d}^{10}w \ \left[ -\frac{1}{4} \mathcal{F}_{MN}^a \mathcal{F}^{a\, MN} - \frac{i}{2} \bar{\Lambda}^a \Gamma^M (D_M \Lambda)^a \right] \, ,
\end{align}
where $M,N = 0, \ldots, 9$. This action is `on-shell' invariant under the supersymmetry transformations \eqref{eq:YMsusy2}
\begin{align}\label{eq:YM10Dsusy}
\delta \mathcal{A}_M^a = - i (\bar{\Lambda}^a \Gamma_M \varepsilon_{10}) \, , \quad \delta \Lambda_\alpha^a = - \frac{1}{2} \left( \Gamma^{MN} \varepsilon_{10} \right)_\alpha \mathcal{F}_{MN}^a \, ,
\end{align}
where $\epsilon_{10}$ is a 10-dimensional Majorana-Weyl spinor (see appendix \ref{app:Spinors}). The first step of the dimensional reduction is to set 
\begin{align}
\partial_{3+i} = \partial_{6+i} = 0 \quad \text{for} \quad i = 1,2,3 \, .
\end{align}
This breaks the $O(1,9)$ Lorentz symmetry of the 10-dimensional theory \eqref{eq:YM10D} down to a $O(1,3) \otimes O(6) \simeq SL(2,\mathbb{C}) \otimes SU(4)$ symmetry. Moreover, it implies the split of the spacetime indices $M=(\mu, i, j)$. Likewise, we decompose the coordinates $w^M = (x^\mu, y^i, z^j)$ and the gauge field
\begin{align}
\mathcal{A}_M^a(x,y,z) = (A_\mu^a(x), A_i^a(x), B_j^a(x)) \, .
\end{align}
Notice that the dependence on the internal coordinates $y^i$ and $z^j$ is dropped. To decompose the spinor field, we choose a particular representation of the 10-dimensional $32 \times 32$ $\Gamma$-matrices
\begin{align}
\begin{aligned}\label{eq:SFTGamma10}
&\Gamma^\mu \coloneqq \gamma^\mu \otimes \mathbbm{1}_8 \, , &&\mu = 0,1,2,3 \, , \\
&\Gamma^{3+i} \coloneqq \gamma_5 \otimes \begin{pmatrix} 0 & i \alpha^i \\ -i \alpha^i & 0 \end{pmatrix} \, , \quad \quad &&i = 1,2,3 \, , \\
&\Gamma^{6+i} \coloneqq \gamma_5 \otimes \begin{pmatrix} 0 & \beta^i \\ \beta^i & 0 \end{pmatrix} \, , &&i = 1,2,3 \, .
\end{aligned}
\end{align}
The $\gamma$-matrices are given by
\begin{align}
\gamma^\mu \coloneqq \begin{pmatrix} 0 & \sigma^\mu \\ \bar{\sigma}^\mu & 0 \end{pmatrix} \, , \quad \gamma^5 = \begin{pmatrix} - \mathbbm{1}_2 & 0 \\ 0 & \mathbbm{1}_2 \end{pmatrix} \, ,
\end{align}
with $\sigma^\mu \coloneqq (\mathbbm{1}_2, \sigma^\mu)$ and $\bar{\sigma}^\mu \coloneqq (\mathbbm{1}_2, - \sigma^i)$. They satisfy the Clifford algebra relation $\{ \gamma^\mu, \gamma^\nu \} = 2 \eta^{\mu\nu}$. The $\alpha^i$ and $\beta^i$ matrices are defined as
\begin{align}
\begin{aligned}\label{eq:AlphaBeta}
&\alpha_{ik}^j \coloneqq \varepsilon_{ijk} \, , &&\alpha_{i4}^j = - \alpha_{4i}^j \coloneqq - \delta_i^j \, , &&\alpha_{44}^i \coloneqq 0 \, , \\
&\beta_{ik}^j \coloneqq - \varepsilon_{ijk} \, , &&\beta_{i4}^j = - \beta_{4i}^j \coloneqq - \delta_i^j \, , &&\beta_{44}^i \coloneqq 0 \, .
\end{aligned}
\end{align}
They satisfy \eqref{eq:AlphaBetaRelations}. All of these definitions are such that the $\Gamma^M$ satisfy the 10-dimensional Clifford algebra relation $\{ \Gamma^M, \Gamma^N \} = 2 \eta^{MN}$. The eleventh gamma matrix is defined as $\Gamma_{11} \coloneqq \Gamma_0 \cdots \Gamma_9$ and the 10-dimensional charge conjugation matrix is
\begin{align}
\mathcal{C}_{10} = \mathcal{C}_4 \otimes \begin{pmatrix} 0 & \mathbbm{1}_4 \\ \mathbbm{1}_4 & 0 \end{pmatrix} \, ,
\end{align}
where $\mathcal{C}_4 = i \gamma_2 \gamma_0$ is the 4-dimensional charge conjugation matrix. $\Gamma_{11}$ commutes with all the other gamma matrices and
\begin{align}
\mathcal{C}_{10} \Gamma_M \mathcal{C}_{10}^{-1} = - \Gamma_M^T \, .
\end{align}
The given representation of the 10-dimensional gamma matrices implies 
\begin{align}\label{eq:SFTLambda10}
\Lambda^a = \begin{pmatrix} \chi^a \\ \bar{\chi}^a \end{pmatrix} \quad \text{with} \quad
\chi^a = \begin{pmatrix} \mathbf{0} \\ \psi_1^a \\ \mathbf{0} \\ \psi_2^a \end{pmatrix} \quad \text{and} \quad 
\psi_i^a = \begin{pmatrix} \omega_{1i}^a \\ \omega_{2i}^a \\ \omega_{3i}^a \\ \omega_{4i}^a \end{pmatrix}  \, ,  \quad 
\mathbf{0} = \begin{pmatrix} 0 \\ 0 \\ 0 \\ 0 \end{pmatrix} \, .
\end{align}
This 32-component spinor satisfies the Majorana condition $\bar{\Lambda}^a = \Lambda^{a \, T} \mathcal{C}_{10}$ as well as the Weyl condition $\Lambda^a = \frac{1}{2} \left( \mathbbm{1}_{32} - \Gamma_{11} \right) \Lambda^a$. It decomposes into four 4-dimensional Majorana spinors 
\begin{align}
\lambda_A^a = \begin{pmatrix} \omega_{A1}^a \\ \omega_{A2}^a \\ - \bar{\omega}_{A2}^a \\ \bar{\omega}_{A1}^a \end{pmatrix} \, .
\end{align}
Subsequently, we find
\begin{align}
\begin{aligned}
&\bar{\Lambda}^a \Gamma^\mu (D_\mu \Lambda)^a = \bar{\lambda}_A^a \gamma^\mu (D_\mu \lambda_A)^a \, , \\
&\bar{\Lambda}^a \Gamma^{3+i} (D_{3+i} \Lambda)^a = i g f^{abc} \bar{\lambda}_A^a \alpha_{AB}^i A_i^b \lambda_B^c \, , \\
&\bar{\Lambda}^a \Gamma^{6+i} (D_{6+i} \Lambda)^a = - g f^{abc} \bar{\lambda}_A^a \gamma_5 \beta_{AB}^i B_i^b \lambda_B^c 
\end{aligned}
\end{align}
and
\begin{align}
\begin{aligned}
\mathcal{F}_{mn}^a \mathcal{F}^{a\, mn} &= F_{\mu\nu}^a F^{a\, \mu\nu} + 2 (D_\mu A_i)^a (D^\mu A^i)^a+ 2 (D_\mu B_i)^a (D^\mu B^i)^a \\
&\quad + g^2 f^{abc} f^{ade} \left(A_i^b A_j^c A^{d \, i} A^{e \, j} + B_i^b B_j^c B^{d \, i} B^{e \, j} + 2 A_i^b B_j^c A^{d \, i} B^{e \, j} \right) \, .
\end{aligned}
\end{align}
Hence the 10-dimensional $\mathcal{N}=1$ action \eqref{eq:YM10D} decomposes into the 4-dimensional $\mathcal{N}=4$ action \eqref{eq:YM4}. Furthermore the 10-dimensional $\mathcal{N}=1$ supersymmetry variations \eqref{eq:YM10Dsusy} imply the 4-dimensional $\mathcal{N}=4$ supersymmetry variations \eqref{eq:susyAlgebraYM3}.

\subsection[Maximally extended \texorpdfstring{$\mathcal{N}=1$}{N=1} super Yang-Mills]{Maximally extended $\boldsymbol{\mathcal{N}}$\:=\;1 super Yang-Mills}
A different way of deriving the $\mathcal{N}=4$ super Yang-Mills action is through maximally extended 4-dimensional $\mathcal{N}=1$ super Yang-Mills. From the superspace formulation of $\mathcal{N}=1$ super Yang-Mills, it is evident that we can couple it to three copies of the 4-dimensional Wess-Zumino model. We will not go into the details here, but in the end, the action reads (see \cite{Grisaru:1988iz} or \cite{Sohnius:1985qm} for a review)
\begin{align}
\begin{aligned}\label{eq:YMMax}
S_\mathrm{inv}^\mathrm{max} = \int \mathrm{d}^4x \ \bigg[ &-\frac{1}{4} F_{\mu\nu}^a F^{a\, \mu\nu} - \frac{i}{2} \bar{\lambda}^a \gamma^\mu (D_\mu \lambda)^a + \frac{1}{2} D^aD^a \\
& + \frac{1}{2} (D_\mu A_i)^a (D^\mu A_i)^a + \frac{1}{2} (D_\mu B_i)^a (D^\mu B_i)^a + \frac{1}{2} F_i^a F_i^a + \frac{1}{2} G_i^a G_i^a \\
& - \frac{i}{2} \bar{\chi}_{i}^a \gamma^\mu (D_\mu \chi_{i})^a - g f^{abc} \left( D^a A_i^b B_i^c + (\bar{\chi}_i^a (A_i^b + i \gamma_5 B_i^b) \lambda^c) \right) \\
& + \varepsilon_{ijk} g f^{abc} \left(\frac{1}{2} \bar{\chi}_i^a (A_j^b - i \gamma_5 B_j^b) \chi_k^c - \frac{1}{2} F_i^a (A_j^b A_k^c - B_j^b B_k^c)- G_i^a A_j^b B_k^c \right) \bigg] \, .
\end{aligned}
\end{align}
This action is invariant under the supersymmetry transformations
\begin{align}
\begin{aligned}\label{eq:YMMaxsusy}
&\begin{aligned}
&\delta A_\mu^a = i \bar{\varepsilon} \gamma_\mu {\lambda}^a \, , &&\delta {\lambda}_\alpha^a = - \frac{1}{2} (\gamma^{\mu\nu} \varepsilon)_\alpha F_{\mu\nu}^a + i (\gamma_5 \varepsilon)_\alpha D^a \, , &&\delta D^a = - \bar{\varepsilon} \gamma_5 \gamma^\mu (D_\mu \lambda)^a \, , \\ 
&\delta A_i^a = \bar{\varepsilon} \chi_i^a \, , &&\delta B_i^a = i \bar{\varepsilon} \gamma_5 \chi_i^a \, , \\
\end{aligned} \\
&\delta F_i^a = i \bar{\varepsilon} \gamma^\mu (D_\mu \chi_i)^a +g f^{abc} \bar{\varepsilon} ( A_i^b + i \gamma_5 B_i^b) \lambda^c \, , \\
&\delta G_i^a = - \bar{\varepsilon}\gamma_5 \gamma^\mu (D_\mu \chi_i)^a - g f^{abc} \bar{\varepsilon} ( B_j^b - i \gamma_5 A_i^b) \lambda^c \, , \\
&\delta \chi_{i\alpha}^a = i (\gamma^\mu \varepsilon)_\alpha (D_\mu A_i)^a + (\gamma_5 \gamma^\mu \varepsilon)_\alpha (D_\mu B_i)^a + F_i^a \varepsilon_\alpha + i (\gamma_5 \varepsilon)_\alpha {G}_i^a \, .
\end{aligned}
\end{align}
In order to turn the $\mathcal{N}=1$ supersymmetry into a $\mathcal{N}=4$ supersymmetry we integrate out the auxiliary fields $D^a$, $F_i^a$ and $G_i^a$. Their algebraic equations of motion are
\begin{align}\label{eq:YMMaxAuxEOM}
&D^a = g f^{abc} A_i^b B_i^c \, , \quad F_i^a = \frac{g}{2} f^{abc} \varepsilon_{ijk} \left(A_j^b A_k^c - B_j^b B_k^c \right) \, , \quad G_i^a = g f^{abc} \varepsilon_{ijk} A_j^b B_k^c \, .
\end{align}
Plugging this back into the action, we use the Jacobi identity \eqref{eq:JacId} to obtain
\begin{align}
\begin{aligned}
S_\mathrm{inv}^4 = \int \mathrm{d}^4x \ \bigg[ &-\frac{1}{4} F_{\mu\nu}^a F^{a\, \mu\nu} + \frac{1}{2} (D_\mu A_i)^a (D^\mu A_i)^a + \frac{1}{2} (D_\mu B_i)^a (D^\mu B_i)^a \\
& - \frac{i}{2} \bar{\lambda}^a \gamma^\mu (D_\mu \lambda)^a - \frac{i}{2} \bar{\chi}_{i}^a \gamma^\mu (D_\mu \chi_{i})^a - g f^{abc} \bar{\chi}_i^a (A_i^b + i \gamma_5 B_i^b) \lambda^c \\
& + \frac{g}{2} \varepsilon_{ijk} f^{abc} \bar{\chi}_i^a (A_j^b - i \gamma_5 B_j^b) \chi_k^c \\
& - \frac{g^2}{4} f^{abc} f^{ade} \left(A_i^b A_j^c A_i^d A_j^e + B_i^b B_j^c B_i^d B_j^e + 2 A_i^b B_j^c A_i^d B_j^e \right) \bigg] \, . \\
\end{aligned}
\end{align}
This is almost the correct result. To write it in the form \eqref{eq:YM4} we introduce the four Majorana spinors $\lambda_A^a$ with
\begin{align}
\lambda_i^a \coloneqq \chi_i^a \quad \text{for } i = 1,2,3 \quad \text{and} \quad \lambda_4^a \coloneqq \lambda^a 
\end{align}
and the matrices $\alpha^i$ and $\beta^i$ defined as in \eqref{eq:AlphaBeta}. These identifications also imply the maximally extended $\mathcal{N}=1$ version of the fermion propagator \eqref{eq:N4FermProp} and Dirac equation \eqref{eq:N4DiracEq}.

\subsection{Gauge Fixing, Correlation Functions and Ward Identities}
Since the 4-dimensional $\mathcal{N}=4$ super Yang-Mills theory can be obtained from 10-dimensional $\mathcal{N}=1$ super Yang-Mills, it also inherits many properties from $\mathcal{N}=1$ super Yang-Mills. For example, the gauge fixing procedure of $\mathcal{N}=4$ super Yang-Mills is identical to that of $\mathcal{N}=1$ super Yang-Mills. Hence the action \eqref{eq:YM4} is supplemented by the same gauge fixing term as before, \emph{i.e.}
\begin{align}
S^4 = S_\mathrm{inv}^4 + S_\mathrm{gf} 
\end{align}
with
\begin{align}
S_\mathrm{gf} =\int \mathrm{d}^4x \ \left[ \frac{1}{2\xi} (\mathcal{G}^\mu A_\mu^a) (\mathcal{G}^\nu A_\nu^a) + \frac{1}{2} \bar{C}^a \mathcal{G}^\mu (D_\mu C)^a \right] \, .
\end{align}
The combined action is invariant under the BRST variations
\begin{align}
\begin{aligned}\label{eq:YMMaxBRST}
&s A_\mu^a = (D_\mu C)^a \, ,   &&s A_i^a = g f^{abc} A_i^b C^c \, ,  &&s B_i^a = g f^{abc} B_i^b C^c \, , \\
& s \lambda_A^a = - g f^{abc} \lambda_A^b C^c \, ,  && s \, \bar{C}^a = - \frac{1}{\xi} \mathcal{G}^a(\tilde{A}) \, , &&s \, C^a = - \frac{g}{2}  f^{abc} C^b C^c \, .
\end{aligned}
\end{align}
Thus, correlation functions in $\mathcal{N}=4$ super Yang-Mills are defined in the same way as in $\mathcal{N}=1$ super Yang-Mills. Moreover, also the Ward identities \eqref{eq:YMWardId2} and \eqref{eq:YMWardId3} remain unchanged. All definitions for maximally extended $\mathcal{N}=1$ super Yang-Mills arise accordingly. 

\subsection[The \texorpdfstring{$\mathcal{N}=4$}{N=4} Action with Weyl Spinors]{The $\boldsymbol{\mathcal{N}}$\:=\;4 Action with Weyl Spinors}
For some applications, it can be useful to write the $\mathcal{N}=4$ super Yang-Mills action with Weyl spinors. We have summarized our notation and conventions for Weyl spinors in appendix \ref{app:Spinors}. First, we combine the $3+3$ scalar and pseudoscalar fields $A_i^a$ and $B_i^a$ into the six scalar fields $\phi_I^a$ via
\begin{align}\label{eq:N4Intermediate1}
\phi_i^a \coloneqq A_i^a \, , \quad \phi_{i+3}^a \coloneqq B_i^a \, , \quad i =1,2,3 \, .
\end{align}
Moreover, we introduce the six anti-symmetric $4 \times 4$ sigma matrices
\begin{align}
\begin{aligned}\label{eq:N4Intermediate2}
&\Sigma_{AB}^i \coloneqq \alpha_{AB}^i \, , &&\bar{\Sigma}_i^{AB} \coloneqq \alpha_i^{AB} \, &&\Sigma_{AB}^{i+3} \coloneqq -i \beta_{AB}^i \, , &&\bar{\Sigma}_{i+3}^{AB} \coloneqq i \beta_i^{AB} \, , &&i =1,2,3 \, .
\end{aligned}
\end{align}
They satisfy the Clifford algebra $\{ \Sigma^I , \bar{\Sigma}^J \} = - 2 \delta^{IJ}$ with $I, J = 1, \ldots, 6$. We write the Majorana spinors and gamma matrices in the Weyl basis, \emph{i.e.}
\begin{align}
\lambda_A^a = \begin{pmatrix} \psi_\alpha^a \\ \bar{\psi}^{\dot{\alpha}} \end{pmatrix} \, , \quad \gamma^\mu = \begin{pmatrix} 0 & \sigma^\mu \\ \bar{\sigma}^\mu & 0 \end{pmatrix} \, , \quad \gamma^5 = \begin{pmatrix} - \mathbbm{1}_2 & 0 \\ 0 & \mathbbm{1}_2 \end{pmatrix}  \, .
\end{align}
Together with the above definitions \eqref{eq:N4Intermediate1} and \eqref{eq:N4Intermediate2} this implies
\begin{align}
\bar{\lambda}_A^\alpha \gamma^\mu (D_\mu\lambda_A)^a = 2 \, \psi^{\alpha A} \sigma_{\alpha\dot{\alpha}}^\mu (D_\mu \bar{\psi}_A^{\dot{\alpha}})^a 
\end{align}
and
\begin{align}
f^{abc} \bar{\lambda}_A^a \left( \alpha_{AB}^i A_i^b + i \gamma_5 \beta_{AB}^i B_i^b \right) \lambda_B^c = f^{abc} \left( \psi^{a \, A} \Sigma_{AB}^I \phi_I^b \psi^{c \, B} \right) + f^{abc} \left( \bar{\psi}_A^a \bar{\Sigma}_I^{AB} \phi^{b \, I} \bar{\psi}_B^c \right) \, .
\end{align}
In contrast to Majorana spinors, the subscript and superscript placement of the spinor indices $A,B = 1, \ldots, 4$ matters for Weyl spinors. 

Furthermore, recall that the gauge group of $\mathcal{N}=4$ super Yang-Mills is $U(N)$ or $SU(N)$. The generators of the fundamental representation of the associated Lie algebra are denoted by $t^a$. Given an element $u$ of  $\mathfrak{u}(N)$ (respectively $\mathfrak{su}(N)$) the unitary transformation $u t^a u^{-1}$ must be a linear combination of the generators $t^b$, \emph{i.e.}
\begin{align}
u t^a u^{-1} = t^b R^{ba}  \, ,
\end{align}
where $R^{ba}$ is a real $N^2 \times N^2$ (respectively $(N^2-1) \times (N^2-1)$) matrix. It can be shown that the $R^{ba}$ are elements of the adjoint representation of $\mathfrak{u}(N)$ (respectively $\mathfrak{su}(N)$). Thus we can write any field $X^a$ as an element of the gauge Lie algebra $\mathfrak{u}(N)$ (respectively $\mathfrak{su}(N)$) via $X = t^a X^a$. For any gauge Lie algebra element $u$ we have $u X u^{-1} = t^b R^{ba} X^a$. Hence the quantity $X^a = 2 \delta^{ab} \tr_c(t^b X)$ transforms in the adjoint representation of the gauge Lie algebra
\begin{align}
X^a \to 2 \delta^{ab} \tr_c(t^b u X u^{-1}) = 2 \delta^{ab} \tr_c(t^b t^c R^{cd} X^d) = R^{ab} X^b \, .
\end{align}
Since all the fields in $\mathcal{N}=4$ super Yang-Mills are in the adjoint representation of the gauge Lie algebra, we can write them as Lie algebra valued objects $\phi = t^a \phi^a$. In the case of the gauge field $A_\mu = t^a A_\mu^a$ is called the gauge potential. 

Subsequently, the action \eqref{eq:YM4}, written in terms of the Lie algebra valued objects with Weyl spinors and the six scalar fields $\phi_I$, becomes
\begin{align}
\begin{aligned}\label{eq:YM5}
S_\mathrm{inv}^4 =  \int \mathrm{d}^4x \ \tr_c \bigg[ &- \frac{1}{2} F_{\mu\nu} F^{\mu\nu} - (D_\mu \phi_I) (D^\mu \phi^I) + \frac{g^2}{4} [\phi_I, \phi_J] [\phi^I, \phi^J] \\
& -  2i  \, \psi^{\alpha A} \sigma_{\alpha\dot{\alpha}}^\mu (D_\mu \bar{\psi}_A^{\dot{\alpha}}) - ig \, \psi^{\alpha A} [ \Sigma_{AB}^I \phi_I, \psi_\alpha^B] - ig \,   \bar{\psi}_{\dot{\alpha}A} [ \bar{\Sigma}_I^{AB} \phi^I, \bar{\psi}_B^{\dot{\alpha}}] \bigg] .
\end{aligned}
\end{align}
We will introduce the supersymmetry variations and equations of motion for this action in chapter \ref{ch:BPS}.

\chapter{The Wess-Zumino Model}\label{ch:WZ}
In this chapter, we introduce the Nicolai map for the 2-dimensional Wess-Zumino model. Most of the content presented here has been developed by Nicolai in a series of papers \cite{Nicolai:1979nr, Nicolai:1980hh,Nicolai:1980jc, Nicolai:1980js} with a comprehensive summary given in \cite{Nicolai:1984jg}. However, we will also expand the previously existing results by providing an explicit expression for the Nicolai map up to and including the fifth order in the coupling constant.

In the first section, we recall the 2-dimensional Wess-Zumino model and state the main theorem capturing the map's properties. Furthermore, we explain the practical application of the theorem. In section \ref{sec:WZROp} we derive the infinitesimal generator of the inverse Nicolai map, called the $\mathcal{R}_\lambda$-operator and compute the inverse Nicolai map to second order. Moreover, we briefly introduce an alternative approach to computing the Nicolai map developed by Lechtenfeld and Rupprecht. In section \ref{sec:WZProof} we prove the main theorem. Section \ref{sec:WZResult} contains the Nicolai map and a discussion of the result. Finally, in the last section, we test the Nicolai map. 
\section{Introduction and Main Theorem}\label{sec:WZIntroduction}
In this chapter, we work in 2-dimensional Euclidean space. Recall the action of the `on-shell' 2-dimensional Wess-Zumino model \eqref{eq:WZ2}
\begin{align}\label{eq:WZAction1}
S_\mathrm{wz} &= \frac{1}{2} \int \mathrm{d}^2x \, \left[ (\partial_\mu A )( \partial^\mu A) + ( m A + \lambda A^3 )^2 + (\bar{\psi} \gamma^\mu \partial_\mu \psi) + ( m + 3 \lambda A^2) (\bar{\psi} \psi) \right] \, ,
\end{align}
where $m$ is the mass and $\lambda$ is the coupling constant. The action is invariant under the supersymmetry variations \eqref{eq:WZsusy2}
\begin{align}\label{eq:WZsusyVar}
\delta A = (\bar{\psi} \varepsilon) \, , \quad \delta \psi_\alpha = (\gamma^\mu \varepsilon)_\alpha \partial_\mu A - \varepsilon_\alpha (mA + \lambda A^3) \, .
\end{align}
The central result is summarized in the following theorem \cite{Nicolai:1980jc,Nicolai:1984jg}.
\begin{theorem}\label{th:WZTheorem}
The 2-dimensional Wess-Zumino model is characterized by the existence of a non-linear and non-local transformation $\mathcal{T}_\lambda$ of the bosonic field
\begin{align*}
\mathcal{T}_\lambda: A(x) \mapsto A^\prime(x,m,\lambda;A) \, , 
\end{align*}
which is invertible, at least in the sense of a formal power series such that
\begin{enumerate}
\item The bosonic Wess-Zumino action is mapped to the abelian action, 
\begin{align*}
S_\mathrm{wz}[\lambda;A] = S_\mathrm{wz}[0;\mathcal{T}_\lambda A] \, , 
\end{align*}
where $S_\mathrm{wz}[\lambda;A]$ is the bosonic part of the action $S_\mathrm{wz}[\lambda;A,\psi]$ and $S_\mathrm{wz}[0;A]$ denotes the free bosonic action, i.e. $S_\mathrm{wz}[\lambda;A]$ at $\lambda = 0$. 
\item The Jacobian determinant of $\mathcal{T}_\lambda$ equals the Matthews-Salam-Seiler determinant, i.e.
\begin{align*}
\mathcal{J}(\mathcal{T}_\lambda A) = \Delta_\text{MSS}[\lambda;A]
\end{align*}
at least order by order in perturbation theory.
\end{enumerate}
\end{theorem}
In the following sections, we give a constructive proof of the main theorem by deriving the Nicolai map $\mathcal{T}_\lambda$ and checking that it satisfies \textit{i)} and \textit{ii)}.

However, first, let us investigate the implication of the theorem. Assume that we have indeed found the map $\mathcal{T}_\lambda$ and its inverse $\mathcal{T}_\lambda^{-1}$. We will later see that the inverse map acts on bosonic monomials $X[A]$ by $\mathcal{T}_\lambda^{-1} X[A] = X[\mathcal{T}_\lambda^{-1} A]$. Computing the free bosonic vacuum expectation value of such a monomial we find
\begin{align}
\big< X[\mathcal{T}_\lambda^{-1} A] \big>_0 &= \int \mathcal{D}_0A \ e^{- S_\mathrm{wz}[0;A]} \ X[\mathcal{T}_\lambda^{-1} A] = \int \mathcal{D}_0A \ \mathcal{J}(\mathcal{T}_\lambda A) \ e^{- S_\mathrm{wz}[0;\mathcal{T}_\lambda A]} \ X[A] \, .
\end{align}
In the second step, we have performed a change of variables from $A$ to $\mathcal{T}_\lambda A$. On the other hand, integrating out the fermionic degrees of freedom in the interacting vacuum expectation value of $X[A]$ yields
\begin{align}
\big< \!\! \big< X[A] \big> \!\!\big>_\lambda = \int \mathcal{D}A \ \mathcal{D}\psi \ e^{- S_\mathrm{wz}[\lambda;A,\psi]} \ X[A] = \int \mathcal{D}_0A \ \Delta_\mathrm{MSS}[\lambda;A] \ e^{- S_\mathrm{wz}[\lambda;A]} \ X[A] \, .
\end{align}
If we now assume that $\mathcal{T}_\lambda$ satisfies \textit{i)} and \textit{ii)} from the main theorem we obtain
\begin{align}\label{eq:WZNMCorr}
\big< \!\! \big< X[A] \big> \!\!\big>_\lambda = \big< X[\mathcal{T}_\lambda^{-1} A] \big>_0 \, .
\end{align}
Notice that this transformation does not render the vacuum expectation value trivial. The complexity is hidden in the perturbative expansion of the non-linear and non-local transformation $\mathcal{T}_\lambda^{-1}$. Using the linearity of the correlation function $\left<\!\!\left< \ldots \right>\!\!\right>_\lambda$ and $\mathcal{T}_\lambda^{-1} X[A] = X[\mathcal{T}_\lambda^{-1} A]$ we can extend \eqref{eq:WZNMCorr} to n-point correlators of bosonic operators $\mathcal{O}_i(x_i)$, \emph{i.e.}
\begin{align}\label{eq:WZNicolaiCorr}
\big<\!\!\big< \mathcal{O}_1(x_1) \ldots \mathcal{O}_n(x_n) \big>\!\!\big>_\lambda = \big< (\mathcal{T}_\lambda^{-1} \mathcal{O}_1)(x_1) \ldots (\mathcal{T}_\lambda^{-1} \mathcal{O}_n)(x_n) \big>_0 \, .
\end{align}
So instead of computing interacting $n$-point correlation functions of the Wess-Zumino model (with fermions), we can simply compute the free correlation function of the transformed operators. After working out the transformations $(\mathcal{T}_\lambda^{-1} \mathcal{O})(x)$ to the desired order in the coupling, we use Wick’s theorem to obtain the free correlator. This property of the Nicolai map was first discovered by Dietz and Lechtenfeld in \cite{Lechtenfeld:1984me,Dietz:1984hf,Dietz:1985hga}.

\section[The \texorpdfstring{$\mathcal{R}_\lambda$}{R}-Operator]{The $\boldsymbol{\mathcal{R}_\lambda}$-Operator}\label{sec:WZROp}
The inverse Nicolai map $\mathcal{T}_\lambda^{-1}$ is obtained from its infinitesimal generator $\mathcal{R}_\lambda$, a non-local functional differential operator, \emph{i.e.}
\begin{align}\label{eq:WZInverseNM}
(\mathcal{T}_\lambda^{-1} A)(x) \coloneqq \sum_{n=0}^\infty \frac{\lambda^n}{n!} (\mathcal{R}_\lambda^n A)(x) \big\vert_{\lambda = 0} \, .
\end{align}
The actual map $(\mathcal{T}_\lambda A)(x)$ is then obtained order by order in $\lambda$ by formally inverting the power series. For an arbitrary monomial of bosonic operators $X[A]$, the linear response of its vacuum expectation value to a change in the coupling constant is given by
\begin{align}
 \frac{\mathrm{d}}{\mathrm{d}\lambda} \left<\!\!\left< X[A] \right>\!\!\right>_\lambda = \left<\!\!\!\!\left< \frac{\mathrm{d} X[A]}{\mathrm{d}\lambda} \right>\!\!\!\!\right>_\lambda - \left<\!\!\!\!\left< \frac{\mathrm{d} S_\mathrm{wz}[\lambda;A,\psi]}{\mathrm{d}\lambda} \, X[A] \right>\!\!\!\!\right>_\lambda \eqqcolon \left< \mathcal{R}_\lambda \, X[A] \right>_\lambda \, . 
\end{align}
At this point, it is crucial that the correlation function is automatically properly normalized due to supersymmetry, and we do not have to divide it by a $\lambda$ dependent constant. If this were not the case, it would spoil the simple form of the $\lambda$ derivative above. Making use of supersymmetry, we want to rewrite the right-hand side in terms of a derivational operator $\mathcal{R}_\lambda$. Luckily even for the `on-shell' version of the Wess-Zumino model, the action still is the top component of a supersymmetry multiplet, which does not rely on the equations of motion to close. Thus we obtain
\begin{align}\label{eq:WZdSdl}
\frac{\mathrm{d}S_\mathrm{wz}[\lambda;A,\psi]}{\mathrm{d}\lambda} = - \delta_\alpha \Delta_\alpha \quad \text{with} \quad \Delta_\alpha \coloneqq \frac{1}{2} \int \mathrm{d}^2x \ \psi_\alpha(x) A^3(x) \, .
\end{align}
Recall that $\delta_\alpha$ anti-commutes with other anti-commuting operators. Using the supersymmetry Ward identity \eqref{eq:WZWardId}
\begin{align}
\left<\!\!\left< \delta_\alpha Y \right>\!\!\right>_\lambda = 0
\end{align}
for an arbitrary string of operators $Y$, we then obtain
\begin{align}
\left<\!\!\!\!\left< \frac{\mathrm{d} S_\mathrm{wz}[\lambda;A,\psi]}{\mathrm{d}\lambda} \, X[A] \right>\!\!\!\!\right>_\lambda = - \left<\!\!\left< \delta_\alpha \Delta_\alpha \, X[A] \right>\!\!\right>_\lambda = - \left<\!\!\left< \Delta_\alpha \, \delta_\alpha X[A] \right>\!\!\right>_\lambda \, .
\end{align}
Subsequently, we integrate out the fermionic degrees of freedom and obtain the $\mathcal{R}_\lambda$-operator
\begin{align}\label{eq:WZRop1}
\mathcal{R}_\lambda \, X[A] = \frac{\mathrm{d} X[A]}{\mathrm{d}\lambda} + \bcontraction{}{\Delta}{_\alpha \, }{\delta} \Delta_\alpha \, \delta_\alpha X[A] \, , 
\end{align}
which is now a purely bosonic expression. The $\mathcal{R}_\lambda$-operator acts distributively
\begin{align}
\mathcal{R}_\lambda(XY) = \mathcal{R}_\lambda(X) Y + X \mathcal{R}_\lambda(Y) \, .
\end{align}
Plugging in the definition for $\Delta_\alpha$ and the supersymmetry variation for $A(x)$ we arrive at
\begin{align}\label{eq:WZRop2}
\mathcal{R}_\lambda = \frac{\mathrm{d}}{\mathrm{d}\lambda} - \frac{1}{2} \int \mathrm{d}^2x \ \mathrm{d}^2y \ \bcontraction{}{\psi}{_\alpha(x) }{\bar{\psi}} \psi_\alpha(x) \bar{\psi}_\alpha(y) \, A^3(y) \, \frac{\delta}{\delta A(x)} \, .
\end{align}
We recognize the fermionic propagator $\bcontraction{}{\psi}{(x) }{\bar{\psi}} \psi(x) \bar{\psi}(y) \equiv S(x,y;A)$. From its definition \eqref{eq:WZFermionProp}
\begin{align}
\big[ \gamma^\mu \partial_\mu + m + 3 \lambda A^2(x) \big]_{\alpha\gamma} 
\bcontraction{}{\psi}{_\gamma(x) }{\bar{\psi}}
\psi_\gamma(x) \bar{\psi}_\beta(y) = \delta_{\alpha\beta} \delta(x-y) 
\end{align}
we obtain the identities
\begin{align}
\frac{\mathrm{d}S(x,y;A)}{\mathrm{d}\lambda} = -3 \, \int \mathrm{d}^2z \ S(x,z;A) A^2(z) S(z,y;A)
\end{align}
and
\begin{align}
\frac{\delta S(x,y;A)}{\delta A(z)} = - 6 \lambda \, \int \mathrm{d}^2z \ S(x,z;A) A(z) S(z,y;A) \, .
\end{align}
Subsequently the $\mathcal{R}_\lambda$-operator \eqref{eq:WZRop2} becomes
\begin{align}\label{eq:WZRop3}
\mathcal{R}_\lambda = \frac{\mathrm{d}}{\mathrm{d}\lambda} - \frac{1}{2} \int \mathrm{d}^2x \ \mathrm{d}^2y \ \tr [ S(x,y;A) ] \, A^3(y) \, \frac{\delta}{\delta A(x)} \, .
\end{align}
In the limit $\lambda = 0$ the propagator $S(x,y;A)$ gets replaced by the free fermion propagator \eqref{eq:WZFreeFermion}
\begin{align}
S_0(x-y) = (- \slashed{\partial} + m) C(x-y) \, .
\end{align}
After iteratively computing $(\mathcal{R}_\lambda^n A)(x)$ to any desired order $n$, we set $\lambda = 0$ and obtain $(\mathcal{T}_\lambda^{-1}A)(x)$ at $\mathcal{O}(\lambda^n)$. The actual map $\mathcal{T}_\lambda$ is obtained from its inverse by formal power series inversion. Let
\begin{align}\label{eq:WZNicolaiMapDef}
(\mathcal{T}_\lambda A) = \sum_{n=0}^\infty \frac{\lambda^n}{n!} (T_n A) \, .
\end{align}
Expanding $\mathcal{T}_\lambda^{-1} \mathcal{T}_\lambda = \mathrm{id}$ in powers of $\lambda$ and matching coefficients we readily obtain
\begin{align}\label{eq:WZNicolaiMapExp}
\begin{aligned}
(T_0 A) &= A \, , \\
(T_n A) &= - \sum_{i=0}^{n-1} \genfrac(){0pt}{}{n}{i} \mathcal{R}_\lambda^n (T_i A) \Big\vert_{\lambda = 0} \, .
\end{aligned}
\end{align}

\subsection{The Inverse Nicolai Map}
To gain a better understanding of the $\mathcal{R}_\lambda$-operator we compute $(\mathcal{T}_\lambda^{-1}A)(x)$ up to order $\lambda^2$. Hence we have to act twice with $\mathcal{R}_\lambda$ on $A(x)$. The first application of the $\mathcal{R}_\lambda$-operator yields
\begin{align}
(\mathcal{R}_\lambda A)(x) = - \frac{1}{2} \int \mathrm{d}^2y \ \tr[ S(x,y;A) ] A^3(y) \, .
\end{align}
When acting again, we must remember that $S(x,y;A)$ depends on both $\lambda$ and $A(x)$, thus, it contributes two terms to the second order. A third term comes from the action of $\mathcal{R}_\lambda$ on $A^3(y)$
\begin{align}
\begin{aligned}
(\mathcal{R}_\lambda^2A)(x) &= \frac{3}{2} \int \mathrm{d}^2y \ \tr[ S(x,z;A) A^2(z) S(z,y;A)] A^3(y) \\
&\quad - \frac{3 \lambda}{2} \int \mathrm{d}^2y \ \mathrm{d}^2z \ \mathrm{d}^2w \ \tr[ S(x,z;A) A^2(z) S(z,y;A)] A^3(y) \tr [ S(y,w;A) ] A^3(w) \\
&\quad + \frac{3}{4} \int \mathrm{d}^2y \ \mathrm{d}^2z \ \tr[ S(x,y;A) ] A^2(y) \tr [S(y,z;A)] A^3(z) \, .
\end{aligned}
\end{align}
We set $\lambda = 0$
\begin{align}
\begin{aligned}
&(\mathcal{R}_\lambda A)(x) \Big\vert_{\lambda = 0} = - \frac{1}{2} \int \mathrm{d}^2y \ \tr[ (m - \gamma^\mu \partial_\mu) C(x-y) ] A^3(y) \, , \\
&\begin{aligned}
(\mathcal{R}_\lambda^2A)(x) \Big\vert_{\lambda = 0} &= \frac{3}{2} \int \mathrm{d}^2y \ \tr[ (m - \gamma^\mu \partial_\mu) C(x-z) A^2(z) (m - \gamma^\nu \partial_\nu) C(z-y) ] A^3(y) \\
&\quad + \frac{3}{4} \int \mathrm{d}^2y \ \mathrm{d}^2z \ \tr[ (m - \gamma^\mu \partial_\mu) C(x-y) ] \\
&\quad \quad \quad \quad \quad \quad \quad \quad \times A^2(y) \tr [ (m - \gamma^\nu \partial_\nu) C(y-z) ] A^3(z) \, .
\end{aligned}
\end{aligned}
\end{align}
Computing the traces (with $\tr \, \mathbf{1} = 2$) and simplifying the results we subsequently obtain
\begin{align}
\begin{aligned}
(\mathcal{T}_\lambda^{-1} A)(x) &= A(x) - m \lambda \int \mathrm{d}^2y \ C(x-y) A^3(y) \\
&\quad + 3 \, m^2 \lambda^2 \int \mathrm{d}^2y \ \mathrm{d}^2z \ C(x-y) A^2(y) C(y-z) A^3(z) \\
&\quad + \frac{3 \lambda^2}{2} \int \mathrm{d}^2y \ \mathrm{d}^2z \ \partial_\mu C(x-y) A^2(y) \partial^\mu C(y-z) A^3(z) \, .
\end{aligned}
\end{align}
This result can now be inverted using the steps explained above. The explicit expression for $(\mathcal{T}_\lambda A)(x)$ up to $\mathcal{O}(\lambda^5)$ is presented in section \ref{sec:WZResult}. 

\subsection{An Alternative Construction of the Nicolai Map}
To close this section, let us remark that there exists an alternative but equivalent construction of the Nicolai map developed by Lechtenfeld and Rupprecht in \cite{Lechtenfeld:2021uvs}. It is based on earlier work of Lechtenfeld (see for example \cite{Lechtenfeld:1984me}). In this construction, one does not need the main theorem \ref{th:WZTheorem} but rather starts with the equation
\begin{align}\label{eq:WZLechtnefeld1}
\big< \!\! \big< X[A] \big> \!\! \big>_\lambda = \big< X[\mathcal{T}_\lambda^{-1} A] \big>_0 
\end{align}
as a defining property of the inverse Nicolai map. Differentiating \eqref{eq:WZLechtnefeld1} with respect to $\lambda$ yields
\begin{align}\label{eq:WZLechtnefeld2}
\frac{\mathrm{d}}{\mathrm{d}\lambda} \big<\!\!\big< X[A] \big>\!\!\big>_\lambda = \left<\!\!\!\!\left< \left( \frac{\mathrm{d}}{\mathrm{d}\lambda} - \frac{\mathrm{d} S_\mathrm{wz}[\lambda;A,\psi]}{\mathrm{d}\lambda} \right) X[A] \right>\!\!\!\!\right>_\lambda = \left< \left( \frac{\mathrm{d}}{\mathrm{d}\lambda} + \check{\mathcal{R}}_\lambda \right) X[A] \right>_\lambda
\end{align}
with the functional differential operator
\begin{align}
\check{\mathcal{R}}_\lambda = \int \mathrm{d}^2x \ \left( \partial_\lambda \mathcal{T}_\lambda^{-1} \circ \mathcal{T}_\lambda \right) A(x) \ \frac{\delta}{\delta A(x)} \, .
\end{align}
The $\check{\mathcal{R}}_\lambda$-operator differs from the $\mathcal{R}_\lambda$-operator introduced above by the $\lambda$ derivative, \emph{i.e.} $\mathcal{R}_\lambda = \frac{\mathrm{d}}{\mathrm{d}\lambda} + \check{\mathcal{R}}_\lambda$. By setting $X[A] = \mathcal{T}_\lambda \, A$ in \eqref{eq:WZLechtnefeld1} we derive the relation
\begin{align}
\left( \frac{\mathrm{d}}{\mathrm{d}\lambda} + \check{\mathcal{R}}_\lambda \right) \mathcal{T}_\lambda \, A = 0 \, .
\end{align}
The solution to this equation is the path-ordered exponential
\begin{align}
(\mathcal{T}_\lambda \, A)(x) = \overset{\rightarrow}{\mathcal{P}} \exp \left[ - \int_0^\lambda \mathrm{d}h \ \check{\mathcal{R}}_h \right] A(x) \, .
\end{align}
The inverse Nicolai map is given by
\begin{align}
(\mathcal{T}_\lambda^{-1} \, A)(x) = \overset{\leftarrow}{\mathcal{P}} \exp \left[ \int_0^\lambda \mathrm{d}h \ \check{\mathcal{R}}_h \right] A(x) \, .
\end{align}
In the end, one finds from \eqref{eq:WZLechtnefeld2}
\begin{align}
\check{\mathcal{R}}_\lambda = - \frac{1}{2} \int \mathrm{d}^2x \ \mathrm{d}^2y \ \tr [ S(x,y;A) ] \, A^3(y) \, \frac{\delta}{\delta A(x)} \, .
\end{align}
Compared to \eqref{eq:WZInverseNM} the $\lambda$ derivatives have now been traded for integrals. However, this construction allows us to obtain the Nicolai map directly without passing through its inverse first. Furthermore, this construction not only works for scalar theories, such as the Wess-Zumino model but also for gauge theories. For more details, see \cite{Lechtenfeld:2021uvs}. 

Finally, Lechtenfeld and Rupprecht have also developed a graphical representation of the Nicolai map, see section 4 of \cite{Ananth:2020lup} and \cite{Lechtenfeld:2022qpa}.

\section{Proof of the Main Theorem}\label{sec:WZProof}
In this section, we prove the main theorem \ref{th:WZTheorem}. Large parts of the proof are universal to all the supersymmetric field theories discussed in this thesis and will thus not be repeated in the subsequent chapters. The existence, non-locality and non-linearity, as well as the inversion property of $\mathcal{T}_\lambda$ are all given by the explicit construction in the previous section. Hence we only need to show that $\mathcal{T}_\lambda$ has the properties \textit{i)} and \textit{ii)}. 
\subsection{Part i)}
For part \textit{i)} we need to show that $S_\mathrm{wz}[\lambda;A] = S_\mathrm{wz}[0;\mathcal{T}_\lambda A]$. This is equivalent to showing that $S_\mathrm{wz}[\lambda;\mathcal{T}_\lambda^{-1} A] = S_\mathrm{wz}[0;A]$. Thus we write
\begin{align}\label{eq:WZProof1}
S_\mathrm{wz}[\lambda;A] = S_0[A] + \lambda S_1[A] + \lambda^2 S_2[A]
\end{align}
and Taylor expand $S_\mathrm{wz}[\lambda;\mathcal{T}_\lambda^{-1} A]$ around $\lambda = 0$
\begin{align}
\begin{aligned}\label{eq:WZProof2}
S_\mathrm{wz}[\lambda;\mathcal{T}_\lambda^{-1} A]&= 
S_\mathrm{wz}[\lambda;\mathcal{T}_\lambda^{-1} A] \Big\vert_{\lambda=0} 
+ \sum_{n=1}^\infty \frac{\lambda^n}{n!} \left[ \frac{\mathrm{d}^n}{\mathrm{d}\lambda^n} S_\mathrm{wz}[\lambda;\mathcal{T}_\lambda^{-1} A] \Big\vert_{\lambda=0} \right] \\
&= S_0[A] 
+ \sum_{n=1}^\infty \frac{\lambda^n}{n!} \left[ \frac{\mathrm{d}^n}{\mathrm{d}\lambda^n} S_0[ \mathcal{T}_\lambda^{-1} A] \Big\vert_{\lambda=0} \right] \\
&\quad + \sum_{n=2}^\infty \frac{\lambda^n}{n!} \left[ \frac{\mathrm{d}^{n-1}}{\mathrm{d}\lambda^{n-1}} S_1[ \mathcal{T}_\lambda^{-1} A] \Big\vert_{\lambda=0} \right]
+ \sum_{n=3}^\infty \frac{\lambda^n}{n!} \left[ \frac{\mathrm{d}^{n-2}}{\mathrm{d}\lambda^{n-2}} S_2[ \mathcal{T}_\lambda^{-1} A] \Big\vert_{\lambda=0} \right] \, .
\end{aligned}
\end{align}
Inserting the definition of $\mathcal{T}_\lambda^{-1}$ into the second term yields
\begin{align}
\begin{aligned}\label{eq:WZProof3}
S_0[ \mathcal{T}_\lambda^{-1} A] &= \int \mathrm{d}^2x \ \left( \sum_{n=0}^\infty \frac{\lambda^n}{n!} \left[ \partial_\mu (\mathcal{R}^n A)(x) \Big\vert_{\lambda = 0} \right] \right) \left( \sum_{n=0}^\infty \frac{\lambda^n}{n!} \left[ \partial^\mu (\mathcal{R}^n A)(x) \Big\vert_{\lambda = 0} \right] \right) \\
&\quad + m \sum_{n=0}^\infty \frac{\lambda^n}{n!} \left[ (\mathcal{R}^n A)(x) \Big\vert_{\lambda = 0} \right] \, .
\end{aligned}
\end{align}
Taking the derivative with respect to $\lambda$ gives $\frac{\mathrm{d}^n}{\mathrm{d}\lambda^n} S_0[\mathcal{T}_\lambda^{-1} A] \vert_{\lambda=0} = \mathcal{R}_\lambda^n (S_0[A])\vert_{\lambda=0}$ and similar relations for the other two terms. Subsequently we conclude
\begin{align}
\begin{aligned}\label{eq:WZProof4}
S_\mathrm{wz}[\lambda;\mathcal{T}_\lambda^{-1} A] &= S_0[A] 
+ \sum_{n=1}^\infty \frac{\lambda^n}{n!} \left[ \mathcal{R}_\lambda^n (S_0[A]) + \mathcal{R}_\lambda^n (\lambda S_1[A]) + \mathcal{R}_\lambda^n (\lambda^2 S_2[A]) \right] \Big\vert_{\lambda=0} \\
&= S_0[A] 
+ \sum_{n=1}^\infty \frac{\lambda^n}{n!} \mathcal{R}_\lambda^n (S_\mathrm{wz}[\lambda;A]) \Big\vert_{\lambda=0} \, .
\end{aligned}
\end{align}
It remains to be shown that $\mathcal{R}_\lambda (S_\mathrm{wz}[\lambda;A])=0$. This requires the explicit form of the $\mathcal{R}_\lambda$-operator given in \eqref{eq:WZRop1}. We find
\begin{align}
\begin{aligned}
\mathcal{R}_\lambda (S_\mathrm{wz}[\lambda;A]) &= \frac{\mathrm{d}}{\mathrm{d}\lambda} \left( \frac{1}{2} \int \mathrm{d}^2x \, \left[ (\partial_\mu A )( \partial^\mu A) + ( m A + \lambda A^3 )^2 \right] \right) \\
&\quad - \frac{1}{4} \int \mathrm{d}^2x \ \mathrm{d}^2y \ \mathrm{d}^2z \ \tr[S(x,y;A)] A^3(y) \, \frac{\delta (\partial_\mu A(z) )( \partial^\mu A(z)) }{\delta A(x)} \\
&\quad - \frac{1}{4} \int \mathrm{d}^2x \ \mathrm{d}^2y \ \mathrm{d}^2z \ \tr[S(x,y;A)] A^3(y) \, \frac{\delta ( m A(z) + \lambda A^3(z) )^2 }{\delta A(x)} \\
&= \int \mathrm{d}^2x \ \left[ m A^4(x) + \lambda A^6(x) \right] \\
&\quad + \frac{1}{2} \int \mathrm{d}^2x \ \mathrm{d}^2y \ \Box A(x) \, \tr[S(x,y;A)] A^3(y) \\
&\quad - \frac{1}{2} \int \mathrm{d}^2x \ \mathrm{d}^2y \ ( m A(x) + \lambda A^3(x) ) ( m + 3 \lambda A^2(x) ) \, \tr[S(x,y;A)] A^3(y) \, , 
\end{aligned}
\end{align}
with the Laplacian $\Box = \partial_\mu \partial^\mu$. In the third term, we use the Dirac equation \eqref{eq:WZFermionProp}
\begin{align}
 ( m + 3 \lambda A^2(x) ) \, \tr[S(x,y;A)] = - \tr[ \slashed{\partial} S(x,y;A)] + 2 \delta(x-y) 
\end{align}
such that
\begin{align}
\begin{aligned}
\mathcal{R}_\lambda (S_\mathrm{wz}[\lambda;A]) &= \color{myblue} \int \mathrm{d}^2x \ \left[ m A^4(x) + \lambda A^6(x) \right] \\
&\quad + \frac{1}{2} \int \mathrm{d}^2x \ \mathrm{d}^2y \ \Box A(x) \, \tr[S(x,y;A)] A^3(y) \\
&\quad + \frac{1}{2} \int \mathrm{d}^2x \ \mathrm{d}^2y \ ( m A(x) + \lambda A^3(x) ) \, \tr[\slashed{\partial} S(x,y;A)] A^3(y) \\
&\quad \color{myblue} - \int \mathrm{d}^2x \ \left[ m A^4(x) + \lambda A^6(x) \right] \color{black} \, .
\end{aligned}
\end{align}
The two {\color{myblue}blue} terms cancel. In the two remaining terms, we use integration by parts and the relation $\Box = \partial_\mu \partial^\mu = \slashed{\partial} \slashed{\partial}$
\begin{align}
\begin{aligned}
\mathcal{R}_\lambda (S_\mathrm{wz}[\lambda;A]) &= - \frac{1}{2} \int \mathrm{d}^2x \ \mathrm{d}^2y \ \tr \left[ \slashed{\partial} A(x) \slashed{\partial} S(x,y;A) \right] A^3(y) \\
&\quad - \frac{1}{2} \int \mathrm{d}^2x \ \mathrm{d}^2y \ \tr \left[ \slashed{\partial} ( m A(x) + \lambda A^3(x) ) \, S(x,y;A) \right] A^3(y) \\
&= - \frac{1}{2} \int \mathrm{d}^2x \ \mathrm{d}^2y \ \tr \left[ \slashed{\partial} A(x) (\slashed{\partial} + m + 3 \lambda A^2(x) ) \, S(x,y;A) \right] A^3(y) \\
&= - \frac{1}{2} \int \mathrm{d}^2x \ \mathrm{d}^2y \ \tr \left[ \slashed{\partial} A(x) \delta(x-y) \right] A^3(y) \\
&= 0 \, .
\end{aligned}
\end{align}
In the second step, we have again used \eqref{eq:WZFermionProp}. The last equality follows from $\tr(\gamma^\mu) = 0$. Thus we have proven that $S_\mathrm{wz}[\lambda; \mathcal{T}_\lambda^{-1} A] = S_\mathrm{wz}[0;A]$. 

Equations \eqref{eq:WZProof1} - \eqref{eq:WZProof4} are universal to all supersymmetric field theories, which are at most quadratic in the coupling. Only the proof of $\mathcal{R}_\lambda (S_\mathrm{wz}[\lambda;A])=0$ was specific to the Wess-Zumino model and has to be repeated for other field theories.

\subsection{Part ii)}
For the proof of part \textit{ii)} consider the Taylor expansion
\begin{align}
\left< \!\! \left< X[A] \right> \!\! \right> _\lambda &= \sum_{n=0}^\infty \frac{\lambda^n}{n!} \left[ \frac{\mathrm{d}^n}{\mathrm{d}\lambda^n} \left< \!\! \left< X[A] \right> \!\! \right>_\lambda \Big\vert_{\lambda = 0} \right] \, .
\end{align}
Inserting the definition of the $\mathcal{R}_\lambda$-operator and the expectation value yields
\begin{align}
\begin{aligned}
\left< \!\! \left< X[A] \right> \!\! \right>_\lambda &= \sum_{n=0}^\infty \frac{\lambda^n}{n!} \left[ \left< \mathcal{R}^n X[A] \right>_\lambda \Big\vert_{\lambda = 0} \right] = \int \mathcal{D}_0A \ e^{-S_\mathrm{wz}[0;A] } \sum_{n=0}^\infty \frac{\lambda^n}{n!} \left[ \mathcal{R}^n X[A] \Big\vert_{\lambda = 0} \right] \\
&= \int \mathcal{D}_0A \ e^{-S_\mathrm{wz}[0;A] } X[ \mathcal{T}_\lambda^{-1} A] = \int \mathcal{D}_0A \ \mathcal{J}(\mathcal{T}_\lambda A)\, e^{-S_\mathrm{wz}[0;\mathcal{T}_\lambda A]} X[A] \, .
\end{aligned}
\end{align}
In the second step, we have used $\Delta_\mathrm{MSS}[\lambda=0;A] = 1$. On the other hand, integrating out the fermionic degrees of freedom in the correlation function yields the Matthews-Salam-Seiler determinant, \emph{i.e.}
\begin{align}
\begin{aligned}
\left< \!\! \left< X[A] \right> \!\! \right>_\lambda &= \int \mathcal{D}A \ \mathcal{D}\psi \ e^{-S_\mathrm{wz}[\lambda;A,\psi] } X[A] = \int \mathcal{D}_0 A \ \Delta_\mathrm{MSS}[\lambda;A] \, e^{-S_\mathrm{wz}[\lambda;A]} X[A] \, .
\end{aligned}
\end{align}
From part \textit{i)} we know that $S_\mathrm{wz}[0;\mathcal{T}_\lambda A] = S_\mathrm{wz}[\lambda;A]$ and since these two equations hold for all bosonic monomials $X[A]$ we can deduce that
\begin{align}
\mathcal{J}(\mathcal{T}_\lambda A) = \Delta_\mathrm{MSS}[\lambda;A] 
\end{align}
at least order by order in perturbation theory. These steps are universal to any scalar supersymmetric field theory. However, for gauge theories, we will obtain an additional term from the Faddeev-Popov determinant. This concludes the proof. 

\section{Result and Discussion}\label{sec:WZResult}
We present the explicit formula for the Nicolai map up to $\mathcal{O}(\lambda^5)$ extending the previously existing result by three orders. The first two lines correspond to the result obtained in \cite{Nicolai:1984jg}\footnote{A different result up to $\mathcal{O}(\lambda^3)$ obtained in \cite{Nicolai:1979nr} via trial and error hints towards an ambiguity in the map $\mathcal{T}_\lambda$. However, we do not expect this result to hold up in higher-order calculations. A similar case for $\mathcal{N}=1$ super Yang-Mills in six dimensions is discussed in section \ref{sec:YM2Ambiguity}.}. In the following section, we will verify that this result satisfies both statements of the main theorem \ref{th:WZTheorem}, providing a highly non-trivial test. The map reads\footnote{As usual, all anti-symmetrizations are with strength one, such that \emph{e.g.} $[ab] = \frac{1}{2!}(ab -ba)$.}
\begin{footnotesize}
\begin{align}\label{eq:WZResult}
\begin{aligned}
(\mathcal{T}_\lambda A)(x) &= A(x) + m \lambda \int \mathrm{d}^2y \ C(x-y) A^3(y) \\
&\quad - \frac{3\lambda^2}{2} \int \mathrm{d}^2 y \ \mathrm{d}^2z \ \partial_\mu C(x-y) A^2(y) \, \partial^\mu C(y-z) A^3(z) \\
&\quad + \frac{m \lambda^3}{2} \int \mathrm{d}^2y \ \mathrm{d}^2z \ \mathrm{d}^2w \ \partial_\mu C(x-y) \ \Big\{ \\
&\quad \quad \quad - 2 \, A(y) \partial^\mu C(y-z) A^3(z) \, C(y-w) A^3(w) \\
&\quad \quad \quad + 3 \, A^2(y) \partial^\mu C(y-z) A^2(z) C(z-w) \, A^3(w) \\
&\quad \quad \quad + 6 \, A^2(y) C(y-z) \, A^2(z) \partial^\mu C(z-w) A^3(w) \Big\} \\
&\quad - \frac{m^2 \lambda^4}{4} \int \mathrm{d}^2y \ \mathrm{d}^2z \ \mathrm{d}^2w \ \mathrm{d}^2v \ \partial_\mu C(x-y) \\ 
&\quad \quad \quad \times \partial^\mu C(y-z) A^3(z) C(y-w) A^3(w) C(y-v) A^3(v) \\
& \quad + \frac{3 \, m^2 \lambda^4}{2} \int \mathrm{d}^2y \ \mathrm{d}^2z \ \mathrm{d}^2w \ \mathrm{d}^2v \ \partial_\mu C(x-y) A(y) \ \Big\{ \\
&\quad \quad \quad + \partial^\mu C(y-z) A^3(z) C(y-w) A^2(w) C(w-v) A^3(v) \\
&\quad \quad \quad + C(y-z) A^3(z) C(y-w) A^2(w) \partial^\mu C(w-v) A^3(v) \Big\} \\
&\quad + \frac{3 \, m^2 \lambda^4}{4} \int \mathrm{d}^2y \ \mathrm{d}^2z \ \mathrm{d}^2w \ \mathrm{d}^2v \ \partial_\mu C(x-y) A^2(y) \ \Big\{ \\
&\quad \quad \quad + \partial^\mu C(y-z) A(z) C(z-w) A^3(w) C(z-v) A^3(v) \\
&\quad \quad \quad + 2 \, C(y-z) A(z) \partial^\mu C(z-w) A^3(w) C(z-v) A^3(v) \\
&\quad \quad \quad + 3 \, \partial^\mu C(y-z) A^2(z) C(z-w) A^2(w) C(w-v) A^3(v) \\
&\quad \quad \quad - 6 \, C(y-z) A^2(z) \partial^\mu C(z-w) A^2(w) C(w-v) A^3(v) \\
&\quad \quad \quad - 9 \, C(y-z) A^2(z) C(z-w) A^2(w) \partial^\mu C(w-v) A^3(v) \Big\} \\
&\quad + \frac{9 \lambda^4}{4} \int \mathrm{d}^2y \ \mathrm{d}^2z \ \mathrm{d}^2w \ \mathrm{d}^2v \ \partial_\mu C(x-y) A(y) \\
&\quad \quad \quad \times \partial^\mu C(y-z) A^3(z) \partial_\nu C(y-w) A^2(w) \partial^\nu C(w-v) A^3(v) \\
&\quad + \frac{27 \lambda^4}{8} \int \mathrm{d}^2y \ \mathrm{d}^2z \ \mathrm{d}^2w \ \mathrm{d}^2v \ \partial_\mu C(x-y) A^2(y) \ \Big\{ \\
&\quad \quad \quad - \partial^\mu C(y-z) A^2(z) \partial_\nu C(z-w) A^2(w) \partial^\nu C(w-v) A^3(v) \\
&\quad \quad \quad + 4 \, \partial_\nu C(y-z) A^2(z) \partial^{[\mu} C(z-w) A^2(w) \partial^{\nu]} C(w-v) A^3(v) \Big\} 
\end{aligned}
\end{align}
\end{footnotesize}
\raggedbottom
\begin{footnotesize}
\begin{align*}
\begin{aligned}
\hphantom{(\mathcal{T}_\lambda A)(x)}&+ \frac{3 \, m^3 \lambda^5}{5} \int \mathrm{d}^2y \ \mathrm{d}^2z \ \mathrm{d}^2w \ \mathrm{d}^2v \ \mathrm{d}^2u \ \partial_\mu C(x-y) \\
&\quad \quad \quad \times \partial^\mu C(y-z) A(z)^3 C(y-w) A(w)^3 C(y-v) A(v)^2 C(v-u) A(u)^3 \\
&+ \frac{3 \, m^3 \lambda^5}{10} \int \mathrm{d}^2y \ \mathrm{d}^2z \ \mathrm{d}^2w \ \mathrm{d}^2v \ \mathrm{d}^2u \ \partial_\mu C(x-y) A(y) \ \Big\{\\
&\quad \quad \quad - 3 \, \partial^\mu C(y-z) A(z)^2 C(z-w) A(w)^3 C(y-v) A(v)^2 C(v-u) A(u)^3 \\ 
&\quad \quad \quad + 6 \, \partial^\mu C(y-z) A(z)^2 C(y-w) A(w)^3 C(y-v) A(v)^2 C(v-u) A(u)^3 \\
&\quad \quad \quad + 2 \, \partial^\mu C(y-z) A(z)^3 C(y-w) A(w) C(w-v) A(v)^3 C(w-u) A(u)^3 \\ 
&\quad \quad \quad - 9 \, \partial^\mu C(y-z)A(z)^3 C(y-w) A(w)^2 C(w-v) A(v)^2 C(v-u) A(u)^3 \\ 
&\quad \quad \quad + C(y-z) A(z)^3 \partial^\mu C(y-w) A(w) C(w-v) A(v)^3 C(w-u) A(u)^3 \\ 
&\quad \quad \quad + 3 \, C(y-z) A(z)^3 \partial^\mu C(y-w) A(w)^2 C(w-v) A(v)^2 C(v-u) A(u)^3 \\ 
&\quad \quad \quad - 9 \, C(y-z) A(z)^2 \partial^\mu C(z-w) A(w)^3 C(y-v) A(v)^2 C(v-u) A(u)^3 \\
&\quad \quad \quad - 3 \, C(y-z) A(z)^3 C(y-w) A(w)^2 \partial^\mu C(w-v) A(v)^2 C(v-u) A(u)^3 \\ 
&\quad \quad \quad + 4 \, C(y-z) A(z)^3 C(y-w) A(w) \partial^\mu C(w-v) A(v)^3 C(w-u) A(u)^3 \\ 
&\quad \quad \quad - 9 \, C(y-z) A(z)^3 C(y-w) A(w)^2 C(w-v) A(v)^2 \partial^\mu C(v-u) A(u)^3 \Big\} \\
&+ \frac{3 \, m^3 \lambda^5}{20} \int \mathrm{d}^2y \ \mathrm{d}^2z \ \mathrm{d}^2w \ \mathrm{d}^2v \ \mathrm{d}^2u \ \partial_\mu C(x-y) A(y)^2 \ \Big\{ \\
&\quad \quad \quad + \partial^\mu C(y-z) C(z-w) A(w)^3 C(y-v) A(v)^3 C(y-u) A(u)^3 \\ 
&\quad \quad \quad + \partial^\mu C(y-z) C(z-w) A(w)^3 C(z-v) A(v)^3 C(z-u) A(u)^3 \\ 
&\quad \quad \quad + 27 \, \partial^\mu C(y-z) A(z)^2 C(z-w) A(w)^2 C(w-v) A(v)^2 C(v-u) A(u)^3 \\ 
&\quad \quad \quad - 6 \, \partial^\mu C(y-z) A(z)^2 C(z-w) A(w) C(w-v) A(v)^3 C(w-u) A(u)^3 \\ 
&\quad \quad \quad + 2 \, C(y-z) \partial^\mu C(z-w) A(w)^3 C(z-v) A(v)^3 C(z-u) A(u)^3 \\ 
&\quad \quad \quad - 18 \, C(y-z) A(z) \partial^\mu C(z-w) A(w)^3 C(z-v) A(v)^2 C(v-u) A(u)^3 \\ 
&\quad \quad \quad - 12 \, C(y-z) A(z)^2 \partial^\mu C(z-w) A(w) C(w-v) A(v)^3 C(w-u) A(u)^3 \\ 
&\quad \quad \quad + 54 \, C(y-z) A(z)^2 \partial^\mu C(z-w) A(w)^2 C(w-v) A(v)^2 C(v-u) A(u)^3 \\ 
&\quad \quad \quad - 6 \, C(y-z) A(z) C(z-w) A(w)^3 \partial^\mu C(z-v) A(v)^2 C(v-u) A(u)^3 \\ 
&\quad \quad \quad - 18 \, C(y-z) A(z)^2 C(z-w) A(w) \partial^\mu C(w-v) A(v)^3 C(w-u) A(u)^3 \\
&\quad \quad \quad + 81 \, C(y-z) A(z)^2 C(z-w) A(w)^2 \partial^\mu C(w-v) A(v)^2 C(v-u) A(u)^3 \\
&\quad \quad \quad - 18 \, C(y-z) A(z) C(z-w) A(w)^3 C(z-v) A(v)^2 \partial^\mu C(v-u) A(u)^3 \\
&\quad \quad \quad + 108 \, C(y-z) A(z)^2 C(z-w) A(w)^2 C(w-v) A(v)^2 \partial^\mu C(v-u) A(u)^3 \Big\} 
\end{aligned}
\end{align*}
\end{footnotesize}
\begin{footnotesize}
\begin{align*}
\begin{aligned}
\hphantom{(\mathcal{T}_\lambda A)(x)}&+ \frac{21 \, m \lambda^5}{20} \int \mathrm{d}^2y \ \mathrm{d}^2z \ \mathrm{d}^2w \ \mathrm{d}^2v \ \mathrm{d}^2u \ \partial_\mu C(x-y) \\
&\quad \quad \quad \times \partial^\mu C(y-z) A(z)^3 C(y-w) A(w)^3 \partial_\nu C(y-v) A(v)^2 \partial^\nu C(v-u) A(u)^3 \\
&+ \frac{9 \, m \lambda^5}{20} \int \mathrm{d}^2y \ \mathrm{d}^2z \ \mathrm{d}^2w \ \mathrm{d}^2v \ \mathrm{d}^2u \ \partial_\mu C(x-y) A(y) \ \Big\{ \\
&\quad \quad \quad - 8 \, \partial^\mu C(y-z) A(z)^3 C(y-w) A(w)^2 \partial_\nu C(w-v) A(v)^2 \partial^\nu C(v-u) A(u)^3 \\
&\quad \quad \quad - 12 \, \partial^\mu C(y-z) A(z)^3 \partial_\nu C(y-w) A(w)^2 C(w-v) A(v)^2 \partial^\nu C(v-u) A(u)^3 \\ 
&\quad \quad \quad - 9 \, \partial^\mu C(y-z) A(z)^3 \partial_\nu C(y-w) A(w)^2 \partial^\nu C(w-v) A(v)^2 C(v-u) A(u)^3 \\ 
&\quad \quad \quad + 2 \, \partial^\mu C(y-z) A(z)^3 \partial_\nu C(y-w) A(w) \partial^\nu C(w-v) A(v)^3 C(w-u) A(u)^3 \\ 
&\quad \quad \quad - \partial^\mu C(y-z) A(z)^2 C(z-w) A(w)^3 \partial_\nu C(y-v) A(v)^2 \partial^\nu C(v-u) A(u)^3 \\ 
&\quad \quad \quad + C(y-z) A(z)^3 \partial^\mu C(y-w) A(w)^2 \partial_\nu C(w-v) A(v)^2 \partial^\nu C(v-u) A(u)^3 \\ 
&\quad \quad \quad + 12 \, C(y-z) A(z)^3 \partial_\nu C(y-w) A(w)^2 \partial^{[\mu} C(w-v) A(v)^2 \partial^{\nu]} C(v-u) A(u)^3 \\ 
&\quad \quad \quad - 8 \, C(y-z) A(z)^2 \partial^\mu C(z-w) A(w)^3 \partial_\nu C(y-v) A(v)^2 \partial^\nu C(v-u) A(u)^3 \Big\} \\
&+ \frac{9 \, m \lambda^5}{40} \int \mathrm{d}^2y \ \mathrm{d}^2z \ \mathrm{d}^2w \ \mathrm{d}^2v \ \mathrm{d}^2u \ \partial_\mu C(x-y) A(y)^2 \ \Big\{ \\
&\quad \quad \quad - 14 \, \partial^\mu C(y-z) A(z) C(z-w) A(w)^3 \partial_\nu C(z-v) A(v)^2 \partial^\nu C(v-u) A(u)^3 \\ 
&\quad \quad \quad + 24 \, \partial^\mu C(y-z) A(v)^2 C(z-w) A(w)^2 \partial_\nu C(w-v) A(v)^2 \partial^\nu C(v-u) A(u)^3 \\ 
&\quad \quad \quad + 27 \, \partial^\mu C(y-z) A(z)^2 \partial_\nu C(z-w) A(w)^2 \partial^\nu C(w-v) A(v)^2 C(v-u) A(u)^3 \\ 
&\quad \quad \quad - 6 \, \partial^\mu C(y-z) A(z)^2 \partial_\nu C(z-w) A(w) \partial^\nu C(w-v) A(v)^3 C(w-u) A(u)^3 \\ 
&\quad \quad \quad + 36 \, \partial^\mu C(y-z) A(z)^2 \partial_\nu C(z-w) A(w)^2 C(w-v) A(v)^2 \partial^\nu C(v-u) A(u)^3 \\ 
&\quad \quad \quad - 16 \, C(y-z) A(z) \partial^\mu C(z-w) A(w)^3 \partial_\nu C(z-v) A(v)^2 \partial^\nu C(v-u) A(u)^3 \\
&\quad \quad \quad + 48 \, C(y-z) A(z)^2 \partial^\mu C(z-w) A(w)^2 \partial_\nu C(w-v) A(v)^2 \partial^\nu C(v-u) A(u)^3 \\ 
&\quad \quad \quad + 12 \, \partial_\nu C(y-z) A(z)^2 \partial^\mu C(z-w) A(w) \partial^\nu C(w-v) A(v)^3 C(w-u) A(u)^3 \\ 
&\quad \quad \quad - 108 \, \partial_\nu C(y-z) A(z)^2 \partial^{[\mu} C(z-w) A(w)^2 \partial^{\nu]} C(w-v)A(v)^2 C(v-u) A(u)^3 \\ 
&\quad \quad \quad - 144 \, \partial_\nu C(y-z) A(z)^2 \partial^{[\mu} C(z-w) A(w)^2 C(w-v) A(v)^2 \partial^{\nu]} C(v-u) A(u)^3 \\ 
&\quad \quad \quad - 144 \, C(y-z) A(z)^2 \partial_\nu C(z-w) A(w)^2 \partial^{[\mu} C(w-v) A(v)^2 \partial^{\nu]} C(v-u) A(u)^3 \\ 
&\quad \quad \quad - 144 \, \partial_\nu C(y-z) A(z)^2 C(z-w) A(w)^2 \partial^{[\mu} C(w-v) A(v)^2 \partial^{\nu]} C(v-u) A(u)^3 \\
&\quad \quad \quad - 13 \, \partial_\nu C(y-z) A(z)^2 \partial^\nu C(z-w) A(w) \partial^\mu C(w-v) A(v)^3 C(w-u) A(u)^3 \\ 
&\quad \quad \quad + 24 \, \partial_\nu C(y-z) A(z) C(z-w) A(w)^3 \partial^{[\mu} C(z-v)A(v)^2 \partial^{\nu]} C(v-u) A(u)^3 \Big\} \\
&\quad + \mathcal{O}(\lambda^6) \, . 
\end{aligned}
\end{align*}
\end{footnotesize}%
\newpage
\flushbottom
Technically, it is possible to compute the Nicolai map to arbitrary order in the coupling constant $\lambda$. However, as can be seen above, the number of terms grows very fast. In particular, we do not expect to obtain a closed-form expression of the map like in supersymmetric quantum mechanics (\emph{c.f.} section \ref{sec:FirstExample} or \cite{Nicolai:1984jg}). 

\subsection{The `Off-Shell' Nicolai Map}
Instead of considering the `on-shell' Wess-Zumino action \eqref{eq:WZ2} and `on-shell' supersymmetry variations \eqref{eq:WZsusy2}, we could have also considered the respective `off-shell' expressions \eqref{eq:WZ1} and \eqref{eq:WZsusy}. Remarkably, we notice that even in the `off-shell' case \eqref{eq:WZdSdl} remains true with the same expression for $\Delta_\alpha$ as in the `on-shell' case. Thus we immediately obtain the `off-shell' $\mathcal{R}_\lambda$-operator
\begin{align}
\begin{aligned}
\mathcal{R}_\lambda = \frac{\mathrm{d}}{\mathrm{d}\lambda} &- \frac{1}{2} \int \mathrm{d}^2x \ \mathrm{d}^2y \ \tr [ S(x,y;A) ] \, A^3(y) \, \frac{\delta}{\delta A(x)} \\
&- \frac{i}{2} \int \mathrm{d}^2x \ \mathrm{d}^2y \ \tr [ \gamma^\mu \partial_\mu S(x,y;A) ] \, A^3(y) \, \frac{\delta}{\delta F(x)} \, .
\end{aligned}
\end{align}
This $\mathcal{R}_\lambda$-operator gives rise to two Nicolai maps. One for $A(x)$ and one for $F(x)$. Since the first line of the `off-shell' $\mathcal{R}_\lambda$-operator is free of the auxiliary field $F(x)$ the `off-shell' Nicolai map for $A(x)$ coincides with its `on-shell' counterpart \eqref{eq:WZResult}.

\subsection{The Radius of Convergence}
A few remarks on the radius of convergence of the map $(\mathcal{T}_\lambda A)$ are in order. For supersymmetric quantum mechanics with a topological $\theta$ term, it has been shown by Lechtenfeld that, at sufficiently small coupling, the Nicolai map converges to a mathematically well-defined functional \cite{Lechtenfeld:2022qed}. Unfortunately, the gamma trace in the $\mathcal{R}_\lambda$-operator \eqref{eq:WZRop3} complicates the situation for the Wess-Zumino model. The trace is similar to an additional loop structure in spinor space attached to the tree diagrams and leads to spacetime index contractions between partial derivatives distributed all over the tree. Consequently, with these additional terms, the estimations in \cite{Lechtenfeld:2022qed} are not good enough to give a finite radius of convergence for the Wess-Zumino model. In \cite{Nicolai:1984jg}, it was speculated, but not proven, that the Wess-Zumino Nicolai map should converge. 

A different approach to finding the radius of convergence of the Nicolai map was given in \cite{Lechtenfeld:2021zgd}. Lechtenfeld and Nicolai demonstrated that in supermembrane theory, the Jacobian of the Nicolai map has a non-zero radius of convergence. Moreover, the authors argued that, with appropriate UV and IR regularization, also the Jacobian of the Nicolai map in super Yang-Mills theories (at least in the axial gauge) has a finite radius of convergence. However, the finite radius of convergence of the Jacobian does not imply a finite radius of convergence of the map itself. But it at least constrains the series expansion. 

So, while still unproven, a finite radius of convergence also for the Nicolai map of the Wess-Zumino model is very likely. Finally, let us remark that in any supersymmetric theory, a finite radius of convergence of the Nicolai map does, of course, not imply a finite radius of convergence for the perturbative expansion of the theory itself. For example, it has been shown by Balian, Itzykson, Parisi and Zuber in a series of papers \cite{Itzykson:1976zu,Itzykson:1977mf,Balian:1977mk} (and also generally argued by Dyson \cite{Dyson:1952tj}), that the perturbative expansion of quantum electrodynamics has a vanishing radius of convergence. However, in our formalism, the usual perturbation expansion is split into two steps. In the first step, the calculation of $(\mathcal{T}_\lambda^{-1} A)$, the radius of convergence is finite and then at the second step, the computation of correlators and thus production of loop diagrams, it is not.

\section{Tests}\label{sec:WZTests}
In this section, we show that the Nicolai map \eqref{eq:WZResult} satisfies part \textit{i)} and \textit{ii)} of the main theorem \ref{th:WZTheorem} order by order in $\lambda$. Thus providing two highly non-trivial tests for our main result \eqref{eq:WZResult}. To keep this section concise, we however only consider the first three orders. The calculations at higher orders follow accordingly and yield the expected results.

\subsection{The Free Action}
By the first statement in the main theorem, the bosonic action is mapped to the abelian action. Hence the transformed bosonic field $A^\prime \equiv (\mathcal{T}_\lambda \, A)$ from \eqref{eq:WZResult} must satisfy
\begin{align}\label{eq:WZFreeAction}
\frac{1}{2} \int \mathrm{d}^2x \ A^\prime (- \Box + m^2 ) A^\prime \overset{!}{=} \frac{1}{2} \int \mathrm{d}^2x \ \left[ (\partial_\mu A)(\partial^\mu A) + ( m A + \lambda A^3 )^2 \right] + \mathcal{O}(\lambda^6) \, .
\end{align}
At the leading order, there is nothing to do. At the first order we plug $A^\prime(x) \vert_{\mathcal{O}(\lambda)}$ into the left-hand side of \eqref{eq:WZFreeAction} and integrate the second term by parts
\begin{align}
\begin{aligned}\label{WZ:FreeActionFirstOrder}
&\quad \frac{m \lambda}{2} \int \mathrm{d}^2x \ \mathrm{d}^2y \ \left[ A(x) (-\Box + m^2) C(x-y) A^3(y) + C(x-y) A^3(y) (- \Box + m^2 ) A(x) \right] \\
&= m \lambda \int \mathrm{d}^2x \ \mathrm{d}^2y \ A(x) (- \Box + m^2 ) C(x-y) A^3(y) \\
&= m \lambda \int \mathrm{d}^2x \ \mathrm{d}^2y \ A(x) \delta(x-y) A^3(y) \\
&= m \lambda \int \mathrm{d}^2x \ A^4(x) \, .
\end{aligned}
\end{align}
In the second step, we have used the definition of the free massive scalar propagator $C(x)$. Collecting the terms of $\mathcal{O}(\lambda)$ on the right-hand side of \eqref{eq:WZFreeAction}, we find agreement with \eqref{WZ:FreeActionFirstOrder}. Similarly, we use integration by parts and the definition of the free scalar propagator at $\mathcal{O}(\lambda^2)$ to obtain
\begin{align}
\begin{aligned}
&\quad - \frac{3 \lambda^2}{4} \int \mathrm{d}^2x \ \mathrm{d}^2y \ \mathrm{d}^2z \ A(x) (-\Box + m^2) \partial_\mu C(x-y) A^2(y) \partial^\mu C(y-z) A^3(z) \\
&\quad - \frac{3 \lambda^2}{4} \int \mathrm{d}^2x \ \mathrm{d}^2y \ \mathrm{d}^2z \ \partial_\mu C(x-y) A^2(y) \partial^\mu C(y-z) A^3(z) (-\Box + m^2)A(x) \\
&\quad + \frac{m^2 \lambda^2}{2} \int \mathrm{d}^2x \ \mathrm{d}^2y \ \mathrm{d}^2z \ C(x-y) A^3(y) (-\Box + m^2) C(x-z) A^3(z) \\
&= \frac{3 \lambda^2}{2} \int \mathrm{d}^2x \ \mathrm{d}^2y \ \mathrm{d}^2z \ \partial_\mu A(x) (-\Box + m^2) C(x-y) A^2(y) \partial^\mu C(y-z) A^3(z) \\
&\quad + \frac{m^2 \lambda^2}{2} \int \mathrm{d}^2x \ \mathrm{d}^2y \ A^3(x) C(x-y) A^3(y) \\
&= \frac{\lambda^2}{2} \int\mathrm{d}^2x \ \mathrm{d}^2y \ \partial_\mu A^3(x) \partial^\mu C(x-y) A^3(y) \\
&\quad + \frac{m^2 \lambda^2}{2} \int \mathrm{d}^2x \ \mathrm{d}^2y \ A^3(x) C(x-y) A^3(y) \\
&= \frac{\lambda^2}{2} \int \mathrm{d}^2x \ \mathrm{d}^2y \ A^3(x) (-\Box + m^2) C(x-y) A^3(y) \\
&= \frac{\lambda^2}{2} \int \mathrm{d}^2x \ A^6(x) \, .
\end{aligned}
\end{align}
Again this agrees with the right-hand side of \eqref{eq:WZFreeAction}. In every subsequent order, the general procedure remains the same. By partial integration, we seek to obtain factors of $(-\Box + m^2)C(x-y) = \delta(x-y)$ to cancel terms with fewer scalar propagators. At $\mathcal{O}(\lambda^3)$ we have for the left-hand side of \eqref{eq:WZFreeAction}
\begin{align}
\begin{aligned}
&\quad \frac{m \lambda^3}{2} \int \mathrm{d}^2x \ \mathrm{d}^2y \ \mathrm{d}^2z \ \mathrm{d}^2w \ \Big\{ \\
&\quad \quad \quad - 3 \, \partial_\mu C(x-y) A^2(y) \partial^\mu C(y-z) A^3(z) (-\Box + m^2) C(x-w) A^3(w) \\
&\quad \quad \quad - 2 \, \partial^\mu C(x-y) A(y) \partial_\mu C(y-z) A^3(z) C(y-w) A^3(w) (-\Box + m^2) A(x) \\
&\quad \quad \quad + 3 \, \partial^\mu C(x-y) A^2(y) \partial_\mu C(y-z) A^2(z) C(z-w) A^3(w) (-\Box + m^2) A(x) \\
&\quad \quad \quad + 6 \, \partial^\mu C(x-y) A^2(y) C(y-z) A^2(z) \partial_\mu C(z-w) A^3(w) (-\Box + m^2) A(x) \Big\} \\
&= \frac{m \lambda^3}{2} \int \mathrm{d}^2x \ \mathrm{d}^2y \ \mathrm{d}^2z \ \Big\{ \\ 
&\quad \quad \quad - A^3(x) \partial_\mu C(x-y) A^2(y) \partial^\mu C(y-z) A^3(z) \\
&\quad \quad \quad - A^2(x) \Box C(x-y) A^3(y) C(x-z) A^3(z)\\
&\quad \quad \quad - A^2(x) \partial_\mu C(x-y) A^3(y) \partial^\mu C(x-z) A^3(z) \\
&\quad \quad \quad + A^3(x) \Box C(x-y) A^2(y) C(y-z) A^3(z) \Big\} \\
&= 0 \, .
\end{aligned}
\end{align}
For the last equality, we have exchanged $x \leftrightarrow y$ in the third and fourth term and used $\Box C(x-y) = \Box C(y-x)$ as well as $\partial_\mu C(y-x) = - \partial_\mu C(x-y)$ to cancel them against the first two terms. A similar calculation at $\mathcal{O}(\lambda^4)$ and $\mathcal{O}(\lambda^5)$ also agrees with \eqref{eq:WZFreeAction}. It is worth pointing out here that the very existence of a non-local field transformation mapping one local action to another local action is a remarkable fact in itself, independently of supersymmetry (but in the absence of supersymmetry, locality would be spoilt by the Jacobian).

\subsection{Jacobian and Fermion Determinant}
We need to show that the Jacobian determinant is equal to the Matthews-Salam-Seiler determinant at least order by order in perturbation theory. This is done by considering the logarithms of the determinants rather than the determinants themselves, \emph{i.e.}
\begin{align}
\log \mathcal{J}(A^\prime) = \log \det \left( \frac{\delta A^\prime(x)}{\delta A(y)} \right) \overset{!}{=} \log (\Delta_\mathrm{MSS}[\lambda;A] ) \, .
\end{align}
Up to $\mathcal{O}(\lambda^3)$ the logarithm of the MSS determinant \eqref{eq:WZMSS} is
\begin{align}\label{eq:WZMSSO3}
\begin{aligned}
\log(\Delta_\mathrm{MSS}[\lambda;A] ) &= \frac{1}{2} \Tr \log \left[ 1 + 3 \lambda S_0 \ast A^2 \right] \\
&= \color{myred} 3 \, m \lambda \int \mathrm{d}^2 x \ C(0) A^2(x) \\
&\quad \color{myblue} - \frac{9 \lambda^2}{2}\int \mathrm{d}^2 x \ \mathrm{d}^2y \ \partial_\mu C(x-y) A^2(y) \partial^\mu C(y-x) A^2(x) \\
&\quad \color{myblue} - \frac{9 \, m^2 \lambda^2}{2} \int \mathrm{d}^2 x \ \mathrm{d}^2y \ C(x-y) A^2(y) C(y-x) A^2(x) \\
&\quad \color{mygreen} + 9 \, m \lambda^3 \, \, \int \mathrm{d}^2x \ \mathrm{d}^2y \ \mathrm{d}^2z \ \Big\{ \\
&\quad \quad \color{mygreen} + 3 \, \partial_\mu C(x-y) A^2(y) \partial^\mu C(y-z) A^2(z) C(z-x) A^2(x) \\
&\quad \quad \color{mygreen} + m^2 C(x-y) A^2(y) C(y-z) A^2(z) C(z-x) A^2(x) \Big\} \\
&\quad + \mathcal{O}(\lambda^4) \, .
\end{aligned}
\end{align}
The coloring is for later convenience. Recall that $C(0)$ can be regulated. The Jacobian determinant is of the form $1 + X$, where $X$ is a power series in the coupling $\lambda$ starting at $\mathcal{O}(\lambda^1)$. Hence, the logarithm of the Jacobian determinant is given by
\begin{align}
\log(1 + X) = - \sum_{n=1}^\infty \frac{(-1)^n}{n} X^n \, .
\end{align}
We must keep this equation in mind when collecting the terms at each order in $\lambda$. In particular, we see that there is no contribution at the leading order.  At $\mathcal{O}(\lambda)$ we find
\begin{align}
\log \det \left( \frac{\delta A^\prime(x)}{\delta A(y)} \right) \bigg\vert_{\mathcal{O}(\lambda^1)} = \Tr \left[ \frac{\delta A^\prime}{\delta A} \bigg\vert_{\mathcal{O}(\lambda^1)} \right] \, .
\end{align}
The final trace is done by setting $y = x$ and integrating over $x$. The computation is straightforward, and we find
\begin{align}
\log \det \left( \frac{\delta A^\prime(x)}{\delta A(y)} \right) \bigg\vert_{\lambda=0} = \color{myred} 3 m \lambda \int \mathrm{d}^2 x \ C(0) A^2(x) \color{black} \, , 
\end{align}
which matches the {\color{myred}red} term in \eqref{eq:WZMSSO3}. In the second order, we find
\begin{align}
\begin{aligned}
\log \det \left( \frac{\delta A^\prime(x)}{\delta A(y)} \right) \bigg\vert_{\mathcal{O}(\lambda^2)} &= \Tr \left[ \frac{\delta A^\prime}{\delta A} \bigg\vert_{\mathcal{O}(\lambda^2)} \right] - \frac{1}{2} \Tr \left[ \frac{\delta A^\prime}{\delta A} \bigg\vert_{\mathcal{O}(\lambda^1)} \frac{\delta A^\prime}{\delta A} \bigg\vert_{\mathcal{O}(\lambda^1)} \right] \\
&= \color{myblue} - \frac{9 \lambda^2}{2}\int \mathrm{d}^2 x \ \mathrm{d}^2y \ \partial_\mu C(x-y) A^2(y) \partial^\mu C(y-x) A^2(x) \\
&\quad \color{myblue} - \frac{9 \, m^2 \lambda^2}{2} \int \mathrm{d}^2 x \ \mathrm{d}^2y \ C(x-y) A^2(y) C(y-x) A^2(x) \color{black} \, .
\end{aligned}
\end{align}
Again this agrees with \eqref{eq:WZMSSO3}. Finally, at $\mathcal{O}(\lambda^3)$ we have
\begin{align}
\begin{aligned}
\log \det \left( \frac{\delta A^\prime(x)}{\delta A(y)} \right) \bigg\vert_{\mathcal{O}(\lambda^3)} &= \Tr \left[ \frac{\delta A^\prime}{\delta A} \bigg\vert_{\mathcal{O}(\lambda^3)} \right] - \left( 2 \cdot \frac{1}{2} \right) \Tr \left[ \frac{\delta A^\prime}{\delta A} \bigg\vert_{\mathcal{O}(\lambda^2)} \frac{\delta A^\prime}{\delta A} \bigg\vert_{\mathcal{O}(\lambda^1)} \right] \\
&\quad + \frac{1}{3} \, \Tr \left[ \frac{\delta A^\prime}{\delta A} \bigg\vert_{\mathcal{O}(\lambda^1)} \frac{\delta A^\prime}{\delta A} \bigg\vert_{\mathcal{O}(\lambda^1)} \frac{\delta A^\prime}{\delta A} \bigg\vert_{\mathcal{O}(\lambda^1)} \right] \, .
\end{aligned}
\end{align}
For the first term, on the right-hand side, we find
\begin{align}
\begin{aligned}
\Tr \left[ \frac{\delta A^\prime}{\delta A} \bigg\vert_{\mathcal{O}(\lambda^3)} \right] &= 9 m \int \mathrm{d}^2x \ \mathrm{d}^2y \ \mathrm{d}^2z \ \Big\{ \\
&\quad \quad + A(x) C(x-y) \partial_\mu C(x-y) A(y)^2 \partial^\mu C(x-z) A(z)^3 \\
&\quad \quad + \frac{3}{2} \, A(x)^2 \partial_\mu C(x-y) A(y)^2 C(y-z) A(z)^2 \partial^\mu C(z-x) \Big\} \, .
\end{aligned}
\end{align}
The second term gives
\begin{align}
\begin{aligned}
&\quad- \left( 2 \cdot \frac{1}{2} \right) \Tr \left[ \frac{\delta A^\prime}{\delta A} \bigg\vert_{\mathcal{O}(\lambda^2)} \frac{\delta A^\prime}{\delta A} \bigg\vert_{\mathcal{O}(\lambda^1)} \right] \\
& = 9 \, m \int \mathrm{d}^2x \ \mathrm{d}^2y \ \mathrm{d}^2z \ \Big\{ \\
&\quad \quad - A(x) C(x-y) \partial_\mu C(x-y) A(y)^2 \partial^\mu C(x-z) A(z)^3 \\ 
&\quad \quad + \frac{3}{2} \, A(x)^2 \partial_\mu C(x-y) A(y)^2 C(y-z) A(z)^2 \partial^\mu C(z-x) \Big\} \, .
\end{aligned}
\end{align}
Finally, the last term gives
\begin{align}
\begin{aligned}
&\quad\frac{1}{3} \, \Tr \left[ \frac{\delta A^\prime}{\delta A} \bigg\vert_{\mathcal{O}(\lambda^1)} \frac{\delta A^\prime}{\delta A} \bigg\vert_{\mathcal{O}(\lambda^1)} \frac{\delta A^\prime}{\delta A} \bigg\vert_{\mathcal{O}(\lambda^1)} \right] \\
&= 9 \, m^3 \int \mathrm{d}^2x \ \mathrm{d}^2y \ \mathrm{d}^2z \ A^2(x) C(x-y) A^2(y) C(y-z) A^3(z) C(z-x) \, .
\end{aligned}
\end{align}
Adding these three terms gives
\begin{align}
\begin{aligned}
\log \det \left( \frac{\delta A^\prime(x)}{\delta A(y)} \right) \bigg\vert_{\mathcal{O}(\lambda^3)} &= \color{mygreen} 9 \, m \int \mathrm{d}^2x \ \mathrm{d}^2y \ \mathrm{d}^2z \ \Big\{ \\
&\quad \quad \color{mygreen} + 3 \, A(x)^2 \partial_\mu C(x-y) A(y)^2 C(y-z) A(z)^2 \partial^\mu C(z-x) \\
&\quad \quad \color{mygreen} + m^2 A^2(x) C(x-y) A^2(y) C(y-z) A^3(z) C(z-x) \Big\} \color{black} \, ,
\end{aligned}
\end{align}
which matches the {\color{mygreen}green} terms in \eqref{eq:WZMSSO3}. The calculation for the fourth and fifth order work accordingly and yield the expected results. Thus we have shown that the map \eqref{eq:WZResult} satisfies \textit{i)} and \textit{ii)} from the main theorem \ref{th:WZTheorem}. This concludes our discussion of the Nicolai map for the 2-dimensional Wess-Zumino model.

\chapter[head={Off-Shell $\mathcal{N}$\:=\;1 super Yang-Mills}]{Off-Shell \texorpdfstring{$\boldsymbol{\mathcal{N}}$\:=\;1}{N=1} super Yang-Mills}\label{ch:YM1}
In this chapter, we discuss the construction of the Nicolai map for `off-shell' $\mathcal{N}=1$ super Yang-Mills in four dimensions. `Off-shell' meaning that the auxiliary field is present in the action and the supersymmetry algebra closes without evoking the equations of motion. The `off-shell' supersymmetry is crucial to the existence of the Nicolai map in general gauges.

In the first section, we recall the relevant notation and introduce rescaled fields. Furthermore, we state the `off-shell' $\mathcal{N}=1$ super Yang-Mills version of the main theorem \ref{th:WZTheorem}. In section \ref{sec:YM1R} we derive the $\tilde{\mathcal{R}}_g$-operator and in section \ref{sec:YM1Proof}, we prove the new main theorem. In the following two sections, we discuss properties of the $\tilde{\mathcal{R}}_g$-operator and give the Nicolai map in axial gauge up to the second order in the coupling constant $g$. In section \ref{sec:YM1Tests} we test the Nicolai map from the previous section. The last section discusses a potential simplification of the Nicolai map by resorting to the `on-shell' formulation of $\mathcal{N}=1$ super Yang-Mills and the Landau gauge. This will motivate the construction presented in the next chapter.

This chapter is mostly based on the author's publication \cite{Malcha:2021ess}. However, the $\tilde{\mathcal{R}}_g$-operator given in section  \ref{sec:YM1R} was first derived by Dietz and Lechtenfeld in \cite{Dietz:1984hf,Lechtenfeld:1984me,Dietz:1985hga}. 

\section{Introduction and Main Theorem}\label{sec:YM1Intro}
Recall the 4-dimensional gauge invariant `off-shell' $\mathcal{N}=1$ super Yang-Mills action \eqref{eq:YM1}
\begin{align}\label{eq:YM11}
S_\mathrm{inv}^1 = \int \mathrm{d}^4x \ \left[ - \frac{1}{4} F_{\mu \nu}^a F^{a \, \mu \nu}  - \frac{i}{2} \bar{\lambda}^a \gamma^\mu (D_\mu \lambda)^a + \frac{1}{2} D^a D^a \right] \, , 
\end{align}
with the standard definitions
\begin{align}
&F_{\mu\nu}^a \coloneqq \partial_\mu A_\nu^a - \partial_\nu A_\mu^a + gf^{abc} A_\mu^b A_\nu^c \, , \\
&(D_\mu \lambda)^a \coloneqq \partial_\mu \lambda^a + gf^{abc} A_\mu^b \lambda^c \, , 
\end{align}
for the field strength tensor and covariant derivative. The action \eqref{eq:YM11} is invariant under the supersymmetry variations \eqref{eq:YMsusy1}
\begin{align}\label{eq:YMsusy11}
&\delta A_\mu^a = - i (\bar{\lambda}^a \gamma_\mu \varepsilon )  \, , 
&&\delta \lambda_\alpha^a = -\frac{1}{2} (\gamma^{\mu\nu} \varepsilon)_\alpha F_{\mu \nu}^a + i (\gamma^5 \varepsilon)_\alpha D^a \, , 
&&\delta D^a = -  ( \bar{\varepsilon} \gamma^5 \gamma^\mu (D_\mu \lambda^a) ) \, .
\end{align}
Thanks to the presence of the auxiliary field, the supersymmetric action \eqref{eq:YM11} can be written as a supervariation
\begin{align}\label{eq:YM1dDelta}
S_\mathrm{inv} = \delta_\alpha \Delta_\alpha
\end{align}
with the gauge invariant functional
\begin{align}\label{eq:YM1Delta}
\Delta_\alpha \coloneqq \int \mathrm{d}^4 x \ \left[ - \frac{1}{16} (\gamma^{\mu\nu} \lambda^a)_\alpha F^a_{\mu\nu}  - \frac{i}{8} (\gamma^5 \lambda^a)_\alpha D^a \right] \, . 
\end{align}
When computing path integrals, we have to add a gauge fixing term to the action \eqref{eq:YM11} to prevent the over-counting of physically equivalent paths
\begin{align}\label{eq:YMgf1}
S_\mathrm{gf} = \int \mathrm{d}^4x \ \left[ \frac{1}{2\xi} (\mathcal{G}^\mu A_\mu^a) (\mathcal{G}^\nu A_\nu^a) + \frac{1}{2} \bar{C}^a \mathcal{G}^\mu (D_\mu C)^a \right] \, .
\end{align}
In the following, we restrict the choice of gauge fixing functionals to $\mathcal{G}^\mu = \partial^\mu$ for the $R_\xi$ type gauges and $\mathcal{G}^\mu = n^\mu$ for the axial type gauges (including the light-cone gauge with $n_\mu n^\mu = 0$). The most general gauge fixing functional compatible with the subsequent discussion is any functional obeying the scaling relation \cite{Malcha:2021ess}
\begin{align}\label{eq:YM1GScaling}
\mathcal{G}^a(A) = g \, \mathcal{G}^a(g^{-1} A) \, .
\end{align}
The complete `off-shell' $\mathcal{N}=1$ super Yang-Mills action $S^1 = S_\mathrm{inv}^1 + S_\mathrm{gf}$ is invariant under the following BRST (or Slavnov) variations \eqref{eq:YMBRST}
\begin{align}
\begin{aligned}\label{eq:BRST11}
&s A_\mu^a = (D_\mu C)^a \, , &&s D^a = g f^{abc} D^b C^c \, , &&s \lambda^a = - g f^{abc} \lambda^b C^c \, , \\
&s \, \bar{C}^a = - \frac{1}{\xi} \mathcal{G}^\mu A_\mu^a \, , &&s\, C^a = - \frac{g}{2} f^{abc} C^b C^c \, .
\end{aligned}
\end{align}
The derivation of the `off-shell' $\mathcal{R}_g$-operator will necessitate a `detour' via a reformulation of the theory in terms of rescaled fields
\begin{align}
\tilde{A}_\mu^a = g A_\mu^a \,, \quad \tilde{\lambda}^a = g\lambda^a \, , \quad\tilde{D}^a = g D^a \, , \quad
 \tilde{C}^a = g \, C^a\, , \quad \bar{\tilde{C}}^a = g  \, \bar{C}^a \, , 
\end{align}
such that the coupling constant appears only as an overall factor outside the action \eqref{eq:YM11}
\begin{align}\label{eq:YM12}
\tilde{S}_\text{inv}^1  =  \frac1{g^2} \int \mathrm{d}^4x \ \left[ - \frac{1}{4} \tilde{F}_{\mu \nu}^a \tilde{F}^{a \, \mu \nu}  - \frac{i}{2} \bar{\tilde{\lambda}}^a \gamma^\mu (D_\mu \tilde{\lambda})^a + \frac{1}{2} \tilde{D}^a \tilde{D}^a \right] \, , 
\end{align}
where now
\begin{align}
\tilde{F}_{\mu\nu}^a & \,\equiv\, \partial_\mu \tilde{A}_\nu^a - \partial_\nu \tilde{A}_\mu^a 
+ f^{abc} \tilde{A}_\mu^b \tilde{A}_\nu^c \, , \\
(D_\mu \tilde{\lambda})^a &\,\equiv\, \partial_\mu \tilde{\lambda}^a + f^{abc} \tilde{A}_\mu^b \tilde{\lambda}^c \, .
\end{align}
The ghost action $\tilde{S}_\text{gf}$ as well as the supersymmetry and the BRST transformations are obtained from \eqref{eq:YMgf1}, \eqref{eq:YMsusy11} and \eqref{eq:BRST11} by dropping $g$ and putting tildes on all fields; idem for \eqref{eq:YM1dDelta} and \eqref{eq:YM1Delta} (it is here that we would need the scaling relation \eqref{eq:YM1GScaling}). For clarity of notation, we always put tildes on all quantities involving rescaled fields. Correlation functions of tilded and untilded monomials of, for example, the gauge field are related by
\begin{align}
\big<\!\! \big< \tilde{A}_{\mu_1}^{a_1}(x_1) \ldots \tilde{A}_{\mu_n}^{a_n}(x_n) \big>\!\!\big>_g = g^n \,
\big< \!\! \big< A_{\mu_1}^{a_1}(x_1) \ldots A_{\mu_n}^{a_n}(x_n) \big>\!\!\big>_g \, .
\end{align}
Finally, let us comment on the limit $g \to 0$ in both the tilded and untilded version of the theory. For the untilded version, the limit of $S_\text{inv}^1 + S_\text{gf}$ is simply the free supersymmetric Maxwell theory. By contrast, the $g \to 0$ limit of $\tilde{S}_\text{inv}^1 + \tilde{S}_\text{gf}$ localizes the bosonic Yang-Mills action on zero curvature configurations. Here we will be concerned with the former case and make use of the tilded formulation only as an intermediate device. 

Subsequently, we can formulate the main theorem \cite{Malcha:2021ess}.
\begin{theorem}\label{th:YM1Theorem}
`Off-shell' 4-dimensional $\mathcal{N}=1$ super Yang-Mills is characterized by the existence of a non-linear and non-local transformation $\mathcal{T}_g$ of the bosonic fields $\Phi = (A_\mu^a, D^a)$
\begin{align*}
\mathcal{T}_g: \Phi(x) \mapsto \Phi^\prime(x,g;\Phi) \, , 
\end{align*}
which is invertible, at least in the sense of a formal power series such that
\begin{enumerate}
\item The gauge-fixing function $\mathcal{G}^\mu A_\mu^a$ is a fixed point of the map $\mathcal{T}_g$. 
\item The bosonic Yang-Mills action without gauge-fixing terms is mapped to the abelian action, 
\begin{align*}
S_\mathrm{inv}^1[g;\Phi] = S_\mathrm{inv}^1[0; \mathcal{T}_g \Phi]  \, .
\end{align*}
\item The Jacobian determinant of $\mathcal{T}_g$ is equal to the product of the Matthews-Salam-Seiler and Faddeev-Popov determinants, i.e.
\begin{align*}
\mathcal{J}(\mathcal{T}_g \, \Phi) = \Delta_\mathrm{MSS}[g;\Phi] \Delta_\mathrm{FP}[g;\Phi] 
\end{align*}
at least order by order in perturbation theory.
\end{enumerate}
\end{theorem}
Similarly to the previous chapter, we give a constructive proof of the theorem by deriving $\mathcal{T}_g$ and showing that it satisfies \textit{i) - iii)}. 

However, first, we show that the relation between free and interacting correlation functions \eqref{eq:WZNicolaiCorr} we have derived for the Wess-Zumino model extends to gauge theories. The $\mathcal{N}=1$ super Yang-Mills Nicolai map has similar properties as the Wess-Zumino Model Nicolai map. In particular its inverse acts on bosonic monomials $X[\Phi]$ by $\mathcal{T}_g^{-1} X[\Phi] = X[\mathcal{T}_g^{-1} \Phi]$. Computing the free bosonic vacuum expectation value of such a monomial and performing a change of variables gives
\begin{align}
\begin{aligned}
\big< X[\mathcal{T}_g^{-1} \Phi] \big>_0 &= \int \mathcal{D}_0 \Phi \ e^{- i S_\mathrm{inv}^1[0;\Phi] - i S_\mathrm{gf}[g;\Phi] } \ X[\mathcal{T}_g^{-1} \Phi] \\
&= \int \mathcal{D}_0 \Phi \ \mathcal{J}(\mathcal{T}_g \Phi)  e^{- i S_\mathrm{inv}^1[0;\mathcal{T}_g \Phi] - i S_\mathrm{gf}[0;\mathcal{T}_g \Phi] } \ X[ \Phi] \, .
\end{aligned}
\end{align}
On the other hand, integrating out the fermionic degrees of freedom in the interacting vacuum expectation value of $X[\Phi]$ yields
\begin{align}
\begin{aligned}
\big<\!\!\big< X[\Phi] \big>\!\!\big>_g &= \int \mathcal{D} \Phi \ \mathcal{D}\Psi \ e^{- i S_\mathrm{inv}^1[g;\Phi,\Psi] - i S_\mathrm{gf}[g;\Phi,\Psi]} \ X[\Phi] \\
&= \int \mathcal{D}_0 \Phi \ \Delta_\mathrm{MSS}[g;\Phi] \Delta_\mathrm{FP}[g;\Phi]  e^{- i S_\mathrm{inv}^1[g;\Phi] - i S_\mathrm{gf}[g;\Phi] } \ X[\Phi] \, .
\end{aligned}
\end{align}
Notice the appearance of the Faddeev-Popov determinant from integrating out the ghost fields. Assuming that $\mathcal{T}_g$ satisfies \textit{i)} - \textit{iii)} of the main theorem above we conclude that
\begin{align}
\big<\!\!\big< X[\Phi] \big>\!\!\big>_g = \big< X[\mathcal{T}_g^{-1} \Phi] \big>_0 \, .
\end{align}
Again the generalization to $n$-point correlation functions of arbitrary bosonic operators $\mathcal{O}_i(x_i)$ is immediate and we conclude
\begin{align}
\big<\!\!\big< \mathcal{O}_1(x_1) \ldots \mathcal{O}_n(x_n) \big>\!\!\big>_g = \big< (\mathcal{T}_g^{-1} \mathcal{O}_1)(x_1) \ldots (\mathcal{T}_g^{-1}  \mathcal{O}_n)(x_n) \big>_0 \, .
\end{align}
Remember that this does not render the correlation function trivial since the complexity lives on in the perturbative expansion of the inverse Nicolai map.

\section[The \texorpdfstring{$\tilde{\mathcal{R}}_g$}{R}-Operator]{The $\boldsymbol{\tilde{\mathcal{R}}_g}$-Operator}\label{sec:YM1R}
Similar to the discussion in the previous chapter, we construct the inverse map $\mathcal{T}_g^{-1}$ via its infinitesimal generator $\tilde{\mathcal{R}}_g$. For `off-shell' super Yang-Mills, this construction was first done in 1984 by Dietz and Lechtenfeld \cite{Dietz:1984hf,Lechtenfeld:1984me,Dietz:1985hga}. However, they were not able to obtain the correct inverse Nicolai map from the $\tilde{\mathcal{R}}_g$-operator since they did not realize at the time that one must act with the $\tilde{\mathcal{R}}_g$-operator on $A_\mu^a(x) = \frac{1}{g} \tilde{A}_\mu^a(x)$ and $D^a(x) = \frac{1}{g} \tilde{D}^a$ rather than $\tilde{A}_\mu^a(x)$ and $\tilde{D}^a$. The correct power series for the inverse Nicolai map of $A_\mu^a(x)$ was first published in \cite{Malcha:2021ess} (and \cite{Lechtenfeld:2021yjb})
\begin{align}\label{eq:YM1InverseNicolaiMap}
\big( \mathcal{T}_g^{-1} A \big)_\mu^a(x)  \equiv \big( \mathcal{\tilde{T}}_g^{-1}  \big({\textstyle\frac{1}{g}} \tilde{A}\big)\big)_\mu^a(x) \coloneqq \sum_{n=0}^\infty \frac{g^n}{n!} \left[ \big( \mathcal{\tilde{R}}_{g}^n \big({\textstyle\frac{1}{g}} \tilde{A} \big) \big)_\mu^a(x)  \, \Big\vert_{\tilde{\Phi}= g\Phi} \Big\vert_{g=0} \right] \, .
\end{align}
By swapping $A_\mu^a(x)$ and $D^a(x)$, we obtain the corresponding map for the auxiliary field. For a bosonic monomial, $X \equiv X[\Phi]$ the linear response of its vacuum expectation value to a change in the coupling constant is given by
\begin{align}
\frac{\mathrm{d}}{\mathrm{d} g} \big< \!\! \big< X \big> \!\! \big>_g = 
\left< \!\!\!\! \left< \frac{\mathrm{d}X}{\mathrm{d} g} \right> \!\!\!\! \right>_g -
i \left< \!\!\!\! \left<  \frac{\mathrm{d}(\tilde{S}_\mathrm{inv} + \tilde{S}_\mathrm{gf})}{\mathrm{d} g} \ X
\right> \!\!\!\! \right>_g \eqqcolon \big< \mathcal{\tilde{R}}_g \, X  \big>_g  \, .
\end{align}
Notice the factor of $i$ from the definition of the correlation function in \eqref{eq:YMCorr1}. Because the $g$ dependence appears only as an overall factor in the action $\tilde{S}^1 =
\tilde{S}_\mathrm{inv}^1 + \tilde{S}_\mathrm{gf}$ we have
\begin{align}
\frac{\mathrm{d}\tilde{S}_\mathrm{inv}}{\mathrm{d} g} &= - \frac{2 \tilde{S}_\mathrm{inv}}{g} = - \frac{2}{g^3} \, \tilde{\delta}_\alpha \tilde{\Delta}_\alpha \, ,
\end{align}
where $\Delta_\alpha$ is defined in \eqref{eq:YM1Delta} and $\tilde{\Delta}_\alpha$ refers to the expression with all fields rescaled. $\tilde{\delta}_\alpha$ is the supersymmetry variation for rescaled fields. So
\begin{align}\label{eq:YM1Intermediate1}
\frac{\mathrm{d}}{\mathrm{d} g}  \big< \!\! \big<  X  \big> \!\! \big>_g =
\left< \!\!\!\! \left< \frac{\mathrm{d}X}{\mathrm{d} g} \right> \!\!\!\! \right>_g + 
\frac{2i}{g^3} \big< \!\! \big< (\tilde{\delta}_\alpha \tilde{\Delta}_\alpha) \,  X  \big> \!\! \big>_g + 
\frac{2i}{g} \big< \!\! \big< \tilde{S}_\mathrm{gf} \, X  \big> \!\! \big>_g  \, .
\end{align}
We want to rewrite
\begin{align}
\big< \!\! \big< (\tilde{\delta}_\alpha \tilde{\Delta}_\alpha)   X \big> \!\! \big>_g =
\big< \!\! \big<\tilde{\delta}_\alpha \big(\tilde{\Delta}_\alpha X \big) \big> \!\! \big>_g   + 
\big< \!\! \big< \tilde{\Delta}_\alpha (\tilde{\delta}_\alpha X) \big> \!\! \big>_g  
\end{align}
using the supersymmetry Ward identity \eqref{eq:YMWardId3}
\begin{align}
\big< \!\! \big< \tilde{\delta}_\alpha Y \big> \!\! \big>_g =  i \left< \!\!\!\! \left<  \frac{1}{g^2} \int \mathrm{d}^4x \ \bar{\tilde{C}}^a \,  \tilde{\delta}_\alpha ( \mathcal{G}^\mu \tilde{A}_\mu^a) \, s(Y)  \right> \!\!\!\! \right> _g
\end{align}
applied to $Y = \tilde{\Delta}_\alpha X$. Because $\tilde{\Delta}_\alpha$ is gauge invariant we have $s(\tilde{\Delta}_\alpha) = 0$ and thus $s(\tilde{\Delta}_\alpha X) = - \tilde{\Delta}_\alpha \, s(X)$ (the minus sign here appears because $s$ anti-commutes with fermionic expressions such as $\tilde{\Delta}_\alpha$).
For the third term on the right-hand side of \eqref{eq:YM1Intermediate1} we use 
\begin{align}
\tilde{S}_\mathrm{gf}  =  s \left( - \frac{1}{2g^2}  \int \mathrm{d}^4 x \ \bar{\tilde{C}}^a \, (\mathcal{G}^\mu \tilde{A}_\mu^a) \right)
\end{align}
and the BRST Ward identity \eqref{eq:YMWardId2} such that
\begin{align}
 \big< \!\! \big< \tilde{S}_\mathrm{gf} \, X  \big> \!\! \big>_g  = \left< \!\!\!\! \left< - \frac{1}{2g^2}  \int \mathrm{d}^4 x \ \bar{\tilde{C}}^a \, (\mathcal{G}^\mu \tilde{A}_\mu^a)  \, s(X) \right> \!\!\!\! \right>_g \, .
\end{align}
Subsequently, we put everything back together and obtain
\begin{align}
\begin{aligned}\label{eq:YM1ROp1}
\frac{\mathrm{d}}{\mathrm{d} g}  \big< \!\! \big<  X  \big> \!\! \big>_g 
&= \left< \!\!\!\! \left< \frac{\mathrm{d}X}{\mathrm{d} g} \right> \!\!\!\! \right>_g
+  \frac{2i}{g^3} \big< \!\! \big<  \tilde{\Delta}_\alpha(\tilde{\delta}_\alpha X)  \big> \!\! \big>_g 
+ \frac{2}{g^5} \left< \!\!\!\! \left<   \int \mathrm{d}^4x \ \bar{\tilde{C}}^a(x) \,  \tilde{\delta}_\alpha (\mathcal{G}^\mu \tilde{A}_\mu^a) \,  \tilde{\Delta}_\alpha \, s(X) \right> \!\!\!\! \right>_g \\
&\quad - \frac{i}{g^3} \left< \!\!\!\! \left<   \int \mathrm{d}^4 x \ \bar{\tilde{C}}^a(x) \, (\mathcal{G}^\mu \tilde{A}_\mu^a) \, s(X) \right> \!\!\!\! \right>_g  \, .
\end{aligned}
\end{align}
Finally, we integrate over the fermionic degrees of freedom. Recall that each spinor field contraction comes with a factor of $i$ and each ghost field contraction comes with a factor of $(-i)$. Due to the rescaling of the fields, both contractions furthermore come with an additional factor of $g^2$. Subsequently, we arrive at
\begin{align}
\begin{aligned}\label{eq:YM1ROp2}
\mathcal{\tilde{R}}_g \, X &= \frac{\mathrm{d}X}{\mathrm{d} g}  - \frac{2}{g} 
\bcontraction{}{\tilde{\Delta}}{_\alpha (}{\tilde{\delta}} \tilde{\Delta}_\alpha (\tilde{\delta}_\alpha X)
+ \frac{2}{g} \int \mathrm{d}^4x \ 
\bcontraction[2ex]{}{\bar{\tilde{C}}}{^a(x) \, \tilde{\delta}_\alpha (\mathcal{G}^\mu \tilde{A}_\mu^a) \, \tilde{\Delta}_\alpha }{s} 
\bar{\tilde{C}}^a(x) \, 
\bcontraction{}{\tilde{\delta}}{_\alpha (\mathcal{G}^\mu \tilde{A}_\mu^a) \, }{\tilde{\Delta}}
\tilde{\delta}_\alpha(\mathcal{G}^\mu \tilde{A}_\mu^a) \, \tilde{\Delta}_\alpha \, s(X) \\
&\quad - \frac{1}{g} \int \mathrm{d}^4x \ 
\bcontraction{}{\bar{\tilde{C}}}{^a(x) \, (\mathcal{G}^\mu \tilde{A}_\mu^a) \,  }{s}
\bar{\tilde{C}}^a(x) \, (\mathcal{G}^\mu \tilde{A}_\mu^a) \,  s(X) \, .
\end{aligned}
\end{align}
This operator is manifestly distributive. We use the definitions of the fermion and ghost propagators
\begin{align}\label{eq:YM1FermGhostProp}
i \bcontraction{}{\tilde{\lambda}}{^a(x) }{\bar{\tilde{\lambda}}} \tilde{\lambda}^a(x) \bar{\tilde{\lambda}}^b(y) \equiv \tilde{S}^{ab}(x,y;\tilde{A}) \, , \quad \bcontraction{}{\tilde{C}}{^a(x) }{\bar{\tilde{C}}} \tilde{C}^a(x) \bar{\tilde{C}}^b(y) \equiv \tilde{G}^{ab}(x,y;\tilde{A}) 
\end{align}
as well as the definition of $\tilde{\Delta}_\alpha$ from \eqref{eq:YM1Delta} and the supersymmetry and BRST transformations \eqref{eq:YMsusy11} and \eqref{eq:BRST11} to obtain the final form
\begin{align}\label{eq:YM1R}
\tilde{\mathcal{R}}_g = \tilde{\mathcal{R}}_\mathrm{inv} + \tilde{\mathcal{R}}_\mathrm{gf} 
\end{align}
with
\begin{align}\label{eq:YM1Rinv}
\begin{aligned}
\mathcal{\tilde{R}}_\mathrm{inv} = \frac{\mathrm{d}}{\mathrm{d} g} 
&+ \frac{1}{8g} \int \mathrm{d}^4x \ \mathrm{d}^4y \ \tr \big( \gamma_\mu \tilde{S}^{ab}(x,y;\tilde{A}) \gamma^{\rho \lambda} \big) \tilde{F}_{\rho\lambda}^b(y) \, \frac{\delta}{\delta \tilde{A}_\mu^a(x)} \\
&+ \frac{i}{4g} \int \mathrm{d}^4x \ \mathrm{d}^4y \ \tr \big( \gamma_5 \gamma_\mu \tilde{S}^{ab}(x,y;\tilde{A}) \big) \tilde{D}^b(y) \, \frac{\delta}{\delta \tilde{A}_\mu^a(x)} \\
&+ \frac{1}{g} \int \mathrm{d}^4x \ \tilde{D}^a(x) \, \frac{\delta}{\delta \tilde{D}^a(x)} 
\end{aligned}
\end{align}
and 
\begin{align}\label{eq:YM1Rgf}
\begin{aligned}
\mathcal{\tilde{R}}_\mathrm{gf} &= - g \int \mathrm{d}^4x \ \mathrm{d}^4y \ (D_\mu \tilde{G})^{ab}(x,y;\tilde{A}) \, \mathcal{\tilde{R}}_\mathrm{inv}\left( \textstyle\frac{1}{g} \, \mathcal{G}^\mu \tilde{A}_\mu^b(y) \right) \, \frac{\delta}{\delta \tilde{A}_\mu^a(x)}  \\
&\quad + g f^{abc} \int \mathrm{d}^4x \ \mathrm{d}^4y \  \tilde{G}^{bd}(x,y;\tilde{A}) \, \mathcal{\tilde{R}}_\mathrm{inv}\left( \textstyle\frac{1}{g} \, \mathcal{G}^\mu \tilde{A}_\mu^b(y) \right) \, \tilde{D}^c(x) \, \frac{\delta}{\delta \tilde{D}^a(x)} \, .
\end{aligned}
\end{align}
Like the super Yang-Mills action, also the $\tilde{\mathcal{R}}_g$-operator is divided into a gauge invariant and a gauge fixing part. Since the gauge fixing part is constructed from the two terms with BRST variations in \eqref{eq:YM1ROp2}, its action on any gauge invariant operator $\mathcal{O}_\mathrm{inv}$ is trivial, \emph{i.e.} $(\tilde{\mathcal{R}}_\mathrm{gf} \, \mathcal{O}_\mathrm{inv}) = 0$. Furthermore, since the other term of the $\tilde{\mathcal{R}}_g$-operator is gauge invariant also $(\tilde{\mathcal{R}}_\mathrm{inv} \, O_\mathrm{inv})$ is gauge invariant. This can significantly simplify the calculation of the Nicolai map for gauge invariant operators. 

From the definition of the $\mathcal{N}=1$ super Yang-Mills fermion propagator \eqref{eq:YMFermProp} we obtain
\begin{align}
\frac{\delta \tilde{S}^{ab}(x,y;\tilde{A})}{\delta \tilde{A}_\mu^m(z)} = - f^{cmd} \tilde{S}^{ac}(x,z;\tilde{A}) \gamma^\mu \tilde{S}^{db}(z,y;\tilde{A}) \, .
\end{align}
Similarly, we obtain from \eqref{eq:YMGhostProp} for the ghost propagator
\begin{align}
\frac{\delta \tilde{G}^{ab}(x,y;\tilde{A})}{\delta \tilde{A}_\mu^m(z)} =  f^{cmd} \tilde{G}^{ac}(x,z;\tilde{A}) \overleftarrow{\partial}{\!\!}_z^\mu \tilde{G}^{db}(z,y;\tilde{A}) \, .
\end{align}
In their rescaled formulations, neither one of these propagators depends on the coupling constant $g$. For practical calculations, it is sometimes useful to write out these equations in Dyson-Schwinger (integrated) form
\begin{align}
\begin{aligned}\label{eq:YM1DysonSchwinger}
\tilde{S}^{ab}(x,y;\tilde{A}) &=  \delta^{ab} S_0(x-y) - f^{acd} \int \mathrm{d}^4z \ S_0(x-z) \gamma^\mu \tilde{A}_\mu^c(z) \tilde{S}^{db}(z,y;\tilde{A}) \, , \\
\tilde{G}^{ab}(x,y;\tilde{A}) &= \delta^{ab} G_0(x-y) - f^{acd} \int \mathrm{d}^4z \ G_0(x-z) \mathcal{G}^\mu \tilde{A}_\mu^c(z) \tilde{G}^{db}(z,y;\tilde{A}) \, .
\end{aligned}
\end{align}
These identities can be obtained by multiplying the equation of motion for $\tilde{S}^{ab}(x,y;\tilde{A})$ (respectively $\tilde{G}^{ab}(x,y;\tilde{A})$) with $S_0(y-z)$ (respectively $G_0(x-y)$) and integrating over $y$. 

Like in the previous chapter, the actual Nicolai map is obtained by power series inversion. The relevant equations are \eqref{eq:WZNicolaiMapDef} and \eqref{eq:WZNicolaiMapExp}. However, one must be careful to rescale the fields in the expansion of $(\mathcal{T}_g^{-1} A)_\mu^a$ before again acting with the $\tilde{\mathcal{R}}_g$-operator. Subsequently, we obtain
\begin{align}
(\mathcal{T}_g A)_\mu^a = \sum_{n=0}^\infty \frac{g^n}{n!} (T_n A)_\mu^a \equiv  \sum_{n=0}^\infty \frac{g^n}{n!} (\tilde{T}_n (\textstyle\frac{1}{g}\tilde{A}))_\mu^a  
\end{align}
with
\begin{align}
\begin{aligned}
(T_0 A)_\mu^a &= A_\mu^a \, , \\
(T_n A)_\mu^a &= - \sum_{i=0}^{n-1}  \genfrac(){0pt}{}{n}{i} (\tilde{\mathcal{R}}_g^n (\tilde{T}_i (\textstyle\frac{1}{g}\tilde{A})))_\mu^a \Big\vert_{\tilde{\Phi}= g \Phi} \Big\vert_{g = 0} \, .
\end{aligned}
\end{align}
The corresponding expressions for $(\mathcal{T}_g D)^a$ follow immediately. 
\section{Proof of the Main Theorem}\label{sec:YM1Proof}
As we have pointed out in the last chapter, large parts of the proof are universal to all the supersymmetric field theories in this thesis. Hence, in this chapter we shall only discuss the steps differing from the previous discussion. The first difference is the new part \textit{i)} in the main theorem \ref{th:YM1Theorem}.

\subsection{Part i)}
The gauge fixing functional being a fixed point of the Nicolai map is equivalent to it being annihilated by the $\tilde{\mathcal{R}}_g$-operator. This follows directly from \eqref{eq:YM1InverseNicolaiMap}. We write
\begin{align}
\begin{aligned}
\tilde{\mathcal{R}}_g &= \tilde{\mathcal{R}}_\mathrm{inv}  - g \int \mathrm{d}^4x \ \mathrm{d}^4y \ (D_\mu \tilde{G})^{ab}(x,y;\tilde{A}) \, \mathcal{\tilde{R}}_\mathrm{inv}\left( \textstyle\frac{1}{g} \, \mathcal{G}^\mu \tilde{A}_\mu^b(y) \right) \, \frac{\delta}{\delta \tilde{A}_\mu^a(x)}  \\
&\quad + g f^{abc} \int \mathrm{d}^4x \ \mathrm{d}^4y \  \tilde{G}^{bd}(x,y;\tilde{A}) \, \mathcal{\tilde{R}}_\mathrm{inv}\left( \textstyle\frac{1}{g} \, \mathcal{G}^\mu \tilde{A}_\mu^b(y) \right)  \tilde{D}^c(x) \, \frac{\delta}{\delta \tilde{D}^a(x)} \, .
\end{aligned}
\end{align}
Hence, we find
\begin{align}
\begin{aligned}
\tilde{\mathcal{R}}_g \left( \textstyle{\frac{1}{g}} \mathcal{G}^\mu\tilde{A}_\mu^a(z) \right) &= \tilde{\mathcal{R}}_\mathrm{inv} \left( \textstyle{\frac{1}{g}}  \mathcal{G}^\mu\tilde{A}_\mu^a(z)  \right) \\
&\quad -  \int \mathrm{d}^4x \ \mathrm{d}^4y \ \frac{\delta ( \mathcal{G}^\mu\tilde{A}_\mu^a(z) )}{\delta \tilde{A}_\mu^b(x)}  (\mathcal{D}_\mu \tilde{G})^{bc}(x,y;\tilde{A}) \, \mathcal{\tilde{R}}_\mathrm{inv}\left( \textstyle\frac{1}{g} \,  \mathcal{G}^\mu\tilde{A}_\mu^a(y)  \right) \\
&= 0 \, .
\end{aligned}
\end{align}

\subsection{Part ii)}
For the proof of part \textit{ii)} we must now only show that $\tilde{\mathcal{R}}_g$-operator annihilates the bosonic part of the gauge invariant $\mathcal{N}=1$ super Yang-Mills action. A direct calculation yields
\begin{align}
\begin{aligned}
 \tilde{\mathcal{R}}_\mathrm{inv} \Big( \tilde{S}_\mathrm{inv}^1[g; \tilde{\Phi} ] \Big) 
  &= \frac{\mathrm{d}}{\mathrm{d}g } \left(   \frac{1}{g^2} \int \mathrm{d}^4x \ \left[ - \frac{1}{4} \tilde{F}_{\mu \nu}^a(x) \tilde{F}^{a \, \mu \nu}(x)  + \frac{1}{2} \tilde{D}^a(x) \tilde{D}^a(x) \right] \right)  \\ 
&\quad - \frac{1}{32g^3} \int \mathrm{d}^4x \ \mathrm{d}^4y \ \mathrm{d}^4z \  \tr \big( \gamma_\mu \tilde{S}^{ab}(x,y;\tilde{A}) \gamma^{\rho \lambda} \big) \tilde{F}_{\rho\lambda}^b(y) \frac{\delta \left( \tilde{F}_{\sigma\tau}^c(z) \tilde{F}^{c \, \sigma\tau}(z) \right)}{\delta \tilde{A}_\mu^a(x)}  \\
&\quad - \frac{i}{16g^3} \int \mathrm{d}^4x \ \mathrm{d}^4y \ \mathrm{d}^4z \  \tr \big( \gamma^5 \gamma_\mu \tilde{S}^{ab}(x,y;\tilde{A}) \big) \tilde{D}^b(y) \,  \frac{\delta \left( \tilde{F}_{\nu\lambda}^c(z) \tilde{F}^{c \, \nu\lambda}(z) \right)}{\delta \tilde{A}_\mu^a(x)} \\
&\quad + \frac{1}{2g^3} \int \mathrm{d}^4x \ \mathrm{d}^4y \ \tilde{D}^a(x) \, \frac{\delta \tilde{D}^b(y) \tilde{D}^b(y)}{\delta \tilde{D}^a(x)} \\
&= - \frac{2}{g^3} \int \mathrm{d}^4x \ \left[ - \frac{1}{4} \tilde{F}_{\mu \nu}^a(x) \tilde{F}^{a \, \mu \nu}(x) \color{green!20!blue}  + \frac{1}{2} \tilde{D}^a(x) \tilde{D}^a(x)   \color{black} \right] \\ 
&\quad \color{green!50!black} + \frac{1}{8g^3} \int \mathrm{d}^4x \ \mathrm{d}^4y \  \tr \big( \gamma^\mu \tilde{S}^{ab}(x,y;\tilde{A}) \gamma^{\rho \lambda} \big) \tilde{F}_{\rho\lambda}^b(y)   (D^\nu \tilde{F}_{\nu\mu})^a(x)  \\
&\quad \color{green!50!black} + \frac{i}{4g^3} \int \mathrm{d}^4x \ \mathrm{d}^4y \  \tr \big( \gamma^5 \gamma^\mu \tilde{S}^{ab}(x,y;\tilde{A}) \big) \tilde{D}^b(y) (D^\nu \tilde{F}_{\nu\mu})^a(x)\\
&\quad \color{green!20!blue} + \frac{1}{g^3} \int \mathrm{d}^4x  \ \tilde{D}^a(x) \tilde{D}^a(x) \color{black} \, .
\end{aligned}
\end{align}
The {\color{green!20!blue}blue} terms cancel. In the two {\color{green!50!black}green} terms we use
\begin{align}
\begin{aligned}\label{eq:YM1DF}
\gamma^\mu  (D^\nu \tilde{F}_{\nu\mu})^a(x) &= \gamma^{[\mu} \eta^{\nu]\rho}  (D_\sigma \tilde{F}_{\nu\mu})^a(x) = \frac{1}{2} \gamma^{\mu\nu}  \gamma^\rho (D_\rho \tilde{F}_{\nu\mu})^a(x) - \frac{1}{2} \gamma^{\mu\nu\rho}  (D_\rho \tilde{F}_{\nu\mu})^a(x)   \, .
\end{aligned}
\end{align}
Due to the Bianchi identity $\gamma^{\mu\nu\rho}  (D_\rho \tilde{F}_{\nu\mu})^a = 0$. Subsequently, we integrate by parts and obtain
\begin{align}
\begin{aligned}
\tilde{\mathcal{R}}_\mathrm{inv} \Big( \tilde{S}_\mathrm{inv}^1[g; \tilde{\Phi} ] \Big)   &=  \frac{1}{2g^3} \int \mathrm{d}^4x \  \tilde{F}_{\mu \nu}^a(x) \tilde{F}^{a \, \mu \nu}(x)  \\ 
&\quad - \frac{1}{16g^3} \int \mathrm{d}^4x \ \mathrm{d}^4y \  \tr \big( \gamma^{\mu\nu}  \gamma^\sigma  (D_\sigma \tilde{S})^{ab}(x,y;\tilde{A}) \gamma^{\rho \lambda} \big) \tilde{F}_{\rho\lambda}^b(y) \tilde{F}_{\nu\mu}^a(x)  \\
&\quad - \frac{i}{8g^3} \int \mathrm{d}^4x \ \mathrm{d}^4y  \  \tr \big( \gamma^5 \gamma^{\mu\nu}  \gamma^\sigma (D_\sigma \tilde{S})^{ab}(x,y;\tilde{A}) \big) \tilde{D}^b(y) \tilde{F}_{\nu\mu}^a(x) \, .
\end{aligned}
\end{align}
We identify the Dirac equation $\gamma^\sigma (D_\sigma \tilde{S})^{ab}(x,y;\tilde{A}) = \delta^{ab} \delta(x-y)$ and thus get
\begin{align}
\begin{aligned}
\tilde{\mathcal{R}}_\mathrm{inv} \Big( \tilde{S}_\mathrm{inv}^1[g; \tilde{\Phi} ] \Big)   &= \color{red}  \frac{1}{2g^3} \int \mathrm{d}^4x \  \tilde{F}_{\mu \nu}^a(x) \tilde{F}^{a \, \mu \nu}(x) - \frac{1}{16g^3} \int \mathrm{d}^4x  \  \tr \big( \gamma^{\mu\nu} \gamma^{\rho \lambda} \big) \tilde{F}_{\rho\lambda}^a(x) \tilde{F}_{\nu\mu}^a(x)  \\ 
&\quad - \frac{i}{8g^3} \int \mathrm{d}^4x \  \tr \big( \gamma^5 \gamma^{\mu\nu} \big) \tilde{D}^b(y) \tilde{F}_{\nu\mu}^a(x)  \\
&= 0\, .
\end{aligned}
\end{align}
The {\color{red}red} terms cancel upon taking the trace. The remaining term vanishes since $\tr(\gamma^5 \gamma^{\mu\nu}) = 0$. Furthermore, we notice that the $\tilde{\mathcal{R}}_g$-operator annihilates the two terms of the gauge invariant bosonic action separately. This concludes the proof of part \textit{i)}. 

\subsection{Part iii)}
The proof of part \textit{iii)} is again largely similar to the discussion in the previous chapter. The only difference is the appearance of the Faddeev-Popov determinant from the integration over the ghost fields, \emph{i.e.}
\begin{align}
\begin{aligned}
\big< \!\! \big< X[\Phi]  \big> \!\! \big>_g &= \int \mathcal{D}\Phi \ \mathcal{D}\Psi \ e^{-i S_\mathrm{inv}^1[g;\Phi,\Psi] - i S_\mathrm{gf}[g;\Phi,\Psi] } X[\Phi] \\
&= \int \mathcal{D}_0\Phi \ \Delta_\mathrm{MSS}[\Phi] \Delta_\mathrm{FP}[\Phi] \, e^{-i S_\mathrm{inv}^1[g;\Phi]-iS_\mathrm{gf}[g,\Phi]} X[\Phi] \, . 
\end{aligned}
\end{align}
Thus we obtain
\begin{align}
\mathcal{J}(\mathcal{T}_\lambda \, \Phi) = \Delta_\mathrm{MSS}[\Phi] \Delta_\mathrm{FP}[\Phi] 
\end{align}
at least order by order in perturbation theory. This concludes the proof. 

\section[Properties of the \texorpdfstring{$\tilde{\mathcal{R}}_g$}{R}-Operator]{Properties of the $\boldsymbol{\tilde{\mathcal{R}}_g}$-Operator}\label{sec:YM1ROpProperties}
In the proof of the main theorem \ref{th:YM1Theorem}, we have seen that $\tilde{\mathcal{R}}_g$ annihilates the two terms of the gauge invariant bosonic action separately. Furthermore, neither the gauge condition nor the Matthews-Salam-Seiler or Faddeev-Popov determinant depend on the auxiliary field $\tilde{D}^a$. Hence, we shall disregard all terms involving the auxiliary field in the remainder of this chapter. Nevertheless the Nicolai map $(\mathcal{T}_g \, A)_\mu^a$ still satisfies \textit{i) - iii)} of the main theorem \ref{th:YM1Theorem} for any gauge with the scaling property \eqref{eq:YM1GScaling}. Finally, let us point out that when considering `on-shell' supersymmetry, the auxiliary field will drop out anyway.

Furthermore, we want to study the well-definedness of \eqref{eq:YM1InverseNicolaiMap}. A priori it is not obvious that the limit $g \to 0$ in $\big( \mathcal{\tilde{R}}_{g}^n \big({\textstyle\frac{1}{g}} \tilde{A} \big) \big)_\mu^a(x)  \, \big\vert_{\tilde{\Phi}= g\Phi} \big\vert_{g=0}$ exists. When computing the inverse Nicolai map $\mathcal{T}_g^{-1}$ to some arbitrary order in the coupling we must act with $\tilde{\mathcal{R}}_g$ on $(\frac{1}{g} \tilde{A}_\mu^a)$, the fermion propagator $\tilde{S}^{ab}(x,y;\tilde{A})$ and the ghost propagator $\tilde{G}^{ab}(x,y;\tilde{A})$. We want to check that each of these calculations gives a finite result when replacing $\tilde{A}_\mu^a = gA_\mu^a$ and taking the limit $g \to 0$. Therefore let us rewrite \eqref{eq:YM1ROp1} and \eqref{eq:YM1ROp2} by means of the identity \cite{Lechtenfeld:1984me}
\begin{align}
\gamma^{\rho\lambda} \tilde{F}_{\rho\lambda}^b = 2 \gamma^\rho \gamma^\lambda (D_\rho \tilde{A}_\lambda)^b - 2 \partial^\lambda \tilde{A}_\lambda^b - 
f^{bde} \gamma^{\rho\lambda} \tilde{A}_\rho^d \tilde{A}_\lambda^e \, ,
\end{align}
leaving the gauge functional arbitrary. Integrating by parts, so $D_\rho$ acts on the fermionic propagator to give a $\delta$-function, then leads to the new representation
\begin{align}\label{eq:YM1R0R1R2}
\mathcal{\tilde{R}}_g = \mathcal{\tilde{R}}_0 + \mathcal{\tilde{R}}_1 + \mathcal{\tilde{R}}_2
\end{align}
with the counting operator (now with $\tilde{D}^a=0$)
\begin{align}\label{eq:YM1CountingOp}
\mathcal{\tilde{R}}_0 \coloneqq \frac{\mathrm{d}}{\mathrm{d} g} + \frac{1}{g} \int \mathrm{d}^4x \ \tilde{A}_\mu^a(x) \, \frac{\delta}{\delta \tilde{A}_\mu^a(x)} \, .
\end{align}
The other two operators are given by
\begin{align}
\begin{aligned}\label{eq:YM1R1}
\mathcal{\tilde{R}}_1 \coloneqq &- \frac{1}{8g} \int \mathrm{d}^4x \ \mathrm{d}^4y \ \tr \left( \gamma_\mu \tilde{S}^{ab}(x,y;\tilde{A}) \gamma^{\rho \lambda} \right) f^{bcd} \tilde{A}_\rho^c(y) \tilde{A}_\lambda^d(y) \frac{\delta}{\delta \tilde{A}_\mu^a(x)} \\
&- \frac{1}{8g} \int \mathrm{d}^4 x \ \mathrm{d}^4 y \ \mathrm{d}^4 z \ (D_\mu \tilde{G})^{ae}(x,z;\tilde{A}) \,
\tr \left( \gamma_\nu  \mathcal{G}^\nu \tilde{S}^{eb}(z,y;\tilde{A}) \gamma^{\rho\lambda} \right) \\
&\quad \quad  \times  f^{bcd} \tilde{A}_\rho^c(y) \tilde{A}_\lambda^d(y) \, \frac{\delta}{\delta \tilde{A}_\mu^a(x)} 
\end{aligned}
\end{align}
and 
\begin{align}
\begin{aligned}\label{eq:YM1R2}
\mathcal{\tilde{R}}_2 \coloneqq &- \frac{1}{4g} \int \mathrm{d}^4 x \ \mathrm{d}^4 y \ \tr \left( \gamma_\mu \tilde{S}^{ab}(x,y;\tilde{A}) \right) \partial^\lambda \tilde{A}_\lambda^b(y) \, \frac{\delta}{\delta \tilde{A}_\mu^a(x)} \\
&+ \frac{1}{4g} \int \mathrm{d}^4 x \ \mathrm{d}^4 y \ \mathrm{d}^4 z \ (D_\mu \tilde{G})^{ab}(x,y;\tilde{A}) \, 
\tr \left( \gamma_\nu  \mathcal{G}^\nu \tilde{S}^{bc}(y,z;\tilde{A}) \right) \partial^\lambda \tilde{A}_\lambda^c(z) \, \frac{\delta}{\delta \tilde{A}_\mu^a(x)} .
\end{aligned}
\end{align}
The counting operator $\mathcal{\tilde{R}}_0$ obeys
\begin{align}
\mathcal{\tilde{R}}_0 (A_\mu^a) \equiv \mathcal{\tilde{R}}_0\left( \textstyle\frac{1}{g}\tilde{A}_\mu^a \right) = 0 
\end{align}
as well as relations like
\begin{align}
\mathcal{\tilde{R}}_0 \left( \tilde{S}^{ab}(x,y;\tilde{A}) \right) = -  f^{cde}  \int \mathrm{d}^4 z \ \tilde{S}^{ac}(x,z;\tilde{A})
\gamma^\mu  \tilde{A}_\mu^d(z) \tilde{S}^{eb}(z,y;\tilde{A}) \qquad \emph{etc.}
\end{align}
Counting the powers of $\tilde{A}$ shows that also $\tilde{\mathcal{R}}_1$ gives finite results for $(\frac{1}{g}\tilde{A}_\mu^a)$, $\tilde{S}^{ab}(x,y;\tilde{A})$ and $\tilde{G}^{ab}(x,y;\tilde{A})$ in the limit $A = g \tilde{A}$ and $g \to 0$. For $\tilde{\mathcal{R}}_2$, we need to be a bit more careful. We consider the potentially singular zeroth order contributions in both integrands of \eqref{eq:YM1R2}, using the Dyson-Schwinger identities \eqref{eq:YM1DysonSchwinger}, 
\begin{align}
\begin{aligned}
\tilde{S}^{ab} (x,y;\tilde{A}) &= - \delta^{ab} \gamma^\rho \partial_\rho C(x-y) + \mathcal{O}(\tilde{A}) \, , \\
\tilde{G}^{ab} (x,y;\tilde{A}) & = \delta^{ab} G_0(x-y) + \mathcal{O}(\tilde{A}) \, ,
\end{aligned}
\end{align} 
where $C(x)$ is the free scalar propagator obeying $\Box C(x) = - \delta(x)$. We can ignore the $\mathcal{O}(\tilde{A})$ terms since they are non-singular as $g\rightarrow 0$. If we had chosen a more general gauge in agreement with \eqref{eq:YM1GScaling}, the second term in \eqref{eq:YM1R2} would potentially have another $\tilde{A}$ dependence in front of the fermion propagator. However, it also follows from \eqref{eq:YM1GScaling} that
\begin{align}
\frac{\delta \mathcal{G}^a[\tilde{A}](x)}{\delta \tilde{A}_\mu^b(y)} = \delta^{ab} \mathcal{G}^\mu \delta(x-y) + \mathcal{O}(\tilde{A}) \, .
\end{align}
Hence, this does not harm the $g \to 0$ limit. For the Landau gauge ($\mathcal{G}^\mu = \partial^\mu$) the cancellation of the singular term follows easily upon use of $\gamma^\mu \partial_\mu S_0(x) =  \delta(x)$ and $G_0(x) = -C(x)$. For the axial gauge ($\mathcal{G}^\mu = n^\mu$), we compute
\begin{align}
\tr \left( \gamma_\mu n^\mu S_0(y-z) \right) = - 4 n^\mu \partial_\mu C(y-z) \, ,
\end{align}
integrate by parts, and use the defining equation for the free ghost propagator $n^\mu \partial_\mu G_0^{ab}(x) = \delta^{ab} \delta(x)$ to show that these contributions cancel again (as we pointed out, higher order terms in the gauge functional do not affect this argument). All remaining terms in \eqref{eq:YM1R2} are at least of order $\tilde{A}$ and therefore possess a well-defined limit for $g \to 0$. 

\newpage
\section{The Nicolai Map in Axial Gauge}\label{eq:YM1Result}
We present the expansion of the Nicolai map $(\mathcal{T}_g \, A)_\mu^a(x)$ in axial gauge $\mathcal{G}^\mu A_\mu^a = n^\mu A_\mu^a$ up to the second order in the coupling \cite{Malcha:2021ess}\footnote{As explained above we have disregarded all terms containing the auxiliary field $D^a$.}
\begin{footnotesize}
\begin{align}
\begin{aligned}\label{eq:YM1TAxial}
\left(\mathcal{T}_g \, A \right)_\mu^a(x) &= A_\mu^a(x) + g f^{abc} \int \mathrm{d}^4y \ \mathrm{d}^4z \ \left( \eta_{\mu\nu} \delta(x-y) - \partial_\mu G_0(x-y) n_\nu \right) \\
&\quad \quad \quad \times \Big\{ A^{b \, \nu}(y) C(y-z) \, \partial \cdot \! A^c(z) + \partial^\lambda C(y-z) A^{b \, \nu}(z) A_\lambda^c(z) \Big\} \\
&\quad + 2 g f^{abc} \int \mathrm{d}^4y \ \mathrm{d}^4z \ \mathrm{d}^4w \ \left( \eta_{\mu\nu} \delta(x-y) - \partial_\mu G_0(x-y) n_\nu \right) \\
&\quad \quad \quad \times \partial_\lambda C(y-z) A^{b \, [\nu}(z) \partial^{\lambda]} C(z-w) \, \partial \cdot \! A^c(w) \\
&\quad + \frac{g^2}{2} f^{abc} f^{bde} \int \mathrm{d}^4y \ \mathrm{d}^4z \ \mathrm{d}^4w \ \left( \eta_{\mu\nu} \delta(x-y) - \partial_\mu G_0(x-y) n_\nu \right) \ \Big\{ \\
&\quad \quad \quad - 2 A^{c \, \nu}(y) C(y-z) A_\lambda^d(z) \partial^\lambda C(z-w) \, \partial \cdot \! A^e(w) \\
&\quad \quad \quad - A^{c \, \nu}(y) C(y-z) \, \partial \cdot \! A^d(z) C(z-w) \, \partial \cdot \! A^e(w) \\ 
&\quad \quad \quad - \frac{1}{2} C(y-z) \, \partial \cdot \! A^c(z) \partial^\lambda C(z-w) A^{d \, \nu}(w) A_\lambda^e(w) \\ 
&\quad \quad \quad + \frac{1}{2} C(y-z) \, \partial \cdot \! A^c(z) \partial^\lambda C(y-w) A^{d \, \nu}(w) A_\lambda^e(w) \\ 
&\quad \quad \quad - \frac{1}{2} C(y-z) A^{d \, \nu}(z) A_\lambda^e(z) \partial^\lambda C(z-w) \, \partial \cdot \! A^c(w) \\ 
&\quad \quad \quad + \frac{1}{2} \partial^\lambda C(y-z) A^{d \, \nu}(z) A_\lambda^e(z) C(z-w) \, \partial \cdot \! A^c(w) \\
&\quad \quad \quad - 2\partial_\lambda C(y-z) A^{c \, [\nu}(z) A^{d\, \lambda]}(z) C(z-w) \, \partial \cdot \! A^e(w) \\
&\quad \quad \quad + 3 \partial_\rho C(y-z) A_\lambda^c(z) \partial^{[\nu} C(z-w) A^{d\, \lambda}(w) A^{e \, \rho]}(w) \Big\} \\ 
&\quad + \frac{g^2}{2} f^{abc} f^{bde} \int \mathrm{d}^4y \ \mathrm{d}^4z \ \mathrm{d}^4w \ \mathrm{d}^4v \ \left( \eta_{\mu\nu} \delta(x-y) - \partial_\mu G_0(x-y) n_\nu \right) \ \Big\{ \\
&\quad \quad \quad - C(y-z) A^{d \, [\nu}(z) \partial^{\lambda]} C(z-w) \, \partial \cdot \! A^e(w) \partial_\lambda C(z-v) \, \partial \cdot \! A^c(v) \\ 
&\quad \quad \quad - C(y-z) \, \partial \cdot \! A^c(z) \partial_\lambda C(z-w) A^{d\, [\nu}(w) \partial^{\lambda]} C(w-v) \, \partial \cdot \! A^e(v) \\
&\quad \quad \quad + C(y-z) \, \partial \cdot \! A^c(z) \partial_\lambda C(y-w) A^{d\, [\nu}(w) \partial^{\lambda]} C(w-v) \, \partial \cdot \! A^e(v) \\
&\quad \quad \quad - \partial_\lambda C(y-z) A^{d \, [\nu}(z) \partial^{\lambda]} C(z-w) \, \partial \cdot \! A^e(w) C(z-v) \, \partial \cdot \! A^c(v) \\ 
&\quad \quad \quad - \partial^\lambda C(y-z) \partial^\nu C(z-w) \, \partial \cdot \! A^c(w) \partial^\rho C(z-v) A_\lambda^d(v) A_\rho^e(v) \\ 
&\quad \quad \quad + 2 \partial_\lambda C(y-z) \partial^{[\nu} A^{d\, \lambda]}(z) C(z-w) \, \partial \cdot \! A^e(w) C(z-v) \, \partial \cdot \! A^c(v) \\
&\quad \quad \quad - 2 \partial_\lambda C(y-z) A^{c \, [\nu}(z) \partial^{\lambda]} C(z-w) \, \partial \cdot \! A^d(w) C(w-v) \, \partial \cdot \! A^e(v) \\ 
&\quad \quad \quad - 4 \partial_\lambda C(y-z) A^{c \, [\nu}(z) \partial^{\lambda]} C(z-w) A_\rho^d(w) \partial^\rho C(w-v) \, \partial \cdot \! A^e(v) \\ 
&\quad \quad \quad + 6 \partial_\rho C(y-z) A_\lambda^c(z) \partial^{[\nu} C(z-w) A^{d \, \lambda}(w) \partial^{\rho]} C(w-v) \, \partial \cdot \! A^e(v) \Big\} \\ 
&\quad - g^2 f^{abc} f^{bde} \int \mathrm{d}^4y \ \mathrm{d}^4z \ \mathrm{d}^4w \ \mathrm{d}^4v \ \mathrm{d}^4u \ \left( \eta_{\mu\nu} \delta(x-y) - \partial_\mu G_0(x-y) n_\nu \right) \\
&\quad \quad \quad \times \partial^\lambda C(y-z) \partial^\nu C(z-w) \, \partial \cdot \! A^c(w) \partial^\rho C(z-v) A_{[\lambda}^d(v) \partial_{\rho]} C(v-u) \, \partial \cdot \! A^e(u) \\
&\quad + \mathcal{O}(g^3) \, . 
\end{aligned}
\end{align}
\end{footnotesize}%
\newpage
We point out that this result is, in principle, valid for all $n^\mu$, regardless of whether they are time-like, space-like or null. It, therefore, applies to the light-cone gauge as well. However, \eqref{eq:YM1TAxial} is substantially more complicated than the corresponding expression for the Wess-Zumino model \eqref{eq:WZResult}. There the first and second order each only consisted of a single term. In section \ref{sec:YM1OnShell} we explain how to simplify this result when accepting some restrictions. 

Recall that the free axial gauge ghost propagator from \eqref{eq:YMaxialGhostProp} satisfies
\begin{align}\label{eq:YM1GhostProp}
n^\mu \partial_\mu G_0(x-y) = \delta(x-y) \, .
\end{align}
In writing the above result, we have regrouped the terms in such a way that they all appear with the axial projector
\begin{align}\label{eq:YM2Projector}
\Pi_{\mu\nu}(x) \coloneqq \eta_{\mu\nu} \delta(x) - \partial_\mu G_0(x) n_\nu
\end{align}
in front. This projector obeys $n^\mu \Pi_{\mu\nu}(x) = 0$ (but $\Pi_{\mu\nu}(x) n^\nu \not= 0$). By \eqref{eq:YM1GScaling} we further have
\begin{align}\label{eq:YM1ProjectorId}
\int \mathrm{d}^4y \ \Pi_{\mu\nu}(x-y) \partial^\nu F(y) = 0 
\end{align}
for any function $F$. Hence, the second order result in axial gauge can be written in such a way that it differs from the result in the $R_\xi$ type gauges only by the insertion of this projector, since all terms of type \eqref{eq:YM1ProjectorId} drop out.

\section{Tests}\label{sec:YM1Tests}
There are three tests we can perform on \eqref{eq:YM1TAxial}. They correspond to the three main properties of the Nicolai map from the theorem \ref{th:YM1Theorem}. In the following let $A_\mu^{\prime \, a} \equiv (\mathcal{T}_g  \, A)_\mu^a$. 
\subsection{The Gauge Condition}
The preservation of the gauge condition 
\begin{align}\label{eq:YM1Test0}
n^\mu A_\mu^{\prime \, a}(x) = n^\mu A_\mu^a(x) + \mathcal{O}(g^3) 
\end{align}
is trivially satisfied up to the order considered. This is because the axial projector obeys $n^\mu \Pi_{\mu\nu}(x) = 0$. So in particular $n^\mu A_\mu^{\prime \, a}(x) = n^\mu A_\mu^a(x)$.

\subsection{The Free Action}
By the second statement in the main theorem, the bosonic Yang-Mills action without gauge-fixing terms is mapped to the abelian action. We integrate the abelian action by parts and obtain
\begin{align}\label{eq:YM1FreeAction}
\frac{1}{2} \int \mathrm{d}^4x \ A_\mu^{\prime \, a} (- \Box \, \eta^{\mu\nu} + \partial^\mu \partial^\nu ) A_\nu^{\prime a} \overset{!}{=} \frac{1}{4} \int \mathrm{d}^4 x \ F_{\mu\nu}^a F^{a\, \mu\nu} + \mathcal{O}(g^3) \, .
\end{align}
In the leading order, the statement is trivial. We notice that any term which can be written as a total derivative $\partial_\mu^x \left( \ldots \right)$ does not contribute by the gauge invariance of the free action. In particular this reduces the axial projector $\Pi_{\mu\nu}(x-y) = \eta_{\mu\nu} \delta(x-y) - \partial_\mu G_0(x-y) n_\nu$ to simply $\eta_{\mu\nu} \delta(x-y)$.  In the first order, we find for the left-hand side of \eqref{eq:YM1FreeAction}
\begin{align}
\begin{aligned}
&\quad \frac{1}{2} \int \mathrm{d}^4x \ A_\mu^{\prime \, a}(x) (- \Box \, \eta^{\mu\nu} + \partial^\mu \partial^\nu ) A_\nu^{\prime \, a}(x) \bigg\vert_{\mathcal{O}(g^1)} \\
& = g f^{abc} \int \mathrm{d}^4x \ \mathrm{d}^4y \  A_\mu^b(x) C(x-y) \, \partial \cdot \! A^c(y) (- \Box \,  \eta^{\mu\nu} + \partial^\mu \partial^\nu ) A_\nu^a(x)  \\
&\quad + g f^{abc} \int \mathrm{d}^4x \ \mathrm{d}^4y \  \partial^\lambda C(x-y) A_\mu^b(y) A_\lambda^c(y) (- \Box \,  \eta^{\mu\nu} + \partial^\mu \partial^\nu ) A_\nu^a(x)  \\
&\quad  + 2 g f^{abc} \int \mathrm{d}^4x \ \mathrm{d}^4y \ \mathrm{d}^4z \ \partial^\lambda C(x-y) A_{[\mu}^b(y) \partial_{\lambda]} C(y-z) \, \partial \cdot \! A^c(z) (- \Box \, \eta^{\mu\nu} + \partial^\mu \partial^\nu ) A_\nu^a(x) \, .
\end{aligned}
\end{align}
We integrate by parts and remove anti-symmetric terms
\begin{align}
\begin{aligned}
&\quad \frac{1}{2} \int \mathrm{d}^4x \ A_\mu^{\prime \, a}(x) (- \Box \, \eta^{\mu\nu} + \partial^\mu \partial^\nu ) A_\nu^{\prime \, a}(x) \bigg\vert_{\mathcal{O}(g^1)} \\
&= - g f^{abc} \int \mathrm{d}^4x \ \partial_\lambda A_\mu^a(x) A^{b \, \mu}(x) A^{c \, \lambda}(x) \\
&\quad + g f^{abc} \int \mathrm{d}^4x \ \mathrm{d}^4y \ \Big\{ \\
&\quad \quad \quad - \Box A_\mu^a(x) A^{b \, \mu}(x) C(x-y) \, \partial \cdot \! A^c(y) \\
&\quad \quad \quad + \partial_\mu \, \partial \cdot \! A^a(x) A^{b \, \mu}(x) C(x-y) \, \partial \cdot \! A^c(y) \\
&\quad \quad \quad  - 2 \partial_\lambda A_\mu^a(x) A^{b \, [\mu}(x) \partial^{\lambda]} C(x-y) \, \partial \cdot \! A^c(y) \Big\} \\
&= g f^{abc} \int \mathrm{d}^4x \ \partial_\mu A_\lambda^a(x) A^{b \, \mu}(x) A^{c \, \lambda}(x) \\
&= \frac{1}{4} \int \mathrm{d}^4x \ F_{\mu\nu}^a(x) F^{a \, \mu\nu}(x) \bigg\vert_{\mathcal{O}(g^1)} \, .
\end{aligned}
\end{align}
In the second order, the steps are generally the same. However, we will need to use the Jacobi identity \eqref{eq:JacId}
\begin{align}
f^{abc} f^{ade} + f^{abd} f^{aec} + f^{abe} f^{acd} = 0 \,.
\end{align}
Furthermore, we can again disregard half the terms because of the axial projector $\Pi_{\mu\nu}(x-y)$. After carefully collecting all the terms contributing to the left-hand side of \eqref{eq:YM1FreeAction} at $\mathcal{O}(g^2)$, we perform similar partial integrations as above and obtain
\begin{align}
\begin{aligned}
&\quad \frac{1}{2} \int \mathrm{d}^4x \ A_\mu^{\prime \, a}(x) (- \Box \, \eta^{\mu\nu} + \partial^\mu \partial^\nu ) A_\nu^{\prime \, a}(x) \bigg\vert_{\mathcal{O}(g^2)} \\
&= \int \mathrm{d}^4x \ A_\mu^{\prime \, a}(x) \big\vert_{\mathcal{O}(g^2)} ( - \Box \, \eta^{\mu\nu} + \partial^\mu \partial^\nu ) A_\nu^{\prime a}(x) \big\vert_{\mathcal{O}(g^0)} \\
&\quad  + \frac{1}{2} \int \mathrm{d}^4x \ A_\mu^{\prime \, a}(x) \big\vert_{\mathcal{O}(g^1)} ( - \Box \, \eta^{\mu\nu} + \partial^\mu \partial^\nu ) A_\nu^{\prime a}(x) \big\vert_{\mathcal{O}(g^1)} \\
&= - \frac{g^2}{4} f^{abc} f^{bde} \int \mathrm{d}^4x \ A_\mu^a(x) A_\lambda^c(x) A^{d\, \mu}(x) A^{e \, \lambda}(x) \\ 
&\quad- \frac{g^2}{2} \int \mathrm{d}^4x \ \mathrm{d}^4y \ A_\mu^a(x) A_\lambda^e(x) \partial^\lambda A^{d \, \mu}(x) C(x-y) \, \partial \cdot \! A^c(y) \\
&\quad \quad \quad  \times \left( f^{abc} f^{bde} + f^{eba} f^{bdc} + f^{cbe} f^{bda} \right) \\
&=  \frac{g^2}{4} f^{abc} f^{ade} \int \mathrm{d}^4x \ A_\mu^b(x) A_\lambda^c(x) A^{d\, \mu}(x) A^{e \, \lambda}(x) \\
& = \frac{1}{4} \int \mathrm{d}^4x \ F_{\mu\nu}^a(x) F^{a \, \mu\nu}(x) \bigg\vert_{\mathcal{O}(g^2)} \, .
\end{aligned}
\end{align}
Thus \eqref{eq:YM1FreeAction} is satisfied. These relations are independent of dimension.

\subsection{Jacobian, Fermion and Ghost Determinant}
Finally, we need to perturbatively show that the Jacobian determinant is equal to the product of the MSS and FP determinants. This is done order by order by considering the logarithms of the determinants rather than the determinants themselves, \emph{i.e.}
\begin{align}\label{eq:YM1Determinants}
\log \mathcal{J}(\mathcal{T}_g A) = \log \det \left( \frac{\delta A_\mu^{\prime \, a}(x)}{\delta A_\nu^b(y)} \right) \overset{!}{=} \log \left( \Delta_\mathrm{MSS}[g;A] \Delta_\mathrm{FP}[g;A] \right) \, .
\end{align}
For the logarithm on the right-hand side, remember that $\log(a \cdot b) = \log(a) + \log(b)$. In section \ref{sec:YMCorrFct} we have discussed how to obtain the Matthews-Salam-Seiler and Faddeev-Popov determinants. Recall the results for the MSS determinant \eqref{eq:YMLogMSS}
\begin{align}
\begin{aligned}\label{eq:YM1LogMSS}
\log(\Delta_\mathrm{MSS}[g;A]) &= g^2 N \int \mathrm{d}^2 x \ \mathrm{d}^4 y \ \Big\{ \\ 
&\quad \quad + 2\, \partial^\mu C(x-y) A_\mu^a(y) \partial^\rho C(y-x) A_\rho^a(x) \\
&\quad \quad - \partial^\mu C(x-y) A^{a \, \rho}(y) \partial_\mu C(y-x) A_\rho^a(x) \Big\} \\
&\quad + \mathcal{O}(g^3) \, .
\end{aligned}
\end{align}
And the FP determinant in axial gauge \eqref{eq:YMLogFPAxial}
\begin{align}
\begin{aligned}\label{eq:YM1LogFPAxial}
\log( \Delta_\mathrm{FP}[g;A]) &=  \frac{g^2 N}{2}  \int \mathrm{d}^4x \ \mathrm{d}^4y \ G_0(x-y) n \cdot A^c(y) G_0(y-x) n \cdot A^d(x) + \mathcal{O}(g^3) \, .
\end{aligned}
\end{align}
In the leading order, there is no contribution from the Jacobian determinant since $\log 1 = 0$. At $\mathcal{O}(g)$, the logarithm of the Jacobian determinant is proportional to $f^{aab} = 0$. Also, the MSS and FP determinants have no contribution in this order. At $\mathcal{O}(g^2)$ the logarithm of the Jacobian determinant consists of two terms 
\begin{align}\label{eq:YM1Intermediate2}
\log \det \left( \frac{\delta A_\mu^{\prime \, a}(x)}{\delta A_\nu^b(y)} \right) \Bigg\vert_{\mathcal{O}(g^2)} = \Tr \left[ \frac{\delta A^\prime}{\delta A} \bigg\vert_{\mathcal{O}(g^2)} \right] - \frac{1}{2} \Tr \left[ \frac{\delta A^\prime}{\delta A} \bigg\vert_{\mathcal{O}(g^1)} \frac{\delta A^\prime}{\delta A} \bigg\vert_{\mathcal{O}(g^1)} \right] \, ,
\end{align}
where the trace is done by setting $\nu=\mu$, $b=a$, $y=x$ and integrating over $x$. The computation is straightforward, but we must be careful with formally divergent terms.
The first term gives
\begin{align}
\begin{aligned}\label{eq:YM1Jac1}
\tr \left[ \frac{\delta A^\prime}{\delta A} \bigg\vert_{\mathcal{O}(g^2)} \right] &= g^2 N \int \mathrm{d}^4x \ \mathrm{d}^4y \ \Big\{ \\ 
&\quad \quad -  \partial_\mu C(x-y) A_\nu^a(y) \partial^\mu C(y-x) A^{a \, \nu}(x) \\ 
&\quad \quad \color{green!20!blue} - C(x-y)  \, \partial^\mu \left( A_\mu^a(y) G_0(y-x) \right)  \, n \! \cdot \! A^a(x)  \\ 
&\quad \quad \color{green!20!blue} + \frac{1}{4} C(x-y) \, \partial \cdot \! A^a(y) \left( C(y-x) - 2 C(0) \right) \, \partial \cdot \! A^a(x)  \color{black} \Big\} \\ 
&\quad+ g^2 N \int \mathrm{d}^4x \ \mathrm{d}^4y \ \mathrm{d}^4z \ \Big\{ \\
&\quad \quad \color{green!20!blue} + \frac{1}{4} G_0(x-z) n^\mu C(z-x) \, \partial \cdot \! A^a(y)  \partial_\mu C(y-x) \, \partial \cdot \! A^a(x) \\ 
&\quad \quad \color{green!20!blue} + 2 G_0(x-z) \partial^\mu C(z-x) \, \partial \cdot \! A^a(y) n^\nu \partial_{ \{\nu} C(y-x) A_{\mu \}}^a(x) \\
&\quad \quad \color{green!20!blue} - 2 G_0(x-z) \partial^\mu C(z-y) A_\nu^a(y) n^\lambda \partial^\nu \partial_{\{\lambda} C(y-x) A_{\mu\}}^a(x) \\
&\quad \quad \color{green!20!blue} + \frac{3}{2} \delta(0) C(z-y) \, \partial \cdot \! A^a(y) C(z-x) \, \partial \cdot \! A^a(x)  \color{black} \Big\} \, .
\end{aligned}
\end{align}
The second term gives
\begin{align}
\begin{aligned}\label{eq:YM1Jac2}
- \frac{1}{2} \tr \left[ \frac{\delta A^\prime}{\delta A} \bigg\vert_{\mathcal{O}(g^1)} \frac{\delta A^\prime}{\delta A} \bigg\vert_{\mathcal{O}(g^1)} \right] 
&= g^2 N \int \mathrm{d}^4x \ \mathrm{d}^4y \ \Big\{ \\
&\quad \quad + 2 \partial^\mu C(x-y) A_\mu^a(y) \partial^\nu C(x-y) A_\nu^a(x) \\
&\quad \quad + \frac{1}{2} G_0(x-y) \, n \! \cdot \! A^a(y) G_0(y-x) \, n \! \cdot \! A^a(x) \\ 
&\quad \quad  \color{green!20!blue} + C(x-y)  \, \partial^\mu \Big( A_\mu^a(y) G_0(y-x) \Big)  \, n \! \cdot \! A^a(x)  \\ 
&\quad \quad  \color{green!20!blue} - \frac{1}{4} C(x-y) \, \partial \cdot \! A^a(y) \left( C(y-x) - 2 C(0) \right) \, \partial \cdot \! A^a(x) \color{black}  \Big\} \\ 
&\quad + g^2 N \int \mathrm{d}^4x \ \mathrm{d}^4y \ \mathrm{d}^4z \ \Big\{ \\
&\quad \quad  \color{green!20!blue} - \frac{1}{4} G_0(x-z) n^\mu C(z-x) \, \partial \cdot \! A^a(y)  \partial_\mu C(y-x) \, \partial \cdot \! A^a(x) \\ 
&\quad \quad  \color{green!20!blue} - 2 G_0(x-z) \partial^\mu C(z-x) \, \partial \cdot \! A^a(y) n^\nu \partial_{ \{\nu} C(y-x) A_{\mu \}}^a(x) \\
&\quad \quad  \color{green!20!blue} + 2 G_0(x-z) \partial^\mu C(z-y) A_\nu^a(y) n^\lambda \partial^\nu \partial_{\{\lambda} C(y-x) A_{\mu\}}^a(x) \\
&\quad \quad  \color{green!20!blue} - \frac{3}{2} \delta(0) C(z-y) \, \partial \cdot \! A^a(y) C(z-x) \, \partial \cdot \! A^a(x)  \color{black} \Big\} \, .
\end{aligned}
\end{align}
Carefully collecting all the {\color{green!20!blue} blue} terms, we see that they cancel. Notice that this also applies to the formally divergent terms, including a factor of $\delta(0)$ (which can be appropriately regularized). The three remaining black terms in \eqref{eq:YM1Jac1} and \eqref{eq:YM1Jac2} match the three terms from \eqref{eq:YM1LogMSS} and \eqref{eq:YM1LogFPAxial}. This concludes the determinant test. 

Notably, the determinant test also works in general dimensions $d$ for which $r = 2(d-2)$. However, we do not expect this lucky coincidence to survive in higher orders. Furthermore, we have also checked that the determinant test is passed in the $R_\xi$ type gauges.

\section{Going `On-Shell'}\label{sec:YM1OnShell}
From the `off-shell' $\mathcal{N}=1$ super Yang-Mills Nicolai map \eqref{eq:YM1TAxial} in axial gauge, we can easily obtain the corresponding result in Landau gauge. We must simply make the substitution
\begin{align}
\partial_\mu G_0(x-y) n_\nu \  \to \ - \partial_\mu C(x-y) \partial_\nu \, .
\end{align}
This is because, in Landau gauge, the free ghost propagator is $G_0(x) = - C(x)$ and $\mathcal{G}^\mu = \partial^\mu$. After the substitution, some terms will drop out as they are now anti-symmetric under the exchange of two spacetime indices. Furthermore, recall that in Landau gauge $\xi \to 0$. Thus the bosonic part of the gauge fixing action behaves like a delta function forcing us on the gauge surface $\partial^\mu A_\mu^a = 0$. Most terms in \eqref{eq:YM1TAxial} are proportional to the gauge condition $\mathcal{G}^a(A) = \partial^\mu A_\mu^a$. Thus when going `on-shell', these terms drop out and we obtain
\begin{align}
\begin{aligned}\label{eq:YM1Landau}
\left(\mathcal{T}_g A \right)_\mu^a(x) &= A_\mu^a(x) + g f^{abc} \int \mathrm{d}^4y \  \partial^\rho C(x-y) A^{b \, \mu}(y) A_\rho^c(y)  \\
&\quad + \frac{3g^2}{2} f^{abc} f^{bde} \int \mathrm{d}^4y \ \mathrm{d}^4z \  \partial^\rho C(x-y) A^{c \, \lambda}(y) \partial_{[\rho} C(y-z) A_\mu^d(z) A_{\lambda]}^e(z)\\
&\quad + \mathcal{O}(g^3) \, . 
\end{aligned}
\end{align}
This result is much simpler than \eqref{eq:YM1TAxial}. However, it only exists in the Landau gauge. Going `on-shell' in the axial or light-cone gauge, \emph{i.e.} considering the limit $\xi \to 0$, does not simplify \eqref{eq:YM1TAxial}. It appears that the Landau gauge is the preferred gauge of super Yang-Mills field theories. In section \ref{sec:YM1ROpProperties} we have reformulated the $\tilde{\mathcal{R}}_g$-operator \eqref{eq:YM1R} using the identity 
\begin{align}
\gamma^{\rho\lambda} \tilde{F}_{\rho\lambda}^b = 2 \gamma^\rho \gamma^\lambda (D_\rho \tilde{A}_\lambda)^b - 2 \partial^\lambda \tilde{A}_\lambda^b - 
f^{bde} \gamma^{\rho\lambda} \tilde{A}_\rho^d \tilde{A}_\lambda^e \, ,
\end{align}
This led to a formulation of the $\tilde{\mathcal{R}}_g$-operator where the term $\tilde{\mathcal{R}}_2$ was proportional to the Landau gauge condition. Thus we suspect that the favoritism of super Yang-Mills field theories for the Landau gauge stems from the form of the field strength tensor. 

In any kind of practical calculation, we will only consider correlation functions of gauge invariant operators $\mathcal{O}_i(x_i)$. Thus for
\begin{align}
\big<\!\!\big< \mathcal{O}_1(x_1) \ldots \mathcal{O}_n(x_n) \big> \!\! \big>_g = \big< (\mathcal{T}_g^{-1} \mathcal{O}_1)(x_1) \ldots (\mathcal{T}_g^{-1} \mathcal{O}_n)(x_n) \big>_0
\end{align}
it does not make a difference whether we use the inverse of \eqref{eq:YM1TAxial} or \eqref{eq:YM1Landau}. However, from a practical point of view, it is much simpler to use \eqref{eq:YM1Landau}. In the next chapter, we will show how to obtain \eqref{eq:YM1Landau} without the detour through a rescaled field formulation. Furthermore, we will be able to obtain the Nicolai map for a wider range of super Yang-Mills field theories, which do not have an `off-shell' formulation with (finitely many) auxiliary fields. Namely, these are $\mathcal{N}=1$ super Yang-Mills in 3, 6 and 10 dimensions and $\mathcal{N}=4$ super Yang-Mills in 4 dimensions. However, we will also see that the `on-shell' formulation requires the Landau gauge.

\chapter[head={On-Shell $\mathcal{N}$\:=\;1 super Yang-Mills}]{On-Shell \texorpdfstring{$\boldsymbol{\mathcal{N}}$\:=\;1}{N=1} super Yang-Mills}\label{ch:YM2}
In this chapter we present the 	`on-shell' $\mathcal{N}=1$ super Yang-Mills Nicolai map in $d=3$, $4$, $6$ and $10$ dimensions up to the third order, with the fourth-order result given in appendix \ref{app:FourthOrder}. We will learn that the `on-shell` construction of the $\mathcal{R}_g$-operator requires the Landau gauge but in return no detour via rescaled fields. However, the `on-shell` version of the $\mathcal{N}=1$ super Yang-Mills Nicolai map is much simpler than the corresponding `off-shell' map. In particular, we will rederive the `on-shell' map \eqref{eq:YM1Landau} in a more direct fashion. 

This chapter is organized as follows. In the first section, we recall the relevant notation and state the main theorem. In particular, we highlight the differences to the `off-shell' formulation of the theorem from the previous chapter. In section \ref{sec:YM2R} we derive the $\mathcal{R}_g$-operator before proving the main theorem in section \ref{sec:YM2Proof}. In section \ref{sec:YM2Result} we give the Nicolai map to third order in Landau gauge and discuss the history of this result. In section \ref{sec:YM2Renormalization} we comment on the renormalization of the $\mathcal{R}_g$-operator and Nicolai map. Section \ref{sec:YM2Tests} contains the tests for the Nicolai map. In section \ref{sec:YM2Ambiguity} we point out an ambiguity in the Nicolai map. We show that a different map up to the third order exists specifically for $d=6$ dimensions, which also passes the three tests from section \ref{sec:YM2Tests}. 

This chapter is heavily based on \cite{Ananth:2020lup} and \cite{Ananth:2020jdr}. The fourth-order Nicolai map was first published in \cite{Malcha:2021ess}.

\section{Introduction and Main Theorem}\label{sec:YM2Intro}
Recall the `on-shell' $d$-dimensional $\mathcal{N}=1$ super Yang-Mills action \eqref{eq:YM2}
\begin{align}\label{eq:YM21}
S_\mathrm{inv}^1 = \int \mathrm{d}^dx \ \left[ -\frac{1}{4} F_{\mu\nu}^a F^{a \, \mu\nu} - \frac{i}{2} 
\bar{\lambda}^a \gamma^\mu (D_\mu \lambda)^a \right] \, .
\end{align}
The action is invariant under the supersymmetry transformations \eqref{eq:YMsusy2}
\begin{align}
\delta A_\mu^a = - i (\bar{\lambda}^a \gamma_\mu \varepsilon ) \, , 
\quad \delta \lambda_\alpha^a = -\frac{1}{2} (\gamma^{\mu\nu} \varepsilon)_\alpha F_{\mu \nu}^a \, .
\end{align}
When computing correlation functions, the gauge-invariant action \eqref{eq:YM21} must be amended by a gauge fixing term \eqref{eq:YMgf}
\begin{align}\label{eq:YM2gf}
S_\mathrm{gf} = \int \mathrm{d}^dx \ \left[ \frac{1}{2\xi} (\mathcal{G}^\mu A_\mu^a) (\mathcal{G}^\nu A_\nu^a) + \frac{1}{2} \bar{C}^a \mathcal{G}^\mu (D_\mu C)^a \right] \, .
\end{align}
The complete action $S = S_\mathrm{inv}^1 + S_\mathrm{gf}$ is invariant under the BRST transformations \eqref{eq:YMBRST}
\begin{align}
\begin{aligned}
&s A_\mu^a = (D_\mu C)^a  \, , \quad s \lambda^a = - g f^{abc} \lambda^b C^c \, , \quad  \, \bar{C}^a = - \frac{1}{\xi} \mathcal{G}^\mu A_\mu^a  \, , \quad s\, C^a = - \frac{g}{2} f^{abc} C^b C^c \, .
\end{aligned}
\end{align}
In this chapter, we present a more direct construction of the Nicolai map with only `on-shell' supersymmetry and without passing through the rescaled fields. This construction necessitates the Landau gauge $\mathcal{G}^\mu A_\mu^a = \partial^\mu A_\mu^a$. However, the `on-shellness' is much less of a restriction for the Nicolai map than, for example, for the supersymmetry algebra. Still, the third part of the main theorem \ref{th:YM1Theorem} from the previous chapter has to be adapted to allow the matching of determinants to be modulo terms proportional to the gauge fixing functional. At the same time we can generalize from 4 spacetime dimensions to $d$ spacetime dimensions.
\begin{theorem}\label{th:YM2Theorem}
`On-shell' $d$-dimensional $\mathcal{N}=1$ super Yang-Mills is characterized by the existence of a non-linear and non-local transformation $\mathcal{T}_g$ of the gauge field
\begin{align*}
\mathcal{T}_g: A_\mu^a(x) \mapsto A_\mu^{\prime \, a}(x,g;A) \, , 
\end{align*}
which is invertible, at least in the sense of a formal power series such that
\begin{enumerate}
\item The Landau gauge-fixing function $\mathcal{G}^\mu A_\mu^a  = \partial^\mu A_\mu^a$ is a fixed point of the map $\mathcal{T}_g$. 
\item The bosonic Yang-Mills action without gauge-fixing terms is mapped to the abelian action, 
\begin{align*}
S_\mathrm{inv}^1[g;A] = S_\mathrm{inv}^1[0; \mathcal{T}_g A]  \, .
\end{align*}
\item Modulo terms proportional to the gauge fixing functional $\mathcal{G}^\mu A_\mu^a = \partial^\mu A_\mu^a$, the Jacobian determinant of $\mathcal{T}_g$ is equal to the product of the Matthews-Salam-Seiler and Faddeev-Popov determinants, i.e.
\begin{align*}
\mathcal{J}(\mathcal{T}_g \, A) = \Delta_\mathrm{MSS}[g;A] \Delta_\mathrm{FP}[g;A] 
\end{align*}
at least order by order in perturbation theory.
\end{enumerate}
\end{theorem}
Regardless of the changes to the theorem, it still implies the important relation
\begin{align}\label{eq:YM2Corr}
\big<\!\!\big< \mathcal{O}_1(x_1) \ldots \mathcal{O}_n(x_n) \big>\!\!\big>_g = \big< (\mathcal{T}_g^{-1} \mathcal{O}_1)(x_1) \ldots (\mathcal{T}_g^{-1}  \mathcal{O}_n)(x_n) \big>_0 \, .
\end{align}
However, this time restricted to the Landau gauge. If the operators $\mathcal{O}_i(x_i)$ are gauge invariant, this restriction is unimportant.

\section[The \texorpdfstring{$\mathcal{R}_g$}{R}-Operator]{The $\boldsymbol{\mathcal{R}_g}$-Operator}\label{sec:YM2R}
We derive the `on-shell' $\mathcal{R}_g$-operator. Since we work with Majorana spinors, this $\mathcal{R}_g$-operator is technically only valid in $d = 3$, 4 and 10 dimensions. However, the Nicolai map itself is independent of the spinor choice and is thus also valid in 6 dimensions\footnote{An $\mathcal{R}_g$-operator with Weyl spinors is, for example, derived in \cite{Lechtenfeld:1984me}.}. Without passing through the rescaled fields, the inverse Nicolai map is defined as
\begin{align}\label{eq:YM2InverseNicolaiMap}
\big( \mathcal{T}_g^{-1} A \big)_\mu^a(x)  \coloneqq \sum_{n=0}^\infty \frac{g^n}{n!} \left[ (\mathcal{R}_g^n \, A)_\mu^a(x) \Big\vert_{g=0} \right] \, .
\end{align}
The Nicolai map is obtained from its inverse by power series inversion, see \eqref{eq:WZNicolaiMapDef} - \eqref{eq:WZNicolaiMapExp}. The $\mathcal{R}_g$-operator was first introduced specifically for $d=4$ in \cite{Flume:1983sx,Dietz:1984hf,Dietz:1985hga}. Only recently the construction has been generalized to $d= 3$, $4$, $6$ and $10$ dimensions \cite{Ananth:2020lup}. As usual, we construct the $\mathcal{R}_g$-operator from the linear response of the vacuum expectation value of a bosonic monomial $X[A]$ to changes in the coupling constant
\begin{align}
\frac{\mathrm{d}}{\mathrm{d} g} \big< \!\! \big< X \big> \!\! \big>_g = 
\left< \!\!\!\! \left< \frac{\mathrm{d}X}{\mathrm{d} g} \right> \!\!\!\! \right>_g -
i \left< \!\!\!\! \left<  \frac{\mathrm{d}(S_\mathrm{inv} + S_\mathrm{gf})}{\mathrm{d} g} \ X
\right> \!\!\!\! \right>_g \eqqcolon \big< \mathcal{R}_g \, X  \big>_g  \, .
\end{align}
In the previous chapter, the `off-shell' supersymmetry allowed us to write the action $S_\mathrm{inv}$ as a supervariation. In the `on-shell' formulation, this is no longer possible. We find
\begin{align}\label{eq:YM2dSdg}
\frac{\mathrm{d}S_\mathrm{inv}^1[g;A,\lambda]}{\mathrm{d}g} = \delta_\alpha \Delta_\alpha - i \left( \frac{1}{2} - \frac{d-1}{r} \right) f^{abc} \int \mathrm{d}^dx \ \left(\bar{\lambda}^a \gamma^\mu A_\mu^b \lambda^c \right)
\end{align}
with
\begin{align}\label{eq:YM2Delta}
\Delta_\alpha = - \frac{1}{2r} f^{abc} \int \mathrm{d}^dx \ \left( \gamma^{\rho\lambda} \lambda^a \right)_\alpha A_\rho^b A_\lambda^c \, .
\end{align}
We want to rewrite
\begin{align}
\big< \!\! \big< (\delta_\alpha \Delta_\alpha)   X \big> \!\! \big>_g =
\big< \!\! \big<\delta_\alpha \big(\Delta_\alpha X \big) \big> \!\! \big>_g   + 
\big< \!\! \big< \Delta_\alpha (\delta_\alpha X) \big> \!\! \big>_g  
\end{align}
using the supersymmetry Ward identity \eqref{eq:YMWardId3}
\begin{align}
\big< \!\! \big< \delta_\alpha Y \big> \!\! \big>_g =  i \left< \!\!\!\! \left<   \int \mathrm{d}^dx \ \bar{C}^a \,  \delta_\alpha ( \partial^\mu A_\mu^a) \, s(Y)  \right> \!\!\!\! \right> _g
\end{align}
applied to $Y = \Delta_\alpha X$. Since $\Delta_\alpha$ is not gauge invariant in the `on-shell' formulation, its BRST variation does not vanish. Using the Jacobi identity \eqref{eq:JacId} we find
\begin{align}
s(\Delta_\alpha) = \frac{1}{r} f^{abc} \int \mathrm{d}^dx \ (\gamma^{\rho\lambda} \lambda^a)_\alpha \partial_\rho C^b A_\lambda^c \, .
\end{align}
For the $g$ derivative of $S_\mathrm{gf}$ (in Landau gauge) we obtain
\begin{align}
\frac{\mathrm{d}S_\mathrm{gf}^1[g;A,\lambda]}{\mathrm{d}g} = f^{abc} \int \mathrm{d}^dx \ \bar{C}^a \partial^\mu (A_\mu^b C^c) \, .
\end{align}
Thus, putting everything back together, we have
\begin{align}\label{eq:YM2Intermediate1}
\frac{\mathrm{d}}{\mathrm{d} g} \big< \!\! \big< X \big> \!\! \big>_g = \left< \!\!\!\! \left< \frac{\mathrm{d}X}{\mathrm{d} g} \right> \!\!\!\! \right>_g - i \left<\!\!\left< \Delta_\alpha \, (\delta_\alpha \, X) \right>\!\!\right> + \left<\!\!\!\!\left< \int \mathrm{d}^dx \ \bar{C}^a \,  \delta_\alpha ( \partial^\mu A_\mu^a) \Delta_\alpha \, s(X) \right>\!\!\!\!\right>_g + \left<\!\!\left< Z \ X \right>\!\!\right>_g
\end{align}
with
\begin{align}
\begin{aligned}\label{eq:YM2Z}
Z &\coloneqq \left( \int \mathrm{d}^dy \ \bar{C}^a(y) \, \delta_\alpha (\partial^\mu A_\mu^a(y)) \right) \frac{1}{r} f^{abc} \int \mathrm{d}^dx \left( \gamma^{\rho\lambda} \lambda^a(x) \right)_\alpha \partial_\rho C^b(x) A_\lambda^c(x) \\
&\quad - \left( \frac{1}{2} - \frac{d-1}{r} \right) f^{abc} \int \mathrm{d}^dx \ \left( \bar{\lambda}^a(x) \gamma^\mu A_\mu^b(x) \lambda^c(x) \right) \\
&\quad - i f^{abc} \int \mathrm{d}^dx \ \bar{C}^a(x) \partial^\mu (A_\mu^b(x) C^c(x)) \, .
\end{aligned}
\end{align}
In the next subsection, we will show that $Z$ vanishes upon integrating out the fermionic degrees of freedom. We show that this is true only in the Landau gauge. The $\mathcal{R}_g$-operator then reads\footnote{We have put a factor of $i$ for each integration over two spinor fields and a factor of $(-i)$ for each integration over two ghost fields.}
\begin{align}\label{eq:YM2ROp}
\mathcal{R}_g X \coloneqq \frac{\mathrm{d} X}{\mathrm{d}g} 
+ \bcontraction{}{\Delta}{_\alpha \, (}{\delta} \Delta_\alpha \, (\delta_\alpha  \, X)
+ \int \mathrm{d}^dx \ 
\bcontraction[1.5ex]{}{\bar{C}}{^a \,  \delta_\alpha ( \partial^\mu A_\mu^a) \Delta_\alpha \,  }{s}
\bar{C}^a \,  
\bcontraction{}{\delta}{_\alpha ( \partial^\mu A_\mu^a) }{\Delta}
\delta_\alpha ( \partial^\mu A_\mu^a) \Delta_\alpha \, s (X)+ 
\bcontraction{}{\, }{\ }{} Z \, X \, .
\end{align}
Without the vanishing of $\bcontraction{}{\, }{\ }{} Z$ this operator would not act distributively. Neglecting the multiplicative term, we can plug in the definition of $\Delta_\alpha$ as well as the fermion and ghost propagators to obtain
\begin{align}
\begin{gathered}\label{eq:YM2ROp1}
\mathcal{R}_g = \frac{\mathrm{d}}{\mathrm{d}g} - \frac{1}{2r} \int \mathrm{d}^dx \  \mathrm{d}^dy \ \tr(\gamma_\mu S^{ab}(x,y;A) \gamma^{\rho\lambda}) f^{bcd} A_\rho^c(y) A_\lambda^d(y) \frac{\delta}{\delta A_\mu^a(x)} \\
+ \frac{1}{2r} \int \mathrm{d}^dx \ \mathrm{d}^dy \ \mathrm{d}^dz \ (D_\mu G)^{ae}(x,z;A) \, \tr( \gamma_\nu \partial^\nu S^{eb}(z,y;A) \gamma^{\rho\lambda}) f^{bcd} A_\rho^c(y) A_\lambda^d(y) \frac{\delta}{\delta A_\mu^a(x)} \, .
\end{gathered}
\end{align}
Compared to the $\tilde{\mathcal{R}}_g$-operator \eqref{eq:YM1ROp1} there is no inherently gauge invariant part (other than the $g$ derivative) in this $\mathcal{R}_g$-operator. Thus, when computing the inverse Nicolai map \eqref{eq:YM2InverseNicolaiMap}, we must always act with the entire $\mathcal{R}_g$-operator even on gauge invariant operators. This is because, from the second application of the $\mathcal{R}_g$-operator onward, it will start to act on itself. Since $\Delta_\alpha$ is not gauge invariant, there will be contributions from the second line in \eqref{eq:YM2ROp1} acting on the first. 

Nevertheless, we might still simplify \eqref{eq:YM2ROp1} by introducing the covariant transversal projector
\begin{align}\label{eq:YM2P}
P_{\mu\nu}^{ab}(x,z;A) = \delta^{ab} \delta_{\mu\nu} \delta(x-z) - (D_\mu G)^{ab}(x,z;A) \partial_\nu^z 
\end{align}
obeying $P \ast P = P$ and $\partial^\mu P_{\mu\nu}^{ab} = 0$. This is the non-abelian version of \eqref{eq:YM2Projector}. They differ by the appearance of the gauge covariant derivative and a non-linear dependence on $A_\mu^a$. It allows for a non-standard (non-linear) separation between transversal and longitudinal degrees of freedom, with
\begin{equation}
A_\mu^{a \, \perp}(x) := \int \mathrm{d}^d y \ P_{\mu\nu}^{ab}(x,y;A)\,A^{b \, \nu}(y) \quad \text{and} \quad
A_\mu^{a \, \parallel}(x) := \int \mathrm{d}^d y \ (D_\mu G)^{ab}(x,y;A)\,\partial^\nu\!A_\nu^b(y) \, ,
\end{equation} 
such that the more standard abelian (linear) split of $A_\mu^a(x)$ into transversal and longitudinal parts is recovered by setting $g=0$. Subsequently, we find
\begin{align}
\begin{gathered}\label{eq:YM2ROp2}
\mathcal{R}_g = \frac{\mathrm{d}}{\mathrm{d}g} - \frac{1}{2r} \int \mathrm{d}^dx \ \mathrm{d}^dy \ \mathrm{d}^dz \ P_{\mu\nu}^{ae}(x,z) \, \tr( \gamma^\nu S^{eb}(z,y;A) \gamma^{\rho\lambda}) f^{bcd} A_\rho^c(y) A_\lambda^d(y) \frac{\delta}{\delta A_\mu^a(x)} \, .
\end{gathered}
\end{align}
This means that the $\mathcal{R}_g$-operator acts only on the `covariantly transversal' part of its argument. Consequently, the map $\mathcal{T}_g$ and its inverse $\mathcal{T}_g^{-1}$ affect only the transverse degrees of freedom of the gauge field, whereas they do not change its longitudinal component, which is therefore effectively the same as in the free theory. 

From the definition of the $\mathcal{N}=1$ super Yang-Mills fermion propagator \eqref{eq:YMFermProp} we obtain
\begin{align}
\frac{\mathrm{d} S^{ab}(x,y;A)}{\mathrm{d}g} = - f^{cde} \int \mathrm{d}^dz \ S^{ac}(x,z;A) A_\mu^d(z) S^{eb}(z,y;A) 
\end{align}
and
\begin{align}
\frac{\delta S^{ab}(x,y;A)}{\delta A_\mu^d(z)} = - g f^{cde} S^{ac}(x,z;A) \gamma^\mu S^{db}(z,y;A) \, .
\end{align}
Similarly, we obtain from \eqref{eq:YMGhostProp} for the ghost propagator
\begin{align}
\frac{\mathrm{d}G^{ab}(x,y;A)}{\mathrm{d}g} = f^{cde} \int \mathrm{d}^dz \ G^{ac}(x,z) \overleftarrow{\partial}{\!\!}_z^\mu A_\mu^d(z) G^{eb}(z,y;A) 
\end{align}
and
\begin{align}
\frac{\delta G^{ab}(x,y;A)}{\delta A_\mu^d(z)} =  f^{cde} G^{ac}(x,z;A) \overleftarrow{\partial}{\!\!}_z^\mu G^{db}(z,y;A) \, .
\end{align}
Finally, recall that in the previous chapter we have expressed the $\tilde{\mathcal{R}}_g$-operator as a sum of three operators $\tilde{\mathcal{R}}_g = \tilde{\mathcal{R}}_0 + \tilde{\mathcal{R}}_1 + \tilde{\mathcal{R}}_2$. The `on-shell' counterpart to $\tilde{\mathcal{R}}_0$ is simply the $g$ derivative. Artificially removing the rescaling from $\tilde{\mathcal{R}}_1$ in \eqref{eq:YM1R1} we see that in 4 dimensions
\begin{align}
\mathcal{R}_g = \frac{\mathrm{d}}{\mathrm{d}g} + \mathcal{R}_1 \, .
\end{align}
This is as expected since the $\mathcal{R}_g$-operator must also be compatible with switching between the rescaled and non-rescaled version of the theory. Furthermore, we argue that $\tilde{\mathcal{R}}_2$ does not have an `on-shell' counterpart since it is proportional to $\partial^\mu A_\mu^a$ and we are in Landau gauge, \emph{i.e.} $\partial^\mu A_\mu^a = 0$.

\subsection[Distributivity of the \texorpdfstring{$\mathcal{R}_g$}{R}-Operator]{Distributivity of the $\boldsymbol{\mathcal{R}_g}$-Operator}\label{sec:Distributivity}
We show that $Z$ vanishes in the Landau gauge upon integrating out the fermionic degrees of freedom. The following calculation was first sketched by Flume and Lechtenfeld in \cite{Flume:1983sx}. A more detailed version was first given in \cite{Ananth:2020lup}. Recall that the Landau gauge does not only require $\mathcal{G}^\mu A_\mu^a  = \partial^\mu A_\mu^a$ but also $\xi \to 0$, \emph{i.e.} forcing us onto the gauge surface. Integrating out the anti-commuting degrees of freedom in \eqref{eq:YM2Z} gives
\begin{align}
\begin{aligned}
\bcontraction{}{\, }{\ }{} Z &=    \frac{i}{r}  \int \mathrm{d}^dy \ 
\bcontraction[3ex]{}{\bar{C}}{^a(y) \, ( \partial_\mu^y \bar{\lambda}^a(y) \gamma^\mu)_\alpha  f^{abc} \int \mathrm{d}^dx \  ( \gamma^{\rho\lambda} \lambda^a(x) )_\alpha \partial_\rho}{C}
\bar{C}^a(y) \, (\partial_\mu^y 
\bcontraction[2ex]{}{\bar{\lambda}}{^a(y) \gamma^\mu)_\alpha  f^{abc} \int \mathrm{d}^dx \  ( \gamma^{\rho\lambda} }{\lambda}
\bar{\lambda}^a(y) \gamma^\mu)_\alpha  f^{abc} \int \mathrm{d}^dx \  ( \gamma^{\rho\lambda} \lambda^a(x) )_\alpha \partial_\rho C^b(x) A_\lambda^c(x) \\
&\quad - i \left( \frac{1}{2} - \frac{d-1}{r} \right) f^{abc} \int \mathrm{d}^dx \ 
\bcontraction{}{\bar{\lambda}}{_\alpha^a(x) \gamma_{\alpha\beta}^\mu A_\mu^b(x)  }{\lambda}
\bar{\lambda}_\alpha^a(x) \gamma_{\alpha\beta}^\mu A_\mu^b(x)  \lambda_\beta^c(x) \\
&\quad  + f^{abc} \int \mathrm{d}^dx \  
\bcontraction{}{\bar{C}}{^a(x)  A_\mu^b(x) \partial^\mu }{C}
\bar{C}^a(x)  A_\mu^b(x) \partial^\mu C^c(x) \, .
\end{aligned}
\end{align}
In the last term, we used the Landau gauge condition $\partial^\mu A_\mu^b(x) = 0$ to move the derivative past the vector field. Moreover, we use the identity $\gamma^{\rho \lambda} = \frac{1}{2} (\gamma^\rho \gamma^\lambda -  \gamma^\lambda \gamma^\rho) =  - \gamma^\lambda \gamma^\rho  + \eta^{\rho \lambda}$ and reorder the contracted terms such that we can identify any contraction with a fermion or ghost propagator (in the presence of the gauge-field background) to get
\begin{align}
\begin{aligned}
\bcontraction{}{\, }{\ }{} Z & =  \frac{1}{r} f^{bcd} \int \mathrm{d}^dx \ \mathrm{d}^dy \  \tr \left(  \partial_\rho^x G^{da}(x,y;A) \gamma^\mu  \gamma^\rho  \gamma^{\lambda}  \partial_\mu^y S^{ba}(x,y;A) \right) A_\lambda^c(x) \\
&\quad - \frac{1}{r} f^{bcd} \int \mathrm{d}^dx \ \mathrm{d}^dy \  \tr \left( \partial_\rho^x G^{da}(x,y;A)   \gamma^\mu \delta^{\rho \lambda} \partial_\mu^y S^{ba}(x,y;A) \right) A_\lambda^c(x) \\
&\quad  +    \left( \frac{1}{2} - \frac{d-1}{r} \right) f^{abc} \int \mathrm{d}^dx \ \tr \left( S^{ca}(x,x;A) \gamma^\mu  \right) A_\mu^b(x)  \\
&\quad - f^{abc} \int \mathrm{d}^dx \ \partial_x^\mu G^{ca}(x,x;A) A_\mu^b(x)   \, .
\end{aligned}
\end{align}
The formally singular terms with coinciding arguments can be appropriately regulated if needed. Then we need the following Landau gauge versions of the Dyson-Schwinger identities \eqref{eq:YM1DysonSchwinger}
\begin{align}
\begin{aligned}\label{eq:YM2DysonSchwinger}
S^{ba}(x,y;A) & = \delta^{ba} S_0(x-y)  +  g f^{bmn}  \int \mathrm{d}^d z \  S_0(x-z)  A_\nu^n(z) \gamma^\nu S^{ma}(z,y;A) \, ,\\
\gamma^\rho \partial_\rho^x G^{da}(x,y;A)  &= \delta^{da} S_0(x-y) +  g f^{dmn} \int \mathrm{d}^d z \  S_0(x-z) A_\rho^n(z) \partial_z^\rho G^{ma}(z,y;A)  \, .
\end{aligned}
\end{align}
Integrating by parts and using $\gamma^\mu \partial_\mu^y S_0(x-y) = -   \delta(x-y)$ in $\bcontraction{}{}{\!\!Z}{}Z$ yields
\begin{align}
\begin{aligned}
\bcontraction{}{\, }{\ }{} Z &= \color{myblue} \frac{1}{r} f^{bca} \int \mathrm{d}^dx \  \tr \left( \gamma^\mu S^{ba}(x,x;A) \right) A_\mu^c(x) \\
&\quad  + \frac{g}{r} f^{bcd} f^{dmn} \int \mathrm{d}^dx \ \mathrm{d}^dy \ \mathrm{d}^dz \\
&\quad \quad \quad \times  \tr \left( S_0(x-z) A_\rho^n(z) \partial_z^\rho G^{ma}(z,y;A)   \gamma^\lambda  \partial_\mu^y S^{ba}(x,y;A) \gamma^\mu \right) A_\lambda^c(x) \\
&\quad \color{mygreen} + f^{acd} \int \mathrm{d}^dx  \  \partial_x^\mu G^{da}(x,x;A) A_\mu^c(x) \\
&\quad - \frac{g}{r} f^{bcd}  f^{bmn} \int \mathrm{d}^dx \ \mathrm{d}^dy \ \mathrm{d}^dz \\
&\quad \quad \quad \times  \tr \left( \partial_x^\rho G^{da}(x,y;A)  S_0(x-z) \gamma^\nu A_\nu^n(z)  \partial_\mu^y S^{ma}(z,y;A)  \gamma^\mu \right) A_\rho^c(x) \\
&\quad  \color{myblue}  +   \left( \frac{1}{2} - \frac{d-1}{r} \right) f^{abc} \int \mathrm{d}^dx \ \tr \left( S^{ca}(x,x;A) \gamma^\mu  \right) A_\mu^b(x)  \\
&\quad \color{mygreen} -  f^{abc} \int \mathrm{d}^dx \ \partial_x^\mu G^{ca}(x,x;A) A_\mu^b(x) \color{black}  \, .
\end{aligned}
\end{align}
The pure fermion loops ({\color{myblue}blue} terms) cancel, provided $r = 2(d-2)$ with $d= 3$, 4, 6 or 10. The pure ghost loops ({\color{mygreen} green} terms) cancel independently of dimension. Finally, we use $S_0(x-z) =- S_0(z-x)$ to cancel the two remaining terms
\begin{align}
\begin{aligned}
\bcontraction{}{\, }{\ }{} Z &=   \frac{g}{r} f^{bcd} f^{dmn} \int \mathrm{d}^dx \ \mathrm{d}^dy \ \mathrm{d}^dz \\
&\quad \quad \quad \times  \tr \left( S_0(x-z) A_\rho^n(z) \partial_z^\rho G^{ma}(z,y;A)   \gamma^\lambda  \partial_\mu^y S^{ba}(x,y;A) \gamma^\mu \right) A_\lambda^c(x) \\
&\quad - \frac{g}{r} f^{bcd}  f^{bmn} \int \mathrm{d}^dx \ \mathrm{d}^dy \ \mathrm{d}^dz \\
&\quad \quad \quad \times  \tr \left( \partial_x^\rho G^{da}(x,y;A)  S_0(x-z) \gamma^\nu A_\nu^n(z)  \partial_\mu^y S^{ma}(z,y;A)  \gamma^\mu \right) A_\rho^c(x) \\
&= 0 \, .
\end{aligned}
\end{align}
Without the relation between the free fermion and the free ghost propagator that only exists in the $R_\xi$ type gauges and without using $\partial^\mu A_\mu^b(x) = 0$ in the first step, this proof would not have worked. Thus, without the Landau gauge, the $\mathcal{R}_g$-operator would not act distributively. 

\section{Proof of the Main Theorem}\label{sec:YM2Proof}
The proof of the theorem \ref{th:YM2Theorem} follows the same steps as the proof in the previous chapter. For part \textit{i)} of the main theorem \ref{th:YM2Theorem} we simply get
\begin{align}
\mathcal{R}_g (\partial^\mu A_\mu^a(x)) =  \frac{1}{2r} \int  \mathrm{d}^dy \ \mathrm{d}^dz \  \partial^\mu P_{\mu\nu}^{ae}(x,z) \, \tr( \gamma_\nu \partial^\nu S^{eb}(z,y;A) \gamma^{\rho\lambda}) f^{bcd} A_\rho^c(y) A_\lambda^d(y) = 0 
\end{align}
since $\partial^\mu P_{\mu\nu}^{ab} = 0$. For part \textit{ii)} we show that $\mathcal{R}_g (S_\mathrm{inv}^1[g;A]) = 0$
\begin{align}
\begin{aligned}
\mathcal{R}_g \left( S_\mathrm{inv}^1[g;A] \right) &= \frac{\mathrm{d}}{\mathrm{d}g} \left(- \frac{1}{4} \int \mathrm{d}^dx \ F_{\mu\nu}^a(x)  F^{a\, \mu\nu}(x) \right) \\
&\quad + \frac{1}{4r} \int \mathrm{d}^dx \  \mathrm{d}^dy \ \mathrm{d}^dz \ \mathrm{d}^dw \   P_{\mu\nu}^{ae}(x,z) \, \tr( \gamma^\nu S^{eb}(z,y;A) \gamma^{\rho\lambda}) \\
&\quad \quad \times f^{bcd} A_\rho^c(y) A_\lambda^d(y) \frac{\delta F_{\sigma\tau}^m(w) F^{m \, \sigma\tau}(w)}{\delta A_\mu^a(x)} \\
&= -\frac{1}{2} f^{abc} \int \mathrm{d}^dx \  F^{a\, \mu\nu}(x) A_\mu^b(x) A_\nu^c(x) \\
&\quad - \frac{1}{r} \int \mathrm{d}^dx \  \mathrm{d}^dy \ \mathrm{d}^dz \  (D_\sigma F^{\sigma \mu})^a(x)    P_{\mu\nu}^{ae}(x,z) \, \tr( \gamma^\nu S^{eb}(z,y;A) \gamma^{\rho\lambda}) \\
&\quad \quad \times f^{bcd} A_\rho^c(y) A_\lambda^d(y) \, .
\end{aligned}
\end{align}
In the second term, the projector $P_{\mu\nu}^{ae}$ can be replaced by the identity since
\begin{align}
\int \mathrm{d}^dx \  (D_\sigma F^{\sigma \mu})^a(x)   (D_\mu G)^{ae}(x,z;A) \ldots =  - \int \mathrm{d}^dx  \ (D_\mu D_\sigma F^{\sigma \mu})^a(x)   G^{ae}(x,z;A) \ldots = 0 \, .
\end{align}
Thus we have
\begin{align}
\begin{aligned}
\mathcal{R}_g \left( S_\mathrm{inv}^1[g;A] \right) &= -\frac{1}{2} f^{abc} \int \mathrm{d}^dx \  F^{a\, \mu\nu}(x) A_\mu^b(x) A_\nu^c(x) \\
&\quad - \frac{1}{r}  f^{bcd} \int \mathrm{d}^dx \  \mathrm{d}^dy \  (D^\nu F_{\nu \mu})^a(x)  \tr( \gamma^\mu S^{ab}(x,y;A) \gamma^{\rho\lambda})  A_\rho^c(y) A_\lambda^d(y) \, .
\end{aligned}
\end{align}
Like in the previous chapter, we use \eqref{eq:YM1DF}
\begin{align}
\begin{aligned}
\gamma^\mu  (D^\nu F_{\nu\mu})^a(x) &= \gamma^{[\mu} \eta^{\nu]\rho}  (D_\sigma F_{\nu\mu})^a(x) = \frac{1}{2} \gamma^{\mu\nu}  \gamma^\rho (D_\rho F_{\nu\mu})^a(x) - \frac{1}{2} \gamma^{\mu\nu\rho}  (D_\rho F_{\nu\mu})^a(x)   \, .
\end{aligned}
\end{align}
and subsequently the Bianchi identity $\gamma^{\mu\nu\rho}  (D_\rho F_{\nu\mu})^a = 0$. Moreover, we integrate by parts and use the Dirac equation \eqref{eq:YMFermProp} to get
\begin{align}
\begin{aligned}
\mathcal{R}_g \left( S_\mathrm{inv}^1[g;A] \right) &= -\frac{1}{2} f^{abc} \int \mathrm{d}^dx \  F^{a\, \mu\nu}(x) A_\mu^b(x) A_\nu^c(x) \\
&\quad + \frac{1}{2r}  f^{bcd} \int \mathrm{d}^dx \  \mathrm{d}^dy \   F_{\nu \mu}^a(x)  \tr( \gamma^{\mu\nu} \gamma^\rho (D_\rho S)^{ab}(x,y;A) \gamma^{\rho\lambda})  A_\rho^c(y) A_\lambda^d(y) \\
&= -\frac{1}{2} f^{abc} \int \mathrm{d}^dx \  F^{a\, \mu\nu}(x) A_\mu^b(x) A_\nu^c(x) \\
&\quad + \frac{1}{2r}  f^{acd} \int \mathrm{d}^dx \   F_{\nu \mu}^a(x)  \tr( \gamma^{\mu\nu} \gamma^{\rho\lambda})  A_\rho^c(x) A_\lambda^d(x) \\
&= 0 \, .
\end{aligned}
\end{align}
The proof of part \textit{iii)} remains largely unchanged from the `off-shell' version. However, the limit $\xi \to 0$ in the path integral implies that
\begin{align}
\mathcal{J}(\mathcal{T}_g \, A) = \Delta_\mathrm{MSS}[g;A] \Delta_\mathrm{FP}[g;A] 
\end{align}
must now only hold up to terms proportional to $\partial^\mu A_\mu^a$. This concludes the proof. 
\newpage
\section{Result and Discussion}\label{sec:YM2Result}
We present the explicit formula for the Nicolai map $\mathcal{T}_g$ to cubic order \cite{Ananth:2020lup}
\begin{align}
\begin{aligned}\label{eq:YM2Result}
(\mathcal{T}_g \, A)_\mu^a(x) &= A_\mu^a(x) + g f^{abc} \int \mathrm{d}^d y \ \partial^\rho C(x-y) A_\mu^b(y) A_\rho^c(y) \\
&\quad +  \frac{3g^2}{2} f^{abc} f^{bde} \int \mathrm{d}^d y \ \mathrm{d}^d z \ \partial^\rho C(x-y) A^{c \,  \lambda}(y) \partial_{[\rho} C(y-z) A_\mu^d(z) A_{\lambda]}^e(z) \\
&\quad + \frac{g^3}{2} f^{a b c} f^{b d e} f^{c f g} \int \mathrm{d}^d y \ \mathrm{d}^d z \ \mathrm{d}^d w \ \partial^\rho C(x-y) \\
&\quad \quad \quad \times \partial^\lambda C(y-z) A_\lambda^d(z) A^{e \, \sigma}(z) \partial_{[\rho} C(y-w) A_\mu^f(w) A_{\sigma]}^g(w) \\
&\quad +  g^3 f^{a b c} f^{b d e} f^{d f g} \int \mathrm{d}^d y \ \mathrm{d}^d z \ \mathrm{d}^d w \ \partial^\rho C(x-y) A^{c \,  \lambda}(y) \ \Big\{ \\
&\quad \quad \quad - \partial^{\sigma} C(y-z)A_{\sigma}^e(z) \partial_{[\rho} C(z-w) A_\mu^f(w) A_{\lambda]}^g(w) \\
&\quad \quad \quad + \partial_{[\rho} C(y-z) A_\mu^e(z) \partial^\sigma C(z-w) A_{\lambda ]}^f(w) A_\sigma^g(w) \Big\} \\
&\quad +  \frac{g^3}{3} f^{abc} f^{bde} f^{dfg} \int \mathrm{d}^d y \ \mathrm{d}^d z \ \mathrm{d}^d w \ \Big\{ \\
&\quad \quad \quad + 6 \, \partial_\rho C(x-y) A^{c \,  \lambda}(y) \partial^{[\rho} C(y-z) A^{e\, \sigma]}(z) \partial_{[\lambda} C(z-w) A_\mu^f(w) A_{\sigma]}^g(w) \\
&\quad \quad \quad - 6 \, \partial^\rho C(x-y) A_\lambda^c(y) \partial^{[\lambda} C(y-z) A^{e \, \sigma]}(z) \partial_{[\rho} C(z-w) A_\mu^f(w) A_{\sigma]}^g(w) \\
&\quad \quad \quad - 6 \, \partial_\rho C(x-y) A_\lambda^c(y) \partial_{[\sigma} C(y-z) A_{\mu]}^e(z) \partial^{[\rho} C(z-w) A^{f \, \lambda}(w) A^{\sigma] \, g}(w) \\
&\quad \quad \quad + 2 \, \partial^\rho C(x-y) A_{[\rho}^c(y) \partial_{\mu]} C(y-z) A^{e \, \lambda }(z) \partial^\sigma C(z-w) A_\lambda^f(w) A_\sigma^g(w) \\
&\quad \quad \quad - \partial_\mu C(x-y) \, \partial^\rho \left( A_\rho^c(y) C(y-z) \right) A^{e \, \lambda}(z) \partial^\sigma C(z-w) A_\lambda^f(w) A_\sigma^g(w) \Big\} \\
&\quad -  \frac{g^3}{3} f^{abc} f^{bde} f^{dfg} \int \mathrm{d}^d y \ \mathrm{d}^d z  \ A_\mu^c(x) C(x-y) A^{e \, \rho}(y) \partial^\lambda C(y-z) A_\rho^f(z) A_\lambda^g(z) \\
&\quad + \mathcal{O}(g^4) \, . 
\end{aligned}
\end{align}
The map up to $\mathcal{O}(g^4)$ is given in appendix \ref{app:FourthOrder}. The first two lines of \eqref{eq:YM2Result} were first obtained for $d=4$ dimensions already in 1980 \cite{Nicolai:1980jc}. Much later, it was shown in \cite{Ananth:2020gkt} that the same result also holds for $d= 3$, $6$ and $10$ dimensions. Shortly after, the third \cite{Ananth:2020lup} and fourth order \cite{Malcha:2021ess} were computed. While the map up to $\mathcal{O}(g^2)$ was originally obtained by trial and error, this becomes tricky at higher orders because the number of terms is significantly larger at $\mathcal{O}(g^3)$ and above. In section \ref{sec:YM2Tests}, we will verify that this result satisfies all three statements of the main theorem \ref{th:YM2Theorem} simultaneously, providing a highly non-trivial test. 

\section{Renormalization}\label{sec:YM2Renormalization}
While the 2-dimensional Wess-Zumino model is finite \cite{Browne:1975xf}, the 4-dimensional Wess-Zumino model and (most) super Yang-Mills theories are not and require renormalization. We restrict the following discussion to 4-dimensional theories since quantum field theories in more than four dimensions are generally not renormalizable. 

For pure supersymmetric field theories, there is a beautiful non-renormalization theorem stating that superpotentials of chiral superfields do not get renormalized \cite{Grisaru:1979wc,Grisaru:1982zh}. For the 4-dimensional Wess-Zumino model, this implies that all fields are renormalized with a single renormalization constant $Z^{1/2}$ and that the coupling constant and the mass do not receive any renormalization besides the wave function renormalization, \emph{i.e.} $\lambda = Z^{-3/2} \lambda_r$ and $m = Z^{-1} m_r$ \cite{Wess:1973kz,Iliopoulos:1974zv}. This remains true even in the `on-shell' version of the theory. Subsequently, $n$-point correlation functions $\Gamma^{(n)}(x_1, \ldots , x_n; m, \lambda)$ of fundamental fields are simply renormalized by
\begin{align}\label{eq:YM2CorrRenormalization}
\Gamma_r^{(n)}(x_1, \ldots, x_n; m_r, \lambda_r) = Z^{-n/2} \Gamma^{(n)}(x_1, \ldots, x_n;m,\lambda) \, .
\end{align}
This is similar to the renormalization in the $\phi^4$ theory (see \emph{e.g.} \cite{Ramond:1981pw}). Moreover, one finds that both the $\mathcal{R}_\lambda$-operator and the Nicolai map $\mathcal{T}_\lambda$ of the 4-dimensional Wess-Zumino model are renormalized by a global factor. For instance, we get
\begin{align}
(\mathcal{T}_{\lambda \, r} A)(x, m_r, \lambda_r;A_r) = Z^{-1/2} (\mathcal{T}_\lambda A)(x,m,\lambda;A) \, .
\end{align}
This is compatible with \eqref{eq:YM2CorrRenormalization}. Moreover, the 4-dimensional Wess-Zumino model can be regularized such that the supersymmetry is preserved \cite{Pauli:1949zm,Wess:1973kz}. Hence, regularization and construction of the $\mathcal{R}_\lambda$-operator are interchangeable. 

Unfortunately, the regularization of super Yang-Mills theories is more complicated. All known regularization procedures of supersymmetric gauge theories break  supersymmetry at least partially. In particular, this applies to the popular dimensional regularization and regularization by dimensional reduction \cite{Siegel:1979wq,Siegel:1980qs,Stockinger:2005gx}. Hence, we must fist construct the $\mathcal{R}_g$-operator and then manually regularize the integrals. 

Moreover, also the renormalization exhibits extra difficulties. The usual Wess-Zumino gauge of the $\mathcal{N}=1$ super Yang-Mills vector superfield breaks the linear realization of supersymmetry by introducing a gauge transformation. Thus the non-renormalization theorem of chiral superfields no longer applies to $\mathcal{N}=1$ super Yang-Mills. In particular, the gauge field and the fermion receive different renormalizations even though they belong to the same supermultiplet \cite{Capri:2014jqa,Maggiore:1995gr}. Only the auxiliary field $D^a$ does not get renormalized because it enters the action \eqref{eq:YM1} only quadratically. Hence, there is no difference in the renormalization of the `off-shell' and `on-shell' $\mathcal{N}=1$ theory. The other renormalization constants are gauge-dependent. In \cite{Capri:2014jqa} it was shown that in Landau gauge only three renormalization constants are needed because $Z_C = Z_{\bar{C}}$ and 
\begin{align}\label{eq:YM2NoRenTh}
Z_g Z_A^{1/2} Z_C = 1 \, .
\end{align}
These renormalization constants have been computed to third order in \cite{Velizhanin:2008rw} and \cite{Capri:2014jqa}. The renormalization of the Nicolai map is best understood by studying the renormalization of the $\mathcal{R}_g$-operator. For simplicity, we consider the `on-shell' Landau gauge. The central building blocks of the $\mathcal{R}_g$-operator \eqref{eq:YM2ROp1} are the fermion and ghost propagator. Recall that (with dimensional regularization)
\begin{align}
\begin{aligned}\label{eq:YM2FermPropExp}
&\quad S^{ab}(x,y;A) \\
&= \delta^{ab} S_0(x-y) + g f^{abc} \int \mathrm{d}^{2\omega} z_1 \ S_0(x-z_1) \gamma^{\mu_1} A_{\mu_1}^c(z_1) S_0(z_1-y) \\
&\quad + g^2 f^{adc} f^{dbe} \int \mathrm{d}^{2\omega} z_1 \, \mathrm{d}^{2\omega} z_2 \ S_0(x-z_1) \gamma^{\mu_1} A_{\mu_1}^c(z_1) S_0(z_1-z_2)\gamma^{\mu_2} A_{\mu_2}^e(z_2) S_0(z_2-y)  + \ldots
\end{aligned}
\end{align}
The free fermion propagator receives its renormalization from the fermion, \emph{i.e.}
\begin{align}
S_0(x-y) = Z_\lambda S_{0\, r}(x-y) \, .
\end{align}
Thus with \eqref{eq:YM2NoRenTh} we find for \eqref{eq:YM2FermPropExp}
\begin{align}
\begin{aligned}\label{eq:YM2FermPropExp2}
&\quad S^{ab}(x,y;A) \\
&= Z_\lambda \delta^{ab} S_{0 \, r}(x-y) + Z_\lambda^2 Z_C^{-1} g_r f^{abc} \int \mathrm{d}^{2\omega} z_1 \ S_{0 \, r}(x-z_1) \gamma^{\mu_1} A_{\mu_1 \, r}^c(z_1) S_{0 \, r}(z_1-y) \\
&\quad + Z_\lambda^3 Z_C^{-2} g_r^2 f^{adc} f^{dbe} \int \mathrm{d}^{2\omega} z_1 \ \mathrm{d}^{2\omega} z_2 \ S_{0 \, r}(x-z_1) \gamma^{\mu_1} A_{\mu_1 \, r}^c(z_1) \\
&\quad \quad \times S_{0 \, r}(z_1-z_2)\gamma^{\mu_2} A_{\mu_2 \, r}^e(z_2) S_{0 \, r}(z_2-y)  + \ldots
\end{aligned}
\end{align}
The renormalization is not homogeneous. For the ghost propagator, on the other hand, we find a homogeneous renormalization to all orders
\begin{align}
\begin{aligned}
&\quad G^{ab}(x,y;A) \\
&= Z_C \,  \delta^{ab} G_{0 \, r}(x-y) + Z_C \, g_r f^{abc} \int \mathrm{d}^{2\omega}z_1 \ G_{0 \, r}(x-z_1) \partial^{\mu_1} (A_{\mu_1 \, r}^c(z_1) G_{0 \, r}(z_1-y)) \\
&\quad + Z_C  \, g_r^2 f^{adc} f^{dbe} \int \mathrm{d}^{2\omega}z_1 \ \mathrm{d}^{2\omega}z_2 \ G_{0 \, r}(x-z_1) \partial^{\mu_1} (A_{\mu_1 \, r}^c(z_1) G_{0 \, r}(z_1-z_2)) \\
&\quad \quad \times \partial^{\mu_2} (A_{\mu_2 \, r}^e(z_2) G_{0 \, r}(z_2-y))  + \ldots \\
&= Z_C \, G_r^{ab}(x,y;A) \, .
\end{aligned}
\end{align}
The renormalization of the $g$ derivative in the $\mathcal{R}_g$-operator \eqref{eq:YM2ROp1} is
\begin{align}
\frac{\mathrm{d}}{\mathrm{d}g} = Z_g^{-1} \frac{\mathrm{d}}{\mathrm{d}g_r} \, .
\end{align}
Similarly, the renormalization of the gauge field term is
\begin{align}
A_\rho^c(y) A_\lambda^d(y) \frac{\delta}{\delta A_\mu^a(x)} = Z_A^{1/2} A_{\rho \, r}^c(y) A_{\lambda \, r}^d(y) \frac{\delta}{\delta A_{\mu \, r}^a(x)}
\end{align}
Finally, also the covariant derivative in the second line of \eqref{eq:YM2ROp1} must be properly renormalized. Altogether this gives the inhomogeneous renormalization of the $\mathcal{R}_g$-operator and subsequently the Nicolai map. 

In $\mathcal{N}=4$ super Yang-Mills the situation is largely similar. It has been shown that the theory exhibits ultraviolet finiteness in the light cone gauge \cite{Mandelstam:1982cb,Brink:1982wv}. However, it was already pointed out by Mandelstam in \cite{Mandelstam:1982cb} that the wave function renormalizations are gauge dependent and, generally, not trivial. Only the vanishing of the beta function is a gauge invariant statement and thus $Z_g =1$ is true in all gauges. Structurally the $\mathcal{N}=4$ $\mathcal{R}_g^4$-operator will be very similar to the $\mathcal{N}=1$ $\mathcal{R}_g$-operator \eqref{eq:YM2ROp1}. Thus, also its renormalization will be similar to the $\mathcal{N}=1$ renormalization discussed above. For the $R_\xi$ type gauges, the $\mathcal{N}=4$ super Yang-Mills renormalization constants have been computed up to three loops in \cite{Velizhanin:2008rw}. However, despite the non-trivial renormalization in general gauges we expect that vacuum expectation values and correlation functions of well-defined gauge invariant operators are finite due to the finiteness of the $\mathcal{N}=4$ theory.

\section{Tests}\label{sec:YM2Tests}
We perform the three tests corresponding to the three parts of the main theorem \ref{th:YM2Theorem} for the `on-shell' Nicolai map \eqref{eq:YM2Result} up to $\mathcal{O}(g^3)$. Let $A_\mu^{\prime \, a} \equiv (\mathcal{T}_g \, A)_\mu^a$.

\subsection{The Gauge Condition}
We first verify that $\partial^\mu A_\mu^{\prime \, a}(x)  = \partial_\mu A_\mu^a(x) + \mathcal{O}(g^4)$. Applying $\partial^\mu$ to \eqref{eq:YM2Result} and removing all terms that are manifestly anti-symmetric under the exchange of two spacetime indices yields
\begin{align}
\partial^\mu A_\mu^{\prime \, a}(x) &= \partial^\mu A_\mu^a(x)  +  \frac{g^3}{3} f^{abc} f^{bde} f^{dfg} \int \mathrm{d}^d y \ \mathrm{d}^d z \ \mathrm{d}^d w \ \Big\{ \nonumber   \\
&\quad \quad \quad \color{myblue} + 6 \, \partial^\mu \partial_\rho C(x-y) A^{c \,  \lambda}(y) \partial^{[\rho} C(y-z) A^{\sigma] \, }(z) \partial_{[\lambda} C(z-w) A_\mu^f(w) A_{\sigma]}^g(w) \nonumber \\
&\quad \quad \quad \color{myblue} - 6 \, \partial^\mu \partial_\rho C(x-y) A_\lambda^c(y) \partial_{[\sigma} C(y-z) A_{\mu]}^e(z) \partial^{[\rho} C(z-w) A^{\lambda \, f}(w) A^{\sigma] \, g}(w) \\
&\quad \quad \quad - \Box C(x-y) \, \partial^\rho \left( A_\rho^c(y) C(y-z) \right) A^{\lambda \, e}(z) \partial^\sigma C(z-w) A_\lambda^f(w) A_\sigma^g(w) \Big\} \nonumber \\
&\quad -  \frac{g^3}{3} f^{abc} f^{bde} f^{dfg}  \int \mathrm{d}^d y \ \mathrm{d}^d z  \ \partial^\mu \left( A_\mu^c(x) C(x-y) \right)  A^{\rho \, e}(y) \partial^\lambda C(y-z) A_\rho^f(z) A_\lambda^g(z) \nonumber \\
&\quad + \mathcal{O}(g^4) \nonumber \, . 
\end{align}
The two {\color{myblue}blue} terms cancel each other. In the second to last term we use $\Box \, C(x-y) = - \delta(x-y)$. It is then easy to see that
\begin{align}
\begin{aligned}
\partial^\mu A_\mu^{\prime \, a}(x) &= \partial^\mu A_\mu^a(x) +  \frac{g^3}{3} f^{abc} f^{bde} f^{dfg} \int \mathrm{d}^d y \ \mathrm{d}^d z  \ \Big\{ \\
&\quad \quad \quad +  \partial^\rho \left( A_\rho^c(x) C(x-y) \right) A^{\lambda \, e}(y) \partial^\sigma C(y-z) A_\lambda^f(z) A_\sigma^g(z)  \\
&\quad \quad \quad -  \partial^\mu \left( A_\mu^c(x) C(x-y) \right) A^{\rho \, e}(y) \partial^\lambda C(y-z) A_\rho^f(z) A_\lambda^g(z) \Big\} \\
&\quad + \mathcal{O}(g^4) \\
&= \partial^\mu A_\mu^a(x) + \mathcal{O}(g^4) \, .
\end{aligned}
\end{align}

\subsection{The Free Action}
By the second statement in the main theorem, the transformed gauge field must satisfy
\begin{align}\label{eq:YM2FreeAction}
\frac{1}{2} \int \mathrm{d}^d x \ A_\mu^{\prime \, a} \left( \Box \, \eta^{\mu\nu} - \partial^\mu \partial^\nu
\right) A_\nu^{\prime \, a} \overset{!}{=} -  \frac{1}{4} \int \mathrm{d}^d x \ F_{\mu \nu}^a F^{a\, \mu \nu}  + \mathcal{O}(g^4) \, . 
\end{align}
We stress that, unlike the matching of determinants, the fulfillment of this condition will not depend on the dimension $d$. Because of the previous subsection, we can ignore the second term on the left-hand side from order $g$ onward. At the leading order, the statement \eqref{eq:YM2FreeAction} is trivial. At order $g$, we find
\begin{align}
\begin{aligned}
&\quad \frac{1}{2} \int \mathrm{d}^d x \ A_\mu^{\prime \, a}(x) \left( \Box \, \eta^{\mu\nu} - \partial^\mu \partial^\nu
\right) A_\nu^{\prime \, a}(x) \Big\vert_{\mathcal{O}(g^1)} \\
&= g f^{abc}  \int \mathrm{d}^d x \ \mathrm{d}^dy \ \partial^\nu C(x-y) A_\mu^b(y) A_\nu^c(y)  \Box A^{a\, \mu}(x) \\
&= g f^{abc} \int \mathrm{d}^dx \ \partial^\nu A^{a\, \mu}(x)  A_\mu^b(x) A_\nu^c(x) \\
&= -  \frac{1}{4} \int \mathrm{d}^d x \ F_{\mu \nu}^a(x) F^{a\, \mu \nu}(x) \Big\vert_{\mathcal{O}(g^1)} \, .
\end{aligned}
\end{align}
At order $g^2$, we find
\begin{align}
\begin{aligned}
&\quad \frac{1}{2} \int \mathrm{d}^d x \ A_\mu^{\prime \, a}(x) \left( \Box \, \eta^{\mu\nu} - \partial^\mu \partial^\nu
\right) A_\nu^{\prime \, a}(x) \Big\vert_{\mathcal{O}(g^2)} \\
&= g^2 f^{abc} f^{ade} \int \mathrm{d}^dx \  \mathrm{d}^dy \ \mathrm{d}^dz \ \partial^\nu C(x-y) A_\mu^b(y) A_\nu^c(y) \partial^\lambda \Box  C(x-z) A^{d \, \mu}(z) A_\lambda^e(z) \\
&\quad + 3 g^2 f^{abc} f^{bde} \int \mathrm{d}^d x \ \mathrm{d}^dy \ \mathrm{d}^dz \ \partial^\nu C(x-y) A^{c \, \lambda}(y) \partial_{[\nu} C(y-z) A_\mu^d(z) A_{\lambda]}^e(z)  \Box A^{a\, \mu}(x) \\
&= g^2 f^{abc} f^{ade} \int \mathrm{d}^dx \  \mathrm{d}^dy \ \partial^\lambda \partial^\nu C(x-y) A_\mu^b(y) A_\nu^c(y)  A^{d \, \mu}(x) A_\lambda^e(x) \\
&\quad + 3 g^2 f^{abc} f^{bde} \int \mathrm{d}^d x \ \mathrm{d}^dy  \ \partial^\nu A^{a\, \mu}(x) A^{c \, \lambda}(x) \partial_{[\nu} C(x-y) A_\mu^d(y) A_{\lambda]}^e(y)  \, .
\end{aligned}
\end{align}
Since the second term is symmetric under the simultaneous exchange of $a \leftrightarrow c$ and $\nu \leftrightarrow \lambda$ we can replace $\partial_\nu A^{a\, \mu}(x)  A^{c \, \lambda}(x)$ by $\frac{1}{2} \partial_\nu (A^{a\, \mu}(x)  A^{c \, \lambda}(x))$. Then we integrate by parts and expand the anti-symmetrization in the second term
\begin{align}
\begin{aligned}
&\quad \frac{1}{2} \int \mathrm{d}^d x \ A_\mu^{\prime \, a}(x) \left( \Box \, \eta^{\mu\nu} - \partial^\mu \partial^\nu
\right) A_\nu^{\prime \, a}(x) \Big\vert_{\mathcal{O}(g^2)} \\
&= g^2 f^{abc} f^{ade} \int \mathrm{d}^dx \  \mathrm{d}^dy \ \partial^\lambda \partial^\nu C(x-y) A_\mu^b(y) A_\nu^c(y)  A^{d \, \mu}(x) A_\lambda^e(x) \\
&\quad - \frac{3}{2} g^2 f^{abc} f^{bde} \int \mathrm{d}^d x \ \mathrm{d}^dy  \  A^{a\, \mu}(x) A^{c \, \lambda}(x) \partial^\nu \partial_{[\nu} C(x-y) A_\mu^d(y) A_{\lambda]}^e(y) \\
&= \color{myblue} g^2 f^{abc} f^{ade} \int \mathrm{d}^dx \  \mathrm{d}^dy \ \partial^\lambda \partial^\nu C(x-y) A_\mu^b(y) A_\nu^c(y)  A^{d \, \mu}(x) A_\lambda^e(x) \\
&\quad - \frac{g^2}{2} f^{abc} f^{bde} \int \mathrm{d}^d x \ \mathrm{d}^dy  \  A^{a\, \mu}(x) A^{c \, \lambda}(x) \Box C(x-y) A_\mu^d(y) A_{\lambda}^e(y) \\
&\quad \color{myblue} + g^2 f^{abc} f^{bde} \int \mathrm{d}^d x \ \mathrm{d}^dy  \  A^{a\, \mu}(x) A^{c \, \lambda}(x) \partial^\nu \partial_{\mu} C(x-y) A_\nu^d(y) A_{\lambda}^e(y)  \\
&= \frac{g^2}{2} f^{abc} f^{bde} \int \mathrm{d}^d x \ \mathrm{d}^dy  \  A^{a\, \mu}(x) A^{c \, \lambda}(x) A_\mu^d(x) A_{\lambda}^e(x) \\ 
&= -  \frac{1}{4} \int \mathrm{d}^d x \ F_{\mu \nu}^a(x) F^{a\, \mu \nu}(x) \Big\vert_{\mathcal{O}(g^2)}  \, .
\end{aligned}
\end{align}
The two {\color{myblue}blue} terms cancel. In the third order, there are many more terms, but the steps of the calculation remain largely the same. We find for the left-hand side of \eqref{eq:YM2FreeAction}
\begin{align}
\begin{aligned}
&\quad \frac{1}{2} \int \mathrm{d}^d x \ A_\mu^{\prime \, a}(x) \left( \Box \, \eta^{\mu\nu} - \partial^\mu \partial^\nu
\right) A_\nu^{\prime \, a}(x) \Big\vert_{\mathcal{O}(g^3)} \\
&= \frac{g^3}{2} f^{a b c} f^{b d e} f^{c f g} \int \mathrm{d}^dx \  \mathrm{d}^d y \ \mathrm{d}^d z \ \mathrm{d}^d w \ \partial^\rho C(x-y) \\
&\quad \quad \quad \times \partial^\lambda C(y-z) A_\lambda^d(z) A^{e \, \sigma}(z) \partial_{[\rho} C(y-w) A_\mu^f(w) A_{\sigma]}^g(w) \Box A^{a\, \mu}(x) \\
&\quad +  g^3 f^{a b c} f^{b d e} f^{d f g} \int \mathrm{d}^dx \ \mathrm{d}^d y \ \mathrm{d}^d z \ \mathrm{d}^d w \ \partial^\rho C(x-y) A^{c \,  \lambda}(y) \ \Big\{ \\
&\quad \quad \quad - \partial^{\sigma} C(y-z)A_{\sigma}^e(z) \partial_{[\rho} C(z-w) A_\mu^f(w) A_{\lambda]}^g(w) \Box A^{a\, \mu}(x) \\
&\quad \quad \quad + \partial_{[\rho} C(y-z) A_\mu^e(z) \partial^\sigma C(z-w) A_{\lambda ]}^f(w) A_\sigma^g(w) \Box A^{a\, \mu}(x) \Big\} \\
&\quad +  \frac{g^3}{3} f^{abc} f^{bde} f^{dfg} \int \mathrm{d}^dx \ \mathrm{d}^d y \ \mathrm{d}^d z \ \mathrm{d}^d w \ \Big\{ \\
&\quad \quad \quad + 6 \, \partial_\rho C(x-y) A^{c \,  \lambda}(y) \partial^{[\rho} C(y-z) A^{e \, \sigma]}(z) \partial_{[\lambda} C(z-w) A_\mu^f(w) A_{\sigma]}^g(w) \Box A^{a\, \mu}(x)  \\
&\quad \quad \quad - 6 \, \partial^\rho C(x-y) A_\lambda^c(y) \partial^{[\lambda} C(y-z) A^{e \, \sigma]}(z) \partial_{[\rho} C(z-w) A_\mu^f(w) A_{\sigma]}^g(w) \Box A^{a\, \mu}(x) \\
&\quad \quad \quad - 6 \, \partial_\rho C(x-y) A_\lambda^c(y) \partial_{[\sigma} C(y-z) A_{\mu]}^e(z) \partial^{[\rho} C(z-w) A^{f \, \lambda}(w) A^{\sigma] \, g}(w) \Box A^{a\, \mu}(x) \\
&\quad \quad \quad + 2 \, \partial^\rho C(x-y) A_{[\rho}^c(y) \partial_{\mu]} C(y-z) A^{e \, \lambda }(z) \partial^\sigma C(z-w) A_\lambda^f(w) A_\sigma^g(w) \Box A^{a\, \mu}(x) \\
&\quad \quad \quad - \partial_\mu C(x-y) \, \partial^\rho \left( A_\rho^c(y) C(y-z) \right) A^{e \, \lambda}(z) \partial^\sigma C(z-w) A_\lambda^f(w) A_\sigma^g(w) \Box A^{a\, \mu}(x) \Big\} \\
&\quad -  \frac{g^3}{3} f^{abc} f^{bde} f^{dfg} \! \int \mathrm{d}^dx \ \mathrm{d}^d y \ \mathrm{d}^d z \,  A_\mu^c(x) C(x-y) A^{e \, \rho}(y) \partial^\lambda C(y-z) A_\rho^f(z) A_\lambda^g(z) \Box A^{a\, \mu}(x) \\
&\quad  + \frac{3 g^3}{2}  f^{a b c} f^{b d e} \int \mathrm{d}^d x \  \mathrm{d}^d y \ \mathrm{d}^d z \  \mathrm{d}^d w \ \partial^\rho C(x-y) A^{c \,\lambda}(y) \partial_{[\mu} C(y-z) A_\lambda^{d}(z) A_{\rho ]}^{e}(z)  \\
&\quad \quad \quad \times \Box \left( f^{a f g}  \partial^\sigma C(x-w) A^{f \, \mu}(w) A_\sigma^{g}(w) \right) \, .
\end{aligned}
\end{align}
We follow the same steps of integrating by parts and using $\Box C(x-y) = - \delta(x-y)$ as above and arrive at
\begin{align}
\begin{aligned}
&\quad \frac{1}{2} \int \mathrm{d}^d x \ A_\mu^{\prime \, a}(x) \left( \Box \, \eta^{\mu\nu} - \partial^\mu \partial^\nu
\right) A_\nu^{\prime \, a}(x) \Big\vert_{\mathcal{O}(g^3)}  \\
&= g^3 f^{a b c} f^{b d e} f^{c f g} \int \mathrm{d}^d x \  \mathrm{d}^d y \ \mathrm{d}^d z \ \Big\{   \\
&\quad \quad \quad \color{myblue} -  \frac{1}{2} A^{a\, \mu}(x) \partial^\rho \partial^\lambda C(x-y) A_\lambda^d(y) A^{e \, \sigma}(y) \partial_{[\rho} C(x-z) A_\mu^f(z) A_{\sigma]}^g(z)   \\
&\quad \quad \quad \color{myred} -  \frac{1}{2} A^{a\, \mu}(x) \partial^\lambda C(x-y) A_\lambda^d(y) A^{e \, \sigma}(y) \partial^\rho \partial_{[\rho} C(x-z) A_\mu^f(z) A_{\sigma]}^g(z) \color{black} \Big\}   \\
&\quad +  g^3 f^{abc} f^{bde} f^{dfg} \int \mathrm{d}^d x \  \mathrm{d}^d y \ \mathrm{d}^d z \ \Big\{ \\
&\quad \quad \quad \color{mygreen} + \frac{1}{2}  A^{a\, \mu}(x) A^{c \,  \lambda}(x) \partial^\rho  \partial^{\sigma} C(x-y)A_{\sigma}^e(y) \partial_{[\rho} C(y-z) A_\mu^f(z) A_{\lambda]}^g(z)  \\
&\quad \quad \quad \color{myred} - \frac{1}{2}  A^{a\, \mu}(x) A^{c \,  \lambda}(x) \partial^\rho \partial_{[\rho} C(x-y) A_\mu^e(y) \partial^\sigma C(y-z) A_{\lambda ]}^f(z) A_\sigma^g(z)   \\
&\quad \quad \quad \color{mygreen} -  A^{a\, \mu}(x) A^{c \,  \lambda}(x) \partial_\rho  \partial^{[\rho} C(x-y) A^{e \, \sigma]}(y) \partial_{[\lambda} C(y-z) A_\mu^f(z) A_{\sigma]}^g(z)  \\
&\quad \quad \quad \color{myblue} + 2 A^{a\, \mu}(x) A_\lambda^c(x)  \partial^\rho \partial^{[\lambda} C(x-y) A^{e \, \sigma]}(y) \partial_{[\rho} C(y-z) A_\mu^f(z) A_{\sigma]}^g(z)   \\
&\quad \quad \quad + \frac{2}{3} \, \partial^\rho A^{a\, \mu}(x) A_{[\rho}^c(x) \partial_{\mu]} C(x-y) A^{e \, \lambda }(y) \partial^\sigma C(y-z) A_\lambda^f(z) A_\sigma^g(z)   \\
&\quad \quad \quad -  \frac{1}{3} \partial_\mu A^{a\, \mu}(x)  \partial^\rho \left( A_\rho^c(x) C(x-y) \right) A^{e \, \lambda}(y) \partial^\sigma C(y-z) A_\lambda^f(z) A_\sigma^g(z)   \\
&\quad \quad \quad - \frac{1}{3} \Box A^{a\, \mu}(x)  A_\mu^c(x) C(x-y) A^{e \, \rho}(y) \partial^\lambda C(y-z) A_\rho^f(z) A_\lambda^g(z) \Big\}  \\
\hphantom{=}&\quad  \color{myblue}  +  \frac{3g^3}{2}  f^{a b c} f^{b d e}  f^{a f g} \int \mathrm{d}^d x \  \mathrm{d}^d y \ \mathrm{d}^d z \  \\
&\quad \quad \quad \color{myblue} \times   A^{f \, \mu}(x) A_\sigma^{g}(x) \partial^\rho \partial^\sigma C(x-y)  A^{c \,\lambda}(y) \partial_{[\mu} C(y-z) A_\lambda^{d}(z) A_{\rho ]}^{e}(z) \color{black} \, . 
\end{aligned}
\end{align}
The {\color{myblue}blue}, {\color{myred}red} and black terms cancel upon renaming some variables. The {\color{mygreen}green} terms combine to give
\begin{align}
\begin{aligned}
&\quad \color{mygreen}  \frac{g^3}{2} f^{abc} f^{bde} f^{dfg} \int \mathrm{d}^dx \ \mathrm{d}^d y \   A^{a\, \mu}(x) A^{c \,  \lambda}(x)  A^{e \, \sigma}(x) \partial_{[\lambda} C(x-y) A_\mu^f(y) A_{\sigma]}^g(y)   \\
&\color{mygreen}  = \frac{g^3}{6} \left( f^{abc} f^{bde} + f^{eba} f^{bdc} + f^{cbe} f^{bda} \right) f^{dfg} \int \mathrm{d}^dx \ \mathrm{d}^d y \\
&\quad \quad \quad \color{mygreen} \times   A^{a\, \mu}(x) A^{c \,  \lambda}(x)  A^{e \, \sigma}(x) \partial_{[\lambda} C(x-y) A_\mu^f(y) A_{\sigma]}^g(y) \\
&\color{mygreen} =  0  \color{black} \, .
\end{aligned}
\end{align}
Here we used the Jacobi identity \eqref{eq:JacId}. Subsequently, we conclude
\begin{align}
\frac{1}{2} \int \mathrm{d}^d x \ A_\mu^{\prime \, a}(x) \left( \Box \, \eta^{\mu\nu} - \partial^\mu \partial^\nu
\right) A_\nu^{\prime \, a}(x) \Big\vert_{\mathcal{O}(g^3)} = 0 \, .
\end{align}
Thus, the condition \eqref{eq:YM2FreeAction} holds up to and including $\mathcal{O}(g^3)$. 

\subsection{Jacobian, Fermion and Ghost Determinant}
Finally, we need to show that the logarithm of the Jacobian determinant matches the logarithm of the product of the Matthews–Salam–Seiler and Faddeev–Popov determinant up to the third order, \emph{i.e.}
\begin{align}\label{eq:YM2Determinants}
\log \mathcal{J}(\mathcal{T}_g \, A) = \log \det \left( \frac{\delta A_\mu^{\prime \, a}(x)}{\delta A_\nu^b(y)} \right) \overset{!}{=} \log( \Delta_\mathrm{MSS}[g;A] \Delta_\mathrm{FP}[g;A] ) \, .
\end{align}
Recall the logarithm of the Matthews–Salam–Seiler determinant \eqref{eq:YMLogMSS} for general Clifford algebra dimension $r$ and spacetime dimension $d$
\begin{align}
\begin{aligned}\label{eq:YM2LogMSS}
\log(\Delta_\mathrm{MSS}[g;A]) &= \frac{r g^2}{4} \int \mathrm{d}^d x \ \mathrm{d}^d y \ \Big\{ \\ 
&\quad \quad + 2\, \partial^\mu C(x-y) A_\mu^a(y) \partial^\rho C(y-x) A_\rho^a(x) \\
&\quad \quad - \partial^\mu C(x-y) A^{a \, \rho}(y) \partial_\mu C(y-x) A_\rho^a(x) \Big\} \\
&\quad + \frac{r g^3}{6} f^{a d m} f^{b e m} f^{c d e} \ \int \mathrm{d}^d x \ \mathrm{d}^d y \ \mathrm{d}^d z \ \Big\{ \\
&\quad \quad \color{myred} - 6 \, \partial^\mu C(x-y) A_\mu^b(y) \partial^\rho C(y-z) A^{c \, \lambda}(z) \partial_\rho C(z-x) A_\lambda^a(x) \\ 
&\quad \quad \color{myblue} + 2 \, \partial^\mu C(x-y) A_\rho^b(y) \partial^\rho C(y-z) A_\lambda^c(z) \partial^\lambda C(z-x) A_\mu^a(x) \\ 
&\quad \quad  \color{myorange} + 3 \, \partial^\mu C(x-y) A_\rho^b(y) \partial^\rho C(y-z) A_\mu^c(z) \partial^\lambda C(z-x) A_\lambda^a(x) \\ 
&\quad \quad \color{mygreen} - \partial^\mu C(x-y) A_\rho^b(y) \partial^\lambda C(y-z) A_\mu^c(z) \partial^\rho C(z-x) A_\lambda^a(x) \\ 
&\quad \quad \color{myviolet} + 3 \, \partial^\mu C(x-y) A^{b \, \rho}(y) \partial^\lambda C(y-z) A_\mu^c(z) \partial_\lambda C(z-x) A_\rho^a(x)\color{black} \Big\} \\
&\quad + \mathcal{O}(g^4) \, .
\end{aligned}
\end{align}
The coloring is for later convenience. Also, recall the logarithm of the Faddeev–Popov determinant \eqref{eq:YMLogFPLandau} (in Landau gauge)
\begin{align}
\begin{aligned}\label{eq:YM2LogFPLandau}
\log( \Delta_\mathrm{FP}[g;A]) &=  \frac{g^2 N}{2} f^{abc} f^{bad} \int \mathrm{d}^dx \ \mathrm{d}^dy \ \partial^\mu C(x-y) A_\rho^a(y) \partial^\rho C(y-x) A_\mu^a(x) \\
&\quad \color{myblue} - \frac{g^3}{3} f^{abc} f^{bde} f^{dam} \int \mathrm{d}^dx \ \mathrm{d}^dy \ \mathrm{d}^dz  \\
&\quad \quad \color{myblue} \times  \partial^\rho C(x-y) A_\mu^c(y) \partial^\mu C(y-z) A_\nu^e(z) \partial^\nu C(z-x) A_\rho^m(x) \\
&\quad + \mathcal{O}(g^4)\, .
\end{aligned}
\end{align}
Like in the previous chapter, there is nothing to compute at order $g$. At order $g^2$, there are two terms contributing to the logarithm of the Jacobian determinant
\begin{align}
\log \det \left( \frac{\delta A_\mu^{\prime \, a}(x)}{\delta A_\nu^b(y)} \right) \Bigg\vert_{\mathcal{O}(g^2)} &= \Tr \left[ \frac{\delta A^\prime}{\delta A} \bigg\vert_{\mathcal{O}(g^2)} \right] - \frac{1}{2} \Tr \left[ \frac{\delta A^\prime}{\delta A} \bigg\vert_{\mathcal{O}(g^1)} \frac{\delta A^\prime}{\delta A} \bigg\vert_{\mathcal{O}(g^1)}  \right]   \, .
\end{align}
As always, the trace $\Tr$ is done by setting $\nu=\mu$, $b=a$, $y=x$ and integrating over $x$. After a brief calculation, we arrive at
\begin{align}
\begin{aligned}
\log \det \left( \frac{\delta A_\mu^{\prime \, a}(x)}{\delta A_\nu^b(y)} \right) \Bigg\vert_{\mathcal{O}(g^2)} 
&=  - \frac{3-2d}{2} g^2 N   \int \mathrm{d}^d x \ \mathrm{d}^d y \  \partial^\mu C(x-y) A_\mu^a(y) \partial^\rho C(y-x) A_\rho^a(x)  \\
&\quad + \frac{2-d}{2} \, g^2 N  \int \mathrm{d}^d x \ \mathrm{d}^d y \  \partial^\mu C(x-y) A^{a \,  \rho}(y) \partial_{\mu} C(y-x)  A_{\rho}^a(x)  \, .
\end{aligned}
\end{align}
This matches the sum of the black terms in \eqref{eq:YM2LogFPLandau} and \eqref{eq:YM2LogMSS} if $r = 2(d-2)$ because then 
\begin{align}
\frac{r}{2} - \frac{1}{2} =  - \frac{3-2d}{2} \quad \text{and} \quad - \frac{r}{4} =  \frac{2-d}{2} \, .
\end{align}
At $\mathcal{O}(g^3)$ the logarithm of the Jacobian determinant schematically consists of three terms  
\begin{align}
\begin{aligned}\label{eq:YM2LogJacDet3}
\log \det \left( \frac{\delta A_\mu^{\prime \, a}(x)}{\delta A_\nu^b(y)} \right) \Bigg\vert_{\mathcal{O}(g^3)} 
&= \tr \left[ \frac{\delta A^\prime}{\delta A} \bigg\vert_{\mathcal{O}(g^3)} \right]  
- \left( 2 \cdot \frac{1}{2} \right) \tr \left[ \frac{\delta A^\prime}{\delta A}  \bigg\vert_{\mathcal{O}(g^2)} 
 \frac{\delta A^\prime}{\delta A} \bigg\vert_{\mathcal{O}(g^1)} \right] \\
&\quad +   \frac{1}{3} \tr \left[ \frac{\delta A^{\prime}}{\delta A} \bigg\vert_{\mathcal{O}(g^1)} 
\frac{\delta A^{\prime}}{\delta A} \bigg\vert_{\mathcal{O}(g^1)} 
\frac{\delta A^{\prime}}{\delta A} \bigg\vert_{\mathcal{O}(g^1)} \right] \, .
\end{aligned}
\end{align} 
The computation of each one of these terms is straightforward. For the third term, we find
\begin{align}
\begin{aligned}\label{eq:YM2JacDet31}
&\quad \frac{1}{3} \tr \left[ \frac{\delta A^{\prime}}{\delta A} \bigg\vert_{\mathcal{O}(g^1)} \frac{\delta A^{\prime}}{\delta A} \bigg\vert_{\mathcal{O}(g^1)} 
\frac{\delta A^{\prime}}{\delta A} \bigg\vert_{\mathcal{O}(g^1)} \right]  \\
&= g^3 f^{a d m} f^{b e m} f^{c d e}  \ \int \mathrm{d}^d x \ \mathrm{d}^d y \ \mathrm{d}^d z \ \Big\{ \\
&\quad \quad \quad \color{myblue} - \frac{3-d}{3} \,   \partial^\mu C(x-y) A_\rho^b(y) \partial^\rho C(y-z) A_\lambda^c(z)  \partial^\lambda C(z-x) A_\mu^a(x) \\
&\quad \quad \quad \color{myorange} +  \partial^\mu C(x-y)  A_\rho^b(y) \partial^\rho C(y-z)  A_\mu^c(z)  \partial^\lambda C(z-x) A_\lambda^a(x)  \\ 
&\quad \quad \quad \color{mygreen} - \frac{1}{3} \,   \partial^\mu C(x-y) A_\rho^b(y) \partial^\lambda C(y-z) A_\mu^c(z)   \partial^\rho C(z-x) A_\lambda^a(x) \color{black}  \Big\} \, .
\end{aligned}
\end{align}
The second term gives
\begin{align}
\begin{aligned}\label{eq:YM2JacDet32}
&\quad - \left( 2 \cdot \frac{1}{2} \right) \tr \left[ \frac{\delta A^\prime}{\delta A}  \bigg\vert_{\mathcal{O}(g^2)} 
 \frac{\delta A^\prime}{\delta A} \bigg\vert_{\mathcal{O}(g^1)} \right] \\
&= g^3 f^{a d m} f^{b e m} f^{c d e}  \ \int \mathrm{d}^d x \ \mathrm{d}^d y \ \mathrm{d}^d z \ \Big\{ \\
&\quad \quad \quad  \color{myred} + \frac{1-d}{2}   \,  \partial^\mu C(x-y)   A_\mu^b(y)  \partial^\rho C(y-z) A^{c\, \lambda}(z)   \partial_\rho C(z-x)  A_\lambda^a(x)  \\ 
&\quad \quad \quad  \color{myblue} + \frac{1}{2}  \,   \partial^\mu C(x-y) A_\rho^b(y) \partial^\rho C(y-z) A_\lambda^c(z)  \partial^\lambda C(z-x) A_\mu^a(x) \\ 
&\quad \quad \quad \color{myorange} - \frac{3-d}{2} \,  \partial^\mu C(x-y)  A_\rho^b(y) \partial^\rho C(y-z)  A_\mu^c(z)  \partial^\lambda C(z-x) A_\lambda^a(x)\\ 
&\quad \quad \quad \color{myviolet} + \frac{1}{2} \,  \partial^\mu C(x-y) A^{b \, \rho}(y) \partial^\lambda C(y-z)  A_\mu^c(z)  \partial_\lambda C(z-x)  A_\rho^a(x)    \color{black} \Big\} \, .
\end{aligned}
\end{align}
Finally, the first term gives
\begin{align}
\begin{aligned}\label{eq:YM2JacDet33}
&\quad \tr \left[ \frac{\delta A^\prime}{\delta A} \bigg\vert_{\mathcal{O}(g^3)} \right] \\
&= g^3 f^{a d m} f^{b e m} f^{c d e}  \ \int \mathrm{d}^d x \ \mathrm{d}^d y \ \mathrm{d}^d z \ \Big\{ \\
&\quad \quad \quad  \color{myred} + \frac{7-3d}{2}   \,  \partial^\mu C(x-y)   A_\mu^b(y)  \partial^\rho C(y-z) A^{c \, \lambda}(z)   \partial_\rho C(z-x)  A_\lambda^a(x)  \\ 
&\quad \quad \quad  \color{myblue} - \frac{3-2d}{6} \,   \partial^\mu C(x-y) A_\rho^b(y) \partial^\rho C(y-z) A_\lambda^c(z)  \partial^\lambda C(z-x) A_\mu^a(x) \\ 
&\quad \quad \quad   \color{myorange} - \frac{3-d}{2}    \,  \partial^\mu C(x-y)  A_\rho^b(y) \partial^\rho C(y-z)  A_\mu^c(z)  \partial^\lambda C(z-x) A_\lambda^a(x)\\  
&\quad \quad \quad   \color{mygreen} + \frac{3-d}{3}  \,  \partial^\mu C(x-y) A_\rho^b(y) \partial^\lambda C(y-z) A_\mu^c(z)   \partial^\rho C(z-x) A_\lambda^a(x) \\ 
&\quad \quad \quad  \color{myviolet}-  \frac{5-2d}{2}  \,  \partial^\mu C(x-y) A^{b\, \rho}(y) \partial^\lambda C(y-z)  A_\mu^c(z)  \partial_\lambda C(z-x)  A_\rho^a(x)   \color{black} \Big\} \\ 
&\quad  - \frac{2g^3}{3}  f^{aem} f^{bde} f^{cdm}  \ \int \mathrm{d}^d x \ \mathrm{d}^d y \  A^{b \, \mu}(x) A_\rho^c(x) C(x-y) \partial^\rho C(x-y) A_\mu^a(y) \\
&\quad  + \frac{g^3}{3}  f^{adm} f^{bce} f^{dem} \int  \mathrm{d}^d x \ \mathrm{d}^d y \ \mathrm{d}^d z  \  A^{a \, \mu}(x) \left(  \partial_\rho C(x-y) \right)^2  \partial^\lambda C(y-z) A_\lambda^b(z) A_\mu^c(z)  \\
&\quad - \frac{g^3}{3}   f^{adm} f^{bce} f^{dem} \int  \mathrm{d}^d x \ \mathrm{d}^d y \ C(0) A^{a\, \mu}(x)   \partial^\rho C(x-y) A_\rho^b(y) A_\mu^c(y) \, .
\end{aligned}
\end{align}
To obtain this last term, we had to use the Landau gauge condition $\partial^\mu A_\mu^a = 0$. Before addressing the black terms, let us show that the colored terms match the respectively colored terms in the MSS and FP determinants. Imposing $r = 2(d-2)$ we obtain
\begin{align}
\begin{aligned}
\color{myred} -r &\color{myred} =  \frac{1-d}{2} + \frac{7-3d}{2} = 4 - 2d \, ,  \\
\color{myblue} \frac{r+1}{3} &\color{myblue} = - \frac{3-d}{3} + \frac{1}{2} - \frac{3-2d}{6} = \frac{2d-3}{3} \color{black} \, , \\
\color{myorange} \frac{r}{2} & \color{myorange} = 1 - \frac{3-d}{2} - \frac{3-d}{2} = d-2 \color{black} \, ,  \\
\color{mygreen} - \frac{r}{6} & \color{mygreen} = - \frac{1}{3} + \frac{3-d}{3} =  \frac{2-d}{3} \color{black} \, ,  \\
\color{myviolet} \frac{r}{2} & \color{myviolet} =  \frac{1}{2} - \frac{5-2d}{2} = d -2 \color{black} \, .
\end{aligned}
\end{align}
All five equations are happily satisfied. We turn to the three black terms in \eqref{eq:YM2JacDet33}. Using the Jacobi identity in the first term 
and $f^{abc} f^{abd} = N  \delta^{cd}$ in the latter two yields
\begin{align}
\begin{aligned}
& - \frac{g^3N}{3} f^{abc}   \int \mathrm{d}^d x \ \mathrm{d}^d y \  A^{b \, \mu}(x) A_\rho^c(x) C(x-y) \partial^\rho C(x-y) A_\mu^a(y) \\
&+ \frac{g^3N}{3} f^{abc} \int  \mathrm{d}^d x \ \mathrm{d}^d y \ \mathrm{d}^d z  \  A^{a \, \mu}(x) \left(  \partial_\rho C(x-y) \right)^2  \partial^\lambda C(y-z) A_\lambda^b(z) A_\mu^c(z)  \\
&- \frac{g^3 N}{3} f^{abc}  \int  \mathrm{d}^d x \ \mathrm{d}^d y \ C(0) A^{a \, \mu}(x)   \partial^\rho C(x-y) A_\rho^b(y) A_\mu^c(y) \, .
\end{aligned}
\end{align} 
The second term is rewritten using the identity
\begin{align}
\Box ( C^2(x-y)) = - 2 \,  C(0) \delta(x-y) + 2 \,  \partial^\rho C(x-y) \partial_\rho C(x-y)  \, .
\end{align}
This simplifies the expression above to
\begin{align}
\begin{aligned}
&\quad - \frac{g^3N}{3} f^{abc}   \int \mathrm{d}^d x \ \mathrm{d}^d y \  A^{b \, \mu}(x) A_\rho^c(x) C(x-y) \partial^\rho C(x-y) A_\mu^a(y) \\
&\quad + \frac{g^3N}{6} f^{abc} \int  \mathrm{d}^d x \ \mathrm{d}^d y \ \mathrm{d}^d z  \  A^{a \, \mu}(x)\Box(C^2(x-y))  \partial^\lambda C(y-z) A_\lambda^b(z) A_\mu^c(z)  \\
&\quad + \frac{g^3N}{3} f^{abc} \int  \mathrm{d}^d x \ \mathrm{d}^d y   \  C(0) A^{a \, \mu}(x)  \partial^\rho C(x-y) A_\rho^b(y) A_\mu^c(y)  \\
&\quad - \frac{g^3 N}{3} f^{abc}  \int  \mathrm{d}^d x \ \mathrm{d}^d y \ C(0) A^{a \, \mu}(x)   \partial^\rho C(x-y) A_\rho^b(y) A_\mu^c(y) \\
& = 0 \, .
\end{aligned}
\end{align} 
Thus, \eqref{eq:YM2Determinants} is satisfied. Showing that the Nicolai map passes these three tests in the fourth order follows the same steps. However, given the size of the fourth-order result, there are, of course, many more terms to consider. 

\section{An Ambiguity in Six Dimensions}\label{sec:YM2Ambiguity}
The first Nicolai maps in \cite{Nicolai:1979nr,Nicolai:1980hh,Nicolai:1980jc,Nicolai:1980js,Nicolai:1984jg} were not obtained by a systematic construction via the $\mathcal{R}_g$-operator but rather by trial and error. For lower orders in perturbation theory, this is possible because here the map is rather constrained. However, at higher orders, the number of terms grows rapidly. In this section, we present another trial and error Nicolai map for 6-dimensional $\mathcal{N}=1$ super Yang-Mills \cite{Ananth:2020jdr}
\begin{align}
\begin{aligned}\label{eq:YM2Result6D}
(\mathcal{\check{T}}_g \, A)_\mu^a(x) &= A_\mu^a(x) + g f^{abc} \int \mathrm{d}^6 y \ \partial^\rho C(x-y) A_\mu^b(y) A_\rho^c(y) \\
&\quad +  \frac{3g^2}{2} f^{abc} f^{bde} \int \mathrm{d}^6 y \ \mathrm{d}^6 z \ \partial^\rho C(x-y) A^{c \,  \lambda}(y) \partial_{[\rho} C(y-z) A_\mu^d(z) A_{\lambda]}^e(z) \\
&\quad +  \frac{3g^3}{2} f^{a b c} f^{b d e} f^{d f g} \int \mathrm{d}^6 y \ \mathrm{d}^6 z \ \mathrm{d}^6 w \ \partial^\rho C(x-y) A^{c \,  \lambda}(y) \ \Big\{ \\
&\quad \quad \quad + \partial_\lambda C(y-z) A^{e \, \sigma}(z) \partial_{[\mu} C(z-w) A_\rho^f(w) A_{\sigma]}^g(w) \\
&\quad \quad \quad - \partial_\mu C(y-z)  A^{e \, \sigma}(z) \partial_{[\lambda} C(z-w) A_\rho^f(w) A_{\sigma]}^g(w) \\
&\quad \quad \quad - \partial_\rho C(y-z)  A^{e \, \sigma}(z) \partial_{[\mu} C(z-w) A_\lambda^f(w) A_{\sigma]}^g(w) \Big\} \\
&\quad - g^3 f^{a b c} f^{b d e} f^{d f g} \int \mathrm{d}^6 y \ \mathrm{d}^6 z \ \mathrm{d}^6 w \ \partial^\rho C(x-y) A^{c \,  \lambda}(y) \ \Big\{ \\
&\quad \quad \quad + \partial^\sigma C(y-z) A_\sigma^e(z) \partial_{[\mu} C(z-w) A_\lambda^f(w) A_{\rho]}^g(w) \\
&\quad \quad \quad - \partial^\sigma C(y-z) A_\rho^e(z) \partial_{[\mu} C(z-w) A_\lambda^f(w) A_{\sigma]}^g(w) \\
&\quad \quad \quad + \partial^\sigma C(y-z) A_\mu^e(z) \partial_{[\rho} C(z-w) A_\lambda^f(w) A_{\sigma]}^g(w) \\
&\quad \quad \quad + \partial^\sigma C(y-z) A_\lambda^e(z) \partial_{[\mu} C(z-w) A_\rho^f(w) A_{\sigma]}^g(w) \Big\} \\
&\quad + \mathcal{O}(g^4) \, . 
\end{aligned}
\end{align}
We found this Nicolai map while working on the calculations presented in section \ref{sec:YM2Tests}. Since this is a trial-and-error expression, there is no systematic way to obtain it. In particular, there is no $\mathcal{R}_g$-operator to generate the inverse Nicolai map. Up to $\mathcal{O}(g^2)$ \eqref{eq:YM2Result6D} agrees with \eqref{eq:YM2Result}. However, in the third order, the two expressions differ. \eqref{eq:YM2Result6D} is shorter than \eqref{eq:YM2Result} and the structure of the terms is more homogeneous in \eqref{eq:YM2Result6D}. In particular, the last and second to last term from \eqref{eq:YM2Result} are absent here. 

There is no guarantee that the map \eqref{eq:YM2Result6D} has an extension to higher orders. The freedom in finding different Nicolai maps at any given order stems from the fact that the main theorem \ref{th:YM2Theorem} only makes statements about the derivatives of the map. Another example of a map that differs from the one obtained via the $\mathcal{R}_g$-operator, is the third-order map for the 2-dimensional Wess-Zumino model proposed in \cite{Nicolai:1979nr}. When computing a Nicolai map via the $\mathcal{R}_g$-operator, the result is always unique. So all order Nicolai maps are expected to be unique.

\section{Tests in Six Dimensions}\label{sec:YM26DTest}
We show that \eqref{eq:YM2Result6D} satisfies the three tests from section \ref{sec:YM2Tests}. We will see that the determinant test only works for $d=6$ dimensions. The other two tests work in any dimension. Since the new result agrees with \eqref{eq:YM2Result} up to $\mathcal{O}(g^2)$, we only have to perform the tests in the third order.

\subsection{The Gauge Condition}
Let $\check{A}_\mu^{\prime \, a} \equiv (\mathcal{\check{T}}_g \, A)_\mu^a$. Acting with a derivative on \eqref{eq:YM2Result6D} yields
\begin{align}
\begin{aligned}
\partial^\mu \check{A}_\mu^{\prime \, a}(x) &= \partial^\mu A_\mu^a(x) +   \frac{3g^3}{2} f^{a b c} f^{b d e} f^{d f g} \int \mathrm{d}^6 y \ \mathrm{d}^6 z \ \mathrm{d}^6 w \ \partial^\mu \partial^\rho C(x-y) A^{c \,  \lambda}(y) \ \Big\{ \\
&\quad \quad \quad + \partial_\lambda C(y-z) A^{e \, \sigma}(z) \partial_{[\mu} C(z-w) A_\rho^f(w) A_{\sigma]}^g(w) \\
&\quad \quad \quad \color{myblue} - \partial_\mu C(y-z)  A^{e \, \sigma}(z) \partial_{[\lambda} C(z-w) A_\rho^f(w) A_{\sigma]}^g(w) \\
&\quad \quad \quad \color{myblue} - \partial_\rho C(y-z)  A^{e \, \sigma}(z) \partial_{[\mu} C(z-w) A_\lambda^f(w) A_{\sigma]}^g(w) \color{black} \Big\} \\
&\quad - g^3 f^{a b c} f^{b d e} f^{d f g} \int \mathrm{d}^6 y \ \mathrm{d}^6 z \ \mathrm{d}^6 w \ \partial^\mu \partial^\rho C(x-y) A^{c \,  \lambda}(y) \ \Big\{ \\
&\quad \quad \quad + \partial^\sigma C(y-z) A_\sigma^e(z) \partial_{[\mu} C(z-w) A_\lambda^f(w) A_{\rho]}^g(w) \\
&\quad \quad \quad \color{myred} - \partial^\sigma C(y-z) A_\rho^e(z) \partial_{[\mu} C(z-w) A_\lambda^f(w) A_{\sigma]}^g(w)  \\
&\quad \quad \quad \color{myred} + \partial^\sigma C(y-z) A_\mu^e(z) \partial_{[\rho} C(z-w) A_\lambda^f(w) A_{\sigma]}^g(w) \\
&\quad \quad \quad + \partial^\sigma C(y-z) A_\lambda^e(z) \partial_{[\mu} C(z-w) A_\rho^f(w) A_{\sigma]}^g(w) \Big\} \\
&\quad + \mathcal{O}(g^4) \\
\hphantom{\partial^\mu \check{A}_\mu^{\prime \, a}(x)}&= \partial^\mu A_\mu^a(x) + \mathcal{O}(g^4)  \, .
\end{aligned}
\end{align}
The {\color{myblue}blue} and {\color{myred}red} terms cancel respectively. The other terms are anti-symmetric in $\mu$ and $\rho$. 

\subsection{The Free Action}
The calculation of the free action \eqref{eq:YM2FreeAction} is also straightforward
\begin{align}
\begin{aligned}
&\quad \frac{1}{2} \int \mathrm{d}^6 x \ \check{A}_\mu^{\prime \, a}(x) \left( \Box \, \eta^{\mu\nu} - \partial^\mu \partial^\nu
\right) \check{A}_\nu^{\prime \, a}(x) \Big\vert_{\mathcal{O}(g^3)}  \\
&= \frac{3g^3}{2} f^{a b c} f^{b d e} f^{d f g} \int \mathrm{d}^6 x \  \mathrm{d}^6 y \ \mathrm{d}^6 z \ \mathrm{d}^6 w \ \partial^\rho C(x-y) A^{c \,  \lambda}(y) \ \Big\{  \\
&\quad \quad \quad + \partial_\lambda C(y-z) A^{e \, \sigma}(z) \partial_{[\mu} C(z-w) A_\rho^f(w) A_{\sigma]}^g(w) \Box A^{a\, \mu}(x)  \\
&\quad \quad \quad - \partial_\mu C(y-z)  A^{e \, \sigma}(z) \partial_{[\lambda} C(z-w) A_\rho^f(w) A_{\sigma]}^g(w) \Box A^{a\, \mu}(x)   \\
&\quad \quad \quad - \partial_\rho C(y-z)  A^{e \, \sigma}(z) \partial_{[\mu} C(z-w) A_\lambda^f(w) A_{\sigma]}^g(w) \Box A^{a\, \mu}(x)  \Big\}   \\
&\quad - g^3 f^{a b c} f^{b d e} f^{d f g} \int  \mathrm{d}^6 x \  \mathrm{d}^6 y \ \mathrm{d}^6 z \ \mathrm{d}^6 w \ \partial^\rho C(x-y) A^{c \,  \lambda}(y) \ \Big\{   \\
&\quad \quad \quad + \partial^\sigma C(y-z) A_\sigma^e(z) \partial_{[\mu} C(z-w) A_\lambda^f(w) A_{\rho]}^g(w) \Box A^{a\, \mu}(x)   \\
&\quad \quad \quad - \partial^\sigma C(y-z) A_\rho^e(z) \partial_{[\mu} C(z-w) A_\lambda^f(w) A_{\sigma]}^g(w) \Box A^{a\, \mu}(x)    \\
&\quad \quad \quad + \partial^\sigma C(y-z) A_\mu^e(z) \partial_{[\rho} C(z-w) A_\lambda^f(w) A_{\sigma]}^g(w) \Box A^{a\, \mu}(x)    \\
&\quad \quad \quad + \partial^\sigma C(y-z) A_\lambda^e(z) \partial_{[\mu} C(z-w) A_\rho^f(w) A_{\sigma]}^g(w) \Box A^{a\, \mu}(x)  \Big\}   \\
&\quad+ \frac{3 g^3}{2}  f^{a b c} f^{b d e} \int \mathrm{d}^6 x \  \mathrm{d}^6 y \ \mathrm{d}^6 z \  \mathrm{d}^6 w \ \partial^\rho C(x-y) A^{c \,\lambda}(y) \partial_{[\mu} C(y-z) A_\lambda^{d}(z) A_{\rho ]}^{e}(z)   \\
&\quad \quad \quad \times \Box \left( f^{a f g}  \partial^\sigma C(x-w) A^{f \, \mu}(w) A_\sigma^{g}(w) \right) \, . 
\end{aligned}
\end{align}
We integrate by parts and obtain
\begin{align}
\begin{aligned}
&\quad \frac{1}{2} \int \mathrm{d}^6 x \ \check{A}_\mu^{\prime \, a}(x) \left( \Box \, \eta^{\mu\nu} - \partial^\mu \partial^\nu
\right) \check{A}_\nu^{\prime \, a}(x) \Big\vert_{\mathcal{O}(g^3)} \\
&=  \frac{3g^3}{2} f^{a b c} f^{b d e} f^{d f g} \int \mathrm{d}^6 x \  \mathrm{d}^6 y \ \mathrm{d}^6 z  \ \Big\{ \\
&\quad \quad \quad \color{myblue} - A_\mu^a(x) A^{c \,  \lambda}(x) \partial^\rho  \partial_\lambda C(x-y) A^{e \, \sigma}(y) \partial_{[\mu} C(y-z) A_\rho^f(z) A_{\sigma]}^g(z)   \\
&\quad \quad \quad + \frac{1}{2} A_\mu^a(x) A^{c \,  \lambda}(x)\Box  C(x-y)  A^{e \, \sigma}(y) \partial_{[\mu} C(y-z) A_\lambda^f(z) A_{\sigma]}^g(z)   \Big\} \\
&\quad - g^3 f^{a b c} f^{b d e} f^{d f g} \int  \mathrm{d}^6 x \  \mathrm{d}^6 y \ \mathrm{d}^6 z \ \Big\{ \\
&\quad \quad \quad \color{myred} - \frac{1}{2} A_\mu^a(x) A^{c \,  \lambda}(x)  \partial^\rho \partial^\sigma C(x-y) A_\sigma^e(y) \partial_{[\mu} C(y-z) A_\lambda^f(z) A_{\rho]}^g(z)   \\
&\quad \quad \quad \color{myred} + \frac{1}{2}  A_\mu^a(x) A^{c \,  \lambda}(x) \partial^\rho \partial^\sigma C(x-y) A_\rho^e(y) \partial_{[\mu} C(y-z) A_\lambda^f(z) A_{\sigma]}^g(z)  \\
&\quad \quad \quad \color{mygreen} -  A_\mu^a(x) A^{c \,  \lambda}(x)  \partial^\rho \partial^\sigma C(x-y) A_\mu^e(y) \partial_{[\rho} C(y-z) A_\lambda^f(z) A_{\sigma]}^g(z)  \color{black}  \Big\} \\
&\quad  \color{myblue} + \frac{3 g^3}{2}  f^{a b c} f^{b d e} f^{a f g} \int \mathrm{d}^6 x \  \mathrm{d}^6 y \ \mathrm{d}^6 z  \\
&\quad \quad \quad \color{myblue}  \times A^{f \, \mu}(x) A_\sigma^{g}(x) \partial^\sigma \partial^\rho C(x-y) A^{c \,\lambda}(y) \partial_{[\mu} C(y-z) A_\lambda^{d}(z) A_{\rho ]}^{e}(z) \color{black} \, .
\end{aligned}
\end{align}
Again the {\color{myblue}blue} and {\color{myred}red} terms cancel respectively. The {\color{mygreen}green} term is anti-symmetric under the exchange of $\rho$ and $\sigma$. In the remaining black term we use $\Box C(x-y) = - \delta(x-y)$ and the Jacobi identity \eqref{eq:JacId}
\begin{align}
\begin{aligned}
&\quad \frac{1}{2} \int \mathrm{d}^6 x \ \check{A}_\mu^{\prime \, a}(x) \left( \Box \, \eta^{\mu\nu} - \partial^\mu \partial^\nu
\right) \check{A}_\nu^{\prime \, a}(x) \Big\vert_{\mathcal{O}(g^3)} \\
&= - \frac{3g^3}{4} f^{a b c} f^{b d e} f^{d f g} \int \mathrm{d}^6 x \  \mathrm{d}^6 y \   A_\mu^a(x) A^{c \,  \lambda}(x) A^{e \, \sigma}(x) \partial_{[\mu} C(x-y) A_\lambda^f(y) A_{\sigma]}^g(y) \\
&= - \frac{g^3}{4} \left( f^{abc} f^{bde} + f^{eba} f^{bdc} + f^{cbe} f^{bda} \right) f^{dfg} \int \mathrm{d}^6x \ \mathrm{d}^6 y \\
&\quad \quad \quad \times   A^{a\, \mu}(x) A^{c \,  \lambda}(x)  A^{e \, \sigma}(x) \partial_{[\mu} C(x-y) A_\lambda^f(y) A_{\sigma]}^g(y) \\
&=  0 \, .
\end{aligned}
\end{align}

\subsection{Jacobian, Fermion and Ghost Determinants}
Let us, for a moment, assume that \eqref{eq:YM2Result6D} exists in $d$ dimensions. Then we compute the logarithm of the Jacobian determinant at $\mathcal{O}(g^3)$ using \eqref{eq:YM2LogJacDet3} 
\begin{align}\label{eq:YM2JacDetD6}
\begin{aligned}
\log \det \left( \frac{\delta \check{A}_\mu^{\prime \, a}(x)}{\delta A_\nu^b(y)} \right) \Bigg\vert_{\mathcal{O}(g^3)} 
&= \tr \left[ \frac{\delta \check{A}^\prime}{\delta A} \bigg\vert_{\mathcal{O}(g^3)} \right]  
- \left( 2 \cdot \frac{1}{2} \right) \tr \left[ \frac{\delta \check{A}^\prime}{\delta A}  \bigg\vert_{\mathcal{O}(g^2)} 
 \frac{\delta \check{A}^\prime}{\delta A} \bigg\vert_{\mathcal{O}(g^1)} \right] \\
&\quad +   \frac{1}{3} \tr \left[ \frac{\delta \check{A}^{\prime}}{\delta A} \bigg\vert_{\mathcal{O}(g^1)} 
\frac{\delta \check{A}^{\prime}}{\delta A} \bigg\vert_{\mathcal{O}(g^1)} 
\frac{\delta \check{A}^{\prime}}{\delta A} \bigg\vert_{\mathcal{O}(g^1)} \right] \, .
\end{aligned}
\end{align}
We find for \eqref{eq:YM2Result6D} (in $d$ dimensions)
\begin{align}
\begin{aligned}
\tr \left[ \frac{\delta \check{A}^\prime}{\delta A} \bigg\vert_{\mathcal{O}(g^3)} \right]  &=  g^3 f^{a d m} f^{b e m} f^{c d e}  \ \int \mathrm{d}^d x \ \mathrm{d}^d y \ \mathrm{d}^d z \ \Big\{ \\
&\quad \quad \quad  \color{myred} + \frac{27-10d}{6}   \,  \partial^\mu C(x-y)   A_\mu^b(y)  \partial^\rho C(y-z) A^{c \, \lambda}(z)   \partial_\rho C(z-x)  A_\lambda^a(x)  \\ 
&\quad \quad \quad  \color{myblue} - \frac{3-d}{2} \,   \partial^\mu C(x-y) A_\rho^b(y) \partial^\rho C(y-z) A_\lambda^c(z)  \partial^\lambda C(z-x) A_\mu^a(x) \\ 
&\quad \quad \quad   \color{myorange} - \frac{3-2d}{6}    \,  \partial^\mu C(x-y)  A_\rho^b(y) \partial^\rho C(y-z)  A_\mu^c(z)  \partial^\lambda C(z-x) A_\lambda^a(x)\\  
&\quad \quad \quad   \color{mygreen} + \frac{3-d}{3}  \,  \partial^\mu C(x-y) A_\rho^b(y) \partial^\lambda C(y-z) A_\mu^c(z)   \partial^\rho C(z-x) A_\lambda^a(x) \\ 
&\quad \quad \quad  \color{myviolet} -  \frac{21-7d}{6}  \,  \partial^\mu C(x-y) A^{b\, \rho}(y) \partial^\lambda C(y-z)  A_\mu^c(z)  \partial_\lambda C(z-x)  A_\rho^a(x)   \color{black} \Big\} \ \, .
\end{aligned}
\end{align}
Compared to \eqref{eq:YM2JacDet33}, there are no additional terms here that require extra work to vanish. Since \eqref{eq:YM2Result6D} is identical to \eqref{eq:YM2Result} up to the second order, the other two terms in \eqref{eq:YM2JacDetD6} give the same result as before, \emph{i.e.} \eqref{eq:YM2JacDet31} and \eqref{eq:YM2JacDet32}. Comparing these terms to the sum of the logarithm of the Matthews-Salam-Seiler determinant \eqref{eq:YM2LogMSS} and the logarithm of the Faddeev-Popov determinant \eqref{eq:YM2LogFPLandau} at $\mathcal{O}(g^3)$ yields
\begin{align}
\begin{aligned}
\color{myred} -r &\color{myred} =  \frac{1-d}{2} + \frac{27-10d}{6} = \frac{30-13d}{6} \, ,  \\
\color{myblue} \frac{r+1}{3} &\color{myblue} = - \frac{3-d}{3} + \frac{1}{2} - \frac{3-d}{2}  = - \frac{12-5d}{6} \color{black} \, , \\
\color{myorange} \frac{r}{2} & \color{myorange} = 1 - \frac{3-d}{2}  - \frac{3-2d}{6}  = - \frac{6-5d}{6} \color{black} \, ,  \\
\color{mygreen} - \frac{r}{6} & \color{mygreen} = - \frac{1}{3}   -  \frac{21-7d}{6} = \frac{2-d}{3} \color{black} \, ,  \\
\color{myviolet} \frac{r}{2} & \color{myviolet} =  \frac{1}{2} -  \frac{21-7d}{6} = - \frac{18-7d}{6}   \color{black} \, .
\end{aligned}
\end{align}
All of these equations are simultaneously satisfied for $d=6$ and $r = 2(6-2) = 8$. Thus we see that \eqref{eq:YM2Result6D} is only valid in six dimensions. Unlike for the Nicolai map \eqref{eq:YM2Result}, we could not find a fourth-order result compatible with this 6-dimensional third-order result. However, it might still exist.

\chapter[head={$\mathcal{N}$\:=\;4 super Yang-Mills}]{\texorpdfstring{$\boldsymbol{\mathcal{N}}$\:=\;4}{N=4} super Yang-Mills}\label{ch:N4YM}
In this chapter, we discuss the $\mathcal{R}_g$-operators and Nicolai maps for maximally extended 4-dimensional $\mathcal{N}=1$ super Yang-Mills and $\mathcal{N}=4$ super Yang-Mills. Both $\mathcal{R}_g$-operators are rather cumbersome compared to the $\mathcal{N}=1$ super Yang-Mills $\mathcal{R}_g$-operator from the previous chapter. Thus, we will show that the $\mathcal{N}=4$ super Yang-Mills Nicolai map can be obtained from the 10-dimensional $\mathcal{N}=1$ super Yang-Mills Nicolai map by the means of dimensional reduction. 

In section \ref{sec:YM4OffShell} we present the `off-shell' $\tilde{\mathcal{R}}_g^\mathrm{max}$-operator for maximally extended 4-dimensional $\mathcal{N}=1$ super Yang-Mills. In section \ref{sec:YM4OnShell}, we give the `on-shell' $\mathcal{R}_g^4$-operator of $\mathcal{N}=4$ super Yang-Mills and discuss the dimensional reduction of $\mathcal{R}_g$-operators and Nicolai maps. Furthermore, we give the first two orders of the $\mathcal{N}=4$ super Yang-Mills Nicolai maps. In the last section, we discuss $\mathcal{N}=4$ correlation functions, BPS operators and the large $N$ limit in the context of the Nicolai map.

Most of this chapter is based on unpublished work by the author of this thesis. However,  the $\mathcal{R}_g$-operators of maximally extended $\mathcal{N}=1$ super Yang-Mills and $\mathcal{N}=4$ super Yang-Mills have first been derived by Rupprecht in \cite{Rupprecht:2021wdj}. The $\mathcal{N}=4$ super Yang-Mills Nicolai map was first given by Nicolai and Plefka in \cite{Nicolai:2020tgo}.

\section{`Off-Shell' Supersymmetry}\label{sec:YM4OffShell}
The closest we can get to an `off-shell' formulation of the Nicolai map in $\mathcal{N}=4$ super Yang-Mills is the `off-shell' Nicolai map of maximally extended $\mathcal{N}=1$ super Yang-Mills in four dimensions. Since this extended super Yang-Mills theory is a gauge theory with `off-shell' supersymmetry, the main theorem governing the properties of the Nicolai map is identical to the main theorem \ref{th:YM1Theorem}. `Off-shell' supersymmetry allows us to write the action \eqref{eq:YMMax} as a supervariation. Moreover, the gauge fixing procedure and the Ward identities are identical to those of regular $\mathcal{N}=1$ super Yang-Mills. Hence, we may adopt the universal form of the Nicolai map from chapter \ref{ch:YM1} and write
\begin{align}
\begin{aligned}\label{eq:YM4ROp1}
\mathcal{\tilde{R}}_g^\mathrm{max} \, X &= \frac{\mathrm{d}X}{\mathrm{d} g}  - \frac{2}{g} 
\bcontraction{}{\tilde{\Delta}}{_\alpha (}{\tilde{\delta}} \tilde{\Delta}_\alpha (\tilde{\delta}_\alpha X)
+ \frac{2}{g} \int \mathrm{d}^4x \ 
\bcontraction[2ex]{}{\bar{\tilde{C}}}{^a(x) \, \tilde{\delta}_\alpha (\mathcal{G}^\mu \tilde{A}_\mu^a) \, \tilde{\Delta}_\alpha }{s} 
\bar{\tilde{C}}^a(x) \, 
\bcontraction{}{\tilde{\delta}}{_\alpha (\mathcal{G}^\mu \tilde{A}_\mu^a) \, }{\tilde{\Delta}}
\tilde{\delta}_\alpha(\mathcal{G}^\mu \tilde{A}_\mu^a) \, \tilde{\Delta}_\alpha \, s(X) \\
&\quad - \frac{1}{g} \int \mathrm{d}^4x \ 
\bcontraction{}{\bar{\tilde{C}}}{^a(x) \, (\mathcal{G}^\mu \tilde{A}_\mu^a) \,  }{s}
\bar{\tilde{C}}^a(x) \, (\mathcal{G}^\mu \tilde{A}_\mu^a) \,  s(X) \, .
\end{aligned}
\end{align}
This $\tilde{\mathcal{R}}_g^\mathrm{max}$-operator inherits all its relevant properties from the $\tilde{R}_g$-operator \eqref{eq:YM1ROp2}. Thus, we will not repeat the proof of theorem \ref{th:YM1Theorem} from section \ref{sec:YM1Proof}. In particular, the $\tilde{\mathcal{R}}_g^\mathrm{max}$-operator preserves the gauge condition, and it maps the bosonic part of the gauge invariant action to the abelian action.  As before, we split the $\tilde{\mathcal{R}}_g^\mathrm{max}$-operator into a gauge invariant term and a gauge fixing term $\mathcal{\tilde{R}}_g^\mathrm{max} = \mathcal{\tilde{R}}_\mathrm{inv}^\mathrm{max} + \mathcal{\tilde{R}}_\mathrm{gf}^\mathrm{max}$ with
\begin{align}
\mathcal{\tilde{R}}_\mathrm{inv}^\mathrm{max} \, X &= \frac{\mathrm{d}X}{\mathrm{d} g}  - \frac{2}{g} 
\bcontraction{}{\tilde{\Delta}}{_\alpha (}{\tilde{\delta}} \tilde{\Delta}_\alpha (\tilde{\delta}_\alpha X)
\end{align}
and
\begin{align}
\mathcal{\tilde{R}}_\mathrm{gf}^\mathrm{max} \, X = g  \int \mathrm{d}^4x \ 
\bcontraction[1.5ex]{}{\bar{\tilde{C}}}{^a(x) \, \mathcal{\tilde{R}}_\mathrm{inv} \left( \textstyle\frac{1}{g} \,  \mathcal{G}^\mu \tilde{A}_\mu^a \right) \,  }{s}
\bar{\tilde{C}}^a(x) \, \mathcal{\tilde{R}}_\mathrm{inv} \left( \textstyle\frac{1}{g} \,  \mathcal{G}^\mu \tilde{A}_\mu^a \right) \,  s(X) \, .
\end{align}
Subsequently, the inverse Nicolai map of the gauge field is obtained by \eqref{eq:YM1InverseNicolaiMap}
\begin{align}
\big( \mathcal{T}_g^{-1} A \big)_\mu^a(x)  \equiv \big( \mathcal{\tilde{T}}_g^{-1}  \big({\textstyle\frac{1}{g}} \tilde{A}\big)\big)_\mu^a(x) \coloneqq \sum_{n=0}^\infty \frac{g^n}{n!} \left[ \big( (\mathcal{\tilde{R}}_{g}^{\mathrm{max}})^n \big({\textstyle\frac{1}{g}} \tilde{A} \big) \big)_\mu^a(x)  \, \Big\vert_{\tilde{\Phi}= g\Phi} \Big\vert_{g=0} \right] \, ,
\end{align}
where $\Phi = (A_\mu^a, A_i^a, B_i^a, D^a, F_i^a, G_i^a)$ are the bosonic fields in the action \eqref{eq:YMMax}. Thus, technically, there are now 14 Nicolai maps for the 14 bosonic fields.

To find the explicit form of \eqref{eq:YM4ROp1}, we need the maximally extended $\mathcal{N}=1$ super Yang-Mills action and its supersymmetry and BRST variations. Recall the action \eqref{eq:YMMax}
\begin{align}
\begin{aligned}\label{eq:YM4Max}
\tilde{S}_\mathrm{inv}^\mathrm{max} = \frac{1}{g^2} \int \mathrm{d}^4x \ \bigg[ &-\frac{1}{4} \tilde{F}_{\mu\nu}^a \tilde{F}^{a\, \mu\nu} - \frac{i}{2} \bar{\tilde{\lambda}}^a \gamma^\mu (D_\mu \tilde{\lambda})^a + \frac{1}{2} \tilde{D}^a \tilde{D}^a \\
&+ \frac{1}{2} (D^\mu \tilde{A}_i)^a (D_\mu \tilde{A}_i)^a + \frac{1}{2} (D^\mu \tilde{B}_i)^a (D_\mu \tilde{B}_i)^a + \frac{1}{2} \tilde{F}_i^a \tilde{F}_i^a + \frac{1}{2} \tilde{G}_i^a \tilde{G}_i^a \\
&- \frac{i}{2} \bar{\chi}_{i}^a \gamma^\mu (D_\mu \chi_{i})^a - g f^{abc} \left( \tilde{D}^a \tilde{A}_i^b \tilde{B}_i^c + (\bar{\tilde{\chi}}_i^a (\tilde{A}_i^b + i \gamma_5 \tilde{B}_i^b) \tilde{\lambda}^c) \right) \\
&+ \varepsilon_{ijk} g f^{abc} \left(\frac{1}{2} \bar{\tilde{\chi}}_i^a (\tilde{A}_j^b - i \gamma_5 \tilde{B}_j^b) \chi_k^c - \frac{1}{2} \tilde{F}_i^a (\tilde{A}_j^b \tilde{A}_k^c - \tilde{B}_j^b \tilde{B}_k^c)- \tilde{G}_i^a \tilde{A}_j^b \tilde{B}_k^c \right) \bigg] \, .
\end{aligned}
\end{align}
It is invariant under the supersymmetry transformations \eqref{eq:YMMaxsusy}
\begin{align}
\begin{aligned}\label{eq:YM4susy1}
&\begin{aligned}
&\tilde{\delta} \tilde{A}_\mu^a = i \bar{\varepsilon} \gamma_\mu \tilde{\lambda}^a \, , &&\tilde{\delta} \tilde{\lambda}_\alpha^a = - \frac{1}{2} (\gamma^{\mu\nu} \varepsilon)_\alpha \tilde{F}_{\mu\nu}^a + i (\gamma_5 \varepsilon)_\alpha \tilde{D}^a \, , &&\tilde{\delta} \tilde{D}^a = - \bar{\varepsilon} \gamma_5 \gamma^\mu (D_\mu \tilde{\lambda})^a \, , \\ 
&\tilde{\delta} \tilde{A}_i^a = \bar{\varepsilon} \tilde{\chi}_i^a \, , &&\tilde{\delta} \tilde{B}_i^a = i \bar{\varepsilon} \gamma_5 \tilde{\chi}_i^a \, , \\
\end{aligned} \\
&\tilde{\delta} \tilde{F}_i^a = i \bar{\varepsilon} \gamma^\mu (D_\mu \tilde{\chi}_i)^a + f^{abc} \bar{\varepsilon} ( \tilde{A}_i^b + i \gamma_5 \tilde{B}_i^b) \tilde{\lambda}^c \, , \\
&\tilde{\delta} \tilde{G}_i^a = - \bar{\varepsilon}\gamma_5 \gamma^\mu (D_\mu \tilde{\chi}_i)^a - f^{abc} \bar{\varepsilon} ( \tilde{B}_j^b - i \gamma_5 \tilde{A}_i^b) \tilde{\lambda}^c \, , \\
&\tilde{\delta} \tilde{\chi}_{i\alpha}^a = i (\gamma^\mu \varepsilon)_\alpha (D_\mu \tilde{A}_i)^a + (\gamma_5 \gamma^\mu \varepsilon)_\alpha (D_\mu \tilde{B}_i)^a + \tilde{F}_i^a \varepsilon_\alpha + i (\gamma_5 \varepsilon)_\alpha \tilde{G}_i^a \, .
\end{aligned}
\end{align}
We utilize the `off-shell' supersymmetry and write
\begin{align}\label{eq:YM4SdD}
\tilde{S}_\mathrm{inv}^\mathrm{max} = \tilde{\delta}_\alpha \tilde{\Delta}_\alpha^\mathrm{max}
\end{align}
with \cite{Rupprecht:2021wdj}
\begin{align}
\begin{aligned} \label{eq:YM4Delta1}
\tilde{\Delta}_\alpha^\mathrm{max} = \int \mathrm{d}^4x \ \bigg[ &- \frac{1}{16} (\gamma^{\mu\nu} \tilde{\lambda}^a)_\alpha \tilde{F}_{\mu\nu}^a - \frac{i}{8} (\gamma^5 \tilde{\lambda}^a)_\alpha \tilde{D}^a - \frac{1}{8} \tilde{\chi}_{i\alpha}^a \tilde{F}_i^a - \frac{i}{8} (\gamma^5 \tilde{\chi}_i)_\alpha \tilde{G}_i^a \\
&- \frac{1}{8} (\gamma^5 \gamma^\mu \tilde{\chi}_i^a)_\alpha (D_\mu \tilde{B}_i)^a - \frac{i}{8} (\gamma^\mu \tilde{\chi}_i^a)_\alpha (D_\mu \tilde{A}_i)^a + \frac{i}{4} f^{abc} (\gamma_5 \tilde{\lambda}^a)_\alpha \tilde{A}_j^b \tilde{B}_j^c \\
&+ \varepsilon_{ijk} f^{abc} \left( - \frac{1}{8} \tilde{\chi}_{i\alpha}^a (\tilde{A}_j^b \tilde{A}_k^c - \tilde{B}_j^b \tilde{B}_k^c) + \frac{i}{4} (\gamma^5 \tilde{\chi}_i^a)_\alpha \tilde{A}_j^b \tilde{B}_k^c \right) \bigg] \, .
\end{aligned}
\end{align}
Given the action \eqref{eq:YM4Max} and the supersymmetry transformations \eqref{eq:YM4susy1}, it is not very complicated to guess the form of \eqref{eq:YM4Delta1} up to the coefficients.\! The first two terms are the same as in the regular $\mathcal{N}=1$ super Yang-Mills version of $\tilde{\Delta}_\alpha^\mathrm{max}$ \eqref{eq:YM1Delta}. The next two terms are simply the analog of the second term for the other two auxiliary fields, and all the remaining terms are the analog of the first term for the new scalar fields $A_i^a$ and $B_i^a$.  The coefficients are then determined from \eqref{eq:YM4SdD}. 

The action \eqref{eq:YM4Max} is subjected to the same gauge fixing procedure as the regular $\mathcal{N}=1$ super Yang-Mills action. Thus, we add to \eqref{eq:YM4Max} the gauge fixing term \eqref{eq:YMgf}
\begin{align}\label{eq:YM4gf}
\tilde{S}_\mathrm{gf} = \frac{1}{g^2} \int \mathrm{d}^4x \ \left[ \frac{1}{2\xi} (\mathcal{G}^\mu \tilde{A}_\mu^a) (\mathcal{G}^\nu \tilde{A}_\nu^a) + \frac{1}{2} \bar{\tilde{C}}^a \mathcal{G}^\mu (D_\mu \tilde{C})^a \right] \, .
\end{align}
The combined action $\tilde{S}^\mathrm{max} = \tilde{S}_\mathrm{inv}^\mathrm{max} + \tilde{S}_\mathrm{gf}$ is invariant under the BRST transformations
\begin{align}
\begin{aligned}\label{eq:YM4BRST}
&s \tilde{A}_\mu^a = (D_\mu \tilde{C})^a \, , && s \tilde{F}_{\mu\nu}^a = f^{abc} \tilde{F}_{\mu\nu}^b \tilde{C}^c \, , && s \tilde{\lambda}^a = - f^{abc} \tilde{\lambda}^b \tilde{C}^c \, , \\
&s \tilde{A}_i^a = f^{abc} \tilde{A}_i^b \tilde{C}^c \, , &&s\tilde{B}_i^a = f^{abc} \tilde{B}_i^b \tilde{C}^c \, , &&s\tilde{\chi}_i^a = - f^{abc} \tilde{C}_i^b \tilde{C}^c \, , \\
&s \tilde{D}^a = f^{abc} \tilde{D}^b \tilde{C}^c \, , &&s\tilde{F}_i^a = f^{abc} \tilde{F}_i^b \tilde{C}^c \, , &&s \, \tilde{G}_i^a = f^{abc} \tilde{G}_i^b \tilde{C}^c \, , \\
&s \, \tilde{C}^a = - \frac{1}{2} f^{abc} \tilde{C}^b \tilde{C}^c \, , &&s \, \tilde{\bar{C}}^a = - \frac{1}{\xi} \mathcal{G}^\mu \tilde{A}_\mu^a \, .
\end{aligned}
\end{align}
The definition of the fermion propagator follows from \eqref{eq:N4FermProp}
\begin{align}
\begin{gathered}\label{eq:YM4FermProp}
i \bcontraction{}{\tilde{\lambda}}{^a(x) }{\bar{\tilde{\lambda}}} \tilde{\lambda}^a(x) \bar{\tilde{\lambda}}^b(y) \equiv \tilde{S}_{44}^{ab}(x,y;\tilde{\mathscr{A}}) \, , \quad 
i \bcontraction{}{\tilde{\lambda}}{^a(x) }{\bar{\tilde{\chi}}} \tilde{\lambda}^a(x) \bar{\tilde{\chi}}_i^b(y) \equiv \tilde{S}_{4i}^{ab}(x,y;\tilde{\mathscr{A}}) \, , \\
i \bcontraction{}{\tilde{\chi}}{_i^a(x) }{\bar{\tilde{\chi}}} \tilde{\chi}_i^a(x) \bar{\tilde{\lambda}}^b(y) \equiv \tilde{S}_{i4}^{ab}(x,y;\tilde{\mathscr{A}}) \, , \quad
i \bcontraction{}{\tilde{\chi}}{_i^a(x) }{\bar{\tilde{\chi}}} \tilde{\chi}_i^a(x) \bar{\tilde{\chi}}_j^b(y) \equiv \tilde{S}_{ij}^{ab}(x,y;\tilde{\mathscr{A}}) \, , 
\end{gathered}
\end{align}
with $\tilde{\mathscr{A}} = (\tilde{A}_\mu^a, \tilde{A}_i^a, \tilde{B}_i^a)$. The propagator obeys the Dirac equation
\begin{align}
\begin{aligned}\label{eq:YM4MaxDirac}
&\gamma^\mu (D_\mu \tilde{S}_{4A})^{ab}(x,y;\tilde{\mathscr{A}}) + i f^{aec} \left( \tilde{A}_i^e(x) + i \gamma_5 \tilde{B}_i^e(x) \right) \tilde{S}_{iA}^{cb}(x,y;\tilde{\mathscr{A}}) = \delta_{4A} \delta^{ab} \delta(x-y) \, , \\
&\gamma^\mu (D_\mu \tilde{S}_{iA})^{ab}(x,y;\tilde{\mathscr{A}}) + i \varepsilon_{ijk} f^{aec} \left( \tilde{A}_j^e(x) - i \gamma_5 \tilde{B}_j^e(x) \right) \tilde{S}_{kA}^{cb}(x,y;\tilde{\mathscr{A}}) = \delta_{iA} \delta^{ab} \delta(x-y) \, ,
\end{aligned}
\end{align}
where $A = 1,2,3,4$. The ghost propagator remains unchanged from the one for $\mathcal{N}=1$ super Yang-Mills in \eqref{eq:YM1FermGhostProp}. Thus, we have collected all the relevant equations to obtain the explicit form of the $\tilde{\mathcal{R}}_g^\mathrm{max}$-operator \eqref{eq:YM4ROp1}. The initial result may be simplified slightly by using the Dirac equation \eqref{eq:YM4MaxDirac}. Subsequently, the gauge invariant part of the $\mathcal{\tilde{R}}_g^\mathrm{max}$-operator reads 

\vspace*{-4ex}
\begin{footnotesize}
\begin{align}
\begin{aligned}\label{eq:YM4OffShellRinv}
\tilde{\mathcal{R}}_\text{inv}^\mathrm{max} &= \frac{\mathrm{d}}{\mathrm{d} g} + \frac{1}{8g} \int \mathrm{d}^4x \ \mathrm{d}^4y \ \tr \left[ \gamma_\mu \tilde{S}_{44}^{ab}(x,y;\tilde{\mathscr{A}}) \gamma^{\rho\lambda} \right] \tilde{F}_{\rho\lambda}^b(y) \frac{\delta}{\delta \tilde{A}_\mu^a(x)} \\
&\quad - \frac{i}{4g} \int \mathrm{d}^4x \ \mathrm{d}^4y \ \tr \left[ \gamma_\mu \tilde{S}_{44}^{ab}(x,y;\tilde{\mathscr{A}}) \gamma_5 \right] \tilde{D}^b(y) \frac{\delta}{\delta \tilde{A}_\mu^a(x)} \\
&\quad - \frac{1}{4g} \int \mathrm{d}^4x \ \mathrm{d}^4y \ \tr \left[ \gamma_\mu \tilde{S}_{4i}^{ab}(x,y;\tilde{\mathscr{A}}) \left( \tilde{F}_i^b(y) + i \gamma_5 \tilde{G}_i^b(y) \right) \right] \frac{\delta}{\delta \tilde{A}_\mu^a(x)} \\
&\quad - \frac{i}{8g} \int \mathrm{d}^4x \ \mathrm{d}^4y \ \tr \left[ \tilde{S}_{i4}^{ab}(x,y) \gamma^{\mu\nu} \tilde{F}_{\mu\nu}^b(y) \left( \frac{\delta}{\delta \tilde{A}_i^a(x)} + i \gamma_5 \frac{\delta}{\delta \tilde{B}_i^a(x)} \right)\right] \\
&\quad + \frac{1}{4g} \int \mathrm{d}^4x \ \mathrm{d}^4y \ \tr \left[\gamma_5 \tilde{S}_{i4}^{ab}(x,y;\tilde{\mathscr{A}}) \tilde{D}^b(y) \left( \frac{\delta}{\delta \tilde{A}_i^a(x)} + i \gamma_5 \frac{\delta}{\delta \tilde{B}_i^a(x)} \right) \right] \\
&\quad - \frac{i}{4g} \int \mathrm{d}^4x \ \mathrm{d^4}y \ \tr \left[ \tilde{S}_{ij}^{ab}(x,y;\tilde{\mathscr{A}}) \left( \tilde{F}_j^b(y) + i \gamma_5 \tilde{G}_j^b(y) \right) \left( \frac{\delta}{\delta \tilde{A}_i^a(x)} + i \gamma_5 \frac{\delta}{\delta \tilde{B}_i^a(x)} \right) \right] \\
&\quad + \frac{1}{g} \int \mathrm{d}^4x \left[ \tilde{A}_i^a(x) \frac{\delta}{\delta \tilde{A}_i^a(x)} + \tilde{B}_i^a(x) \frac{\delta}{\delta \tilde{B}_i^a(x)} \right] \\
&\quad - \frac{i}{8g} \int \mathrm{d}^4x \ \mathrm{d}^4y \ \tr \left[\gamma_5 \gamma_\mu (D^\mu \tilde{S}_{44})^{ab}(x,y;\tilde{\mathscr{A}}) \gamma^{\rho\lambda} \right] \tilde{F}_{\rho\lambda}^b(y) \frac{\delta}{\delta \tilde{D}^a(x)} \\
&\quad + \frac{1}{4g} \int \mathrm{d}^4x \ \mathrm{d}^4y \ \tr \left[ \gamma_\mu (D^\mu \tilde{S}_{44})^{ab}(x,y;\tilde{\mathscr{A}}) \right] \tilde{D}^b(y) \frac{\delta}{\delta \tilde{D}^a(x)} \\
&\quad - \frac{i}{4g} \int \mathrm{d}^4x \ \mathrm{d^4}y \ \tr \left[ \gamma_5 \gamma^\mu (D_\mu \tilde{S}_{4i})^{ab}(x,y;\tilde{\mathscr{A}}) \left( \tilde{F}_i^b(y) + i \gamma_5 \tilde{G}_i^b(y) \right) \right] \frac{\delta}{\delta \tilde{D}^a(x)} \\
&\quad + \frac{1}{8g} \int \mathrm{d}^4x \ \mathrm{d^4}y \ \tr \left[ \gamma^\mu (D_\mu \tilde{S}_{i4})^{ab}(x,y) \gamma^{\rho\lambda} \tilde{F}_{\rho\lambda}^b(y) \left( \frac{\delta}{\delta \tilde{F}_i^a(x)} + i \gamma_5 \frac{\delta}{\delta \tilde{G}_i^a(x)} \right)\right] \\
&\quad - \frac{i}{4g} \int \mathrm{d}^4x \ \mathrm{d}^4y \ \tr \left[ \gamma_5 \gamma^\mu (D_\mu \tilde{S}_{i4})^{ab}(x,y;\tilde{\mathscr{A}}) \tilde{D}^b(y) \left( \frac{\delta}{\delta \tilde{F}_i^a(x)} + i \gamma_5 \frac{\delta}{\delta \tilde{G}_i^a(x)} \right)\right] \\
&\quad + \frac{1}{4g} \int \mathrm{d}^4x \ \mathrm{d^4}y \ \tr \left[ \gamma^\mu (D_\mu \tilde{S}_{ij})^{ab}(x,y;\tilde{\mathscr{A}}) \left( \tilde{F}_j^b(y) + i \gamma_5 \tilde{G}_j^b(y) \right) \left( \frac{\delta}{\delta \tilde{F}_i^a(x)} + i \gamma_5 \frac{\delta}{\delta \tilde{G}_i^a(x)} \right)\right] \\
&\quad + \frac{i}{8g} f^{abc} \int \mathrm{d}^4x \ \mathrm{d}^4y \ \tr \left[ (\tilde{A}_i^c(x) + i \gamma_5 \tilde{B}_i^c(x)) \tilde{S}_{44}^{bd}(x,y;\tilde{\mathscr{A}}) \gamma^{\rho\lambda} \tilde{F}_{\rho\lambda}^d(y) \left( \frac{\delta}{\delta \tilde{F}_i^a(x)} + i \gamma_5 \frac{\delta}{\delta \tilde{G}_i^a(x)} \right)\right] \\
&\quad - \frac{1}{4g} f^{abc} \int \mathrm{d}^4x \ \mathrm{d}^4y \ \tr\left[ (\tilde{A}_i^c(x) + i \gamma_5 \tilde{B}_i^c(x)) \tilde{S}_{44}^{ad}(x,y;\tilde{\mathscr{A}}) \gamma_5 \tilde{D}^d(y) \left( \frac{\delta}{\delta \tilde{F}_i^a(x)} + i \gamma_5 \frac{\delta}{\delta \tilde{G}_i^a(x)} \right)\right] \\
&\quad + \frac{i}{4g} f^{abc} \int \mathrm{d}^4x \ \mathrm{d^4}y \\
&\quad \quad \times \tr \left[ (\tilde{A}_i^c(x) + i \gamma_5 \tilde{B}_i^c(x)) \tilde{S}_{4j}^{ae}(x,y;\tilde{\mathscr{A}}) \left( \tilde{F}_j^e(y) - i \gamma_5 \tilde{G}_j^e(y) \right) \left( \frac{\delta}{\delta \tilde{F}_i^a(x)} + i \gamma_5 \frac{\delta}{\delta \tilde{G}_i^a(x)} \right)\right] \, .
\end{aligned}
\end{align}
\end{footnotesize}%

\newpage
Likewise, the gauge fixing term is given by
\begin{footnotesize}
\begin{align}
\begin{aligned}\label{eq:YM4OffShellRg}
\tilde{\mathcal{R}}_\text{gf}^\mathrm{max} &= - g \int \mathrm{d}^4x \ \mathrm{d}^4y \ (D_\mu \tilde{G})^{ab}(x,y;\tilde{A}) \tilde{\mathcal{R}}_\text{inv}\left( \textstyle\frac{1}{g} \, \mathcal{G}^\mu A_\mu^b(y) \right) \frac{\delta}{\delta \tilde{A}_\mu^a(x)} \\
&\quad + g f^{abc} \int \mathrm{d}^4x \ \mathrm{d}^4y \ \tilde{G}^{bd}(x,y;\tilde{A}) \tilde{\mathcal{R}}_\text{inv}\left( \textstyle\frac{1}{g} \, \mathcal{G}^\mu A_\mu^d(y) \right) \tilde{A}_i^c(x) \frac{\delta}{\delta \tilde{A}_i^a(x)} \\
&\quad + g f^{abc} \int \mathrm{d}^4x \ \mathrm{d}^4y \ \tilde{G}^{bd}(x,y;\tilde{A}) \tilde{\mathcal{R}}_\text{inv}\left( \textstyle\frac{1}{g} \, \mathcal{G}^\mu A_\mu^d(y) \right) \tilde{B}_i^c(x) \frac{\delta}{\delta \tilde{B}_i^a(x)} \\
&\quad + g f^{abc} \int \mathrm{d}^4x \ \mathrm{d}^4y \ \tilde{G}^{bd}(x,y;\tilde{A}) \tilde{\mathcal{R}}_\text{inv}\left( \textstyle\frac{1}{g} \, \mathcal{G}^\mu A_\mu^d(y) \right) \tilde{D}^c(x) \frac{\delta}{\delta \tilde{D}^a(x)} \\
&\quad + g f^{abc} \int \mathrm{d}^4x \ \mathrm{d}^4y \ \tilde{G}^{bd}(x,y;\tilde{A}) \tilde{\mathcal{R}}_\text{inv}\left( \textstyle\frac{1}{g} \, \mathcal{G}^\mu A_\mu^d(y)\right) \tilde{F}_i^c(x) \frac{\delta}{\delta \tilde{F}_i^a(x)} \\
&\quad + g f^{abc} \int \mathrm{d}^4x \ \mathrm{d}^4y \ \tilde{G}^{bd}(x,y;\tilde{A}) \tilde{\mathcal{R}}_\text{inv}\left( \textstyle\frac{1}{g} \, \mathcal{G}^\mu A_\mu^d(y) \right) \tilde{G}_i^c(x) \frac{\delta}{\delta \tilde{G}_i^a(x)} \, .
\end{aligned}
\end{align}
\end{footnotesize}%
Without any explicit calculations, we can already tell that the Nicolai maps corresponding to this $\mathcal{\tilde{R}}_g^\mathrm{max}$-operator will be highly complicated. Even though physical observables do not depend on the auxiliary fields, these will be present in the Nicolai maps of all bosonic fields. 

\section{`On-Shell' Supersymmetry}\label{sec:YM4OnShell}
We consider the `on-shell' formulation of maximally extended $\mathcal{N}=1$ super Yang-Mills, \emph{i.e.} $\mathcal{N}=4$ super Yang-Mills, in the hope of a more concise $\mathcal{R}_g$-operator and Nicolai map. The $\mathcal{R}_g^4$-operator of $\mathcal{N}=4$ super Yang-Mills is derived exactly like the `on-shell' $\mathcal{N}=1$ super Yang-Mills $\mathcal{R}_g$-operator in the previous chapter. Due to the similarity between $\mathcal{N}=1$ super Yang-Mills and $\mathcal{N}=4$ super Yang-Mills, we can once again immediately predict the structure of the $\mathcal{N}=4$ super Yang-Mills $\mathcal{R}_g^4$-operator
\begin{align}\label{eq:YM4ROp3}
\mathcal{R}_g^4 X \coloneqq \frac{\mathrm{d} X}{\mathrm{d}g} 
+ \bcontraction{}{\Delta}{_{\alpha A} \, (}{\delta} \Delta_{\alpha A} \, (\delta_{\alpha A}  \, X)
+ \int \mathrm{d}^4x \ 
\bcontraction[1.5ex]{}{\bar{C}}{^a \,  \delta_{\alpha A} ( \partial^\mu A_\mu^a) \Delta_{\alpha A} \,  }{s}
\bar{C}^a \,  
\bcontraction{}{\delta}{_{\alpha A} ( \partial^\mu A_\mu^a) }{\Delta}
\delta_{\alpha A} ( \partial^\mu A_\mu^a) \Delta_{\alpha A} \, s (X)+ 
\bcontraction{}{\, }{\ }{} Z_4 \, X \, .
\end{align}
Notice that the supersymmetry variation and $\Delta_{\alpha A}$ carry an additional index $A= 1,2,3,4$ like the four Majorana fermions $\lambda_{\alpha A}^a$. The multiplicative contribution $Z_4$ again only vanishes in the Landau gauge. This can be seen from a direct calculation or by dimensionally reducing the 10-dimensional $Z$ from \eqref{eq:YM2Z}. 

So, as before, we must now only figure out the supersymmetry and BRST variations as well as the explicit expression of $\Delta_{\alpha A}$ to give the explicit form of \eqref{eq:YM4ROp3}. To this end, recall the gauge invariant $\mathcal{N}=4$ super Yang-Mills action with Majorana spinors \eqref{eq:YM4}
\begin{align}
\begin{aligned}\label{eq:YM41}
S_\mathrm{inv}^4 = \int \mathrm{d}^4x \ \bigg[ &-\frac{1}{4} F_{\mu\nu}^a F^{a\, \mu\nu} - \frac{1}{2} (D_\mu A_i)^a (D^\mu A^i)^a - \frac{1}{2} (D_\mu B_i)^a (D^\mu B^i)^a \\
&- \frac{i}{2} \bar{\lambda}_A^a \gamma^\mu (D_\mu \lambda_A)^a + \frac{g}{2} f^{abc} \bar{\lambda}_A^a (\alpha_{AB}^i A_i^b + i \gamma_5 \beta_{AB}^i B_i^b) \lambda_B^c \\
&- \frac{g^2}{4} f^{abc} f^{ade} \left(A_i^b A_j^c A^{d \, i} A^{e \, j} + B_i^b B_j^c B^{d \, i} B^{e \, j} + 2 A_i^b B_j^c A^{d \, i} B^{e \, j} \right) \bigg] \, .
\end{aligned}
\end{align}
The action is invariant under the supersymmetry transformations \eqref{eq:susyAlgebraYM3}
\begin{align}
\begin{aligned}\label{eq:YM4susy2}
&\begin{aligned}
&\delta A_\mu^a = i (\bar{\varepsilon}_A \gamma_\mu \lambda_A^a) \, , &&\delta A_i^a = - (\bar{\varepsilon}_A \alpha_{i AB} \lambda_B^a) \, , &&\delta B_i^a = - i (\bar{\varepsilon}_A \gamma_5 \beta_{iAB} \lambda_B^a) \, , \\
\end{aligned} \\
&\delta \lambda_{\alpha A}^a = - \frac{1}{2} (\gamma^{\mu\nu} \varepsilon_A)_\alpha F_{\mu\nu}^a - i (\gamma^\mu \alpha_{AB}^i \varepsilon_B)_\alpha (D_\mu A_i)^a - (\gamma_5 \gamma^\mu \beta_{AB}^i \varepsilon_B)_\alpha (D_\mu B_i)^a \\
&\quad \quad \quad - \frac{g}{2} f^{abc} \left( \left(\alpha_{AB}^i A_i^b - i \gamma_5 \beta_{AB}^i B_i^b \right) \left( \alpha_{BC}^j A_j^c + i \gamma_5 \beta_{BC}^j B_j^c \right) \varepsilon_C \right)_\alpha \, .
\end{aligned}
\end{align}
The gauge fixing term is the non-rescaled version of \eqref{eq:YM4gf} and the BRST transformations are the non-rescaled version of \eqref{eq:YM4BRST} without the auxiliary fields. Subsequently, we find that the $\mathcal{N}=4$ super Yang-Mills versions of \eqref{eq:YM2Delta} and \eqref{eq:YM2Z} are given by \cite{Rupprecht:2021wdj}
\begin{align}
\begin{aligned}
\Delta_{\alpha A} = - \frac{1}{16} f^{abc} \int \mathrm{d}^4x \ \bigg[ &\frac{1}{2} (\gamma^{\mu\nu} \lambda_A^a)_\alpha A_\mu^b A_\nu^c + i (\gamma^\mu A_\mu^a (\alpha_{AB}^i A_i^b + i \gamma_5 \beta_{AB}^i B_i^b) \lambda_B^c)_\alpha \\
&+ \frac{1}{2} \left[ \left(\alpha_{AB}^i A_i^a - i \gamma_5 \beta_{AB}^i B_i^a \right) \left(\alpha_{BC}^j A_j^b + i \gamma_5 \beta_{BC}^j B_j^b \right) \lambda_{C}^c \right]_\alpha \bigg] \
\end{aligned}
\end{align}
and
\begin{align}
\begin{aligned}
Z &= \frac{1}{16} \int \mathrm{d}^4y \ \bar{C}^a(y) \delta_{A \, \alpha} \left( \partial^\rho A_\rho^a(y) \right) f^{abc} \int \mathrm{d}^4x \ (\gamma^{\mu\nu} \lambda_A^a(x))_\alpha\partial_\mu C^b(x) A_\nu^c(x) \\
&\quad - \frac{i}{16} \int \mathrm{d}^4y \ \bar{C}^a(y) \delta_{A \, \alpha} \left( \partial^\rho A_\rho^a(y) \right) \\
&\quad \quad \times f^{abc} \int \mathrm{d}^4x \ (\gamma^\mu \partial_\mu C^a(x) (\alpha_{AB}^i A_i^b(x) - i \gamma_5 \beta_{AB}^i B_i^b(x)) \lambda_B^c(x))_\alpha \\
&\quad + \frac{1}{16} f^{abc} \int \mathrm{d}^4 x \ \bar{\lambda}_A^a(x) \gamma^\mu A_\mu^b(x) \lambda_A^c(x) \\
&\quad + \frac{i}{16} f^{abc} \int \mathrm{d}^4 x \ \bar{\lambda}_A^a(x) (\alpha_{AB}^i A_i^b(x) + i \gamma_5 \beta_{AB}^i B_i^b(x)) \lambda_B^c(x) \\
&\quad - i f^{abc} \int \mathrm{d}^4 x \ \bar{C}^a(x) \partial^\mu \left( A_\mu^b(x) C^c(x) \right) \, .
\end{aligned}
\end{align}
The fermion propagator $i \bcontraction{}{\lambda}{_A^a(x) }{\bar{\lambda}} \lambda_A^a(x) \bar{\lambda}_B^b(y) \equiv S_{AB}^{ab}(x,y;\mathscr{A})$ and Dirac equation
\begin{align}\label{eq:SAB2}
\left(\delta_{AC} \delta^{ac} \gamma^\mu D_\mu + i g f^{aec} \alpha_{AC}^i A_i^e(x) - g f^{aec} \gamma_5 \beta_{AC}^i B_i^e(x) \right) S_{CB}^{cb}(x,y;\mathscr{A}) = \delta_{AB} \delta^{ab} \delta(x-y) 
\end{align}
for $\mathcal{N}=4$ super Yang-Mills with $\mathscr{A} = (A_\mu^a, A_i^a, B_i^a)$ were defined in \eqref{eq:N4FermProp} and \eqref{eq:N4DiracEq}.

To obtain a compact expression for the $\mathcal{R}_g$-operator, we would like to write the gauge fixing term as a function of the remaining $\mathcal{R}_g^4$-operator, \emph{i.e.} the `on-shell' analog of \eqref{eq:YM4OffShellRg}. In chapter \ref{ch:YM2}, we have obtained a more compact form by introducing the transversal projector \eqref{eq:YM2P}. In $\mathcal{N}=4$ super Yang-Mills, it appears that it is advantageous to write the $\mathcal{R}_g$-operator as
\begin{align}\label{eq:YM4ROp2}
\mathcal{R}_g^4 = \frac{\mathrm{d}}{\mathrm{d}g } + \mathcal{R}_\mathrm{inv}^4 + \mathcal{R}_\mathrm{gf}^4
\end{align}
with
\begin{footnotesize}
\begin{align}\label{eq:YM4ROp4}
\begin{aligned}
\mathcal{R}_\text{inv}^4 &= - \frac{1}{32} f^{bde} \int \mathrm{d}^4x \ \mathrm{d}^4y \ \tr \Big[ \gamma_\mu S_{AA}^{ab}(x,y;\mathscr{A}) \gamma^{\rho\lambda} \Big] A_\rho^d(y) A_\lambda^e(y) \frac{\delta}{\delta A_\mu^a(x)} \\
&\quad - \frac{i}{16} f^{bde} \int \mathrm{d}^4x \ \mathrm{d}^4y \ \tr \Big[ \gamma_\mu S_{AB}^{ab}(x,y;\mathscr{A}) \left( \alpha_{BA}^i A_i^d(y) + i \gamma_5 \beta_{BA}^i B_i^d(y) \right) \gamma^\rho A_\rho^e(y) \Big] \frac{\delta}{\delta A_\mu^a(x)} \\
&\quad - \frac{i}{32} \int \mathrm{d}^4x \ \mathrm{d}^4y \ \tr \Big[ \gamma_\mu S_{AB}^{ab}(x,y;\mathscr{A}) \left(\alpha_{BA}^j (D_\rho A_j)^b(y)+ i \gamma_5 \beta_{BA}^j (D_\rho B_j)^b(y) \right) \gamma^\rho \Big] \frac{\delta}{\delta A_\mu^a(x)} \\
&\quad + \frac{i}{32} f^{bde} \int \mathrm{d}^4x \ \mathrm{d}^4y \ \tr \bigg[ S_{AB}^{ab}(x,y,\mathscr{A}) \gamma^{\mu\nu} A_\mu^d(y) A_\nu^e(y) \left( \alpha_{BA}^i \frac{\delta}{\delta A_i^a(x)} + i \gamma_5 \beta_{BA}^i \frac{\delta}{\delta B_i^a(x)} \right)\bigg] \\
&\quad - \frac{1}{16} f^{bde} \int \mathrm{d}^4x \ \mathrm{d}^4y \ \tr \bigg[ S_{AB}^{ab}(x,y;\mathscr{A}) \left( \alpha_{BC}^j A_j^d(y) + i \gamma_5 \beta_{BC}^j B_j^d(y) \right) \gamma^\rho A_\rho^e(y) \\
&\quad \quad \times \left( \alpha_{CA}^i \frac{\delta}{\delta A_i^a(x)} + i \gamma_5 \beta_{CA}^i \frac{\delta}{\delta B_i^a(x)} \right) \bigg] \\
&\quad - \frac{1}{32} \int \mathrm{d}^4x \ \mathrm{d}^4y \ \tr \bigg[ S_{AB}^{ab}(x,y;\mathscr{A}) \left(\alpha_{BC}^k (D_\rho A_k)^b(y) + i \gamma_5 \beta_{BC}^k (D_\rho B_k)^b(y) \right) \gamma^\rho \\
&\quad \quad \times \left( \alpha_{CA}^i \frac{\delta}{\delta A_i^a(x)} + i \gamma_5 \beta_{CA}^i \frac{\delta}{\delta B_i^a(x)} \right) \bigg] \\ 
&\quad - \frac{1}{2} \int \mathrm{d}^4x \ \tr \left[ A_i^a(x) \frac{\delta}{\delta A_i^a(x)} + B_i^a(x) \frac{\delta}{\delta B_i^a(x)} \right] 
\end{aligned}
\end{align}
\end{footnotesize}%
and
\begin{align}\label{eq:YM4ROp5}
\begin{aligned}
\mathcal{R}_\text{gf}^4 &= - \int \mathrm{d}^4x \ \mathrm{d}^4y \ (D_\mu G)^{ab}
(x,y;\mathscr{A}) \mathcal{R}_\text{inv} \left( \partial^\rho A_\rho^b(y) \right) \frac{\delta}{\delta A_\mu^a(x)} \\
&\quad + g f^{abc} \int \mathrm{d}^4x \ \mathrm{d}^4y \ G^{bd}(x,y;\mathscr{A}) \mathcal{R}_\text{inv} \left( \partial^\rho A_\rho^d(y) \right) A_i^c(x) \frac{\delta}{\delta A_i^a(x)} \\
&\quad + g f^{abc} \int \mathrm{d}^4x \ \mathrm{d}^4y \ G^{bd}(x,y;\mathscr{A}) \mathcal{R}_\text{inv} \left( \partial^\rho A_\rho^d(y) \right) B_i^c(x) \frac{\delta}{\delta B_i^a(x)} \, .
\end{aligned}
\end{align}
These expressions are already much more concise than \eqref{eq:YM4OffShellRinv} - \eqref{eq:YM4OffShellRg}. However, compared to the $\mathcal{N}=1$ super Yang-Mills $\mathcal{R}_g$-operator \eqref{eq:YM2ROp2}, they are still rather complicated. Hence, we want to show that the construction of the $\mathcal{R}_g$-operator is compatible with the dimensional reduction of 10-dimensional $\mathcal{N}=1$ super Yang-Mills to 4-dimensional $\mathcal{N}=4$ super Yang-Mills. This would allow us to circumvent the $\mathcal{N}=4$ super Yang-Mills $\mathcal{R}_g^4$-operator altogether. Instead, we could simply apply the dimensional reduction to the 10-dimensional $\mathcal{N}=1$ super Yang-Mills Nicolai map \eqref{eq:YM2Result}.

\subsection[The Dimensional Reduction of the 10-dimensional \texorpdfstring{$\mathcal{R}_g$}{R}-Operator]{The Dimensional Reduction of the 10-dimensional $\boldsymbol{\mathcal{R}_g}$-Operator}
We compute the dimensional reduction of the 10-dimensional $\mathcal{R}_g$-operator \eqref{eq:YM2ROp}. Recall that
\begin{align}
\mathcal{R}_g X \coloneqq \frac{\mathrm{d} X}{\mathrm{d}g} 
+ \bcontraction{}{\Delta}{_\tau^{10} \, (}{\delta} \Delta_\tau^{10} \, (\delta_\tau  \, X)
+ \int \mathrm{d}^{10}x \ 
\bcontraction[1.5ex]{}{\bar{C}}{^a \,  \delta_\tau ( \partial^M \mathcal{A}_\mu^a) \Delta_\tau^{10} \,  }{s}
\bar{C}^a \,  
\bcontraction{}{\delta}{_\tau ( \partial^M \mathcal{A}_M^a) }{\Delta}
\delta_\tau ( \partial^M \mathcal{A}_M^a) \Delta_\tau^{10} \, s (X)+ 
\bcontraction{}{\, }{\ }{} Z \, X \, .
\end{align}
with $M= 0, \ldots, 9$, $\tau = 1, \ldots, 32$ and
\begin{align}
\Delta_\tau^{10} = - \frac{1}{64} f^{abc} \int \mathrm{d}^{10}x \ \left( \Gamma^{MN} \Lambda^a \right)_\tau \mathcal{A}_M^b \mathcal{A}_N^c \, .
\end{align}
In section \ref{sec:SFTDimRed}, we have explained how to dimensionally reduce the 10-dimensional $\mathcal{N}=1$ super Yang-Mills action to the 4-dimensional $\mathcal{N}=4$ super Yang-Mills action. The cost of this calculation was that we had to choose an explicit spinor representation \eqref{eq:SFTGamma10}
\begin{align}
\begin{aligned}
&\Gamma^\mu \coloneqq \gamma^\mu \otimes \mathbbm{1}_8 \, , &&\mu = 0,1,2,3 \, , \\
&\Gamma^{3+i} \coloneqq \gamma_5 \otimes \begin{pmatrix} 0 & i \alpha^i \\ -i \alpha^i & 0 \end{pmatrix} \, , \quad \quad &&i = 1,2,3 \, , \\
&\Gamma^{6+i} \coloneqq \gamma_5 \otimes \begin{pmatrix} 0 & \beta^i \\ \beta^i & 0 \end{pmatrix} \, , &&i = 1,2,3 \, .
\end{aligned}
\end{align}
These gamma matrices fix the form of a general 10-dimensional Majorana-Weyl spinor to \eqref{eq:SFTLambda10}
\begin{align}
\Lambda^a = \begin{pmatrix} \chi^a \\ \bar{\chi}^a \end{pmatrix} \quad \text{with} \quad
\chi^a = \begin{pmatrix} \mathbf{0} \\ \psi_1^a \\ \mathbf{0} \\ \psi_2^a \end{pmatrix}  \quad \text{and}  \quad 
\psi_i^a = \begin{pmatrix} \omega_{1i}^a \\ \omega_{2i}^a \\ \omega_{3i}^a \\ \omega_{4i}^a \end{pmatrix} \, , \quad 
\mathbf{0} = \begin{pmatrix} 0 \\ 0 \\ 0 \\ 0 \end{pmatrix} \, .
\end{align}
During the dimensional reduction, this 32-component Majorana-Weyl spinor decomposes into four 4-component Majorana spinors
\begin{align}
\lambda_A^a = \begin{pmatrix} \omega_{A1}^a \\ \omega_{A2}^a \\ - \bar{\omega}_{A2}^a \\ \bar{\omega}_{A1}^a \end{pmatrix} \, .
\end{align}
We already know from section \ref{sec:SFTDimRed} that the dimensional reduction of the 10-dimensional supersymmetry variations implies the 4-dimensional supersymmetry variations \eqref{eq:YM4susy2}. Thus, it remains to be shown that $\Delta_\tau^{10}$ is related to the $\Delta_{\alpha A}$ in the same way that $\Lambda_\tau^a$ is related to the $\lambda_{\alpha A}^a$. A brief calculation indeed reveals that
\begin{align}
\Delta_{4+A}^{10} = \Delta_{1A} \, , \quad \Delta_{12+A}^{10} = \Delta_{2A} \, , \quad \Delta_{16+A}^{10} = - \Delta_{3A} \, , \quad \Delta_{24+A}^{10} = - \Delta_{4A} 
\end{align}
for all $A = 1,2,3,4$. Hence, we can conclude that $\mathcal{R}_g^{10}$ is equivalent to $\mathcal{R}_g^4$ upon dimensional reduction. Thus, instead of computing the $\mathcal{N}=4$ super Yang-Mills Nicolai map via the $\mathcal{R}_g$-operator \eqref{eq:YM4ROp2}, we can take the 10-dimensional $\mathcal{N}=1$ super Yang-Mills Nicolai map from \eqref{eq:YM2Result} and imply the dimensional reduction. Furthermore, this allows us to obtain the map directly for the six scalar fields $\phi_I^a$ instead of the $3+3$ scalar and pseudoscalar fields $A_i^a$ and $B_i^a$, without passing through a Weyl spinor formulation of the $\mathcal{R}_g^4$-operator. Up to the second order in the coupling, we find \cite{Nicolai:2020tgo}
\begin{align}
\begin{aligned}\label{eq:N4NicMap1}
(\mathcal{T}_g \, A)_\mu^a(x) &= A_\mu^a(x) + g f^{abc} \int \mathrm{d}^4 y \ \partial^\rho C(x-y) A_\mu^b(y) A_\rho^c(y) \\
&\quad +  \frac{3g^2}{2} f^{abc} f^{bde} \int \mathrm{d}^4 y \ \mathrm{d}^4 z \ \partial^\rho C(x-y) A^{c \,  \lambda}(y) \partial_{[\rho} C(y-z) A_\mu^d(z) A_{\lambda]}^e(z) \\
&\quad +  g^2 f^{abc} f^{bde} \int \mathrm{d}^4 y \ \mathrm{d}^4 z \ \partial^\rho C(x-y) \phi^{c \, I}(y) \partial_{[\rho} C(y-z) A_{\mu]}^d(z) \phi_I^e(z) \\
&\quad + \mathcal{O}(g^3) 
\end{aligned}
\end{align}
and
\begin{align}
\begin{aligned}\label{eq:N4NicMap2}
(\mathcal{T}_g \, \phi)_I^a(x) &= A_\mu^a(x) + g f^{abc} \int \mathrm{d}^4 y \ \partial^\rho C(x-y) \phi_I^b(y) A_\rho^c(y) \\
&\quad -  g^2 f^{abc} f^{bde} \int \mathrm{d}^4 y \ \mathrm{d}^4 z \ \partial^\rho C(x-y) A^{c \,  \lambda}(y) \partial_{[\rho} C(y-z) A_{\lambda]}^d(z) \phi_I^e(z) \\
&\quad -  \frac{g^2}{2} f^{abc} f^{bde} \int \mathrm{d}^4 y \ \mathrm{d}^4 z \ \partial^\rho C(x-y) \phi^{c \, J}(y) \partial_{\rho} C(y-z) \phi_J^d(z) \phi_I^e(z) \\
&\quad + \mathcal{O}(g^3) \, .
\end{aligned}
\end{align}
In the next chapter, we will see yet another approach to the Nicolai map in $\mathcal{N}=4$ super Yang-Mills. We will compute the vacuum expectation value of a $\mathcal{N}=4$ Wilson loop by expressing it in terms of 10-dimensional $\mathcal{N}=1$ fields. This is the most effective method of computing $\mathcal{N}=4$ super Yang-Mills correlation functions or vacuum expectation values with the Nicolai map since it allows us to directly use the $\mathcal{N}=1$ Nicolai map \eqref{eq:YM2Result}. A Nicolai map which is simpler than any map we have found in this chapter. 

\section{Correlation Functions and the Nicolai Map}
In \cite{Nicolai:2020tgo} Nicolai and Plefka have used the Nicolai maps \eqref{eq:N4NicMap1} and \eqref{eq:N4NicMap2} to compute all scalar 2-, 3-, and 4-point functions in $\mathcal{N}=4$ super Yang-Mills up to $\mathcal{O}(g^2)$. As expected, their results agree with the previous calculation from \emph{e.g.} \cite{Drukker:2009sf}. Furthermore, they constructed the one-loop dilatation operator matching the result from \citep{Beisert:2003tq}.

In the last chapter of \cite{Nicolai:2020tgo}, the authors also speculate about a possible invariance of (certain) Wilson loops under the action of the $\mathcal{R}_g^4$-operator \eqref{eq:YM4ROp3}. However, it turns out that no Wilson loop is invariant under the action of the $\mathcal{R}_g^4$-operator. This can be seen by a direct calculation. More generally, the statement that the vacuum expectation value of some operator $\mathcal{O}$ does not receive quantum corrections is much weaker than the operator being invariant under the action of the $\mathcal{R}_g^4$-operator as this would imply that also all $n$-point functions $\left<\!\left< \mathcal{O} \ldots \mathcal{O}\right>\!\right>_g$ of the operator $\mathcal{O}$ do not receive any quantum corrections. The statement follows from the simple observation that $\mathcal{R}_g^4 \,  \mathcal{O} = 0$ implies $\mathcal{T}_g^{-1} \mathcal{O} = \mathcal{O}$ and thus 
\begin{align}
\left<\!\left< \mathcal{O} \ldots \mathcal{O}\right>\!\right>_g = \big< \mathcal{T}_g^{-1} \mathcal{O} \ldots \mathcal{T}_g^{-1} \mathcal{O}\big>_0 =\left< \mathcal{O} \ldots \mathcal{O}\right>_0 \, .
\end{align}
This is clearly not the case for Wilson loops, as has been demonstrated in \cite{Erickson:2000af} for straight anti-parallel lines and in \cite{Plefka:2001bu} for circles. 

So far, the examples of operators annihilated by the $\mathcal{R}_g$-operator are scarce. In \cite{Nicolai:1984jg} Nicolai showed that if the $\mathcal{R}_g$-operator \eqref{eq:YM2ROp} is restricted to 4 dimensions and modified by letting
\begin{align}
\Delta_\alpha \to \Delta_\alpha^\pm = \frac{1}{2} (1 \pm \gamma^5)_{\alpha\beta} \Delta_\beta
\end{align}
it annihilates the (anti) self-dual field strength
\begin{align}
F_{\mu\nu}^{a \, \pm}(x) = \frac{1}{2} F_{\mu\nu}^a(x) \pm \frac{1}{4} \epsilon_{\mu\nu\rho\lambda} F^{a\, \rho\lambda}(x)
\end{align}
on the gauge surface. For the $\mathcal{N}=4$ super Yang-Mills $\mathcal{R}_g^4$-operator, this is no longer true. The next best prospect are operators which are not annihilated by the $\mathcal{R}_g^4$-operator but whose Nicolai map gives rise to some simplifications. A natural candidate in $\mathcal{N}=4$ super Yang-Mills would be BPS operators. These are operators corresponding to states which are annihilated by some of the supersymmetry generators (see chapters \ref{ch:SCA} and \ref{ch:BPS} for details). In $\mathcal{N}=4$ super Yang-Mills, some of these BPS operators are so strongly protected that their 2- and 3-point functions do not receive any quantum corrections. Unfortunately, however, the Nicolai maps of these operators are not significantly simpler than the maps for similar non-BPS operators. For example compare the Nicolai map of the unprotected Konishi operator $\mathcal{K} = \tr_c(\phi_I \phi^I)$ to the Nicolai map of the $\frac{1}{2}$-BPS operator $\varphi_{IJ} \coloneqq \tr_c( \phi_I \phi_J) - \frac{\eta_{IJ}}{6} \tr_c(\phi_K \phi^K)$ by using \eqref{eq:N4NicMap2}. We suspect this is because the 4-point functions of BPS operators are not protected. Nevertheless, the 2- and 3-point functions computed with the Nicolai map reveal the expected simplifications. Thus possibly a focus on correlation functions of BPS operators via the Nicolai map rather than the study of the Nicolai maps of the operators themselves is more fruitful. However, this is left for the future.

\subsection[The Large \texorpdfstring{$N$}{N} Limit]{The Large $\boldsymbol{N}$ Limit}
In \cite{tHooft:1973alw} `t Hooft showed that the $SU(N)$ symmetry structure of $\mathcal{N}=4$ super Yang-Mills allows us to interpret $N$ as an additional coupling constant and consider correlation functions as expansions in powers of $1/N$. A particularly interesting case arises in the large $N$ (or `t Hooft) limit $N \to \infty$, $g \to 0$ and $g^2 N$ fixed. In this limit, only planar Feynman diagrams survive, and the theory is believed to be integrable \cite{Minahan:2002ve,Beisert:2003tq,Beisert:2003jj,Beisert:2003yb}. Moreover, the planar limit is important in the context of the AdS/CFT correspondence \cite{Maldacena:1997re} (see also \cite{Beisert:2010jr} for a review). In its strongest version, the correspondence claims that there is an exact equivalence between 4-dimensional $\mathcal{N}=4$ super Yang-Mills and type IIB superstring theory on the $\mathrm{AdS}_5 \times S^5$ background. However, profound tests of the correspondence, such as comparing the anomalous dimension of the dilatation operator \cite{Beisert:2003xu} to certain solitonic closed-string solutions \cite{Arutyunov:2003uj}, are constrained to the large $N$ limit of the gauge theory.

Thus, one might wonder about the large $N$ limit in the context of the Nicolai map. After all, the image of the map is a free, \emph{i.e.} integrable field theory. However, the Nicolai map itself \eqref{eq:N4NicMap1} - \eqref{eq:N4NicMap2} does not depend on $N$ but merely the structure constants $f^{abc}$. Thus, the large $N$ expansion becomes visible only when computing vacuum expectation values or correlation functions. We obtain factors of $N$ via contractions of the type $f^{abc} f^{abd} = N \delta^{cd}$. In general, all terms in the Nicolai map contribute to all orders of the $1/N$ expansion through their Wick contractions with each other. This appears to rule out the existence of a large $N$ Nicolai map. Moreover, neither the $\mathcal{N}=4$ super Yang-Mills action nor the supersymmetry variations are altered by resorting to the large $N$ limit. Thus, also the construction of $\mathcal{R}_g^4$ will not change in this limit.

\chapter{Wilson Loops and the Nicolai Map}\label{ch:WilsonLoop}
In this chapter, we demonstrate an application of the Nicolai map by computing the vacuum expectation value of the infinite straight line Maldacena-Wilson loop in $\mathcal{N}=4$ super Yang-Mills to order $g^6$ (for all $N$). Thus, we extend the previous perturbative result by one order. The results of this chapter are twofold. The perturbative cancellations of the different contributions to the Maldacena-Wilson loop are by no means trivial and seem to resemble those of the circular Maldacena-Wilson loop at order $g^4$. Furthermore, we argue that our approach to computing quantum correlation functions is competitive with more standard diagrammatic techniques.

The chapter is organized as follows. In the first section, we introduce the Maldacena-Wilson loop and briefly review previous results and their importance in the AdS/CFT correspondence. In the second section, we outline our strategy for the perturbative calculation of the vacuum expectation value. In section \ref{sec:Divergences}, we discuss the possible divergences arising in the calculation. Section \ref{sec:PerturbationTheory} contains the details of the perturbative calculation of the vacuum expectation value. In particular, we show that all quantum corrections vanish up to $\mathcal{O}(g^6)$. 

This chapter is heavily based on the author's publication \cite{Malcha:2022fuc}.

\section{Introduction}
Wilson loops are gauge invariant operators describing the parallel transport of a gauge field around a closed loop \cite{Wilson:1974sk}. Thus, they are essentially path dependent phase factors. In a pure gauge theory, they form an over-complete basis of gauge invariant operators \cite{Giles:1981ej}. In gauge theories containing matter, open Wilson loops (or Wilson lines) can be thought of as parallel transport operators used to compare two quark fields $\psi(x)$ and $\psi(y)$, transforming under the fundamental representation of the gauge group, at two different points $x$ and $y$ (see \emph{e.g.} \cite{Peskin:1995ev} for details). 

In super Yang-Mills field theories, the Wilson line along some curve $\mathcal{C}$ from $a$ to $b$ is defined as
\begin{align}\label{eq:WL}
\mathcal{W}_\mathcal{C}(a,b) = \mathcal{P} \, \exp \left( i g \int_\mathcal{C} \mathrm{d}x^\mu \, A_\mu(x) \right) \, , 
\end{align}
where $A_\mu(x) = t^a A_\mu^a(x)$ is the Lie algebra valued gauge potential, with $t^a$ the traceless hermitian generators of the fundamental representation of $\mathfrak{su}(N)$. $\mathcal{P}$ denotes the path ordering. Given a parametrization $x^\mu(\tau)$ of the curve $\mathcal{C}$ with $x(0) = a$ and $x(1) = b$ \eqref{eq:WL} becomes
\begin{align}
\mathcal{W}_\mathcal{C}(a,b) = \mathcal{P} \, \exp \left( i g \int_0^1 \mathrm{d}\tau  \, A_\mu(x) \dot{x}^\mu \right) \, , 
\end{align}
where we have abbreviated $\dot{x}^\mu \equiv \frac{\mathrm{d}x(\tau)}{\mathrm{d}\tau}$ and $x \equiv x(\tau)$. The path ordering is such that larger values of $\tau$ stand to the right. When we take the trace of a Wilson line along a closed curve $\mathcal{C}$, we obtain the gauge invariant Wilson loop
\begin{align}
\mathcal{W}(\mathcal{C}) = \frac{1}{N} \tr_c \, \mathcal{P} \, \exp \left( i g \oint_\mathcal{C} \mathrm{d}\tau \, A_\mu(x) \dot{x}^\mu \right) \, .
\end{align}
For the conventions on the trace, see section \ref{sec:Notation}. The factor of $\frac{1}{N}$ is to normalize the large $N$ expansion of the vacuum expectation value.  

In \cite{Maldacena:1998im}, Maldacena generalized the definition of the Wilson loop for $\mathcal{N}=4$ super Yang-Mills by additionally coupling it to the scalar fields. The (gauge invariant) Euclidean Maldacena-Wilson loop is defined as
\begin{align}\label{eq:MWL}
\mathcal{W}^M(\mathcal{C}) = \frac{1}{N} \tr_c \, \mathcal{P} \, \exp \left( i g \oint_{\mathcal{C}} \mathrm{d}\tau \left( A_\mu(x) \dot{x}^\mu + i \phi_I(x) |\dot{x}| \theta^I \right) \right) \, ,
\end{align}
where $\theta^I$ describes a point on the unit 5-sphere, \emph{i.e.} $\theta_I \theta^I = 1$, and $x(\tau)$ parametrizes the curve $\mathcal{C}$. In general $\theta^I$ can depend on $\tau$. However, in the following, we will restrict ourselves to the simpler case of constant $\theta^I$. The Maldacena-Wilson loop plays an important role in the context of the AdS/CFT correspondence as its vacuum expectation value is believed to be dual to the area of the minimal surface of a disk in supergravity \cite{Maldacena:1998im,Drukker:2000rr}. However, the instances of good tests for the AdS/CFT correspondence are scarce. Usually, the problem is that the gauge theory limit of large $N$ and large $g^2N$ is not attainable in perturbation theory. Thus, for a long time, most tests of the AdS/CFT correspondence have been restricted to operators so protected by supersymmetry that they do not receive any quantum corrections (see chapters \ref{ch:SCA} and \ref{ch:BPS}). However, in a series of papers, Erickson, Drukker, Gross, Pestun, Semenoff and Zarembo found an exact expression for the vacuum expectation value of a circular Maldacena-Wilson loop to all orders in the coupling constant \cite{Erickson:2000af,Drukker:2000rr,Pestun:2007rz}. After an initial conjecture by Erickson, Semenoff and Zarembo, who summed only certain types of Feynman diagrams to obtain the vacuum expectation value, Drukker, Gross and Pestun used the technique of localization to obtain the vacuum expectation value of the circular Maldacena-Wilson loop from that of the infinite straight line. The idea behind the proof is that the infinite straight line and the circle are related by a special conformal transformation and that the difference of the vacuum expectation values for Maldacena-Wilson loops of these two shapes depends merely on the one point of the circle that gets sent to infinity during the special conformal transformation turning it into an infinite straight line. The all-order gauge theory result has been successfully tested up to the second order in string theory \cite{Drukker:1999zq,Semenoff:2001xp,Zarembo:2002an,Semenoff:2002kk}. 

Unfortunately, however, there is a problem with this very nice result. While it is widely believed that the vacuum expectation value of the infinite straight line Maldacena-Wilson loop is exactly equal to one, \emph{i.e.}
\begin{align}\label{eq:WLVEV}
\big<\!\!\big< \mathcal{W}(-) \big>\!\!\big>_g = 1 
\end{align}
this has not been proven. In \cite{Zarembo:2002an}, Zarembo showed that the infinite straight line Maldacena-Wilson loop is a $\frac{1}{2}$-BPS object, \emph{i.e.} it preserves $\frac{1}{2}$ of the $\mathcal{N}=4$ supersymmetries. But contrary to the type of BPS operators we will meet in the next two chapters, this does not imply anything regarding the vacuum expectation value of the Maldacena-Wilson loop. Thus, for now, we are limited to perturbation theory in which the vacuum expectation value of the infinite straight line Maldacena-Wilson loop has previously only been computed to $\mathcal{O}(g^4 N^2)$ \cite{Zarembo:2002an}. In the following, we want to show how to use our fourth-order $\mathcal{N}=1$ super Yang-Mills Nicolai map \eqref{eq:FourthOrderResult} to obtain the vacuum expectation value of the infinite straight line Maldacena-Wilson loop up to $\mathcal{O}(g^6)$ (for all $N$). 

\section{The Wilson Loop and the Nicolai Map}
In this chapter we use the Euclidean metric. We parametrize the infinite straight line Maldacena-Wilson loop by $x^\mu(\tau) = (\tau,0,0,0)$
\begin{align}\label{eq:MWLLine}
\mathcal{W}^M(-) = \frac{1}{N} \tr_c \, \mathcal{P} \, \exp \left( i g \int_{-\infty}^\infty \mathrm{d}\tau \left( A_\mu(x) \dot{x}^\mu + i \phi_I(x) |\dot{x}| \theta^I \right) \right) \, .
\end{align}
Due to the work of Dietz and Lechtenfeld, we know that the Nicolai map provides a ghost and fermion free quantization of supersymmetric Yang-Mills theories \cite{Lechtenfeld:1984me, Dietz:1984hf, Dietz:1985hga}. Recall that the vacuum expectation value of any bosonic monomial $X[\Phi]$ is given by
\begin{align}\label{eq:WLVEVX}
\left<\!\!\left< X[\Phi] \right>\!\!\right>_g = \big< X[\mathcal{T}_g^{-1} \, \Phi] \big>_0 \, ,
\end{align}
where $\mathcal{T}_g^{-1}$ is the inverse Nicolai map. Using the linearity of $\big<\!\!\big<\ldots\big>\!\!\big>_g$ and $\mathcal{T}_g^{-1} X[\Phi] = X[\mathcal{T}_g^{-1} \Phi]$ we can extend \eqref{eq:WLVEVX} to $n$-point correlators of bosonic operators $\mathcal{O}_i(x_i)$, \emph{i.e.} 
\begin{align}\label{eq:WLVEVO}
\big<\!\!\big< \mathcal{O}_1(x_1) \ldots \mathcal{O}_n(x_n) \big>\!\!\big>_g = \big< (\mathcal{T}_g^{-1}\mathcal{O}_1)(x_1) \ldots (\mathcal{T}_g^{-1}\mathcal{O}_n)(x_n) \big>_0 \, .
\end{align} 
A priori, there are several ways to compute the vacuum expectation value of the infinite straight line Maldacena-Wilson loop \eqref{eq:MWLLine} with the Nicolai map. The naive approach is to first compute the inverse Nicolai transformation of \eqref{eq:MWLLine} and then use \eqref{eq:WLVEVX} to find the vacuum expectation value. However, unfortunately, the Nicolai transform of \eqref{eq:MWLLine} is by no means a trivial expression and thus, the intermediate expressions in the calculation would be unnecessarily cumbersome. 

So instead of the 4-dimensional $\mathcal{N}=4$ super Yang-Mills Maldacena-Wilson loop \eqref{eq:MWLLine} we consider the 10-dimensional $\mathcal{N}=1$ super Yang-Mills Wilson loop 
\begin{align}\label{eq:WL10D}
\mathcal{W}^M(-) = \frac{1}{N} \tr_c \, \mathcal{P} \, \exp \left( i g \int_{-\infty}^\infty \mathrm{d}\tau \ A_M(z) \dot{z}^M \right) \, ,
\end{align}
with the 10-dimensional gauge field $A_M(z) = t^a A_M^a(z)$ and $\dot{z}^M = ( \dot{x}^\mu, \dot{y}^I) = ( \dot{x}^\mu, i |\dot{x}| \theta^I)$. We abbreviate $z_i \equiv z(\tau_i)$. For an infinite straight line $\dot{z}_i^M$ satisfies
\begin{align}
\delta_{MN} \,  \dot{z}_i^M \dot{z}_j^N = \dot{x}_i \cdot \dot{x}_j - |\dot{x}_i| |\dot{x}_j| = 0 \, .
\end{align}
Furthermore, we will not apply the inverse Nicolai map to \eqref{eq:WL10D} directly but rather first expand the vacuum expectation value of \eqref{eq:WL10D} in powers of the coupling $g$. Then we obtain simple $n$-point correlation functions of the 10-dimensional gauge field $A_M^a$, which are computed by the means of \eqref{eq:WLVEVO} and Wick's theorem. Up to the second order, the inverse of the 10-dimensional Nicolai map \eqref{eq:FourthOrderResult} is given by
\begin{align}
\begin{aligned}\label{eq:WLInverseNicolaiMap}
(\mathcal{T}_g^{-1} A)_M^a(z) &= A_M^a(z) - g f^{abc} \int \mathrm{d}^{10}v \ \partial^N C(z-v) A_M^b(v) A_N^c(v) \\
&\quad + \frac{g}{2} f^{abc} f^{bde} \int \mathrm{d}^{10}v \ \mathrm{d}^{10}w \ \Big\{ \\
&\quad \quad \quad + 3 \partial^N C(z-v) A^{c \, L}(v) \partial_{[M} C(v-w) A_N^d(w) A_{L]}^e(w) \\
&\quad \quad \quad - 4 \partial^N C(z-v) A_{[M}^c(v) \partial^L C(v-w) A_{N]}^d(w) A_L^e(w) \Big\} \\
&\quad + \mathcal{O}(g^3) \, .
\end{aligned}
\end{align}
For $n$-point quantum correlation functions, we define
\begin{align}
\big<\!\!\big< \mathcal{O}_1(x_1) \ldots \mathcal{O}_n(x_n) \big>\!\!\big>_m \coloneqq \big<\!\!\big< \mathcal{O}_1(x_1) \ldots \mathcal{O}_n(x_n) \big>\!\!\big>_g \bigg\vert_{\mathcal{O}(g^m)} \, , 
\end{align}
with $\big<\!\!\big< \mathcal{O}_1(x_1) \ldots \mathcal{O}_n(x_n) \big>\!\!\big>_0 = \big< \mathcal{O}_1(x_1) \ldots \mathcal{O}_n(x_n) \big>_0$. 

\section{Divergences and Dimensional Reduction}\label{sec:Divergences}
Starting at order $g^2$, the vacuum expectation value of a general Maldacena-Wilson loop is divergent when two or more space-time arguments approach each other. However, in \cite{Drukker:1999zq}, it was argued that these linear divergences cancel for loops of the type \eqref{eq:MWL}, which are parametrized by a four-vector $x^\mu$ and a point on the unit 5-sphere $\theta^I$. For an explicit proof at $\mathcal{O}(g^4)$ see \cite{Erickson:2000af}. In the case of the infinite straight line, the situation is even simpler since up to $\mathcal{O}(g^4)$ all divergent terms are proportional to $\dot{x}_\mu \dot{x}^\mu - |\dot{x}| |\dot{x}| = 0$. However, we will see that at order $g^6$, this simplicity ceases to exist as the internal structure of, in particular, the 2- and 3-point correlation functions becomes more involved. We expect to obtain two UV divergent contributions from these correlation functions, which cancel each other when they are summed up. 

Moreover, partial contributions of, for example, fermion or ghost loops to any $n$-point function are generally highly divergent. Luckily, the Nicolai map completely sidesteps the use of fermion and ghost fields in the computation of bosonic correlation functions. Thus, we will not see any divergences related to such loops. 

Since intermediate results in our calculations are UV divergent, regularization by dimensional reduction is in order. Our starting point is $\mathcal{N}=1$ super Yang-Mills in 10 dimensions, where we denote the spacetime indices by $M, N = 0, \ldots , 9$. Dimensionally reducing the 10-dimensional $\mathcal{N}=1$ theory to $\mathcal{N}=4$ super Yang-Mills in $2\omega$ dimensions, we split the spacetime indices $M = (\mu, I)$, where $\mu,\nu = 0, \ldots, 2\omega - 1$ and $I,J = 1, \ldots, 10 - 2\omega$. Likewise we decompose the coordinates $z^M = (x^\mu, y^I)$ and the gauge field
\begin{align}
A_M^a(x,y) = \big( A_\mu^a(x), \phi_I^a(x) \big) \, .
\end{align}
Notice that the dependence on the internal coordinates $y^I$ is dropped. The scalar propagator in $2\omega$ dimensions is (with the Laplacian $\Box \equiv \partial_\mu \partial^\mu$)
\begin{align}
C(x) = \int \frac{\mathrm{d}^{2\omega}k}{(2\pi)^{2\omega}} \frac{e^{ikx}}{k^2} \, .
\end{align}
It satisfies $- \Box \, C(x) = \delta(x)$ with the $2\omega$-dimensional delta function $\delta(x) \equiv \delta^{2\omega}(x)$. In $2\omega$ dimensions, we have
\begin{align}
C(x) = \frac{\Gamma(\omega-1)}{4\pi^\omega} \frac{1}{[x^2]^{\omega-1}} \, .
\end{align}
In 10 dimensions, the gauge field propagator is \eqref{eq:YMFreePropagator}
\begin{align}
\big< A_M^a(x) A_N^b(y) \big>_0 = \delta^{ab} \left( \delta_{MN} - (1-\xi) \frac{\partial_M \partial_N}{\Box} \right) C(x-y) \, .
\end{align}
Here $\xi$ is the gauge parameter. We argue that we can compute the inverse Nicolai map in Landau gauge ($\xi = 0$) whilst using the Feynman gauge ($\xi =1$) for the propagator because the Wilson loop is gauge invariant. When computing its vacuum expectation value, all terms coming from the gauge parameter dependent term in the propagator must vanish. Thus, without loss of generality, we choose $\xi =1$ and the propagator becomes
\begin{align}\label{eq:WLAProp}
\big< A_M^a(x) A_N^b(y) \big>_0 = \delta^{ab} \delta_{MN} C(x-y) \, .
\end{align}

\section{Perturbation Theory}\label{sec:PerturbationTheory}
In perturbation theory, the vacuum expectation value of \eqref{eq:WL10D} is given by
\begin{align}\label{eq:WLPertVEV}
\begin{aligned}
\big<\!\!\big< \mathcal{W}(-) \big>\!\!\big>_g = 1  &+ \frac{ig}{N} \int_{-\infty}^\infty \mathrm{d}\tau_1 \ \dot{z}_1^M \ \tr_c \, \big<\!\!\big< A_M(z_1) \big>\!\!\big>_g \\
& + \frac{i^2g^2}{2!N} \int_{-\infty}^\infty \mathrm{d}\tau_1 \ \mathrm{d}\tau_2 \ \dot{z}_1^M \dot{z}_2^N \ \tr_c \, \mathcal{P} \, \left<\!\!\left< A_M(z_1) A_N(z_2) \right>\!\!\right>_g  \\
& + \frac{i^3g^3}{3!N} \int_{-\infty}^\infty \mathrm{d}\tau_1 \ \mathrm{d}\tau_2 \ \mathrm{d}\tau_3 \ 
\dot{z}_1^M \dot{z}_2^N \dot{z}_3^L \ \tr_c \, \mathcal{P} \, \big<\!\!\big< A_M(z_1) A_N(z_2) A_L(z_3) \big>\!\!\big>_g \\
& + \ldots \, .
\end{aligned}
\end{align}
The expectation value has been computed perturbatively up to order $g^4N^2$ by Erickson, Semenoff and Zarembo in \cite{Erickson:2000af, Zarembo:2002an}. We have checked that their result also holds for all $N$. In the following, we show how to compute the next nontrivial order of \eqref{eq:WLPertVEV} by the means of the Nicolai map. Expanding the vacuum expectation value at order $g^6$, we obtain
\begin{align}
\begin{aligned}\label{eq:WLW}
&\quad \big<\!\!\big< \mathcal{W}(-) \big>\!\!\big>_6 \\
& = \frac{ig}{N} \int_{-\infty}^\infty \mathrm{d}\tau_1 \ \dot{z}_1^M \ \tr_c \, \big<\!\!\big< A_M(z_1) \big>\!\!\big>_5 \\
& \quad + \frac{i^2g^2}{2!N} \int_{-\infty}^\infty \mathrm{d}\tau_1 \ \mathrm{d}\tau_2 \ \dot{z}_1^M \dot{z}_2^N \ \tr_c \, \mathcal{P} \, \big<\!\!\big< A_M(z_1) A_N(z_2) \big>\!\!\big>_4 \\
& \quad + \frac{i^3g^3}{3!N} \int_{-\infty}^\infty \mathrm{d}\tau_1 \ \mathrm{d}\tau_2 \ \mathrm{d}\tau_3 \ 
\dot{z}_1^M \dot{z}_2^N \dot{z}_3^L \ \tr_c \, \mathcal{P} \, \big<\!\!\big< A_M(z_1) A_N(z_2) A_L(z_3) \big>\!\!\big>_3 \\
& \quad + \frac{i^4g^4}{4!N} \int_{-\infty}^\infty \mathrm{d}\tau_1 \ \mathrm{d}\tau_2 \ \mathrm{d}\tau_3 \ \mathrm{d}\tau_4 \ \dot{z}_1^M \dot{z}_2^N \dot{z}_3^L \dot{z}_4^P 
\ \tr_c \, \mathcal{P} \, \big<\!\!\big< A_M(z_1) A_N(z_2) A_L(z_3) A_P(z_4) \big>\!\!\big>_2 \\
& \quad + \frac{i^5g^5}{5!N} \int_{-\infty}^\infty \mathrm{d}\tau_1 \ \mathrm{d}\tau_2 \ \mathrm{d}\tau_3 \ \mathrm{d}\tau_4 \ \mathrm{d}\tau_5 \\
& \quad \quad \quad \times \dot{z}_1^M \dot{z}_2^N \dot{z}_3^L \dot{z}_4^P \dot{z}_5^Q 
\ \tr_c \, \mathcal{P} \, \big<\!\!\big< A_M(z_1) A_N(z_2) A_L(z_3) A_P(z_4) A_Q(z_5) \big>\!\!\big>_1 \\
& \quad + \frac{i^6g^6}{6!N} \int_{-\infty}^\infty \mathrm{d}\tau_1 \ \mathrm{d}\tau_2 \ \mathrm{d}\tau_3 \ \mathrm{d}\tau_4 \ \mathrm{d}\tau_5 \ \mathrm{d}\tau_6 \\
& \quad \quad \quad \times \dot{z}_1^M \dot{z}_2^N \dot{z}_3^L \dot{z}_4^P \dot{z}_5^Q \dot{z}_6^R 
\ \tr_c \, \mathcal{P} \, \big<\!\!\big< A_M(z_1) A_N(z_2) A_L(z_3) A_P(z_4) A_Q(z_5) A_R(z_6) \big>\!\!\big>_0 \, .
\end{aligned}
\end{align}
We briefly discuss the terms which vanish more or less trivially. The trace over the 1-point function is zero since for $t^a \in \mathfrak{su}(N)$
\begin{align}
 \tr_c \, \big<\!\!\big< A_M(z_1) \big>\!\!\big>_5 = \tr_c(t^a) \, \big<\!\!\big< A_M^a(z_1) \big>\!\!\big>_5 = 0 \, .
\end{align}
For the 4-point function, we need to expand the inverse Nicolai map \eqref{eq:WLInverseNicolaiMap} up to $\mathcal{O}(g^2)$. Then we use \eqref{eq:WLVEVO} and collect all terms of $\mathcal{O}(g^2)$. Computing the Wick contractions, we obtain several non-vanishing terms. However, once we multiply the correlation function with $\dot{z}_1^M \dot{z}_2^N \dot{z}_3^L \dot{z}_4^P$ and insert the parametrization of the straight line, everything cancels. The vanishing of the last two terms is rather simple. In both cases, there are Wick contractions of two untransformed fields. These produce terms which are proportional to $\dot{z}_{M \, i} \dot{z}_j^M = 0$. Thus, only the 2- and 3-point functions need to be discussed in detail.

\subsection{2-Point Function}
To compute the two-loop correction to the 2-point function, we need to expand the inverse Nicolai map \eqref{eq:WLInverseNicolaiMap} up to $\mathcal{O}(g^4)$\footnote{When computing the inverse map in Landau gauge it is necessary to explicitly enforce the gauge condition $\partial^\mu A_\mu = 0$ in all terms. This is similar to the determinant test for the Nicolai map in Landau gauge (see \cite{Ananth:2020lup}).}. At the fourth-order $(\mathcal{T}_g^{-1} A)_M^a$ has about 500 terms. We apply \eqref{eq:WLVEVO} to the 2-point function and collect all terms of $\mathcal{O}(g^4)$, \emph{i.e.}
\begin{align}
\begin{aligned}
\Sigma_1 &= \frac{i^2g^2}{2!N} \int_{-\infty}^\infty \mathrm{d}\tau_1 \ \mathrm{d}\tau_2 \ \dot{z}_1^M \dot{z}_2^N \ \tr_c \, \mathcal{P} \, \big<\!\!\big< A_M(z_1) A_N(z_2) \big>\!\!\big>_4 \\
 &= - \frac{g^6}{2N} \, \tr_c(t^a t^b) \int_{-\infty}^\infty \mathrm{d}\tau_1 \ \mathrm{d}\tau_2 \ \dot{z}_1^M \dot{z}_2^N \ \big< (\mathcal{T}_g^{-1} A)_M^a (z_1) (\mathcal{T}_g^{-1} A)_N^b (z_2) \big>_0 \ \Big\vert_{\mathcal{O}(g^4)} \, .
\end{aligned}
\end{align}
Because $\tr_c(t^a t^b) = \tr_c(t^b t^a)$ the path ordering is trivial. After computing the free field expectation value of the transformed fields and some basic simplifications, such as enforcing $f^{aab} = 0$, we obtain roughly 650 terms. Approximately a third of them are proportional to $\delta_{MN} \dot{z}_1^M \dot{z}_2^N = 0$. We observe that most of the remaining terms are proportional to
\begin{align}
\begin{aligned}\label{eq:WL2PointExample}
&\int_{-\infty}^\infty \ \mathrm{d}\tau_1 \ \mathrm{d}\tau_2 \ \dot{z}_1^M \dot{z}_2^N  \int \mathrm{d}^{10}y_1 \ \mathrm{d}^{10}y_2 \ \mathrm{d}^{10}y_3 \ \mathrm{d}^{10}y_4 \ \\
& \times C(z_1 - y_1) \partial_M C(y_1-y_3) \partial^P C(y_1-y_4) C(y_3-y_4) \partial_P C(y_4 - y_2) \partial_N C(y_3-y_2) C(y_2-z_2) \, ,
\end{aligned}
\end{align}
where the four derivatives may sit at any of the seven propagators. Using integration by parts, it is always possible to rearrange the contracted derivatives such that they act on two propagators both, depending on either $y_1$, $y_3$ or $y_4$. In this situation, we use
\begin{align}
\begin{aligned}
\int \mathrm{d}^{10}y_4 \ &\partial^P C(y_1-y_4) C(y_3-y_4) \partial_P C(y_4 - y_2) \\
&\quad \quad 
\begin{aligned}
 = \frac{1}{2} \int \mathrm{d}^{10}y_4 \ \Big\{ &- C(y_1-y_4) \Box \, C(y_3-y_4) C(y_4 - y_2) \\
&+ \Box \, C(y_1-y_4) C(y_3-y_4)  C(y_4 - y_2) \\
&+ C(y_1-y_4) C(y_3-y_4) \Box \, C(y_4 - y_2) \Big\} 
\end{aligned}
\end{aligned}
\end{align}
and $\Box \, C(x-y) = - \delta(x-y)$. Thus, \eqref{eq:WL2PointExample} becomes
\begin{align}
\begin{aligned}
&\frac{1}{4} \int_{-\infty}^\infty \ \mathrm{d}\tau_1 \ \mathrm{d}\tau_2 \ \dot{z}_1^M \dot{z}_2^N \int \mathrm{d}^{10}y_1 \ \mathrm{d^{10}}y_2 \ \mathrm{d}^{10}y_3 \ \Big\{ \\
&\quad \quad \quad + \frac{1}{2} C(z_1 - y_1) \partial_M C(y_1-y_3)^2  \partial_N C(y_3-y_2)^2 C(y_2-z_2) \\
&\quad \quad \quad - C(z_1 - y_1) \partial_M C(y_1-y_3)^2  C(y_1 - y_2) \partial_N C(y_3-y_2) C(y_2-z_2) \\
&\quad \quad \quad - C(z_1 - y_1) \partial_M C(y_1-y_3)  C(y_1-y_2)   \partial_N C(y_3-y_2)^2 C(y_2-z_2) \Big\}  \, .
\end{aligned}
\end{align}
The first term turns out to be a total derivative. Integrating by parts we obtain
\begin{align}
\int_{-\infty}^\infty \mathrm{d}\tau_1 \ \dot{z}_1^M \partial_M C(z_1-y_1) \ [\ldots] = \int_{-\infty}^\infty \mathrm{d}\tau_1 \ \frac{\partial}{\partial \tau_1} \, C(z_1-y_1) \ [\ldots] = 0 \, .
\end{align}
The other two terms can be combined using the observation
\begin{align}
\int \mathrm{d}^{10}y_3 \ \partial_M C(y_1-y_3)^2 \partial_N C(y_3-y_2) = \int \mathrm{d}^{10}y_3 \ \partial_M C(y_1-y_3) \partial_N C(y_3-y_2)^2 \, .
\end{align}
We repeat these steps on the other 400 non-vanishing terms. Subsequently, we perform the dimensional reduction and obtain the now very simple expression
\begin{align}
\begin{aligned}
\Sigma_1 &= \frac{i^2g^2}{2!N} \int_{-\infty}^\infty \mathrm{d}\tau_1 \ \mathrm{d}\tau_2 \ \dot{z}_1^M \dot{z}_2^N \ \tr_c \, \mathcal{P} \, \big<\!\!\big< A_M(z_1) A_N(z_2) \big>\!\!\big>_4 \\
&= g^6 N(N^2-1) \int_{-\infty}^\infty \mathrm{d}\tau_1 \ \mathrm{d}\tau_2 \ \dot{x}_1^\mu \dot{x}_2^\nu \int \mathrm{d}^{2\omega}y_1 \ \mathrm{d}^{2\omega}y_2 \ \mathrm{d}^{2\omega}y_3 \ \Big\{ \\
&\quad \quad \quad + \partial_\mu \partial_\nu C(x_1 - y_1) C(x_1-y_2) C(x_2 - y_1) C(y_1-y_3) C(y_2-y_3)^2 \\
&\quad \quad \quad + \frac{3}{2} \partial_\mu C(x_1-y_1) C(x_1-y_2) C(x_2-y_3) C(y_1-y_2) C(y_1-y_3) \partial_\nu C(y_2-y_3) \Big\} \, .
\end{aligned}
\end{align}
Neither of these terms is a total derivative as there are two $x_1$ dependencies in each. Thus, we must cancel $\Sigma_1$ against the 3-point function.

\subsection{3-Point Function}
For the 3-point function, the procedure is much the same as for the 2-point function. For the trace and path ordering, we find
\begin{align}
\begin{aligned}
\Sigma_2 &= \frac{i^3g^3}{3!N} \int_{-\infty}^\infty \mathrm{d}\tau_1 \ \mathrm{d}\tau_2 \ \mathrm{d}\tau_3 \ 
\dot{z}_1^M \dot{z}_2^N \dot{z}_3^L \ \tr_c \, \mathcal{P} \, \big<\!\!\big< A_M(z_1) A_N(z_2) A_L(z_3) \big>\!\!\big>_3 \\
&= - \frac{ig^3}{24N} \, d^{abc} \int_{-\infty}^\infty \mathrm{d}\tau_1 \ \mathrm{d}\tau_2 \ \mathrm{d}\tau_3 \ 
\dot{z}_1^M \dot{z}_2^N \dot{z}_3^L \, \big<\!\!\big< A_M^a(z_1) A_N^b(z_2) A_L^c(z_3) \big>\!\!\big>_3 \\
&\quad+ \frac{g^3}{24N} \, f^{abc} \int_{-\infty}^\infty \mathrm{d}\tau_1 \ \mathrm{d}\tau_2 \ \mathrm{d}\tau_3 \ \epsilon(\tau_1,\tau_2,\tau_3) \, 
\dot{z}_1^M \dot{z}_2^N \dot{z}_3^L \, \big<\!\!\big< A_M^a(z_1) A_N^b(z_2) A_L^c(z_3) \big>\!\!\big>_3 \, , 
\end{aligned}
\end{align}
where $d^{abc}$ is totally symmetric and
\begin{align}
\epsilon(\tau_1, \tau_2, \tau_3) = \left[ \theta(\tau_1 -\tau_2) - \theta(\tau_2 - \tau_1) \right] \left[ \theta(\tau_1 -\tau_3) - \theta(\tau_3 - \tau_1) \right] \left[ \theta(\tau_2 -\tau_3) - \theta(\tau_3- \tau_2) \right] \, .
\end{align}
So $\epsilon(\tau_1, \tau_2, \tau_3) = 1$ for $\tau_1 > \tau_2 > \tau_3$ and anti-symmetric under the transposition of any two $\tau_i$. The first term will cancel because the 3-point correlation function at $\mathcal{O}(g^3)$ is anti-symmetric in $a$, $b$ and $c$. This time we only need the inverse Nicolai map up to $\mathcal{O}(g^3)$. However, since we now compute a 3-point function instead of a 2-point function, we end up with about the same number of terms as before. But two-thirds of the terms are proportional to $\delta_{MN}$, $\delta_{ML}$ or $\delta_{NL}$ and thus cancel. The remaining terms are simplified using the same integration by parts relations as above. However, for the 3-point function, there are no total derivatives. Subsequently, we perform the dimensional reduction and obtain the 15 terms
\begin{align}
\begin{aligned}
\Sigma_2 &= - \frac{g^6 N(N^2-1)}{12} \int_{-\infty}^\infty \mathrm{d}\tau_1 \ \mathrm{d}\tau_2 \ \mathrm{d}\tau_3 \ \epsilon(\tau_1,\tau_2,\tau_3) \ \dot{x}_1^\mu \dot{x}_2^\nu \dot{x}_3^\rho \int \mathrm{d}^{2\omega}y_1 \ \mathrm{d}^{2\omega}y_2 \ \mathrm{d}^{2\omega}y_3 \ \Big\{ \\
&\quad \quad \quad + \partial_\mu \partial_\nu \partial_\lambda C(x_1-y_1) C(x_2-y_1) C(x_3-y_2) C(y_1-y_3) C(y_2-y_3)^2 \\
&\quad \quad \quad + \partial_\mu \partial_\nu C(x_1-y_1) \partial_\lambda C(x_2-y_1) C(x_3-y_2) C(y_1-y_3) C(y_2-y_3)^2 \\
&\quad \quad \quad + \text{permutations} \hspace*{251pt} \Big\} \\
&\quad + \frac{g^6 N(N^2-1)}{8} \int_{-\infty}^\infty \mathrm{d}\tau_1 \ \mathrm{d}\tau_2 \ \mathrm{d}\tau_3 \ \epsilon(\tau_1,\tau_2,\tau_3) \ \dot{x}_1^\mu \dot{x}_2^\nu \dot{x}_3^\rho \int \mathrm{d}^{2\omega}y_1 \ \mathrm{d}^{2\omega}y_2 \ \mathrm{d^{2\omega}}y_3 \ \Big\{ \\
&\quad \quad \quad + \partial_\mu \partial_\nu C(x_1-y_1) C(x_2-y_2) C(x_3-y_3) C(y_1-y_2) C(y_1-y_3) \partial_\lambda C(y_2-y_3) \\
&\quad \quad \quad - C(x_1-y_1) \partial_\nu \partial_\lambda C(x_2-y_2) C(x_3-y_3) C(y_1-y_2) \partial_\mu C(y_1-y_3) C(y_2-y_3) \\
&\quad \quad \quad + C(x_1-y_1) C(x_2-y_2) \partial_\lambda \partial_\mu C(x_3-y_3) \partial_\nu C(y_1-y_2) C(y_1-y_3) C(y_2-y_3) \Big\} \, .
\end{aligned}
\end{align}
All these terms have a factor of the form
\begin{align}
\dot{x}_i^\mu \partial_\mu C(x_i-y) = \frac{\partial}{\partial \tau_i} \, C(x_i-y) 
\end{align}
and this is their only dependence on $x_i$. Thus, we can integrate by parts and use
\begin{align}
\frac{\partial}{\partial \tau_1 } \epsilon(\tau_1,\tau_2,\tau_3) &= 2 \delta(\tau_1-\tau_2) - 2 \delta(\tau_1-\tau_3) \, .
\end{align}
After integrating over the delta functions and renaming the variables, we obtain
\begin{align}
\begin{aligned}
\Sigma_2 &= g^6 N(N^2-1) \int_{-\infty}^\infty \mathrm{d}\tau_1 \ \mathrm{d}\tau_2 \ \dot{x}_1^\mu \dot{x}_2^\nu \int \mathrm{d}^{2\omega}y_1 \ \mathrm{d}^{2\omega}y_2 \ \mathrm{d}^{2\omega}y_3 \ \Big\{ \\
&\quad \quad \quad + \partial_\mu \partial_\nu C(x_1-y_1) C(x_1-y_1) C(x_2-y_2) C(y_1-y_3) C(y_2-y_3)^2 \\
&\quad \quad \quad - \partial_\mu \partial_\nu C(x_1-y_1) C(x_1-y_2) C(x_2-y_1) C(y_1-y_3) C(y_2-y_3)^2 \\
&\quad \quad \quad + \partial_\nu C(x_1-y_1) \partial_\mu C(x_1-y_1) C(x_2-y_2) C(y_1-y_3) C(y_2-y_3)^2 \\
&\quad \quad \quad - \partial_\mu C(x_1-y_1) C(x_1-y_2) \partial_\nu C(x_2-y_1) C(y_1-y_3) C(y_2-y_3)^2 \\
&\quad \quad \quad - \frac{3}{2} \partial_\mu C(x_1-y_1) C(x_1-y_2) C(x_2-y_3) C(y_1-y_2) C(y_1-y_3) \partial_\nu C(y_2-y_3) \Big\} \, .
\end{aligned}
\end{align}
The first and third term can be combined to give a total derivative. Also, the fourth term is a total derivative. Subsequently, we conclude
\begin{align}
\begin{aligned}
\Sigma_2 &= -g^6 N(N^2-1) \int_{-\infty}^\infty \mathrm{d}\tau_1 \ \mathrm{d}\tau_2 \ \dot{x}_1^\mu \dot{x}_2^\nu \int \mathrm{d}^{2\omega}y_1 \ \mathrm{d}^{2\omega}y_2 \ \mathrm{d}^{2\omega}y_3 \ \Big\{ \\
&\quad \quad + \partial_\mu \partial_\nu C(x_1-y_1) C(x_2-y_1) C(x_1-y_2) C(y_1-y_3) C(y_2-y_3)^2 \\
&\quad \quad + \frac{3}{2} \partial_\mu C(x_1-y_1) C(x_1-y_2) C(x_2-y_3) C(y_1-y_2) C(y_1-y_3) \partial_\nu C(y_2-y_3) \Big\} \, .
\end{aligned}
\end{align}
We see that $\Sigma_1$ and $\Sigma_2$ cancel
\begin{align}
\Sigma_1 + \Sigma_2 = 0 \, .
\end{align}
Thus, we have shown that for a Maldacena-Wilson loop operator of an infinite straight line
\begin{align}
\big<\!\!\big< \mathcal{W}(-) \big>\!\!\big>_g = 1+ \mathcal{O}(g^8)
\end{align}
for all $N$. Although the infinite straight line Maldacena-Wilson loop is a $\frac{1}{2}$-BPS operator, the cancellation of the perturbative corrections at the sixth order is far from trivial. They seem to resemble the cancellations of the fourth-order perturbative corrections for the expectation value of the circular Maldacena-Wilson loop (see \cite{Erickson:2000af}). All correlation functions have been computed using the Nicolai map. Despite the complexity of intermediate results, such as the non-linear and non-local transformation of the gauge field to fourth order in \cite{Malcha:2021ess}, the general procedure is rather simple as it completely circumvents the use of anti-commuting variables. In the future, it will be interesting to see if the Nicolai map can also be used to obtain non-perturbative results for certain Wilson loop operators.

\chapter{The Superconformal Algebra and its Unitary Representations}\label{ch:SCA}
In this chapter, we introduce the superconformal algebra $\mathfrak{psu}(2,2|4)$ and its unitary representations. $\mathfrak{psu}(2,2|4)$ is the symmetry algebra of $\mathcal{N}=4$ super Yang-Mills. Due to the vanishing of the beta function to all orders, the superconformal symmetry is preserved even at the quantum level. Moreover, the superconformal symmetry puts very strong restrictions on the 2- and 3-point functions of certain BPS operators protecting them from receiving any quantum corrections. This chapter aims to establish the technical foundation for the subsequent chapter, where we discuss the stress-tensor multiplet.

The chapter is organized as follows. In section \ref{sec:SCA} we introduce the superconformal algebra $\mathfrak{psu}(2,2|4)$. Then, in section \ref{sec:SU4Rep}, we discuss the representation theory of the $R$-symmetry algebra $\mathfrak{su}(4)$ and subsequently, in section \ref{sec:psurep}, the unitary irreducible representations of the superconformal algebra. In section \ref{sec:Unitarity}, we study the implications of unitarity on the $\mathfrak{psu}(2,2|4)$ representations and then, in section \ref{sec:ShortMultiplet}, the shortening of supermultiplets. Finally, in the last two sections, we briefly discuss the anomalous dimension and 2- and 3-point correlation functions in conformal field theories.

The representation theory of the superconformal algebra has been studied by Dolan and Osborn in \cite{Dolan:2002zh}. In this chapter, we closely follow their work and notation. See also \cite{Eberhardt:2020cxo} for a more pedagogical and less technical introduction to the representation theory of the superconformal algebra. For a general mathematical introduction to the representation theory of Lie algebras, see \cite{Hall2015} and for everything on conformal field theory, including the representation theory of the conformal algebra see \cite{DiFrancesco:1997nk}.

\section{The Superconformal Algebra}\label{sec:SCA}
We first introduce the algebra $\mathfrak{su}(2,2|4)$. Its bosonic subalgebras are the conformal algebra $\mathfrak{su}(2,2)$ and the $R$-symmetry algebra $\mathfrak{u}(4)$. As before, we work in 4-dimensional Minkowski space, where the metric $\eta^{\mu\nu}$ ($\mu, \nu = 0,1,2,3$) has mostly minus signature $(+,-,-,-)$. The 4-dimensional conformal algebra $\mathfrak{su}(2,2) \simeq \mathfrak{so}(2,4)$ is an extension of the Poincaré algebra. We have already introduced the Poincaré algebra in chapter \ref{ch:SFT}. Recall that it consists of the four-momentum generators $P_\mu$ and the six anti-symmetric Lorenz generators $M_{\mu\nu}$
\begin{align}
\begin{aligned}\label{eq:SCAPoincare1}
&[P_\mu, P_\nu] = 0 \, , \\
&[M_{\mu\nu}, M_{\rho\lambda}] = i (\eta_{\mu\rho} M_{\nu\lambda} - \eta_{\mu\lambda} M_{\nu\rho} - \eta_{\nu\rho} M_{\mu\lambda} + \eta_{\nu\lambda} M_{\mu\rho}) \, ,\\
&[M_{\mu\nu},P_\lambda] = i(\eta_{\mu\lambda} P_\nu - \eta_{\lambda\nu} P_\mu) \, .
\end{aligned}
\end{align}
To obtain the conformal algebra, $\mathfrak{su}(2,2)$, the Poincaré algebra is supplemented by the dilatation generator $D$ and the generators of special conformal transformations $K_\mu$. The additional relations are
\begin{align}
\begin{aligned}\label{eq:SCACA1}
&[D, P_\mu] = i P_\mu \, , \quad [D, M_{\mu\nu}] = 0 \, , \quad [D,K_\mu] = - i K_\mu \, , \\
&[P_\mu, K_\nu] = - 2i(M_{\mu\nu} - \eta_{\mu\nu} D) \, , \quad [K_\mu, K_\nu] = 0 \, , \\
&[M_{\mu\nu},K_\lambda] = i(\eta_{\mu\lambda} K_\nu - \eta_{\nu\lambda} K_\mu) \, .
\end{aligned}
\end{align}
For the subsequent discussion, it is advantageous to switch to a spinor basis, eliminating all spacetime indices. We choose to work with Weyl spinors. The spinor indices are $\alpha = 1,2$ and $\dot{\alpha} = 1,2$. For all other conventions and some useful relations involving Weyl spinors, see appendix \ref{app:Spinors}. We define
\begin{align}
\begin{aligned}
&P_{\alpha\dot{\alpha}} \coloneqq \sigma_{\alpha \dot{\alpha}} ^\mu P_\mu \, , 
&& K^{\dot{\alpha}\alpha} \coloneqq (\bar{\sigma}^\mu)^{\dot{\alpha}\alpha} K_\mu \, , \\
&M_\alpha^{\ \beta} \coloneqq - \frac{i}{4} (\sigma^\mu \bar{\sigma}^\nu)_\alpha^{\ \beta} M_{\mu\nu} \, , 
&&\bar{M}_{\ \dot{\beta}}^{\dot{\alpha}} \coloneqq - \frac{i}{4} (\bar{\sigma}^\mu \sigma^\nu)_{\ \dot{\beta}}^{\dot{\alpha}} M_{\mu\nu} \, .
\end{aligned}
\end{align}
Subsequently, the Poincaré algebra \eqref{eq:SCAPoincare1} becomes
\begin{align}
\begin{aligned}
&[P_{\alpha\dot{\alpha}}, P_{\beta\dot{\beta}}] = 0 \, , \\
&[M_\alpha^{\ \beta}, M_\gamma^{\ \delta}] = \delta_\gamma^{\ \beta} M_\alpha^{\ \delta} - \delta_\alpha^{\ \delta} M_\gamma^{\ \beta} \, , && [\bar{M}_{\ \dot{\beta}}^{\dot{\alpha}}, \bar{M}_{\ \dot{\delta}}^{\dot{\gamma}}] = - \delta_{\ \dot{\delta}}^{\dot{\alpha}} \bar{M}_{\ \dot{\beta}}^{\dot{\gamma}} + \delta_{\ \dot{\beta}}^{\dot{\gamma}} \bar{M}_{\ \dot{\delta}}^{\dot{\alpha}} \, , \\
&[M_\alpha^{\ \beta}, P_{\gamma\dot{\gamma}}] = \delta_\gamma^{\ \beta} P_{\alpha\dot{\gamma}} - \frac{1}{2} \delta_\alpha^{\ \beta} P_{\gamma\dot{\gamma}} \, , 
&&[\bar{M}_{\ \dot{\beta}}^{\dot{\alpha}}, P_{\gamma\dot{\gamma}}] = \delta_{\ \dot{\gamma}}^{\dot{\alpha}} P_{\gamma\dot{\beta}} - \frac{1}{2} \delta_{\ \dot{\beta}}^{\dot{\alpha}} P_{\gamma\dot{\gamma}} \, .
\end{aligned}
\end{align}
Similarly, the rest of the conformal algebra \eqref{eq:SCACA1} now reads
\begin{align}
\begin{aligned}
&[D, P_{\alpha\dot{\alpha}}] = i P_{\alpha\dot{\alpha}} \, , \quad [D, M_\alpha^{\ \beta}] = 0 \, , \quad [D, \bar{M}_{\ \dot{\beta}}^{\dot{\alpha}}] = 0 \, , \quad [D,K_{\dot{\alpha}\alpha}] = - i K_{\alpha\dot{\alpha}} \, , \\
&[P_{\alpha\dot{\alpha}}, K^{\dot{\beta}\beta}] = -4 (\delta_\alpha^{\ \beta} \bar{M}_{\ \dot{\alpha}}^{\dot{\beta}} + \delta_{\ \dot{\alpha}}^{\dot{\beta}} M_\alpha^{\ \beta}+ \delta_\alpha^{\ \beta} \delta_{\ \dot{\alpha}}^{\dot{\beta}} D) \, , \quad [K^{\dot{\alpha}\alpha}, K^{\dot{\beta}\beta}] = 0 \, , \\
&[M_\alpha^{\ \beta},K^{\dot{\gamma}\gamma}] = - \delta_\alpha^{\ \gamma} K^{\dot{\gamma}\beta} + \frac{1}{2} \delta_\alpha^{\ \beta} K^{\dot{\gamma}\gamma} \, , \quad [\bar{M}_{\ \dot{\beta}}^{\dot{\alpha}}, K^{\dot{\gamma}\gamma}] = - \delta_{\ \dot{\beta}}^{\dot{\gamma}} K^{\dot{\alpha}\gamma} + \frac{1}{2} \delta_{\ \dot{\beta}}^{\dot{\alpha}} K^{\dot{\gamma}\gamma} \, .
\end{aligned}
\end{align}
Then the 16 supercharges $Q_\alpha^A$ and $\bar{Q}_{\dot{\alpha}A}$ ($A=1,2,3,4$) as well as the 16 superconformal charges $S_A^\alpha$ and $\bar{S}^{\dot{\alpha}A}$ are introduced via
\begin{align}
\begin{aligned}
&\{ Q_\alpha^A, \bar{Q}_{\dot{\beta}B} \} = 2 \delta_{\ B}^A P_{\alpha\dot{\beta}} \, , 
&&\quad \{ \bar{S}^{\dot{\alpha} A}, S_B^\beta \} = 2 \delta_{\ B}^A K^{\dot{\alpha}\beta} \, , \\
&\{ Q_\alpha^A, Q_\beta^B \} = \{ \bar{Q}_{\dot{\alpha}A}, \bar{Q}_{\dot{\beta}B} \} = 0 \, , 
&&\quad \{ S_A^\alpha, S_B^\beta \} = \{ \bar{S}^{\dot{\alpha} A}, \bar{S}^{\dot{\beta} B} \} = 0 \, .
\end{aligned}
\end{align}
The Lie brackets with the generators of the conformal algebra are 
\begin{align}
\begin{aligned}
&[P_{\alpha\dot{\alpha}}, Q_\beta^A] = 0 \, , 
&&\quad[P_{\alpha\dot{\alpha}}, S_A^\beta] = - 2 \delta_\alpha^{\ \beta} \bar{Q}_{\dot{\alpha}A} \, , \\
&[P_{\alpha\dot{\alpha}}, \bar{Q}_{\dot{\beta}A} ] = 0 \, , 
&&\quad[P_{\alpha\dot{\alpha}}, \bar{S}^{\dot{\beta} A}] = 2 \delta_{\ \dot{\alpha}}^{\dot{\beta}} Q_\alpha^A \, , \\
&[M_\alpha^{\ \beta}, Q_\gamma^A] = \delta_\gamma^{\ \beta} Q_\alpha^A - \frac{1}{2} \delta_\alpha^{\ \beta} Q_\gamma^A \, , 
&&\quad [M_\alpha^{\ \beta}, S_A^\gamma] = - \delta_\alpha^{\ \gamma} S_A^\beta + \frac{1}{2} \delta_\alpha^{\ \beta} S_A^\gamma \, , \\
&[\bar{M}_{\ \dot{\beta}}^{\dot{\alpha}}, \bar{Q}_{\dot{\gamma}A}] = - \delta_{\ \dot{\gamma}}^{\dot{\alpha}} \bar{Q}_{\dot{\beta}A} + \frac{1}{2} \delta_{\ \dot{\beta}}^{\dot{\alpha}} \bar{Q}_{\dot{\gamma}A} \, , 
&&\quad [\bar{M}_{\ \dot{\beta}}^{\dot{\alpha}}, \bar{S}^{\dot{\gamma}A}] = \delta_{\ \dot{\beta}}^{\dot{\gamma}} \bar{S}^{\dot{\alpha}A} - \frac{1}{2} \delta_{\ \dot{\beta}}^{\dot{\alpha}} \bar{S}^{\dot{\gamma}A} \, , \\
&[M_\alpha^{\ \beta}, \bar{Q}_{\dot{\gamma}A}] = 0 \, , 
&&\quad [M_\alpha^{\ \beta}, \bar{S}^{\dot{\gamma}A}] = 0 \, , \\
&[\bar{M}_{\ \dot{\beta}}^{\dot{\alpha}}, Q_\gamma^A] = 0 \, , 
&&\quad [\bar{M}_{\ \dot{\beta}}^{\dot{\alpha}}, S_A^\gamma] = 0 \, \\
&[D, Q_\alpha^A] = \frac{i}{2} Q_\alpha^A \, , 
&&\quad [D, S_A^\alpha] = - \frac{i}{2} S_A^\alpha \, , \\
&[D, \bar{Q}_{\dot{\alpha}A}] = \frac{i}{2} \bar{Q}_{\dot{\alpha}A} \, , 
&&\quad [D, \bar{S}^{\dot{\alpha} A}] = - \frac{i}{2} \bar{S}^{\dot{\alpha} A} \, , \\
&[K^{\dot{\alpha}\alpha}, Q_\beta^A] = 2 \delta_\beta^{\ \alpha} \bar{S}^{\dot{\alpha} A} \, , 
&&\quad [K^{\dot{\alpha}\alpha}, S_{\beta A}] = 0 \, , \\
&[K^{\dot{\alpha}\alpha}, \bar{Q}_{\dot{\beta}A}] = - 2 \delta_{\ \dot{\beta}}^{\dot{\alpha}} S_A^\alpha \, , 
&&\quad [K^{\dot{\alpha}\alpha}, \bar{S}_{\dot{\beta}}^A] = 0 \, .
\end{aligned}
\end{align}
The two columns are anti-symmetric under the simultaneous exchange of $P_{\alpha\dot{\alpha}}$ and $K^{\dot{\alpha}\alpha}$ as well as $Q_\alpha^A$ and $S_A^\alpha$. The supercharges anti-commute with the superconformal charges as follows
\begin{align}
\begin{aligned}\label{eq:SCAQS}
&\{ Q_\alpha^A, S_B^\beta \} = 4 \left( \delta_{\ B}^A M_\alpha^{\ \beta} - \frac{i}{2} \delta_{\ B}^A \delta_\alpha^{\ \beta} D - \delta_\alpha^{\ \beta} R_{\ B}^A \right) \, , \\
&\{\bar{S}^{\dot{\alpha} A} , \bar{Q}_{\dot{\beta} B} \} = 4 \left( \delta_{\ B}^A \bar{M}_{\ \dot{\beta}}^{\dot{\alpha}} + \frac{i}{2} \delta_{\ B}^A \delta_{\ \dot{\beta}}^{\dot{\alpha}} D - \delta_{\ \dot{\beta}}^{\dot{\alpha}} R_{\ B}^A \right) \, , \\
&\{ Q_\alpha^A, \bar{S}^{\dot{\beta} B} \} = \{ S_A^\alpha, \bar{Q}_{\dot{\beta} B} \} = 0 \, .
\end{aligned}
\end{align}
The $R_{\ B}^A$ are the generators of the $R$-symmetry. They form the bosonic subalgebra $\mathfrak{u}(4)$
\begin{align}\label{eq:SCAu(4)}
[R_{\ B}^A, R_{\ D}^C] = \delta_{\ B}^C R_{\ D}^A - \delta_{\ D}^A R_{\ B}^C \, .
\end{align}
The $R$-symmetry generators commute with all generators of the conformal subalgebra
\begin{align}
&[R_{\ B}^A, P_\mu] = 0 \, , &&[R_{\ B}^A, M_{\mu\nu}] = 0 \, , &&[R_{\ B}^A, D] = 0 \, , &&[R_{\ B}^A, K_\mu] = 0 \, .
\end{align}
Finally, the Lie brackets of $R_{\ B}^A$ with the supercharges and superconformal charges are
\begin{align}
\begin{aligned}\label{eq:SCAQR}
&[R_{\ B}^A, Q_\alpha^C] = \delta_{\ B}^C Q_\alpha^A - \frac{1}{4} \delta_{\ B}^A Q_\alpha^C \, , 
&&\quad [R_{\ B}^A, S_C^\alpha] = - \delta_{\ C}^A S_B^\alpha + \frac{1}{4} \delta_{\ B}^A S_C^\alpha \, , \\
&[R_{\ B}^A, \bar{Q}_{\dot{\alpha}C}] = - \delta_{\ B}^C \bar{Q}_{\dot{\alpha}A} + \frac{1}{4} \delta_{\ B}^A \bar{W}_{\dot{\alpha}C} \, , 
&&\quad [R_{\ B}^A, \bar{S}^{\dot{\alpha} C}] = \delta_{\ B}^C \bar{S}^{\dot{\alpha} A} - \frac{1}{4} \delta_{\ B}^A \bar{S}^{\dot{\alpha} C} \, . 
\end{aligned}
\end{align}
We notice that $R_{\ A}^A$ (summation over $A$) is in the center of $\mathfrak{su}(2,2|4)$. Upon setting $R_{\ A}^A = 0$ the $R$-symmetry algebra $\mathfrak{u}(4)$ becomes $\mathfrak{su}(4) \simeq \mathfrak{so}(6)$ and thus the superconfromal algebra $\mathfrak{su}(2,2|4)$ becomes $\mathfrak{psu}(2,2|4)$. In other words the superconformal algebra $\mathfrak{psu}(2,2|4)$ is defined through the short exact sequence
\begin{align}
0 \ \longrightarrow \ \mathfrak{u}(1) \ \longrightarrow \ \mathfrak{su}(2,2|4) \ \longrightarrow \ \mathfrak{psu}(2,2|4) \ \longrightarrow 0 \, .
\end{align}
This concludes the introduction of the superconformal algebra and its 56 generators
\begin{align}
(P_{\alpha\dot{\alpha}}, M_\alpha^{\ \beta}, \bar{M}_{\ \dot{\beta}}^{\dot{\alpha}}, K^{\dot{\alpha}\alpha}, D \ | \ Q_\alpha^A, \bar{Q}_{\dot{\alpha}A}, S_A^\alpha, \bar{S}^{\dot{\alpha}A},R_{\ B}^A) \, .
\end{align}
Demanding hermiticity of the operators associated to the generators imposes the following additional conditions
\begin{align}
(M_\alpha^{\ \beta})^\dagger = \bar{M}_{\ \dot{\alpha}}^{\dot{\beta}} \, , \quad
(R_{\ B}^A)^\dagger = R_{\ A}^B \, , \quad
(Q_\alpha^A)^\dagger = \bar{Q}_{\dot{\alpha}A} \, , \quad
(S_A^\alpha)^\dagger = \bar{S}^{\dot{\alpha}A} \, .
\end{align}

\section[Representations of \texorpdfstring{$\mathfrak{su}(4)$}{su(4)}]{Representations of $\boldsymbol{\mathfrak{su}(4)}$}\label{sec:SU4Rep}
Before discussing the representation theory of the superconformal algebra, we briefly study the representation theory of the $R$-symmetry algebra $\mathfrak{su}(4)$. Recall that $\mathfrak{su}(4)$ is a complex simple Lie algebra of rank 3. Thus, all of its irreducible finite-dimensional representations are uniquely classified by a highest weight. We decompose the $\mathfrak{su}(4)$ generators into the Chevalley basis to find these highest weight representations. For each of the three simple roots of $\mathfrak{su}(4)$ there is a $\mathfrak{su}(2)$ algebra generated by $H_i$ and $E_i^\pm$ ($i=1,2,3)$ with
\begin{align}\label{eq:SCAChevalley}
[H_i, H_j] = 0 \, , && [E_i^+, E_j^-] = \delta_{ij} H_j \, , && [H_i, E_j^\pm] = \pm K_{ji} E_j^\pm \quad \text{no sum over j} \, ,
\end{align}
where $K_{ji}$ are the elements of the Cartan matrix
\begin{align}
K = \begin{pmatrix} 2 & -1 & 0 \\ -1 & 2 & -1 \\ 0 & -1 & 2 \end{pmatrix} \, .
\end{align}
The Chevalley basis is the Cartan-Weyl basis but with a different normalization. The $H_i$ are the Cartan generators. Furthermore, the generators $E_i^\pm$ satisfy the Serre relations
\begin{align}\label{eq:SCASerre}
(\mathrm{ad}_{E_i^\pm})^{1-K_{ji}}(E_j^\pm) = \underbrace{[E_i^\pm, [ \ldots [E_i^\pm}_{(1-K_{ji}) \ \text{times}}, E_j^\pm] \ldots ]] = 0 \, , \quad i \not= j \, .
\end{align}
All remaining commutation relations are determined using \eqref{eq:SCAChevalley}, \eqref{eq:SCASerre} and the Jacobi identity. Since $\mathfrak{su}(4)$ is also a compact Lie algebra, we may further impose the hermiticity conditions
\begin{align}
H_i^\dagger = H_i \, , \quad E_i^{+ \, \dagger} = E_i^- \, , \quad i = 1, 2, 3 \, .
\end{align}
Subsequently, we find that the following decomposition of the $\mathfrak{su}(4)$ generators $R_{\ B}^A$ satisfies \eqref{eq:SCAu(4)} and $R_{\ A}^A = 0$
\begingroup
\vspace*{10pt}
\begin{footnotesize}
\begin{align}\label{eq:SCARMatrix}
R = \begin{pmatrix}
\frac{1}{4} (3H_1 + 2H_2 + H_3) & E_1^+ & [E_1^+, E_2^+] & [E_1^+, [E_2^+, E_3^+]] \\
E_1^- & \frac{1}{4} (-H_1 + 2H_2 + H_3) & E_2^+ & [E_2^+, E_3^+] \\
-[E_1^-, E_2^-] & E_2^- & - \frac{1}{4} (H_1 + 2 H_2 - H_3) & E_3^+ \\
[E_1^-, [E_2^-, E_3^-]] & - [E_2^-, E_3^-] & E_3^- & - \frac{1}{4} (H_1 + 2H_2 + 3H_3)
\end{pmatrix} \, .
\end{align}
\end{footnotesize}%
\endgroup%
Let $\mathfrak{h} =\{ H_1, H_2, H_3\}$ be the Cartan subalgebra of $\mathfrak{su}(4)$ and let $\mathfrak{n}^\pm = \{E_1^\pm, E_2^\pm, E_3^\pm\}$. From the discussion above it follows that $\mathfrak{su}(4) = \mathfrak{h} \oplus \mathfrak{n}^- \oplus \mathfrak{n}^+$. Further let $V$ be a representation of $\mathfrak{su}(4)$. Then $v \in V$ is called of weight $\lambda \in H^*$ if $v \not= 0$ and $h.v = \lambda(h) v$ for all $h \in \mathfrak{h}$. The weight space for $\lambda$ is
\begin{align}
V[\lambda] \coloneqq \{ v \in V \ | \ \text{$v$ is a vector of weight $\lambda$} \} \, .
\end{align}
If $V[\lambda] \not= \{ 0 \}$, then $\lambda$ is called weight of $V$. Subsequently, $V \not= \{0\}$ is a highest weight representation of weight $\lambda$ if $V$ is generated by some $v \in V[\lambda]$ ($v \not = 0$) and $\mathfrak{n}^+ v = \{ 0 \}$. As stated above, every irreducible finite-dimensional representation of $\mathfrak{su}(4)$ is a highest weight representation. For the following discussion, we choose the eigenvectors of the $H_i$ as generators of the weight space. In common physics notation, the eigenvectors are labeled by their three eigenvalues with respect to the action of the three $H_i$, \emph{i.e.}
\begin{align}
H_i \ket{\lambda_1, \lambda_2, \lambda_3} = \lambda_i \ket{\lambda_1, \lambda_2, \lambda_3} \, .
\end{align}
In this notation we write $\ket{\lambda_1, \lambda_2, \lambda_3}^\mathrm{hw}$ for the non-zero highest weight vector annihilated by the $E_i^+$
\begin{align}
E_i^+ \ket{\lambda_1, \lambda_2, \lambda_3}^\mathrm{hw} = 0 \, , \quad \lambda_i \ge 0 \, .
\end{align}
Every highest weight representation, \emph{i.e.} every irreducible finite-dimensional representation of $\mathfrak{su}(4)$, is uniquely characterized by its highest weight vector $\ket{\lambda_1, \lambda_2, \lambda_3}^\mathrm{hw}$ and thus by its Dynkin labels $[\lambda_1, \lambda_2, \lambda_3]$.
The remaining basis vectors for the given representation are obtained by the successive action of $E_i^-$ on the highest weight vector. In particular 
\begin{align}\label{eq:SCAEminushw}
E_i^- \ket{\lambda_1, \lambda_2, \lambda_3}^\mathrm{hw} = 0 \quad \text{if} \quad \lambda_i = 0 \, .
\end{align}
In general an irreducible finite dimensional $\mathfrak{su}(4)$ representation with Dynkin labels $[\lambda_1, \lambda_2, \lambda_3]$ has complex dimensions \cite{Hall2015}
\begin{align}\label{eq:SCADim}
d_{(\lambda_1, \lambda_2, \lambda_3)} = \frac{1}{12} (\lambda_1 + \lambda_2 + \lambda_3 + 3)(\lambda_1 + \lambda_2 + 2)(\lambda_2 + \lambda_3 + 2)(\lambda_1+1)(\lambda_2+1) (\lambda_3+1) \, .
\end{align}
This result is obtained using the Weyl dimension formula. For some representations, the dimension will reduce by a factor of two since they are real. 

\section[Unitary Representations of \texorpdfstring{$\mathfrak{psu}(2,2|4)$}{psu(224)}]{Unitary Representations of $\boldsymbol{\mathfrak{psu}(2,2|4)}$}\label{sec:psurep}
Building on the discussion above, we construct all unitary irreducible representations of the superconformal algebra $\mathfrak{psu}(2,2|4)$. Since the superconformal algebra contains the connected, simple, non-compact Poincaré algebra, it has no non-trivial finite-dimensional unitary representations \cite{Dobrev:1985qv}. Thus, all unitary representations are necessarily infinite-dimensional. 

In the previous section, the starting point to finding all the irreducible finite-dimensional representations of $\mathfrak{su}(4)$ was the Cartan subalgebra $\mathfrak{h} = \{H_1, H_2, H_3 \}$, where by \eqref{eq:SCARMatrix}
\begin{align}
H_1 = R_{\ 1}^1 - R_{\ 2}^2 \, , \quad H_2 = R_{\ 2}^2 + R_{\ 3}^3 \, , \quad H_3 = - R_{\ 3}^3 + R_{\ 4}^4 \, .
\end{align}
Thus, the Cartan subalgebra $\mathfrak{h}$ is spanned by the diagonal elements of $R_{\ B}^A$. For any super Lie algebra, the Cartan subalgebra coincides with that of its bosonic subalgebra. The bosonic subalgebras of the superconformal algebra $\mathfrak{psu}(2,2|4)$ are $\mathfrak{su}(2,2)$ and $\mathfrak{su}(4)$. The representation theory of the conformal algebra has extensively been studied in \cite{DiFrancesco:1997nk}. It is built on Wigner's classification of unitary representations of the Poincaré group \cite{Wigner:1939cj} (see also \cite{Weinberg:1995mt}). The Cartan subalgebra of the conformal algebra is generated by $M_\alpha^{\ \beta}$ $(\alpha = \beta)$, $\bar{M}_{\ \dot{\beta}}^{\dot{\alpha}}$ $(\dot{\alpha} = \dot{\beta})$ and $D$. We combine this with the Cartan subalgebra of $\mathfrak{su}(4)$ and find the Cartan subalgebra of $\mathfrak{psu}(2,2|4)$
\begin{align}
\mathfrak{j}^0 = \mathrm{span}(M_\alpha^{\ \beta} (\alpha = \beta) , \bar{M}_{\ \dot{\beta}}^{\dot{\alpha}} (\dot{\alpha} = \dot{\beta}) , R_{\ B}^A (A = B) , D ) \, .
\end{align}
For a super Lie algebra, there is some freedom to distribute the simple fermionic roots in the Dynkin diagram. Here we follow the convention of \cite{Dobrev:1985qz} and obtain the following associations of generators to positive and negative roots
\begin{align}
\begin{aligned}
&\mathfrak{j}^+ = \mathrm{span}(K^{\dot{\alpha}\alpha}, S_A^\alpha, \bar{S}^{\dot{\alpha}A}, M_\alpha^{\ \beta} (\alpha < \beta), \bar{M}_{\ \dot{\beta}}^{\dot{\alpha}} (\dot{\alpha} < \dot{\beta}), R_{\ B}^A (A < B) ) \, , \\
&\mathfrak{j}^- = \mathrm{span}( P_{\alpha\dot{\alpha}}, Q_\alpha^A, \bar{Q}_{\dot{\alpha}A}, M_\alpha^{\ \beta} (\alpha > \beta), \bar{M}_{\ \dot{\beta}}^{\dot{\alpha}} (\dot{\alpha} > \dot{\beta}), R_{\ B}^A (A > B) ) \, .
\end{aligned}
\end{align}
We denote the eigenvectors of the elements of the Cartan subalgebra $\mathfrak{j}^0$ by $\ket{\Delta; j, \bar{j}; \lambda_1, \lambda_2, \lambda_3}$, where $\Delta$ is the scaling dimension, \emph{i.e.} the eigenvalue of the dilatation operator $D$. $j$ and $\bar{j}$ are the two Lorenz spins and $\lambda_1$, $\lambda_2$ and $\lambda_3$ are the three $\mathfrak{su}(4)$ Dynkin labels from before. Acting with the dilatation generator on $\ket{\Delta; j, \bar{j}; \lambda_1, \lambda_2, \lambda_3}$ yields
\begin{align}\label{eq:SCADhw}
D \ket{\Delta, j, \bar{j}; \lambda_1, \lambda_2, \lambda_3} = i \Delta \ket{\Delta, j, \bar{j}; \lambda_1, \lambda_2, \lambda_3} \, .
\end{align}
For the action of the Lorenz generators, we write the spin dependence explicitly, \emph{i.e.}
\begin{align}
\ket{\Delta, j, \bar{j}; \lambda_1, \lambda_2, \lambda_3} = \ket{\Delta; \lambda_1, \lambda_2, \lambda_3}_{\alpha_1 \ldots \alpha_{2j}, \dot{\alpha}_1 \ldots \dot{\alpha}_{2j}}
\end{align}
and obtain
\begin{align}
\begin{aligned}\label{eq:SCAMhw}
M_\alpha^{\ \beta} \ket{\Delta; \lambda_1, \lambda_2, \lambda_3}_{\alpha_1 \ldots \alpha_{2j}, \dot{\alpha}_1 \ldots \dot{\alpha}_{2j}} 
&= 2 j \delta_{(\alpha_1}^{\ \ \ \beta} \ket{\Delta; \lambda_1, \lambda_2, \lambda_3}_{\alpha_1 \ldots \alpha_{2j)} \alpha, \dot{\alpha}_1 \ldots \dot{\alpha}_{2j}} \\
&\quad - j \delta_\alpha^{\ \beta} \ket{\Delta; \lambda_1, \lambda_2, \lambda_3}_{\alpha_1 \ldots \alpha_{2j}, \dot{\alpha}_1 \ldots \dot{\alpha}_{2j}} \, , \\
\bar{M}_{\ \dot{\alpha}}^{\dot{\beta}} \ket{\Delta; \lambda_1, \lambda_2, \lambda_3}_{\alpha_1 \ldots \alpha_{2j}, \dot{\alpha}_1 \ldots \dot{\alpha}_{2j}} 
&= - 2 \bar{j} \ket{\Delta; \lambda_1, \lambda_2, \lambda_3}_{\alpha_1 \ldots \alpha_{2j} \dot{\alpha}, (\dot{\alpha}_1 \ldots \dot{\alpha}_{2j-1}} \delta_{\ \dot{\alpha}_{2j)}}^{\dot{\beta}} \\
&\quad - j \delta_{\ \dot{\alpha}}^{\dot{\beta}} \ket{\Delta; \lambda_1, \lambda_2, \lambda_3}_{\alpha_1 \ldots \alpha_{2j}, \dot{\alpha}_1 \ldots \dot{\alpha}_{2j}} \, .
\end{aligned}
\end{align}
The round brackets indicate a symmetrization of indices. Finally, \eqref{eq:SCARMatrix} gives the action of the diagonal $R$-symmetry generators
\begin{align}
\begin{aligned}\label{eq:SCARhw}
&R_{\ 1}^1 \ket{\Delta; j, \bar{j}; \lambda_1, \lambda_2, \lambda_3} = \frac{1}{4} (3r_1 + 2r_2 + \lambda_3) \ket{\Delta; j, \bar{j}; \lambda_1, \lambda_2, \lambda_3} \, , \\
&R_{\ 2}^2 \ket{\Delta; j, \bar{j}; \lambda_1, \lambda_2, \lambda_3} = \frac{1}{4} (-\lambda_1 + 2r_2 + \lambda_3) \ket{\Delta; j, \bar{j}; \lambda_1, \lambda_2, \lambda_3} \, , \\
&R_{\ 3}^3 \ket{\Delta; j, \bar{j}; \lambda_1, \lambda_2, \lambda_3} = - \frac{1}{4} (\lambda_1 + 2r_2 - \lambda_3) \ket{\Delta; j, \bar{j}; \lambda_1, \lambda_2, \lambda_3} \, , \\
&R_{\ 4}^4 \ket{\Delta; j, \bar{j}; \lambda_1, \lambda_2, \lambda_3} = - \frac{1}{4} (\lambda_1 + 2r_2 + 3r_3) \ket{\Delta; j, \bar{j}; \lambda_1, \lambda_2, \lambda_3} \, .
\end{aligned}
\end{align}
By $\ket{\Delta; j, \bar{j}; \lambda_1, \lambda_2, \lambda_3}^\mathrm{hw}$ we denote the highest weight vectors which are annihilated by all generators of $\mathfrak{j}^+$. All unitary irreducible representations of the superconformal algebra $\mathfrak{psu}(2,2|4)$ are then constructed by iteratively acting with the generators of $\mathfrak{j}^-$ on a highest weight vector $\ket{\Delta; j, \bar{j}; \lambda_1, \lambda_2, \lambda_3}^\mathrm{hw}$. As opposed to the highest weight representations of the simple Lie algebra $\mathfrak{su}(4)$, these representations are infinite-dimensional. Physically a highest weight vector corresponds to the lowest energy state. 

\section{Unitarity}\label{sec:Unitarity}
In the following, we are not interested in the entire infinite-dimensional unitary irreducible representation associated to a highest weight vector $\ket{\Delta; j, \bar{j}; \lambda_1, \lambda_2, \lambda_3}^\mathrm{hw}$ but rather only the supersymmetry multiplet of a highest weight state, which is obtained by the action of the supersymmetry generators $Q_\alpha^A$ and $\bar{Q}_{\dot{\alpha}A}$ on $\ket{\Delta; j, \bar{j}; \lambda_1, \lambda_2, \lambda_3}^\mathrm{hw}$. When no further restrictions are applied, we call these multiplets long and denote them by $\mathcal{A}_{[\lambda_1, \lambda_2, \lambda_3](j, \bar{j})}^\Delta$, where the indices correspond to the eigenvalues of the highest weight states. Since the supersymmetry generators are fermionic operators, the supersymmetry multiplet is finite-dimensional with \cite{Dolan:2002zh}
\begin{align}
\mathrm{dim} \, \mathcal{A}_{[\lambda_1, \lambda_2, \lambda_3](j, \bar{j})}^\Delta = 2^{16} d_{(\lambda_1, \lambda_2, \lambda_3)} (2j + 1) (2 \bar{j} + 1) \, .
\end{align}
Because there are only 16 supersymmetry generators, we can only act 16 times on the highest weight state before one operator appears twice and we obtain zero. 

Using \eqref{eq:SCAQR}, we can express the action of the supercharges $Q_\alpha^A$ and $\bar{Q}_{\dot{\alpha}A}$ on a state $[\lambda_1, \lambda_2, \lambda_3]_{(j, \bar{j})}$ with conformal dimension $\Delta$ in terms of the change in the weight and spin. We have \cite{Dolan:2002zh}
\begin{align}\label{eq:SCAQDynkin1}
\begin{aligned}
&Q_\alpha^1 \sim [+1,0,0]_{(\pm\textstyle\frac12,0)} \, , &&Q_\alpha^2 \sim [-1,+1,0]_{(\pm\textstyle\frac12,0)} \, , \\
&Q_\alpha^3 \sim [0,-1,+1]_{(\pm\textstyle\frac12,0)} \, , &&Q_\alpha^4 \sim [0,0,-1]_{(\pm\textstyle\frac12,0)}
\end{aligned}
\end{align}
and
\begin{align}\label{eq:SCAQDynkin2}
\begin{aligned}
&\bar{Q}_{\dot{\alpha}1} \sim [-1,0,0]_{(0,\pm\textstyle\frac12)} \, , &&\bar{Q}_{\dot{\alpha}2} \sim [+1,-1,0]_{(0,\pm\textstyle\frac12)} \, , \\
&\bar{Q}_{\dot{\alpha}3} \sim [0,+1,-1]_{(0,\pm\textstyle\frac12)} \, , &&\bar{Q}_{\dot{\alpha}4} \sim [0,0,+1]_{(0,\pm\textstyle\frac12)} \, .
\end{aligned}
\end{align}
Every application of $Q_\alpha^A$ or $\bar{Q}_{\dot{\alpha}A}$ increases the conformal dimension by $\frac{1}{2}$. Representations with negative Dynkin labels are zero. Thus, not every supercharge can act on every state. 

In unitary representations, all states have non-negative norm. This restriction gives rise to a so-called unitarity bound \cite{Dobrev:1985qv}\footnote{See \cite{Eberhardt:2020cxo} for instructions on how to compute the unitarity bound.}
\begin{align}\label{eq:SCAUBound}
\Delta \ge \Delta_\mathcal{A} = \max \left( 2 + 2 j + \frac{1}{2} (3 \lambda_1 + 2 \lambda_2 + \lambda_3), \ 2 + 2 \bar{j} + \frac{1}{2} (\lambda_1 + 2 \lambda_2 + 3\lambda_3) \right) \, .
\end{align}
Multiplets that saturate the unitarity bound, \emph{i.e.} where \eqref{eq:SCAUBound} is an equality, have states with zero norm. These are called null states. They form a closed subrepresentation and hence can be consistently removed from the multiplet \cite{Cordova:2016emh}. In particular, the trivial representation saturates the unitarity bound. Multiplets above the unitarity bound do not have any null states. Consequently, multiplets saturating the unitarity bound are shorter than multiplets with $\Delta > \Delta_\mathcal{A}$ and we call them short $\mathcal{A}$-type multiplets. 

\section{Multiplet Shortening}\label{sec:ShortMultiplet}
\begin{figure}
\begin{center}
\begin{tikzpicture}
\fill[myred,opacity=.2] (-.5,-1) rectangle (4,3);
\fill[green!80!black,opacity=.2] (-.5,3) rectangle (4,5);
\draw[very thick,mygreen] (-.8,3) -- (4.5,3) node[right, black] {$\Delta=\Delta_\mathcal{A}$};
\draw[very thick, mygreen] (-.8,1) -- (4.5,1) node[right,black] {$\Delta=\Delta_\mathcal{B}$};
\draw[very thick,->] (0,-1.2) -- (0,5.5) node[right] {$\ \Delta$};
\node at (2,2) {non-unitary};
\node at (2,0) {non-unitary};
\node at (2,4) {unitary};
\end{tikzpicture}
\end{center}
\caption{Unitarity structure of superconformal multiplets. Long multiplets exist for $\Delta > \Delta_\mathcal{A}$ in the green region. At the unitarity bound $\Delta = \Delta_\mathcal{A}$ are the short $\mathcal{A}$-type multiplets. Below the unitarity bound, there is the forbidden region in red. However, at discrete values of the conformal scaling dimension $\Delta = \Delta_\mathcal{B}$, individual short (and semi-short) multiplets exist. In particular these are the BPS multiplets \cite{Eberhardt:2020cxo,Cordova:2016emh}.}\label{fig:SCAUnitarity}
\end{figure}
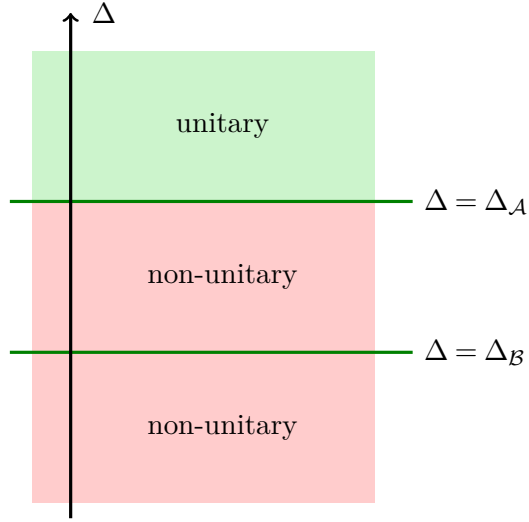

Long multiplets can also be shortened by imposing BPS conditions on the highest weight state, \emph{i.e.} by demanding that
\begin{align}\label{eq:SCAShort1}
Q_\alpha^A \ket{\Delta; j, \bar{j}; \lambda_1, \lambda_2, \lambda_3}^\mathrm{hw} = 0 \quad \text{for} \quad \alpha \in \{1,2 \} 
\end{align}
for some of the $Q_\alpha^A$. The operators corresponding to short multiplets, annihilated by some of the supercharges, are called chiral. Since any highest weight state is also annihilated by the superconformal charges $S_A^\alpha$, \eqref{eq:SCAQS} implies that $\ket{\Delta; j, \bar{j}; \lambda_1, \lambda_2, \lambda_3}^\mathrm{hw}$ must be annihilated by all $M_\alpha^{\ \beta}$. Then \eqref{eq:SCAMhw} implies that $j = 0$ and we get
\begin{align}\label{eq:SCAShort2}
\{ Q_\alpha^A, S_B^\beta \} \ket{\Delta; 0, \bar{j}; \lambda_1, \lambda_2, \lambda_3}^\mathrm{hw} = - 4 \delta_\alpha^{\ \beta} \left( \frac{i}{2} \delta_{\ B}^A D + R_{\ B}^A \right) \ket{\Delta; 0, \bar{j}; \lambda_1, \lambda_2, \lambda_3}^\mathrm{hw} = 0
\end{align}
for $B = 1,2,3,4$. We use \eqref{eq:SCADhw} and \eqref{eq:SCARhw} to solve this equation. For $A=1$ and $B=1$ we obtain
\begin{align}
\Delta = \frac{1}{2} (3 \lambda_1 + 2 \lambda_2 + \lambda_3) \, .
\end{align}
There are no further implications because all $R_{\ B}^1$ for $B > 1$ are in $\mathfrak{j}^+$ and thus annihilate $\ket{\Delta; 0, \bar{j}; \lambda_1, \lambda_2, \lambda_3}^\mathrm{hw}$. Next we solve \eqref{eq:SCAShort2} for $A=1,2$. There are three equations to satisfy
\begin{align}
\begin{gathered}
\Delta_\mathcal{B} = \frac{1}{2} ( 3 \lambda_1 + 2 \lambda_2 + \lambda_3 ) \, , \\
\Delta_\mathcal{B} = \frac{1}{2} ( - \lambda_1 + 2 \lambda_2 + \lambda_3 ) \, , \\
R_{\ 1}^2 \, \ket{\Delta; 0, \bar{j}; \lambda_1, \lambda_2, \lambda_3}^\mathrm{hw} = E_1^- \ket{\Delta; 0, \bar{j}; \lambda_1, \lambda_2, \lambda_3}^\mathrm{hw} = 0 \, .
\end{gathered}
\end{align}
From the first two equations we obtain $\lambda_1 = 0$ and $\Delta_\mathcal{B} = \frac{1}{2} ( 2 \lambda_2 + \lambda_3)$. By \eqref{eq:SCAEminushw} the third equation is satisfied if $\lambda_1 = 0$. The index $\mathcal{B}$ indicates that the scaling dimension $\Delta_\mathcal{B}$ belongs to a short multiplet as opposed to a long mulitplet denoted $\mathcal{A}$. We continue like this for the cases $A=1,2,3$ and $A=1,2,3,4$. Let $s$ be the fraction of supercharges for which \eqref{eq:SCAShort1} holds. Summarizing the results we have
\begin{align}
\begin{aligned}\label{eq:SCACond1}
&A = 1 &&s= \frac{1}{4} &&\Delta_\mathcal{B} = \frac{1}{2} (3r_1 + 2r_2 + \lambda_3) \, , && \\
&A = 1,2 &&s = \frac{1}{2} &&\Delta_\mathcal{B} = \frac{1}{2} (2r_2 + \lambda_3) \, , && \lambda_1 = 0 \, , \\
&A= 1,2,3 &&s = \frac{3}{4} &&\Delta_\mathcal{B} = \frac{1}{2}\lambda_3 \, , && \lambda_1 = \lambda_2 = 0 \, , \\
&A = 1,2,3,4 \quad &&s=1 \quad &&\Delta_\mathcal{B} = 0 \, , &&\lambda_1=\lambda_2=\lambda_3=0 \, .
\end{aligned}
\end{align}
Similar to \eqref{eq:SCAShort1} we can also impose the condition
\begin{align}\label{eq:SCAShort3}
\bar{Q}_{\dot{\alpha}A} \ket{\Delta, j, \bar{j}; \lambda_1, \lambda_2, \lambda_3}^\mathrm{hw} = 0 \quad \text{for} \quad \dot{\alpha} \in \{1,2 \} \, .
\end{align}
This yields the following constraints (with $\bar{j}=0$)
\begin{align}
\begin{aligned}\label{eq:SCACond2}
&A= 4 &&\bar{s}=\frac{1}{4} &&\Delta_\mathcal{B}=\frac{1}{2}(\lambda_1 + 2r_2 + 3r_3) \, , && \\
&A=3,4 &&\bar{s}=\frac{1}{2} &&\Delta_\mathcal{B} =\frac{1}{2}(\lambda_1+2r_2) \, , &&\lambda_3 = 0 \, , \\
&A=2,3,4 &&\bar{s}=\frac{3}{4} &&\Delta_\mathcal{B} = \frac{1}{2} \lambda_1 \, , &&\lambda_2=\lambda_3=0 \, , \\
&A= 1,2,3,4 \quad &&\bar{s}=1 \quad &&\Delta_\mathcal{B} = 0 \, , &&\lambda_1=\lambda_2=\lambda_3=0 \, .
\end{aligned}
\end{align}
To construct all conformal primary states for complete supermultiplets, combining the results for the $Q$ and $\bar{Q}$ supercharges is necessary. If \eqref{eq:SCACond1} and \eqref{eq:SCACond2} are both applied, we must have $j = \bar{j} = 0$ and we denote these short multiplets by $\mathcal{B}_{[\lambda_1,\lambda_2,\lambda_3](0,0)}^{s,\bar{s}}$. In general, there are only three possible cases. The first is $s=\bar{s}=1$, $\lambda_1=\lambda_2=\lambda_3=0$. This case corresponds to the trivial vacuum representation since $P_{\alpha\dot{\alpha}} \ket{0,0,0;0,0,0}^\mathrm{hw} = 0$. The next case are the $\frac{1}{4}$-BPS multiplets
\begin{align}
\mathcal{B}_{[\lambda_1,\lambda_2,\lambda_3](0,0)}^{\frac{1}{4},\frac{1}{4}} \, , \quad \Delta_\mathcal{B} = \lambda_1 + 2\lambda_2 \, .
\end{align}
And finally, the third case are the $\frac{1}{2}$-BPS multiplets
\begin{align}\label{eq:SCA12BPS}
\mathcal{B}_{[0,\lambda_2,0](0,0)}^{\frac{1}{2},\frac{1}{2}} \, , \quad \Delta_\mathcal{B} = \lambda_2 \, .
\end{align}
Besides the short $\mathcal{B}$-type multiplets there are also semi-short multiplets obtained by imposing \cite{Dolan:2002zh}
\begin{align}
\left( \bar{Q}_{1A} + \frac{1}{2 \bar{j} + 1} \bar{M}_{\ \dot{\beta}}^{\dot{\alpha}} \bar{Q}_{2A} \right) \ket{\Delta; j, \bar{j}; \lambda_1, \lambda_2, \lambda_3}^\mathrm{hw} = 0 \quad \text{with} \ (\dot{\alpha} > \dot{\beta}) \, .
\end{align}
Long multiplets can be decomposed into direct sums of semi-short and sometimes also short multiplets. Comparing the scaling dimensions of the long and short multiplets, we notice that $\Delta_\mathcal{A} > \Delta_\mathcal{B}$. Thus, by imposing the BPS conditions on the highest weight state, we have found supermultiplets with lower than initially allowed scaling dimensions which still belong to unitary representations. We have summarized this in figure \ref{fig:SCAUnitarity}. The short and semi-short representations of the superconformal algebra for $\mathcal{N}=2$ and $\mathcal{N}=4$ in four dimensions have been extensively studied in \cite{Dolan:2002zh}. Unitary superconformal multiplets for $1 \le \mathcal{N} \le 8$ supersymmetries in $d \ge 3$ dimensions have also been studied in \cite{Cordova:2016emh}. In particular, the authors of \cite{Cordova:2016emh} have analyzed the operator content of the various multiplets. 

\section{The Anomalous Dimension}
In any supermultiplet, the scaling dimensions of the operators increase by a factor of $\frac{1}{2}$ with each level. When the primary operator has scaling dimension $\Delta$ its first descendants have scaling dimension $\Delta + \frac{1}{2}$. The second descendants have scaling dimension $\Delta + 1$ and so on. A priori, the scaling dimension of any (primary) operator receives quantum corrections. These corrections are called the anomalous dimension. However, since the superconformal symmetry of $\mathcal{N}=4$ super Yang-Mills holds even at the quantum level, all operators in one supermultiplet have the same anomalous dimension.

A particularly special kind of multiplets are the $\frac{1}{2}$-BPS multiplets $\mathcal{B}_{[0,\lambda_2,0](0,0)}^{\frac{1}{2},\frac{1}{2}} $ with $\Delta_\mathcal{B} = \lambda_2$. It has been shown that the operators in these multiplets are subjected to a non-renormalization theorem and do not receive any anomalous dimension \cite{Baggio:2012rr}. This agrees with the results of perturbative calculations \cite{Arutyunov:2000ku, Arutyunov:2000im, Ferrara:1999ed, Bianchi:2001cm, Penati:2001sv}. In particular, the vanishing of the anomalous dimension implies that the $\frac{1}{2}$-BPS multiplets $\mathcal{B}_{[0,\lambda_2,0](0,0)}^{\frac{1}{2},\frac{1}{2}} $ do not recombine into long multiplets \cite{Dolan:2002zh}.

\section{Conformal Correlation Functions}\label{sec:CorFct}
In conformal field theories such as $\mathcal{N}=4$ super Yang-Mills, the form of the 2- and 3-point functions of (quasi) primary operators $\mathcal{O}_i(x_i)$ are completely fixed up to a constant, the scaling dimensions $\Delta_i \equiv \Delta_i(g)$ and its quantum corrections. For the 2-point function the scaling dimensions must agree, $\Delta = \Delta_1 = \Delta_2$ and the correlator is
\begin{align}
\left< \! \left< \mathcal{O}_1(x_1) \mathcal{O}_2(x_2) \right> \! \right>_g = \frac{C_{12}}{|x_1-x_2|^{2\Delta}} \, .
\end{align}
with a constant $C_{12}= C_{12}(g)$ that can be normalized to 1. Similarly, the 3-point function reads
\begin{align}
\left< \! \left< \mathcal{O}_1(x_1) \mathcal{O}_2(x_2) \mathcal{O}_2(x_3) \right> \! \right>_g = \frac{C_{123}}{x_{12}^{\Delta_1+\Delta_2- \Delta_3} x_{23}^{\Delta_2+\Delta_3- \Delta_1} x_{13}^{\Delta_1+\Delta_3- \Delta_2} } \, ,
\end{align}
where $x_{ij} = |x_i-x_j|$ and $C_{123} = C_{123}(g)$ is also a constant. Thus, for operators in the $\frac{1}{2}$-BPS multiplets $\mathcal{B}_{[0,\lambda_2,0](0,0)}^{\frac{1}{2},\frac{1}{2}}$, the exact expressions for the 2- and 3-point function are given by their classical results. 4-point functions are not protected.

\chapter{The Stress Tensor Multiplet}\label{ch:BPS}
We present the explicit field content of the stress tensor multiplet. The stress tensor multiplet is the simplest non-trivial $\frac{1}{2}$-BPS multiplet in $\mathcal{N}=4$ super Yang-Mills. It contains the energy-momentum tensor and the $R$-symmetry currents. Remarkably, its constituents do not acquire an anomalous dimension. The main result of this chapter, \emph{i.e.} the explicit field content of the stress tensor multiplet, has not been published before. Abstractly the field content was studied by Dolan and Osborn in \cite{Dolan:2001tt}.

In the first section, we briefly introduce the stress tensor multiplet and highlight some older results. Furthermore, we recall some important equations. In section \ref{sec:BPSMultiplet}, we give a graphical representation of the states in the stress tensor multiplet. Section \ref{sec:FieldContent} contains the main result of this chapter, \emph{i.e.} the explicit field content of the stress tensor multiplet. In section \ref{sec:BPSCorr} we explain how to compute correlation functions of general operators containing spinor fields with the Nicolai map. Finally, in the last section, we outline the derivation of the field content in the stress tensor multiplet by computing successive supersymmetry transformations of the primary field.

\section{Introduction}
By Noether's theorem to every continuous global symmetry of the action corresponds a conserved current. The symmetry group of $\mathcal{N}=4$ super Yang-Mills is of course $PSU(2,2|4)$. Thus, for every one of its generators, there is a conserved current. The conserved current associated with translation invariance is the energy-momentum tensor (or stress-energy tensor) $T_{\mu\nu}$. In classical 4-dimensional conformal field theories, the energy-momentum tensor is symmetric and traceless. Since the superconformal symmetry of $\mathcal{N}=4$ super Yang-Mills survives even at the quantum level, the energy-momentum tensor remains traceless beyond the classical theory. This is not the case in general field theories with classical conformal symmetry. 

All conserved currents in $\mathcal{N}=4$ super Yang-Mills are part of the same supermultiplet \cite{Sohnius:1985qm}. This is the $\frac{1}{2}$-BPS multiplet $\mathcal{B}_{[0,2,0](0,0)}^{\frac{1}{2},\frac{1}{2}}$ \cite{Dolan:2002zh}. In the following, we call it the stress tensor multiplet. It is the best-studied multiplet in $\mathcal{N}=4$ superconformal symmetries. It has connections to many current topics in high energy physics such as scattering amplitudes, integrability and conformal bootstrap (see \cite{Heslop:2022xgp} for a review). We have already listed some of the properties of this multiplet in the previous chapter. In particular, recall that its constituents do not acquire an anomalous dimension. Thus, the first correlation function to receive quantum corrections is the 4-point function. It has been studied in a variety of ways. For example, it was shown that the conformal symmetry relates the 4-point function of any four operators in the stress tensor multiplet to the 4-point function of the primary field \cite{Eden:2000bk}. Superconformal symmetry has also been used to express the 4-point correlation functions in terms of a single (yet unknown) function of the two conformal invariants \cite{Dolan:2001tt}. This result has, in particular, been applied to the operator product expansion. Another direction of research are perturbative calculations at weak coupling. The 4-point function has been computed to two-loops \cite{Bianchi:2000hn}. The integrand was even given up to ten-loops \cite{Bourjaily:2016evz}. Finally, the 4-point functions of the stress tensor multiplet have also been studied in numerical bootstrap, see \emph{e.g.} \cite{Beem:2016wfs}, where upper bounds on the scaling dimensions and the operator product expansion coefficients have been computed.

In the future, we would like to study the stress tensor multiplet with the Nicolai map and hopefully derive some non-perturbative results for the 4-point functions or operator product expansions. However, this requires us to first know the explicit field content of the entire multiplet and not only its primary field. Thus, in the following, we derive the entire field content of the stress tensor multiplet and show how to compute correlation functions of mixed bosonic and fermionic operators with the Nicolai map. 

\subsection[The \texorpdfstring{$\mathcal{N}=4$}{N=4} super Yang-Mills Action and Equations of Motion]{The $\boldsymbol{\mathcal{N}}$\:=\;4 super Yang-Mills Action and Equations of Motion}
Recall the $\mathcal{N}=4$ super Yang-Mills action with Weyl spinors \eqref{eq:YM5}
\begin{align}
\begin{aligned}\label{eq:BPSAction}
S_\mathrm{inv}^4 = \int \mathrm{d}^4x \ \tr_c \bigg[ &- \frac{1}{2} F_{\mu\nu} F^{\mu\nu} - (D_\mu \phi_I) (D^\mu \phi^I) + \frac{g^2}{4} [\phi_I, \phi_J] [\phi^I, \phi^J] \\
& - 2i \, \psi^{\alpha A} \sigma_{\alpha\dot{\alpha}}^\mu (D_\mu \bar{\psi}_A^{\dot{\alpha}}) - ig \, \psi^{\alpha A} [ \Sigma_{AB}^I \phi_I, \psi_\alpha^B] - ig \,  \bar{\psi}_{\dot{\alpha}A} [ \bar{\Sigma}_I^{AB} \phi^I, \bar{\psi}_B^{\dot{\alpha}}] \bigg] .
\end{aligned}
\end{align}
The expression has been simplified by writing the fields as Lie algebra valued objects, \emph{i.e.} $A_\mu = t^a A_\mu^a$, and taking the trace over the representation space, as in \eqref{eq:ColorTrace}. Recall that $\mu,\nu = 0,\ldots,3$ are the spacetime indices, $\alpha =1,2$ and $\dot{\alpha}=1,2$ are the Weyl spinor indices, $A,B = 1,\ldots,4$ are the $SU(4)$ $R$-symmetry indices, counting the 4 supersymmetries, and $I,J= 1, \ldots , 6$ are the $SO(6)$ indices. For more details on the spinor conventions and notation, see appendix \ref{app:Spinors}.

The $\mathcal{N}=4$ super Yang-Mills action is invariant under the action of the superconformal algebra $\mathfrak{psu}(2,2|4)$, introduced in the previous chapter. The action of the supersymmetry transformations on the fields of the fundamental multiplet is given by
\begin{align}
\begin{aligned}\label{eq:BPSsusy1}
&\delta A_\mu^a = i \varepsilon^{\alpha A} \sigma_{\mu \, \alpha\dot{\alpha}} \bar{\psi}_A^{a\, \dot{\alpha}} - i \psi^{a \, \alpha A} \sigma_{\mu \, \alpha\dot{\alpha}} \bar{\varepsilon}_A^{\dot{\alpha}} \ \, , \\
&\delta \psi_\alpha^{a \, A} = - (\sigma^{\mu\nu})_\alpha^{\ \beta} \varepsilon_\beta^A F_{\mu\nu}^a - i \, \bar{\Sigma}_I^{AB} (D_{\alpha\dot{\alpha}} \phi^I)^a \bar{\varepsilon}_B^{\dot{\alpha}} - \frac{g}{2} f^{abc} \bar{\Sigma}_I^{AB} \Sigma_{BC}^J \varepsilon_{\alpha}^C \phi^{b \, I} \phi_J^c \, , \\
&\delta \bar{\psi}_{\dot{\alpha} A}^a = \bar{\varepsilon}_{\dot{\beta} A} (\bar{\sigma}^{\mu\nu})_{\ \dot{\alpha}}^{\dot{\beta}} F_{\mu\nu}^a + i \, \Sigma_{AB}^I \varepsilon^{\alpha B} (D_{\alpha\dot{\alpha}} \phi_I)^a - \frac{g}{2} f^{abc} \Sigma_{AB}^I \bar{\Sigma}_J^{ BC}\bar{\varepsilon}_{\dot{\alpha} C} \phi_I^b \phi^{c \, J} \, , \\
&\delta \phi_I^a = - \varepsilon^{\alpha A} \Sigma_{I \, AB} \psi_{\alpha}^{a B} - \bar{\varepsilon}_{\dot{\alpha}A} \bar{\Sigma}_I^{AB} \bar{\psi}_B^{a \, \dot{\alpha}} \, ,
\end{aligned}
\end{align}
where $\varepsilon_\alpha^A$ and $\bar{\varepsilon}_{\dot{\alpha}A}$ are anti-commuting Weyl spinors. Since the supersymmetry of $\mathcal{N}=4$ super Yang-Mills is realized only `on-shell,' the action of the supersymmetry algebra closes only up to terms proportional to the equations of motion
\begin{align}
\begin{aligned}\label{eq:BPSeom1}
D_\nu F^{\mu\nu} &= i g [ \phi_I, D^\mu \phi^I] + g [\psi^{\alpha A} \sigma_{\alpha\dot{\alpha}}^\mu, \bar{\psi}_A^{\dot{\alpha}}] \, , \\
D_\mu D^\mu \phi_I &= - g^2 \, [ \phi_J, [ \phi_I, \phi^J]] - \frac{ig}{2} [\psi^{\alpha A}, \Sigma_{I \, AB} \psi_{\alpha}^B] - \frac{ig}{2} [\bar{\psi}_{\dot{\alpha}A}, \bar{\Sigma}_I^{AB} \bar{\psi}_B^{\dot{\alpha}}] \, , \\
D^{\dot{\alpha}\alpha} \psi_\alpha^A &= - g \, [\bar{\Sigma}_I^{AB} \phi^I, \bar{\psi}_B^{\dot{\alpha}}] \, , \\
D_{\alpha\dot{\alpha}} \bar{\psi}_A^{\dot{\alpha}} &= - g \, [ \Sigma_{AB}^I \phi_I, \psi_\alpha^B] \, .
\end{aligned}
\end{align}

\section{The Stress Tensor Multiplet}\label{sec:BPSMultiplet}
We give a diagrammatic representation of all states in the stress tensor multiplet. The highest weight state of the stress tensor multiplet $\mathcal{B}_{[0,2,0](0,0)}^{\frac{1}{2},\frac{1}{2}}$ is of course $[0,2,0]_{(0,0)}$. The irreducible $\mathfrak{su}(4)$ representation associated to the Dynkin label $[0,2,0]$ has real dimension 20 and is denoted by $\mathbf{20}^\prime$. The prime is conventional since the irreducible representation associated to $[2,0,0]$ is also 20-dimensional. 

The descendants of the highest weight state are found by acting with the supercharges. By construction, the highest weight state is annihilated by the action of $Q_\alpha^1$, $Q_\alpha^2$, $\bar{Q}_{\dot{\alpha}3}$ and $\bar{Q}_{\dot{\alpha}4}$ (see section \ref{sec:ShortMultiplet}). In \eqref{eq:SCAQDynkin1} - \eqref{eq:SCAQDynkin2} we have given the changes in the Dynkin labels and spin generated by the action of the supercharges. For example acting with $Q_\alpha^3$ on the highest weight state $[0,2,0]_{(0,0)}$ gives
\begin{align}
Q_\alpha^3 [0,2,0]_{(0,0)} = [0,1,1]_{(\textstyle\frac12,0)} \, .
\end{align} 
Acting again with $Q_\beta^3$ (with $\beta \not = \alpha$) yields
\begin{align}\label{eq:BPSQ3Q3}
Q_\beta^3 Q_\alpha^3 [0,2,0]_{(0,0)} = [0,0,2]_{(0,0)} \, .
\end{align}
Alternatively, acting with $Q_\alpha^4$ gives
\begin{align}\label{eq:BPSQ3Q4}
Q_\beta^4 Q_\alpha^3 [0,2,0]_{(0,0)} = [0,1,0]_{(1,0)} \oplus \color{myred} [0,1,0]_{(0,0)} \color{black} \, .
\end{align}
The {\color{myred}red} term vanishes. It would correspond to the trace over a single Lie algebra valued scalar field $\phi_I$, which is zero for gauge group $SU(N)$. On \eqref{eq:BPSQ3Q3}, we can only act with $Q_\alpha^4$ since acting three times with $Q_\alpha^3$ gives zero. Similarly, on \eqref{eq:BPSQ3Q4} we must act with $Q_\beta^3$. In both cases, we obtain
\begin{align}
Q_\gamma^4 Q_\beta^3 Q_\alpha^3 [0,2,0]_{(0,0)} = [0,0,1]_{(\textstyle\frac12,0)} \, .
\end{align}
Finally, we act one last time with $Q_\delta^4$ and obtain
\begin{align}\label{eq:BPSQQQQ}
Q_\delta^4 Q_\gamma^4 Q_\beta^3 Q_\alpha^3 [0,2,0]_{(0,0)} = [0,0,0]_{(0,0)} \, .
\end{align}
Any further action with a supercharge $Q_\alpha^A$ would give zero. Thus, we have obtained all states in one branch of the $\mathcal{B}_{[0,2,0](0,0)}^{\frac{1}{2},\frac{1}{2}}$ multiplet. 

The entire stress tensor multiplet structure is given in the following diagram \cite{Dolan:2002zh}.
\begin{figure}[H]
\begin{center}
\begin{tikzpicture}
\node (A) at (0,0) {\scriptsize{$[0,2,0]_{(0,0)}$}};
\node (B1) at (-1.4,-1.4) {\scriptsize{$[0,1,1]_{(\frac12,0)}$}}; \node (B2) at (1.4,-1.4) {\scriptsize{$[1,1,0]_{(0,\frac12)}$}};
\node (C1) at (-2.8,-2.8) {\scriptsize{$[0,0,2]_{(0,0)}$}}; \node (C2) at (-2.8,-3.2) {\scriptsize{$[0,1,0]_{(1,0)}$}}; 
\node (C3) at (0,-3) {\scriptsize{$[1,0,1]_{(\frac12,\frac12)}$}}; \node (C4) at (2.8,-2.8) {\scriptsize{$[2,0,0]_{(0,0)}$}}; 
\node (C5) at (2.8,-3.2) {\scriptsize{$[0,1,0]_{(0,1)}$}};
\node (D1) at (-4.3,-4.6) {\scriptsize{$[0,0,1]_{(\frac12,0)}$}}; \node (D2) at (-1.4,-4.6) {\scriptsize{$[1,0,0]_{(1,\frac12)}$}}; 
\node (D3) at (1.4,-4.6) {\scriptsize{$[0,0,1]_{(\frac12,1)}$}}; \node (D4) at (4.3,-4.6) {\scriptsize{$[1,0,0]_{(0,\frac12)}$}};
\node (E1) at (-5.7,-6) {\scriptsize{$[0,0,0]_{(0,0)}$}}; \node (E2) at (0,-6) {\scriptsize{$[0,0,0]_{(1,1)}$}}; \node (E3) at (5.7,-6) {\scriptsize{$[0,0,0]_{(0,0)}$}};
\draw[->] (A) to (B1); \draw[->] (A) to (B2);
\draw[->] (B1) to (C1); \draw[->] (B1) to (C3); \draw[->] (B2) to (C3); \draw[->] (B2) to (C4); 
\draw[->] (C2) to (D1); \draw[->] (C2) to (D2); \draw[->] (C3) to (D2); \draw[->] (C3) to (D3); 
\draw[->] (C5) to (D3); \draw[->] (C5) to (D4);
\draw[->] (D1) to (E1); \draw[->] (D2) to (E2); \draw[->] (D3) to (E2); \draw[->] (D4) to (E3); 
\node (00) at (-7.5,0.8) {\scriptsize{$\Delta$}};
\node (A0) at (-7.5,0) {\scriptsize{$2$}};
\node (B0) at (-7.5,-1.4) {\scriptsize{$\frac{5}{2}$}};
\node (C0) at (-7.5,-3) {\scriptsize{$3$}};
\node (D0) at (-7.5,-4.6) {\scriptsize{$\frac{7}{2}$}};
\node (E0) at (-7.5,-6) {\scriptsize{$4$}};
\end{tikzpicture}
\end{center}
\caption{\label{fig:BPSStressTensorMultipletDynkin}
Diagrammatic representation of the stress tensor multiplet using Dynkin labels.}
\end{figure}
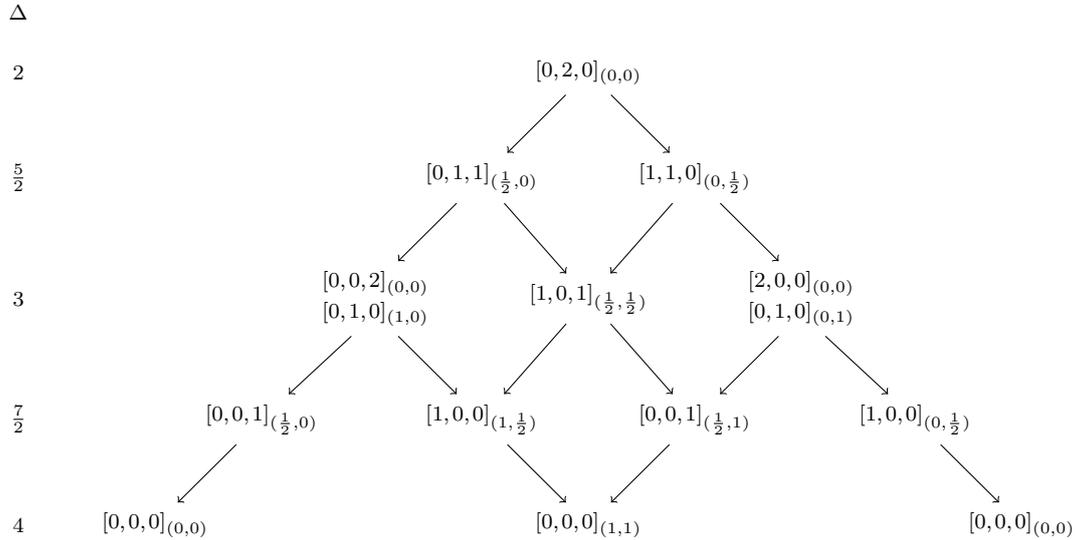
The $\swarrow$ arrows denotes the action of $Q_\alpha^A$ and the $\searrow$ arrows denotes the action of $\bar{Q}_{\dot{\alpha}A}$. The first row has conformal dimension $\Delta = 2$. With every row, the conformal dimension increases by $\frac{1}{2}$. Moreover, the spin increases (or decreases) by $\frac{1}{2}$ with every action of a supercharge. 
\section{The Field Content}\label{sec:FieldContent}
We give the entire field content of the stress tensor multiplet in terms of the fundamental fields. The primary field of the multiplet corresponding to the highest weight state $[0,2,0]_{(0,0)}$ is given by
\begin{align}\label{eq:BPSPrimary}
\varphi_{IJ} \coloneqq \tr_c( \phi_I \phi_J) - \frac{\eta_{IJ}}{6} \tr_c(\phi_K \phi^K) \, . 
\end{align}
The field is symmetric, \emph{i.e.} $\varphi_{IJ} = \varphi_{JI}$, and traceless, \emph{i.e.} $\varphi_I^{\ I} = 0$. Because the fundamental scalar field $\phi_I$ has conformal dimension 1, $\varphi_{IJ}$ has conformal dimension 2. Similarly, there is a field corresponding to each state in the $\frac{1}{2}$-BPS multiplet. Up to hermitian conjugates, the fields in the stress tensor multiplet are\footnote{As usual $(ab) = \frac{1}{2}(ab + ba)$.}
\begin{align}\label{eq:BPSFields}
\begin{aligned}
&\varphi_{IJ} \coloneqq \tr_c (\phi_I \phi_J) - \frac{\eta_{IJ}}{6} \tr_c(\phi_K \phi^K) \, , \\[1.5ex]
&\Psi_{\alpha I}^A \coloneqq \tr_c ( \phi_I \psi_{\alpha}^A ) - \frac{\bar{\Sigma}_I^{AB} \Sigma_{BC}^K }{6} \, \tr_c( \phi_K \psi_{\alpha}^C) \, , \\[1.5ex]
&\rho^{AB} \coloneqq \frac{1}{2} \tr_c \left( \psi^{\alpha A} \psi_\alpha^B \right) - \frac{ig}{12} \, \tr_c \left( (\bar{\Sigma}_I \Sigma_J \bar{\Sigma}_K)^{AB} \phi^I [ \phi^J , \phi^K ] \right) \, , \\[1.5ex]
&f_{\alpha\beta}^I \coloneqq \frac{1}{2} \tr_c\left( \psi_{\alpha}^A \Sigma_{AB}^I \psi_{\beta}^B \right) - \tr_c \left( \phi^I (\sigma^{\mu\nu})_{\alpha}^{\ \gamma} \epsilon_{\gamma \beta} F_{\mu\nu} \right) \, , \\[1.5ex]
&J_{\alpha\dot{\alpha} IJ} \coloneqq \tr_c \left( \psi_\alpha^A (\Sigma_{IJ})_A^{\ B} \bar{\psi}_{\dot{\alpha}B} \right) - i \, \tr_c \left( (D_{\alpha\dot{\alpha}} \phi_{[I}) \phi_{J]} \right) \, , \\[1.5ex]
&\lambda_\alpha^A \coloneqq \tr_c \left( (\sigma^{\mu\nu})_\alpha^{\ \beta} \psi_\beta^A F_{\mu\nu} \right) 
 + \frac{ig}{2} \, \tr_c \left( \bar{\Sigma}_I^{AB} \Sigma_{BC}^J \psi_{\alpha}^C [\phi^I, \phi_J] \right) \, , \\[1.5ex]
&\begin{aligned}
\chi_{\alpha\beta\dot{\beta}A} &\coloneqq - \frac{1}{2} \tr_c \left( (\sigma^{\mu\nu})_\alpha^{\ \gamma} \epsilon_{\gamma\beta} F_{\mu\nu} \bar{\psi}_{\dot{\beta}A} \right) - \frac{i}{3} \, \tr_c \left( \Sigma_{AB}^I \psi_{(\alpha}^B (D_{\beta)\dot{\beta}} \phi_I) \right) \\
&\quad + \frac{i}{6} \, \tr_c \left( \Sigma_{AB}^I (D_{(\beta|\dot{\beta}} \psi_{\alpha)}^B) \phi_I \right) \, , 
\end{aligned} \\[1.5ex]
&\begin{aligned}
\Phi &\coloneqq \frac{1}{2} \, \tr_c \left( F_{\mu\nu} F^{\mu\nu} \right) + \frac{i}{4} \epsilon^{\mu\nu\rho\lambda} \, \tr_c\left( F_{\mu\nu} F_{\rho\lambda} \right) \\
&\quad + i \, \tr_c \left( \psi^{\alpha A} (D_{\alpha\dot{\alpha}} \bar{\psi}_A^{\dot{\alpha}}) \right) 
+ \frac{g^2}{2} \tr_c \left( [\phi_I, \phi_J] [\phi^I, \phi^J] \right) \, , 
\end{aligned} \\[1.5ex]
&\begin{aligned}
T_{\alpha\beta\dot{\alpha}\dot{\beta}} &\coloneqq \frac{1}{2} \, \tr_c \left( (\sigma^{\mu\nu})_{{\color{myred}(\alpha}}^{\ \gamma} \epsilon_{|\gamma|\color{myred}\beta)} \epsilon_{{\color{myblue}(\dot{\alpha}}|\dot{\gamma}|} (\bar{\sigma}^{\rho\lambda})_{\ {\color{myblue}\dot{\beta})}}^{\dot{\gamma}} F_{\mu\nu} F_{\rho\lambda} \right) \\
&\quad + \frac{i}{2} \, \tr_c \left( (D_{\color{myred}(\alpha\color{myblue}(\dot{\alpha}} \psi_{{\color{myred}\beta)}}^A) \bar{\psi}_{{\color{myblue}\dot{\beta})}A} \right) - \frac{i}{2} \, \tr_c \left( \psi_{{\color{myred}(\alpha}}^A (D_{\color{myred}\beta)\color{myblue}(\dot{\alpha}} \bar{\psi}_{{\color{myblue}\dot{\beta})}A}) \right) \\
&\quad + \frac{1}{6} \, \tr_c \left((D_{\color{myred}(\alpha\color{myblue}(\dot{\alpha}} (D_{\color{myred}\beta)\color{myblue}\dot{\beta})} \phi_I)) \phi^I \right) 
- \frac{1}{3} \, \tr_c \left((D_{\color{myred}(\alpha\color{myblue}(\dot{\alpha}} \phi_I) (D_{\color{myred}\beta)\color{myblue}\dot{\beta})} \phi^I) \right) \, .
\end{aligned}
\end{aligned}
\end{align}
This is the main result of this chapter and has not been published before. The paper \cite{Dolan:2001tt} only discusses the field constraints and supersymmetry transformations of the fields \eqref{eq:BPSFields} but does not provide the explicit expressions \eqref{eq:BPSFields}. In section \ref{sec:BPSsusyTrafo}, we show how to derive \eqref{eq:BPSFields} from the primary field \eqref{eq:BPSPrimary}. 

The last field in \eqref{eq:BPSFields} is, of course, the energy-momentum tensor $T_{\alpha\beta\dot{\alpha}\dot{\beta}}$. We colored the indices to distinguish the two symmetrizations over the dotted and undotted indices $\color{myblue}(\dot{\alpha}\dot{\beta})$ and $\color{myred}(\alpha\beta)$. The fact that $T_{\alpha\beta\dot{\alpha}\dot{\beta}}$ is separately symmetric in $\color{myblue}(\dot{\alpha}\dot{\beta})$ and $\color{myred}(\alpha\beta)$ follows from the symmetry and tracelessness of $T_{\mu\nu}$. $\Phi$ is the $SU(4)$ $R$-symmetry current. 

Replacing the Dynkin labels in \ref{fig:BPSStressTensorMultipletDynkin} with the respective fields gives \cite{Dolan:2001tt}:
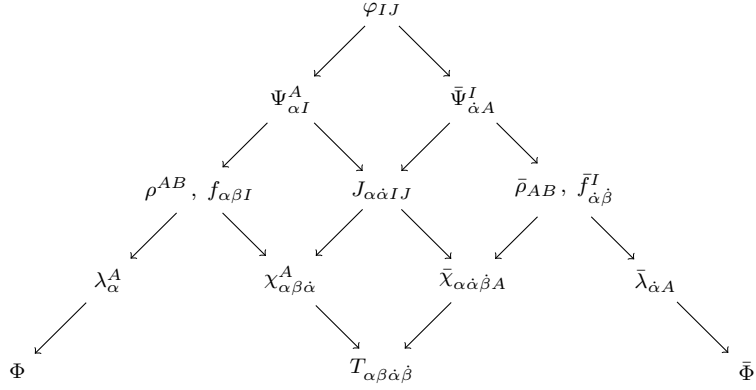
\begin{figure}[H]
\begin{center}
\begin{tikzpicture}
\node (A1) at (0,0) {\scriptsize{$\varphi_{IJ}$}};
\node (B1) at (-1.2,-1.2) {\scriptsize{$\Psi_{\alpha I}^A$}}; \node (B2) at (1.2,-1.2) {\scriptsize{$\bar{\Psi}_{\dot{\alpha} A}^I$}};
\node (C1) at (-2.4,-2.4) {\scriptsize{$\rho^{AB} \, , \ f_{\alpha\beta I}$}}; 
\node (C2) at (0,-2.4) {\scriptsize{$J_{\alpha\dot{\alpha} IJ}$}}; \node (C3) at (2.4,-2.4) {\scriptsize{$\bar{\rho}_{AB} \, , \ \bar{f}_{\dot{\alpha}\dot{\beta}}^I$}}; 
\node (D1) at (-3.6,-3.6) {\scriptsize{$\lambda_\alpha^A$}}; \node (D2) at (-1.2,-3.6) {\scriptsize{$\chi_{\alpha\beta\dot{\alpha}}^A$}}; 
\node (D3) at (1.2,-3.6) {\scriptsize{$\bar{\chi}_{\alpha\dot{\alpha}\dot{\beta} A}$}}; \node (D4) at (3.6,-3.6) {\scriptsize{$\bar{\lambda}_{\dot{\alpha} A}$}};
\node (E1) at (-4.8,-4.8) {\scriptsize{$\Phi$}}; \node (E2) at (0,-4.8) {\scriptsize{$T_{\alpha\beta\dot{\alpha}\dot{\beta}}$}}; \node (E3) at (4.8,-4.8) {\scriptsize{$\bar{\Phi}$}};
\draw[->] (A1) to (B1); \draw[->] (A1) to (B2);
\draw[->] (B1) to (C1); \draw[->] (B1) to (C2); \draw[->] (B2) to (C2); \draw[->] (B2) to (C3); 
\draw[->] (C1) to (D1); \draw[->] (C1) to (D2); \draw[->] (C2) to (D2); \draw[->] (C2) to (D3); 
\draw[->] (C3) to (D3); \draw[->] (C3) to (D4);
\draw[->] (D1) to (E1); \draw[->] (D2) to (E2); \draw[->] (D3) to (E2); \draw[->] (D4) to (E3); 
\end{tikzpicture}
\end{center}
\caption{\label{fig:BPSStressTensorMultipletFields}
Diagrammatic representation of the stress tensor multiplet in terms of the fields.}
\end{figure}
The supersymmetry variations of the fields in the stress tensor multiplet are \cite{Dolan:2001tt}
\begin{align}\label{eq:BPSsusy2}
\begin{aligned}
&\delta \varphi_{IJ} = - 2 \varepsilon^{\alpha A} \Sigma_{AB (I} \Psi_{J) \alpha}^B - 2 \bar{\varepsilon}_{\dot{\alpha}A} \bar{\Sigma}_{(I}^{AB} \bar{\Psi}_{J) B}^{\dot{\alpha}} \, , \\[1.5ex]
&\begin{aligned}
\delta \Psi_{\alpha I}^A &= - \frac{i}{2} (D_{\alpha\dot{\alpha}} \varphi_{IJ}) \bar{\Sigma}^{J \, AB} \bar{\varepsilon}_B^{\dot{\alpha}} + f_{\alpha\beta I} \varepsilon^{\beta A} - \frac{1}{6} f_{\alpha\beta J} \bar{\Sigma}_I^{AB} \Sigma_{BC}^J \varepsilon^{\beta C} \\
&\quad - \rho^{AB} \Sigma_{I \, BC} \varepsilon_{\alpha}^C - J_{\alpha\dot{\alpha} IJ} \bar{\Sigma}^{J \, AB} \bar{\varepsilon}_B^{\dot{\alpha}} + \frac{1}{6} J_{\alpha\dot{\alpha} JK} \bar{\Sigma}_I^{AB} \Sigma_{BC}^J \bar{\Sigma}^{K \, CD} \bar{\varepsilon}_D^{\dot{\alpha}} \, , 
\end{aligned} \\[1.5ex]
&\delta \rho^{AB} = \varepsilon^{\alpha (A} \lambda_\alpha^{B)} - i \bar{\varepsilon}_{\dot{\alpha} C} \bar{\Sigma}_I^{C(A} D^{\dot{\alpha}\alpha} \Psi_{\alpha}^{B) I} \, , \\[1.5ex]
&\delta f_{\alpha\beta}^I = - 2i \Psi_{(\alpha}^{I \, A} \overset{\leftarrow}{D}_{\beta)\dot{\beta}} \bar{\varepsilon}_A^{\dot{\beta}} + 2 \chi_{\alpha\beta\dot{\beta} A} \bar{\Sigma}^{I \, AB} \bar{\varepsilon}_B^{\dot{\beta}} - \varepsilon_{(\alpha}^A \Sigma_{AB}^I \lambda_{\beta)}^B \, , \\[1.5ex]
&\begin{aligned}
\delta J_{\alpha\dot{\alpha}IJ} &= 2i \varepsilon^{\beta A} \Sigma_{AB [I} \Psi_{J]\alpha}^B \overset{\leftarrow}{D}_{\beta\dot{\alpha}}
-i \varepsilon^{\beta A} \Sigma_{AB [I} \Psi_{J]\beta}^B \overset{\leftarrow}{D}_{\alpha\dot{\alpha}}
+ 2 \varepsilon^{\beta A} (\Sigma_{IJ})_A^{\ B} \chi_{\beta\alpha\dot{\alpha}B} \\
&\quad + 2i D_{\alpha\dot{\beta}} \bar{\Psi}_{\dot{\alpha}[I}^A \bar{\Sigma}_{J] \, AB} \bar{\varepsilon}^{\dot{\beta}B} 
- i D_{\alpha\dot{\alpha}} \bar{\Psi}_{\dot{\beta}[I}^A \bar{\Sigma}_{J] \, AB} \bar{\varepsilon}^{\dot{\beta} B}
+ 2 \bar{\chi}_{\alpha\dot{\alpha}\dot{\beta}}^A (\bar{\Sigma}_{IJ})_{\ A}^B \bar{\varepsilon}_B^{\dot{\beta}} \, . 
\end{aligned}\\[1.5ex]
&\delta \lambda_\alpha^A = \varepsilon_\alpha^A \Phi + i ( f_{\alpha\beta}^I \epsilon^{\beta\gamma} \overset{\leftarrow}{D}_{\gamma\dot{\alpha}}) \bar{\Sigma}_I^{AB} \bar{\varepsilon}_B^{\dot{\alpha}} + 2i D_{\alpha\dot{\alpha}} \rho^{AB} \bar{\varepsilon}_B^{\dot{\alpha}} \, , \\[1.5ex]
&\begin{aligned}
\delta \chi_{\alpha\beta\dot{\alpha}A} &= - \frac{i}{2} \varepsilon^{\gamma B} \Sigma_{I \, BA} D_{\gamma\dot{\alpha}} f_{\alpha\beta}^I + \frac{i}{3} \varepsilon^{\gamma B} \Sigma_{I \, BA} D_{(\beta|\dot{\alpha}|} f_{\alpha)\gamma}^I 
- \frac{i}{4} D_{(\alpha|\dot{\beta}|} J_{\beta)\dot{\alpha}IJ} \Sigma_{AB}^I \bar{\Sigma}^{J\, BC} \bar{\varepsilon}_C^{\dot{\beta}} \\
&\quad + \frac{i}{12} D_{(\alpha|\dot{\alpha}|} J_{\beta)\dot{\beta}IJ} \Sigma_{AB}^I \bar{\Sigma}^{J\, BC} \bar{\varepsilon}_C^{\dot{\beta}} 
+ T_{\alpha\beta\dot{\alpha}\dot{\beta}} \bar{\varepsilon}^{\dot{\beta}A} \, ,
\end{aligned} \\[1.5ex]
&\delta \Phi = 2i \bar{\varepsilon}_{\dot{\alpha}A} D^{\dot{\alpha}\alpha} \lambda_\alpha^A \, , \\[1.5ex]
&\delta T_{\alpha\beta\dot{\alpha}\dot{\beta}} =  2 i \varepsilon^{\delta A} D_{\delta\color{myblue}(\dot{\alpha}} \chi_{\alpha\beta{\color{myblue}\dot{\beta})}A} 
- i \varepsilon^{\delta A} D_{{\color{myred}(\alpha}\color{myblue}(\dot{\alpha}} \chi_{|\delta|{\color{myred}\beta)}{\color{myblue}\dot{\beta})}A} - 2i  \bar{\chi}_{{\color{myred}(\alpha}\dot{\alpha}\dot{\beta}} \overset{\leftarrow}{D}_{{\color{myred}\beta)}\dot{\delta}} \bar{\varepsilon}^{\dot{\delta}A}
+ i  \bar{\chi}_{{\color{myred}(\alpha}{\color{myblue}(\dot{\alpha}}|\dot{\delta}} \overset{\leftarrow}{D}_{{\color{myred}\beta)}{\color{myblue}\dot{\beta})}} \bar{\varepsilon}^{\dot{\delta}A} \, .
\end{aligned} 
\end{align}
These can be obtained without knowing the field representations. We know that under a supersymmetry transformation, a field of conformal dimension $\Delta$ must transform into a linear combination of fields of conformal dimension $\Delta + \frac{1}{2}$ and derivatives of fields of conformal dimension $\Delta - \frac{1}{2}$. The coefficients and field constraints are found by demanding that the supersymmetry algebra closes on the fields of the stress tensor multiplet up to terms proportional to the equations of motion. This analysis was first done in Dolan and Osborn. The equations of motion for the fields are \cite{Dolan:2001tt}
\begin{align}
D^{\alpha\dot{\alpha}} J_{\alpha\dot{\alpha} IJ} = 0 \, , \quad
D^{\dot{\alpha}\alpha} \chi_{\alpha\beta\dot{\alpha}}^A = 0 \, , \quad
D^{\dot{\alpha}\alpha} \bar{\chi}_{\alpha\dot{\alpha}\dot{\beta} A} = 0 \, , \quad 
D^{\dot{\alpha}\alpha} T_{\alpha\beta\dot{\alpha}\dot{\beta}} = 0 \, .
\end{align}
They can be verified using the field expressions \eqref{eq:BPSFields} and the $\mathcal{N}=4$ super Yang-Mills equations of motion \eqref{eq:BPSeom1}. The main properties of the fields are summarized in the following table taken from \cite{Dolan:2001tt}.
\begin{table}[H]
\renewcommand{\arraystretch}{1.0}
\begin{small}
\begin{tabular}{|c|c|c|c|r|c|}
\hline 
$\boldsymbol{SU(4)}$ \textbf{Rep} & $\boldsymbol{SU(4)}$ \textbf{Dim} & $\boldsymbol{(j, \bar{j})}$ & \textbf{Field} & \textbf{Field Constraints} & \textbf{Field Dim} \TBspace \\ 
\hline
$[0,2,0]$ & $20_\mathbbm{R}$ & $(0,0)$ & $\varphi_{IJ}$ & $\varphi_{IJ} = \varphi_{(IJ)} \, , \ \varphi_I^{\ I} = 0$ & 20 \\
$[0,1,1]$ & $20_\mathbbm{C}$ & $(\frac12,0)$ & $\Psi_{\alpha I}^A$ & $\Sigma_{AB}^I \Psi_{\alpha I}^B = 0 $ & 40 \\
$[1,1,0]$ & $20_\mathbbm{C}$ & $(0,\frac12)$ & $\bar{\Psi}_{\dot{\alpha} A}^I$ & $\bar{\Sigma}_I^{AB} \bar{\Psi}_{\dot{\alpha} B}^I = 0 $ & 40 \\
$[0,1,0]$ & $6_\mathbbm{R}$ & $(1,0)$ & $f_{\alpha\beta I}$ & $ f_{\alpha\beta I} = f_{(\alpha \beta) I}$ & 18 \\
$[0,1,0]$ & $6_\mathbbm{R}$ & $(0,1)$ & $\bar{f}_{\dot{\alpha}\dot{\beta}}^I$ & $ \bar{f}_{\dot{\alpha}\dot{\beta}}^I = \bar{f}_{(\dot{\alpha}\dot{\beta})}^I$ & 18 \\
$[0,0,2]$ & $10_\mathbbm{C}$ & $(0,0)$ & $\rho^{AB}$ & $\rho^{AB} = \rho^{(AB)}$ & 10 \\
$[2,0,0]$ & $10_\mathbbm{C}$ & $(0,0)$ & $\bar{\rho}_{AB}$ & $\bar{\rho}_{AB} = \bar{\rho}_{(AB)}$ & 10 \\
$[1,0,1]$ & $15_\mathbbm{C}$ & $(\frac12,\frac12)$ & $J_{\alpha\dot{\alpha} IJ}$ & $J_{\alpha\dot{\alpha} IJ} = J_{\alpha\dot{\alpha} [IJ]} \, , \ D^{\alpha\dot{\alpha}} J_{\alpha\dot{\alpha} IJ} = 0 $ & 45 \\
$[0,0,1]$ & $4_\mathbbm{C}$ & $(\frac12,0)$ & $\lambda_\alpha^A$ & & 8 \\
$[1,0,0]$ & $4_\mathbbm{C}$ & $(0,\frac12)$ & $\bar{\lambda}_{\dot{\alpha} A}$ & & 8 \\
$[1,0,0]$ & $4_\mathbbm{C}$ & $(1,\frac12)$ & $\chi_{\alpha\beta\dot{\alpha}}^A$ & $\chi_{\alpha\beta\dot{\alpha}}^A = \chi_{(\alpha\beta)\dot{\alpha}}^A \, , \ D^{\dot{\alpha}\alpha} \chi_{\alpha\beta\dot{\alpha}}^A = 0$ & 16 \\
$[0,0,1]$ & $4_\mathbbm{C}$ & $(\frac12,1)$ & $\bar{\chi}_{\alpha\dot{\alpha}\dot{\beta} A}$ & $\bar{\chi}_{\alpha\dot{\alpha}\dot{\beta} A} = \bar{\chi}_{\alpha(\dot{\alpha}\dot{\beta}) A} \, , \ D^{\dot{\alpha}\alpha} \bar{\chi}_{\alpha\dot{\alpha}\dot{\beta} A} = 0$ & 16 \\
$[0,0,0]$ & $1_\mathbbm{R}$ & $(0,0)$ & $\Phi$ & & 1 \\
$[0,0,0]$ & $1_\mathbbm{R}$ & $(0,0)$ & $\bar{\Phi}$ & & 1 \\
$[0,0,0]$ & $1_\mathbbm{R}$ & $(1,1)$ & $T_{\alpha\beta\dot{\alpha}\dot{\beta}}$ & $T_{\alpha\beta\dot{\alpha}\dot{\beta}} = T_{(\alpha\beta)(\dot{\alpha}\dot{\beta})} \, , \ D^{\dot{\alpha}\alpha} T_{\alpha\beta\dot{\alpha}\dot{\beta}} = 0$ & 5  \\[.6ex]
\hline
\end{tabular}
\end{small}
\end{table}
The index structure of the fields can be inferred from their field dimension and spin. The field constraints are such that the field dimension matches the $SU(4)$ dimension. 

\section{General Correlation Functions with the Nicolai Map}\label{sec:BPSCorr}
The inverse Nicolai map $\mathcal{T}_g^{-1}$ maps quantum correlation functions of bosonic operators to free correlation functions 
\begin{align}\label{eq:BPSCorr}
\big< \!\! \big< \mathcal{O}_1(x_1) \ldots \mathcal{O}_n(x_n) \big> \!\! \big>_g = \big< (\mathcal{T}_g^{-1} \mathcal{O}_1)(x_1) \ldots (\mathcal{T}_g^{-1}\mathcal{O}_n)(x_n) \big>_0 \, .
\end{align}
However, except for the primary field $\varphi_{IJ}$ all fields in the stress tensor multiplet \eqref{eq:BPSFields} contain spinors. Hence, we must adjust \eqref{eq:BPSCorr} to compute correlation functions of these operators. Recall that we can always integrate out the fermions in a correlation function since they appear only quadratically in the action \eqref{eq:BPSAction}. Similar to \eqref{eq:SpinorCorr} we have for example (with Majorana spinors)
\begin{align}
\big<\!\!\big< A_\mu^a(x) \lambda_A^b(y) \bar{\lambda}_B^c(z) \big>\!\!\big>_g = i \big< A_\mu^a(x) \bcontraction{}{\lambda}{_A^b(y) }{\bar{\lambda}} \lambda_A^b(y) \bar{\lambda}_B^c(z) \big>_g =   \big< A_\mu^a(x)  S_{AB}^{bc}(y,z;\mathscr{A}) \big>_g 
\end{align}
where $\mathscr{A} = (A_\mu^a, A_i^a, B_i^a)$. We then find
\begin{gather}
\begin{gathered}
\big<\!\!\big< A_\mu^a(x) \lambda_A^b(y) \bar{\lambda}_B^c(z) \big>\!\!\big>_g =   \big< A_\mu^a(x)  S_{AB}^{bc}(y,z;\mathscr{A}) \big>_g
\\
= \int \mathcal{D}_0 \mathscr{A} \ \Delta_\mathrm{MSS}[g;\mathscr{A}] \Delta_\mathrm{FP}[g;\mathscr{A}] \ e^{- i S_\mathrm{inv}^4[g;\mathscr{A}] - i S_\mathrm{gf}[g,\mathscr{A}]}  \ A_\mu^a(x)  S_{AB}^{bc}(y,z;\mathscr{A}) \, .
\end{gathered}
\end{gather}
The fermion propagator only depends on the free bosonic propagator and the bosonic fields. Hence, we find
\begin{align}
\begin{aligned}
&\quad \big< (\mathcal{T}_g^{-1} A)_\mu^a(x)  S_{AB}^{bc}(y,z; \mathcal{T}_g^{-1} \mathscr{A}) \big>_0 \\
&= \int \mathcal{D}_0 \mathscr{A}  \  e^{- i S_\mathrm{inv}^4[0;\mathscr{A}] - i S_\mathrm{gf}[0,\mathscr{A}]}  \ (\mathcal{T}_g^{-1} A)_\mu^a(x)  S_{AB}^{bc}(y,z; \mathcal{T}_g^{-1} \mathscr{A}) \\
&= \int \mathcal{D}_0 \mathscr{A}  \ \mathcal{J}(\mathcal{T}_g \, \mathscr{A}) \  e^{- i S_\mathrm{inv}^4[0; \mathcal{T}_g \mathscr{A}] - i S_\mathrm{gf}[0, \mathcal{T}_g \mathscr{A}]}  \  A_\mu^a(x)  S_{AB}^{bc}(y,z; \mathscr{A}) \, .
\end{aligned}
\end{align}
Thus, with the main theorem \ref{th:YM2Theorem} we can conclude
\begin{align}
\big<\!\!\big< A_\mu^a(x) \lambda_A^b(y) \bar{\lambda}_B^c(z) \big>\!\!\big>_g = i  \, \big< (\mathcal{T}_g^{-1} A)_\mu^a(x)  S_{AB}^{bc}(y,z; \mathcal{T}_g^{-1} \mathscr{A}) \big>_0 \, .
\end{align}
The generalization of this equality follows from Wick's theorem and \eqref{eq:BPSCorr}. We leave the calculation of correlation functions for future study. Notice that we can only compute correlation functions of operators containing spinors but we cannot compute their Nicolai map.

\section{Supersymmetry Transformations of the Chiral Primary Field}\label{sec:BPSsusyTrafo}
The descendant fields of the stress tensor multiplet are obtained by successively computing the supersymmetry transformations of the chiral primary field $\varphi_{IJ}$. In the following, we outline the details of these calculations and give some examples. Our starting point is \eqref{eq:BPSsusy2}. The first two descendants are found by
\begin{align}
\begin{aligned}
\delta \varphi_{IJ} &= - \tr_c ( \varepsilon^{\alpha A} \Sigma_{I \, AB} \psi_{\alpha}^B \phi_J ) - \tr_c (\phi_I \varepsilon^{\alpha A} \Sigma_{J \, AB} \psi_{\alpha }^B ) + \frac{\eta_{IJ}}{3} \tr_c (\varepsilon^{\alpha A} \Sigma_{AB}^K \psi_{\alpha}^B \phi_K ) \\
&\quad - \tr_c ( \bar{\varepsilon}_{\dot{\alpha} A} \bar{\Sigma}_I^{AB} \bar{\psi}_B^{\dot{\alpha}} \phi^J ) - \tr_c (\phi^I \bar{\varepsilon}_{\dot{\alpha} A} \bar{\Sigma}_J^{AB} \bar{\psi}_B^{\dot{\alpha}} ) + \frac{\eta_{IJ}}{3} \tr_c (\bar{\varepsilon}_{\dot{\alpha} A} \bar{\Sigma}_K^{AB} \bar{\psi}_B^{\dot{\alpha}} \phi^K ) \\
&\overset{!}{=} - 2 \varepsilon^{\alpha A} \Sigma_{AB ( I} \Psi_{J) \alpha}^B - 2 \bar{\varepsilon}_{\dot{\alpha}A} \bar{\Sigma}_{(I}^{AB} \bar{\Psi}_{J) B}^{\dot{\alpha}} \, .
\end{aligned}
\end{align}
We can read off the expressions for $\Psi_{\alpha I}^A$ and $\bar{\Psi}_{\dot{\alpha}A}^I$. We continue down the left branch of the multiplet (see fig. \ref{fig:BPSStressTensorMultipletFields}). The supersymmetry variation of $\Psi_{\alpha I}^A$ is
\begin{align}
\begin{aligned}\label{eq:BPSsusy3}
\delta \Psi_{\alpha I}^A &= 
 - \tr_c \left( \varepsilon^{\beta B} \Sigma_{I \, BC} \psi_{\beta}^C \psi_\alpha^A \right) 
- \tr_c \left( \bar{\varepsilon}_{\dot{\beta}B} \bar{\Sigma}_I^{BC} \bar{\psi}_C^{\dot{\beta}} \psi_\alpha^A \right) 
- \tr_c \left( \phi_I \sigma_{\alpha}^{\mu\nu \, \beta} \varepsilon_\beta^A F_{\mu\nu} \right) \\
&\quad - i \, \tr_c \left( \phi_I \sigma_{\alpha\dot{\alpha}}^\mu \bar{\Sigma}_J^{AB} \bar{\varepsilon}_B^{\dot{\alpha}} (D_\mu \phi^J) \right) 
+ \frac{ig}{2} \tr_c \left( \phi_I \bar{\Sigma}_J^{AB} \Sigma_{BC}^K \varepsilon_{\alpha}^C [\phi^J, \phi_K] \right) \\
&\quad + \frac{\bar{\Sigma}_I^{AB} \Sigma_{BC}^K }{6} \, \tr_c \left( \varepsilon^{\beta D} \Sigma_{K \, DE} \psi_{\beta}^E \psi_{\alpha}^C \right) 
+ \frac{\bar{\Sigma}_I^{AB} \Sigma_{BC}^K }{6} \, \tr_c \left(\bar{\varepsilon}_{\dot{\beta}D} \bar{\Sigma}_K^{DE} \bar{\psi}_E^{\dot{\beta}} \psi_{\alpha}^C \right) \\
&\quad + \frac{\bar{\Sigma}_I^{AB} \Sigma_{BC}^K }{6} \, \tr_c \left( \phi_K \sigma_{\alpha}^{\mu\nu \, \beta} \varepsilon_\beta^C F_{\mu\nu} \right)
+ i \, \frac{\bar{\Sigma}_I^{AB} \Sigma_{BC}^K }{6} \, \tr_c \left( \phi_K \sigma_{\alpha\dot{\alpha}}^\mu \bar{\Sigma}_J^{CD} \bar{\varepsilon}_D^{\dot{\alpha}} (D_\mu \phi^J) \right) \\
&\quad - \frac{ig}{2} \frac{\bar{\Sigma}_I^{AB} \Sigma_{BC}^K }{6} \, \tr_c \left( \phi_K \bar{\Sigma}_{CD}^J \Sigma^{L \, DE} \varepsilon_{\alpha E} [\phi_J, \phi_L] \right) \, .
\end{aligned}
\end{align}
By \eqref{eq:BPSsusy2} this must be equivalent to
\begin{align}
\begin{aligned}\label{eq:BPSsusy4}
\delta \Psi_{\alpha I}^A &= - \frac{i}{2} (D_{\alpha\dot{\alpha}} \varphi_{IJ}) \bar{\Sigma}^{J \, AB} \bar{\varepsilon}_B^{\dot{\alpha}} + f_{\alpha\beta I} \varepsilon^{\beta A} - \frac{1}{6} f_{\alpha\beta J} \bar{\Sigma}_I^{AB} \Sigma_{BC}^J \varepsilon^{\beta C} \\
&\quad - \rho^{AB} \Sigma_{I \, BC} \varepsilon_{\alpha}^C - J_{\alpha\dot{\alpha} IJ} \bar{\Sigma}^{J \, AB} \bar{\varepsilon}_B^{\dot{\alpha}} + \frac{1}{6} J_{\alpha\dot{\alpha} JK} \bar{\Sigma}_I^{AB} \Sigma_{BC}^J \bar{\Sigma}^{K \, CD} \bar{\varepsilon}_D^{\dot{\alpha}} \, .
\end{aligned}
\end{align}
First we focus on all the terms proportional to $\varepsilon_\alpha^A$ and compare the respective terms on the right-hand sides of \eqref{eq:BPSsusy3} and \eqref{eq:BPSsusy4}
\begin{align}
\begin{aligned}\label{eq:BPSsusy5}
&\quad  f_{\alpha\beta I} \varepsilon^{\beta A} - \frac{1}{6} f_{\alpha\beta J} \bar{\Sigma}_I^{AB} \Sigma_{BC}^J \varepsilon^{\beta C} - \rho^{AB} \Sigma_{I \, BC} \varepsilon_{\alpha}^C \\
&\overset{!}{=} - \tr_c \left( \varepsilon^{\beta B} \Sigma_{I \, BC} \psi_{\beta}^C \psi_\alpha^A \right) 
- \tr_c \left( \phi_I \sigma_{\alpha}^{\mu\nu \, \beta} \varepsilon_\beta^A F_{\mu\nu} \right) 
+ \frac{ig}{2} \tr_c \left( \phi_I \bar{\Sigma}_J^{AB} \Sigma_{BC}^K \varepsilon_{\alpha}^C [\phi^J, \phi_K] \right) \\
&\quad + \frac{\bar{\Sigma}_I^{AB} \Sigma_{BC}^K }{6} \, \tr_c \left( \varepsilon^{\beta D} \Sigma_{K \, DE} \psi_{\beta}^E \psi_{\alpha}^C \right) 
\color{myblue} + \frac{\bar{\Sigma}_I^{AB} \Sigma_{BC}^K }{6} \, \tr_c \left( \phi_K \sigma_{\alpha}^{\mu\nu \, \beta} \varepsilon_\beta^C F_{\mu\nu} \right) \\
&\quad \color{mygreen} - \frac{ig}{2} \frac{\bar{\Sigma}_I^{AB} \Sigma_{BC}^K }{6} \, \tr_c \left( \phi_K \bar{\Sigma}_{CD}^J \Sigma^{L \, DE} \varepsilon_{\alpha E} [\phi_J, \phi_L] \right) \color{black} \, .
\end{aligned}
\end{align}
The coloring is for later. We want to solve this equation for $\rho^{AB}$ and $f_{\alpha\beta}^I$. From the table in the previous section we know that $f_{\alpha\beta}^I$ must be symmetric in $\alpha$ and $\beta$ and $\rho^{AB}$ must be symmetric in $A$ and $B$. The only combinations of fields appearing on the right-hand side of \eqref{eq:BPSsusy5}, which are symmetric in $\alpha$ and $\beta$ and match the index structure of $f_{\alpha\beta}^I$ are
\begin{align}\label{eq:BPSguess1}
 \tr_c\left( \psi_{\alpha}^A \Sigma_{AB}^I \psi_{\beta}^B \right) \, , \quad \tr_c \left( \phi^I (\sigma^{\mu\nu})_{\alpha}^{\ \gamma} \epsilon_{\gamma \beta} F_{\mu\nu} \right) \, , 
\end{align}
where $\epsilon_{\alpha\beta}$ is the anti-symmetric tensor and $(\sigma^{\mu\nu})_{\alpha}^{\ \gamma} \epsilon_{\gamma \beta}$ is symmetric in $\alpha$ and $\beta$. Similarly,
\begin{align}\label{eq:BPSguess2}
 \tr_c \left( \psi^{\alpha A} \psi_\alpha^B \right) \, , \quad \tr_c \left( (\bar{\Sigma}_I \Sigma_J \bar{\Sigma}_K)^{AB} \phi^I [ \phi^J , \phi^K ] \right)
\end{align}
are the only combinations symmetric in $A$ and $B$. A term like $\tr_c( \Sigma_I^{AB} \phi^I (\sigma^{\mu\nu})_{\alpha}^{\ \beta} \epsilon_{\beta \alpha} F_{\mu\nu})$ cannot be part of $\rho^{AB}$ as it is anti-symmetric in $A$ and $B$. Thus, we assume that $f_{\alpha\beta}^I$ is a linear combination of \eqref{eq:BPSguess1} and $\rho^{AB}$ is a linear combination of \eqref{eq:BPSguess2}. We compute the left-hand side of \eqref{eq:BPSsusy5} to fix the coefficients $X_i$
\begin{align}
\begin{aligned}\label{eq:BPSRho}
&\quad - \rho^{AB} \Sigma_{I\, BC} \varepsilon_{\alpha}^C \\
&= X_1 \tr_c \left( \psi^{\beta A} \psi_{\beta}^B \right) \Sigma_{I\, BC} \varepsilon_{\alpha}^C + X_2 \tr_c \left( (\bar{\Sigma}_J \Sigma_K \bar{\Sigma}_L)^{AB} \Sigma_{I\, BC} \varepsilon_{\alpha}^C \phi^J [ \phi^K , \phi^L ] \right) \\
&= X_1  \tr_c \left(\varepsilon_{\alpha}^B \Sigma_{I\, BC} \psi^{\beta C} \psi_\beta^A \right) 
- 6 X_2  \tr_c \left( (\bar{\Sigma}_J \Sigma^K)_{\ B}^A \varepsilon_\alpha^B \phi_I [\phi^J, \phi_K] \right) \\
&\color{mygreen}+ X_2 \tr_c \left( (\bar{\Sigma}_I\Sigma^J \bar{\Sigma}_K \Sigma^L)_{\ B}^A \varepsilon_\alpha^B \phi_J [\phi^K, \phi_L] \right) \color{black} \, .
\end{aligned}
\end{align}
The {\color{mygreen}green} term must match the {\color{mygreen}green} term in \eqref{eq:BPSsusy5}. So we obtain $X_2 = - \frac{ig}{12}$. To find $X_1$ we must first compute the terms with $f_{\alpha\beta}^I$ in \eqref{eq:BPSsusy5}
\begin{align}
\begin{aligned}\label{eq:BPSf}
&\quad f_{\alpha\beta I} \varepsilon^{\beta A} - \frac{1}{6} f_{\alpha\beta J} \bar{\Sigma}_I^{AB} \Sigma_{BC}^J \varepsilon^{\beta C} \\
&= X_3 \tr_c\left( \psi_{\alpha}^B \Sigma_{I\, BC} \psi_{\beta}^C \varepsilon^{\beta A} \right) + X_4 \tr_c \left( \phi_I (\sigma^{\mu\nu})_\alpha^{\ \gamma} \epsilon_{\gamma \beta} \varepsilon^{\beta A} F_{\mu\nu} \right) \\
&\quad - \frac{X_3}{6} \tr_c\left( \psi_{\alpha}^D \Sigma_{J \, DE} \psi_{\beta}^E \bar{\Sigma}_I^{AB} \Sigma_{BC}^J \varepsilon^{\beta C} \right) \\
&\quad  - \frac{X_4}{6} \tr_c \left( \phi_J (\sigma^{\mu\nu})_\alpha^{\ \gamma}\epsilon_{\gamma \beta} \bar{\Sigma}_I^{AB} \Sigma_{BC}^J \varepsilon^{\beta C} F_{\mu\nu} \right) \\
&=  \left( X_3 - \frac{X_3}{3} \right) \tr_c\left( \varepsilon^{\beta A} \psi_{\beta}^B \Sigma_{I \, BC} \psi_{\alpha}^C \right) 
- \frac{X_3}{3} \tr_c\left( \varepsilon^{\beta B} \Sigma_{I\, BC} \psi_{\beta}^C \psi_{\alpha}^A \right)
 \\
&\quad + \frac{X_3}{3} \tr_c\left( \varepsilon^{\beta B} \Sigma_{I\, BC} \psi_{\beta}^A \psi_{\alpha}^C \right) + X_4 \tr_c \left( \phi_I \sigma_{\alpha}^{\mu\nu \, \beta} \varepsilon_{\beta}^A F_{\mu\nu} \right) \\
&\quad \color{myblue} - X_4 \frac{\bar{\Sigma}_I^{AB} \Sigma_{BC}^J}{6} \tr_c \left( \phi_J \sigma_{\alpha}^{\mu\nu \, \beta} \varepsilon_{\beta}^C F_{\mu\nu} \right) \color{black} \, . 
\end{aligned}
\end{align}
For the second equality we used $\Sigma_{AB}^I \bar{\Sigma}_I^{CD} = - 4 \delta_A^{\ [C} \delta_B^{\ D]}$. The {\color{myblue}blue} term matches the {\color{myblue}blue} term in \eqref{eq:BPSsusy5} if $X_4 = -1$. To find the last two coefficients $X_1$ and $X_3$ we write out the remaining terms in \eqref{eq:BPSsusy5}, \eqref{eq:BPSRho} and \eqref{eq:BPSf}
\begin{align}
\begin{aligned}\label{eq:BPSsusy6}
&\quad X_1  \tr_c \left(\varepsilon_{\alpha}^B \Sigma_{I\, BC} \psi^{\beta C} \psi_\beta^A \right) + \left( X_3 - \frac{X_3}{3} \right) \tr_c\left( \varepsilon^{\beta A} \psi_{\beta}^B \Sigma_{I \, BC} \psi_{\alpha}^C \right) \\
&\quad - \frac{X_3}{3} \tr_c\left( \varepsilon^{\beta B} \Sigma_{I\, BC} \psi_{\beta}^C \psi_{\alpha}^A \right)
+ \frac{X_3}{3} \tr_c\left( \varepsilon^{\beta B} \Sigma_{I\, BC} \psi_{\beta}^A \psi_{\alpha}^C \right) \\
&\overset{!}{=} - \tr_c \left( \varepsilon^{\beta B} \Sigma_{I \, BC} \psi_{\beta}^C \psi_\alpha^A \right) 
 + \frac{\bar{\Sigma}_I^{AB} \Sigma_{BC}^K }{6} \, \tr_c \left( \varepsilon^{\beta D} \Sigma_{K \, DE} \psi_{\beta}^E \psi_{\alpha}^C \right) \, .
\end{aligned}
\end{align}
In the second term on the right-hand side, we first commute $\bar{\Sigma}_I^{AB}$ and $\Sigma_{BC}^K$ and subsequently use $\Sigma_{AB}^K \bar{\Sigma}_K^{CD} = - 4 \delta_A^{\ [C} \delta_B^{\ D]}$. Thus, \eqref{eq:BPSsusy6} becomes
\begin{align}
\begin{aligned}
&\quad X_1  \tr_c \left(\varepsilon_{\alpha}^B \Sigma_{I\, BC} \psi^{\beta C} \psi_\beta^A \right) \color{myred} + \left( X_3 - \frac{X_3}{3} \right) \tr_c\left( \varepsilon^{\beta A} \psi_{\beta}^B \Sigma_{I \, BC} \psi_{\alpha}^C \right) \\
&\quad - \frac{X_3}{3} \tr_c\left( \varepsilon^{\beta B} \Sigma_{I\, BC} \psi_{\beta}^C \psi_{\alpha}^A \right)
+ \frac{X_3}{3} \tr_c\left( \varepsilon^{\beta B} \Sigma_{I\, BC} \psi_{\beta}^A \psi_{\alpha}^C \right) \\
&\overset{!}{=} - \tr_c \left( \varepsilon^{\beta B} \Sigma_{I \, BC} \psi_{\beta}^C \psi_\alpha^A \right) 
 + \frac{1}{3} \, \tr_c \left( \varepsilon^{\beta B} \Sigma_{I \, BC} \psi_{\beta}^C \psi_{\alpha}^A \right) 
\color{myred} + \frac{1}{3} \, \tr_c \left( \varepsilon^{\beta A} \psi_{\beta}^B \Sigma_{I \, BC} \psi_{\alpha}^C \right) \\
&\quad - \frac{1}{3} \, \tr_c \left( \varepsilon^{\beta B} \Sigma_{I \, BC} \psi_{\beta}^A \psi_{\alpha}^C \right) \, .
\end{aligned}
\end{align}
The {\color{myred}red} terms match if $X_3 = \frac{1}{2}$. Similarly the remaining black term match if $X_1 = \frac{1}{2}$ since 
\begin{align}
 \tr_c \left( \varepsilon^{\beta B} \Sigma_{I\, BC}\psi_{\beta}^C \psi_\alpha^A \right) 
+ \tr_c \left( \varepsilon^{\beta B} \Sigma_{I\, BC} \psi_\beta^A \psi_{\alpha}^C \right) 
+ \tr_c \left(\varepsilon_{\alpha}^B \Sigma_{I\, BC} \psi^{\beta C} \psi_\beta^A \right) = 0 \, .
\end{align}
This can be seen by spelling out the left-hand side for all possible combinations of the indices $\alpha$ and $\beta$. Thus, we have obtained the following expressions for $\rho^{AB}$ and $f_{\alpha\beta}^I$
\begin{align}
\begin{aligned}
&\rho^{AB} \coloneqq \frac{1}{2} \tr_c \left( \psi^{\alpha A} \psi_\alpha^B \right) - \frac{ig}{12} \, \tr_c \left( (\bar{\Sigma}_I \Sigma_J \bar{\Sigma}_K)^{AB} \phi^I [ \phi^J , \phi^K ] \right) \, , \\
&f_{\alpha\beta}^I \coloneqq \frac{1}{2} \tr_c\left( \psi_{\alpha}^A \Sigma_{AB}^I \psi_{\beta}^B \right) - \tr_c \left( \phi^I (\sigma^{\mu\nu})_{\alpha}^{\ \gamma} \epsilon_{\gamma \beta} F_{\mu\nu} \right) \, .
\end{aligned}
\end{align}
Similarly, we can find $J_{\alpha\dot{\alpha}IJ}$ by collecting all terms proportional to $\bar{\varepsilon}_{\dot{\alpha}A}$ in \eqref{eq:BPSsusy3} and \eqref{eq:BPSsusy4}. 

Following this same procedure, we work through the entire multiplet to determine the field content at every level. The calculations become more tedious at every level and the field structure becomes more complicated. For conciseness, we will not provide any further details of these calculations here. Finally, the supersymmetry variations of the $R$-symmetry current $\Phi$ and the energy-momentum tensor $T_{\alpha\beta\dot{\alpha}\dot{\beta}}$ do not reveal any new fields but give only derivatives of the previously obtained fields. Thus, these variations act as tests for the entire calculation.

\chapter{Conclusion and Outlook}\label{ch:Outlook}
In this thesis, we have studied the Nicolai map in various supersymmetric field theories. Besides a formal introduction to the subject, we have used the Nicolai map to compute the vacuum expectation value of the infinite straight line Maldacena-Wilson loop in $\mathcal{N}=4$ super Yang-Mills to order $g^6$. Furthermore, we have calculated the entire field content of the $\mathcal{N}=4$ super Yang-Mills stress tensor multiplet. In the following, we review the new results obtained in this work and name open questions and possible directions for future research.

\section{Summary of Results}
We have learned that supersymmetric theories can be characterized by the existence of a non-linear and non-local transformation of the bosonic fields, called the Nicolai map, which maps the interacting functional measure to that of a free theory such that the Jacobian determinant of the map is equal to the product of the fermionic determinants. The Nicolai map is made possible by the supersymmetric Ward identities relating bosonic and fermionic correlation functions. 

Briefly summarized, the main results of this work are:
\begin{itemize}[noitemsep]
\item The calculation of the Nicolai map \eqref{eq:WZResult} for the 2-dimensional Wess-Zumino model up to the fifth order in the coupling.
\item The introduction of `on- and off-shell' Nicolai maps and $\mathcal{R}_g$-operators. 
\item The calculation of the Nicolai map \eqref{eq:YM1Result} for `off-shell' $\mathcal{N}=1$ super Yang-Mills in axial gauge up to the second order in the coupling.
\item The calculation of the Nicolai map \eqref{eq:FourthOrderResult} for `on-shell' $\mathcal{N}=1$ super Yang-Mills in $d=3$, 4, 6 and 10 dimensions and Landau gauge up to the fourth order in the coupling.
\item The calculation of the vacuum expectation value of the infinite straight line Maldacena-Wilson loop in $\mathcal{N}=4$ super Yang-Mills to order $g^6$.
\item The derivation of the entire field content of the $\mathcal{N}=4$ super Yang-Mills stress tensor multiplet.
\end{itemize}
We will now give some further details to these results.

The Nicolai map of the 2-dimensional Wess-Zumino model was first studied by Nicolai in \cite{Nicolai:1979nr,Nicolai:1980hh,Nicolai:1980jc,Nicolai:1980js,Nicolai:1984jg}. Nicolai derived the corresponding $\mathcal{R}_\lambda$-operator, \emph{i.e.} the infinitesimal generator of the inverse Nicolai map, and provided the first three orders of the perturbative expansion of the Nicolai map. However, at the time the map was not systemically computed. For lower-order results, this can work, but as we have seen already Nicolai's third-order result differs from the one we obtained by a systematic calculation. In chapter \ref{ch:WZ}, we have explained how to derive the $\mathcal{R}_\lambda$-operator and how to compute the Nicolai map. Our results show that the map can be obtained to arbitrary order in perturbation theory. More than anything, the fifth-order result \eqref{eq:WZResult} is a non-trivial proof of concept for the Nicolai map. Furthermore, we have given a pedagogical proof of the main theorem governing the properties of the Nicolai map.

The most important result of this work was given in chapter \ref{ch:YM1}. Namely, the introduction of the `off-shell' Nicolai map \cite{Malcha:2021ess}. For the first time, we were able to obtain a Nicolai map of a supersymmetric gauge theory in general gauges. Despite countless attempts (see \emph{e.g.} \cite{Lechtenfeld:1984me,Dietz:1985hga,deAlfaro:1984gw, deAlfaro:1984hb,deAlfaro:1986mti}) all previous Nicolai maps of gauge theories were constrained to the Landau gauge (see \emph{e.g.} \cite{Nicolai:1984jg}). The crucial observation was to consider the `off-shell' formulation of the theory and rescaled fields $\tilde{A}_\mu^a = g A_\mu^a$. When using rescaled fields, the dependence of the action on the coupling constant $g$ factors out and also the supersymmetry transformations of the fields no longer depend on the coupling. Hence, it becomes possible to write the `off-shell' action of a gauge theory as a supervariation without any gauge constraints. Subsequently, we obtained an $\tilde{\mathcal{R}}_g$-operator not restricted by a gauge choice. The second crucial observation was to act with this $\tilde{\mathcal{R}}_g$-operator on $(\frac{1}{g} \tilde{A}_\mu^a)$ rather than $\tilde{A}_\mu^a$. This led us to the correct expression for the inverse Nicolai map $(\mathcal{T}_g^{-1} A)_\mu^a = (\mathcal{T}_g (\frac{1}{g} \tilde{A}))_\mu^a$. This construction is universal to all supersymmetric gauge theories with `off-shell' supersymmetry. We have applied it to 4-dimensional $\mathcal{N}=1$ super Yang-Mills and computed the Nicolai map in axial gauge \eqref{eq:YM1Result} to the second order in the coupling. 

Due to the presence of the field strength tensor in the `off-shell' $\tilde{\mathcal{R}}_g$-operator, the favored gauge of the super Yang-Mills Nicolai maps is the Landau gauge. For a detailed explanation, see section \ref{sec:YM1OnShell}. Thus, for our next result in this thesis, we constrained ourselves to the Landau gauge and computed the Nicolai map \eqref{eq:FourthOrderResult} for `on-shell' $\mathcal{N}=1$ super Yang-Mills in $d=3$, 4, 6 and 10 dimensions up to the fourth order in the coupling \cite{Ananth:2020lup,Malcha:2021ess}. Thus, we have extended the previously existing result from \cite{Nicolai:1984jg} by two orders. We also found an ambiguity specifically in six dimensions \cite{Ananth:2020jdr} where up to the third order in the coupling, a simpler version of the Nicolai map exists. 

After extensively studying the Nicolai map in $\mathcal{N}=1$ super Yang-Mills, we finally turned to $\mathcal{N}=4$ super Yang-Mills in chapter \ref{ch:N4YM}. Since the $\mathcal{N}=4$ super Yang-Mills theory does not have an `off-shell' formulation with finitely many auxiliary fields and manifest $\mathcal{N}=4$ supersymmetry, the closest we can come to an `off-shell' Nicolai map for the $\mathcal{N}=4$ theory is the `off-shell' Nicolai map of maximally extended 4-dimensional $\mathcal{N}=1$ super Yang-Mills. In \eqref{eq:YM4OffShellRinv} - \eqref{eq:YM4OffShellRg} we have given the `off-shell' $\tilde{\mathcal{R}}_g$-operator of maximally extended 4-dimensional $\mathcal{N}=1$ super Yang-Mills and in \eqref{eq:YM4ROp4} - \eqref{eq:YM4ROp5} we have given the `on-shell' $\mathcal{R}_g^4$-operator of $\mathcal{N}=4$ super Yang-Mills. Since both expressions are much more complicated than any of the $\mathcal{R}_g$-operators derived before, we proved that the Nicolai map of $\mathcal{N}=4$ super Yang-Mills can be obtained from the Nicolai map of 10-dimensional $\mathcal{N}=1$ super Yang-Mills by the means of dimensional reduction. This gave us automatic access to the fourth-order Nicolai map in $\mathcal{N}=4$ super Yang-Mills. In \cite{Rupprecht:2021wdj} it was shown that, when one dimensionally reduces the $\mathcal{N}=1$ $\mathcal{R}_g$-operator to the $\mathcal{N}=4$ $\mathcal{R}_g^4$-operator, there is some freedom in the realization of the $SU(4)$ $R$-symmetry. Thus, there are several possible Nicolai maps for $\mathcal{N}=4$ super Yang-Mills. Our result corresponds to the case of full $SU(4)$ symmetry. Correlation functions do not depend on the choice of map.

In chapter \ref{ch:WilsonLoop}, we have demonstrated a possible perturbative application of the Nicolai map. We have computed the vacuum expectation value of the infinite straight line Maldacena-Wilson loop in $\mathcal{N}=4$ super Yang-Mills to order $g^6$. It is widely believed that the vacuum expectation value of the infinite straight line Maldacena-Wilson loop is exactly equal to 1 \cite{Zarembo:2002an}. However, this has actually never been proven. In this work, we have extended the previous perturbative calculation of \cite{Zarembo:2002an} by one order. This result has first been published in \cite{Malcha:2022fuc}. We found that the cancellations occurring at $\mathcal{O}(g^6)$ are by no means trivial. 

In this thesis, we have also commented on some general questions related to the Nicolai map. It is well known that the perturbative expansion of most quantum field theories has vanishing radius of convergence. The Nicolai map, however, is believed to have a non-zero radius of convergence in all theories. However, thus far, it has only been proven for supersymmetric quantum mechanics in \cite{Lechtenfeld:2022qed}. Similar proofs for other supersymmetric field theories remain a challenge for future research. Concerning the renormalization of Nicolai maps, we have shown that it fundamentally hinges on the renormalization properties of the fermion and ghost propagators in the gauge field background. In super Yang-Mills theories, this yields a non-linear renormalization of the Nicolai map.  

In the second part of this thesis, we have derived the entire field content of the $\mathcal{N}=4$ super Yang-Mills stress tensor multiplet. This is a particularly interesting short multiplet as it contains all the $\mathcal{N}=4$ currents and, in particular, the energy-momentum tensor. Moreover, it has been shown \cite{Baggio:2012rr} that the operators in the stress tensor multiplet do not acquire an anomalous dimension. The structure of the stress tensor multiplet and the 2-, 3- and 4-point correlation functions of its constituents have been extensively studied in \cite{Dolan:2001tt}. However, the explicit expressions for the fields in the multiplet were not provided. We have filled this gap in chapter \ref{ch:BPS}. With the explicit field content at hand we can in the future investigate its correlation functions with the Nicolai map. It will be interesting to see if and how the simplifications in the 2-, 3- and 4-point functions of this mulitplet are visible through the Nicolai map.

Thus, to summarize, we now have a comprehensive understanding of the Nicolai map in a wide range of supersymmetric field theories. In particular, we have addressed many open questions from earlier research in the 1980s. Hopefully, this framework will assist us in the future when we tackle some of the more fundamental questions in supersymmetric gauge theories. 

\section{Outlook}
There are several possible lines of future research on the Nicolai map and related topics. In the following, we highlight a selected few.

The Nicolai maps presented in the chapters \ref{ch:WZ}, \ref{ch:YM1} and \ref{ch:YM2} have been computed using Mathematica. However, not much thought has gone into optimizing the calculations. A more efficient approach could yield results far beyond the fourth- respectively fifth-order Nicolai maps presented in this thesis. It would be interesting to see if calculations such as that of the five-loop anomalous dimension of the Konishi operator \cite{Bajnok:2009vm} are also possible with the Nicolai map. 

The Nicolai map in axial gauge \eqref{eq:YM1TAxial} is significantly more complicated than its Landau gauge counterpart \eqref{eq:YM2Result}. At the moment, there seem to be three possible paths to a more tractable result. The first would be along the lines of section \ref{sec:YM2Ambiguity}, \emph{i.e.} to simply guess a simpler version of the second-order Nicolai map in axial gauge. In \cite{Ananth:2020jdr} an algorithm is outlined which could produce such a result. The advantage of this approach is that it would be relatively simple to implement. However, it is not very likely that we could produce higher-order results this way. 

A second, more promising approach is the introduction of a topological term in the action similar to \cite{Lechtenfeld:2022lvb}. In Landau gauge, special values of $\theta$ make the second order of the Nicolai map vanish altogether and drastically reduce the number of terms in higher orders. We have checked that the exact approach presented in \cite{Lechtenfeld:2022lvb} does not yield any simplifications in the axial gauge. Furthermore, it is only permissible in Euclidean signature and thus not possible for the light-cone gauge. Nevertheless, we still hope that the formalism can be appropriately modified.

A third possibility could be a modification of the gauge field along the lines of \cite{Bassetto:1985mm,Bassetto:1987uh}. Mandelstam's finiteness statements \cite{Mandelstam:1982cb,Leibbrandt:1984yb} seem to only concern the transversal degrees of freedom of the fields. Hence, there is room to modify the longitudinal degrees of freedom in accordance with the finiteness in the light-cone gauge. The hope is that such a modification in the $\mathcal{N}=1$ theory would cancel some of the terms in \eqref{eq:YM1Result}.

Another possibility for future research is to search for non-perturbative results in $\mathcal{N}=4$ super Yang-Mills with the Nicolai map. The most straightforward type of result would be to find operators which are in the kernel of the $\mathcal{R}_g$-operator. A first study of the kernel of the $\mathcal{R}_g$-operator in $\mathcal{N}=1$ super Yang-Mills was done by Lechtenfeld in his doctoral thesis \cite{Lechtenfeld:1984me}. He showed that the (anti) self-dual field strength tensor is in the kernel of the `off-shell' $\tilde{\mathcal{R}}_g$-operator. More generally, also operators for which the Nicolai map takes a closed form could potentially be very interesting. These could give rise to all-order non-trivial perturbative results for vacuum expectation values or correlation functions. A related goal is to give a non-perturbative proof for the vanishing of quantum corrections in the vacuum expectation value of the infinite-straight line Maldacena-Wilson loop. 

Furthermore, studying the Nicolai map of correlation functions rather than individual operators could also yield some interesting results. With the explicit field content of the stress tensor multiplet \eqref{eq:BPSFields} and the instructions for computing general correlation functions with the Nicolai map from section \ref{sec:BPSCorr}, it is now possible to study the multiplet from a new angle. The hope is to derive some non-perturbative statements for the 4-point functions or operator product expansions. 

And finally, there is string theory. So far, all the research on the Nicolai map has been done for ordinary quantum field theories. However, supersymmetry is also present in theories of gravity, such as string theory. It would be interesting to see if it is possible to derive a Nicolai map for the superstring similar to the ones presented here. Maybe even new connections in the context of the AdS/CFT correspondence are possible.

\appendix

\chapter{Spinors}\label{app:Spinors}
In this appendix, we list important properties of sigma and gamma matrices as well as spinors in various dimensions. In the first two sections, we introduce sigma matrices in four and six spacetime dimensions. In section \ref{sec:SGammaMatrices} we introduce gamma matrices in general dimension and provide some explicit representations in four and ten dimensions. In the last four sections, we introduce Weyl, Dirac, Majorana and Majorana-Weyl spinors. In particular, we give the Fierz identities for Weyl and Majorana spinors in two and four dimensions.

\section{Sigma Matrices in Four Dimensions}
Four-dimensional sigma matrices carry two types of spinor indices $\alpha = 1,2$ and $\dot{\alpha} = 1,2$. The three Pauli matrices are
\begin{align}
\sigma^1 \coloneqq \begin{pmatrix} 0 & 1 \\ 1 & 0 \end{pmatrix} \, , \quad 
\sigma^2 \coloneqq \begin{pmatrix} 0 & -i \\ i & 0 \end{pmatrix} \, , \quad
\sigma^3 \coloneqq \begin{pmatrix} 1 & 0 \\ 0 & -1 \end{pmatrix} 
\end{align}
and the two times four sigma matrices are defined as $\sigma^\mu \coloneqq (\mathbbm{1}_2, \sigma^i)$ and $\bar{\sigma}^\mu \coloneqq (\mathbbm{1}_2, - \sigma^i)$. Furthermore, we introduce the anti-symmetric tensors $\epsilon^{\alpha\beta}$ and $\epsilon^{\dot{\alpha}\dot{\beta}}$ with 
\begin{align}
\epsilon_{21} = \epsilon^{12} = 1 \, , \quad \epsilon_{12} = \epsilon^{21} = -1 \, , \quad \epsilon_{11} = \epsilon_{22} = 0 \, .
\end{align}
The barred and unbarred sigma matrices are related by
\begin{align}
\bar{\sigma}^{\mu \, \dot{\alpha}\alpha} = \epsilon^{\dot{\alpha}\dot{\beta}} \epsilon^{\alpha\beta} \sigma_{\beta\dot{\beta}}^\mu \, .
\end{align}
The Fierz identities for the $\sigma$-matrices are
\begin{align}
\sigma_{\alpha\dot{\alpha}}^\mu \bar{\sigma}_\mu^{\dot{\beta}\beta} = 2 \delta_\alpha^{\ \beta} \delta_{\ \dot{\alpha}}^{\dot{\beta}} \, , \quad \sigma_{\mu \, \alpha\dot{\alpha}} \sigma_{\beta\dot{\beta}}^\mu = 2 \epsilon_{\alpha\beta} \epsilon_{\dot{\alpha}\dot{\beta}} \, , \quad \bar{\sigma}_\mu^{\dot{\alpha}\alpha} \bar{\sigma}^{\mu \, \dot{\beta}\beta} = 2 \epsilon^{\alpha\beta} \epsilon^{\dot{\alpha}\dot{\beta}} \, .
\end{align} 
Usually, we will suppress the spinor indices. The metric tensor is $\eta^{\mu\nu}$ with mostly minus signature $(+,-,-,-)$. The sigma matrices satisfy the Clifford algebra
\begin{align}
\{ \sigma^\mu, \bar{\sigma}^\nu \} = 2 \eta^{\mu\nu} \, .
\end{align}
We define the anti-symmetric products
\begin{gather}
\begin{gathered}
(\sigma^{\mu\nu})_\alpha^{\ \beta} \coloneqq \frac{1}{4} (\sigma_{\alpha\dot{\alpha}}^\mu \bar{\sigma}^{\nu \dot{\alpha}\beta} - \sigma_{\alpha\dot{\alpha}}^\nu \bar{\sigma}^{\mu \dot{\alpha}\beta} ) \, , \\
(\bar{\sigma}^{\mu\nu})_{\ \dot{\alpha}}^{\dot{\beta}} \coloneqq \frac{1}{4} ( \bar{\sigma}^{\mu \dot{\alpha}\alpha} \sigma_{\alpha\dot{\beta}}^\nu - \bar{\sigma}^{\nu \dot{\alpha}\alpha} \sigma_{\alpha\dot{\beta}}^\mu ) \, .
\end{gathered}
\end{gather}
These relations and definitions imply
\begin{gather}
\begin{gathered}
\bar{\sigma}^\rho \sigma^{\mu\nu} = \bar{\sigma}^{\rho\mu\nu} + \eta^{\rho[\mu} \bar{\sigma}^{\nu]} \, ,  \\
\sigma^{\mu\nu} \sigma^\rho = \sigma^{\mu\nu\rho} + \sigma^{[\mu} \eta^{\nu]\rho} \, , \\
\frac{1}{2} \left( \sigma^\mu \bar{\sigma}^\nu \sigma^\rho - \sigma^\rho \bar{\sigma}^\nu \sigma^\mu \right) = i \epsilon^{\mu \nu \rho \lambda} \sigma_\lambda 
\end{gathered}
\end{gather}
with the totally anti-symmetric symbol $\epsilon^{0123} = 1$.

\section{Sigma Matrices in Six Dimensions}
In six dimensions, there is one type of spinor index $A= 1,2,3,4$ and two times six sigma matrices $\Sigma_{AB}^I$ and $\bar{\Sigma}_I^{AB}$. The barred and unbarred sigma matrices are related by 
\begin{align}
\Sigma_{AB}^I = \frac{1}{2} \epsilon_{ABCD} \bar{\Sigma}^{I \, CD} \, , \quad \bar{\Sigma}_I^{AB} = \frac{1}{2} \epsilon^{ABCD} \Sigma_{I\, CD} \, ,
\end{align}
with the totally anti-symmetric symbol $\epsilon^{1234} = 1$. The six-dimensional sigma matrices satisfy the Clifford algebra
\begin{align}
\{ \Sigma^I, \bar{\Sigma}^J \} = - 2 \delta^{IJ} 
\end{align}
and the Fierz identities
\begin{align}
\Sigma_{AB}^I \bar{\Sigma}_I^{CD} = - 4 \delta_A^{\ [C} \delta_B^{\ D]} \, , \quad \Sigma_{I \, AB} \Sigma_{CD}^I = - 2 \epsilon_{ABCD} \, , \quad \bar{\Sigma}_I^{AB} \bar{\Sigma}^{I \, CD} = - 2 \epsilon^{ABCD} \, .
\end{align}
Moreover, let
\begin{gather}
\begin{gathered}
(\Sigma^{IJ})_A^{\ B} \coloneqq \frac{1}{4} (\Sigma_{AC}^I \bar{\Sigma}^{J \, CB} - \Sigma_{AC}^J \bar{\Sigma}^{I \, CB} ) \, , \\
(\bar{\Sigma}^{IJ})_{\ A}^B \coloneqq \frac{1}{4} ( \bar{\Sigma}^{I \, AC} \Sigma_{CB}^J - \bar{\Sigma}^{J \, AC} \Sigma_{CB}^I ) \, .
\end{gathered}
\end{gather}
These relations and definitions imply
\begin{gather}
\begin{gathered}
\bar{\Sigma}^I \Sigma^{JK} = \bar{\Sigma}^{IJK} - \delta^{I[J} \bar{\Sigma}^{K]} \, , \\
\Sigma^{IJ} \Sigma^K = \Sigma^{IJK} -  \Sigma^{[I} \delta^{J]K} \, .
\end{gathered}
\end{gather}
Important trace relations are
\begin{gather}
\begin{gathered}
\tr(\Sigma^I \bar{\Sigma}^J) = - 4 \delta^{IJ} \, , \\
\tr(\Sigma^I \bar{\Sigma}^J \Sigma^K \bar{\Sigma}^L) = - 4 \left( \delta^{IJ} \delta^{KL} - \delta^{IK} \delta^{JL} + \delta^{IL} \delta^{JK} \right) \, .
\end{gathered}
\end{gather}
All these relations are independent of the sigma matrix representation. However, some applications require us to choose a representation. In that case, we define the two times six sigma matrices via the real anti-symmetric matrices  $\alpha_{AB}^i$ and $\beta_{AB}^i$ (with $i = 1,2,3$)
\begin{align}\label{eq:SAlphaBeta1}
\begin{aligned}
&\alpha_{ik}^j \coloneqq \varepsilon_{ijk} \, , &&\alpha_{i4}^j = - \alpha_{4i}^j \coloneqq - \delta_i^j \, , &&\alpha_{44}^i \coloneqq 0 \, , \\
&\beta_{ik}^j \coloneqq - \varepsilon_{ijk} \, , &&\beta_{i4}^j = - \beta_{4i}^j \coloneqq - \delta_i^j \, , &&\beta_{44}^i \coloneqq 0 \, ,
\end{aligned}
\end{align}
which satisfy
\begin{align}
\begin{aligned}\label{eq:SAlphaBeta2}
&\{ \alpha^i, \alpha^j \} = - 2 \delta^{ij} \, , \quad [\alpha^i , \alpha^j ] = 2 \varepsilon^{ijk} \alpha^k \, , \\
&\{ \beta^i, \beta^j \} = - 2 \delta^{ij} \, , \quad [\beta^i , \beta^j ] = - 2 \varepsilon^{ijk} \beta^k \, , \quad [\alpha^i, \beta^j] = 0 \, .
\end{aligned}
\end{align}
Subsequently,
\begin{align}
\Sigma_{AB}^i \coloneqq \alpha_{AB}^i \, , &&\bar{\Sigma}_i^{AB} \coloneqq \alpha_{AB}^i \, &&\Sigma_{AB}^{i+3} \coloneqq -i \beta_{AB}^i \, , &&\bar{\Sigma}_{i+3}^{AB} \coloneqq i \beta_{AB}^i  \, .
\end{align}

\section{Gamma Matrices}\label{sec:SGammaMatrices}
Gamma matrices exist in every dimension $d \ge 2$. They carry one type of spinor index $\alpha = 1, \ldots, r$, where $r$ is the dimension of the gamma matrix representation in question. For the majority of this thesis, it is not necessary to specify a gamma matrix representation. In contrast to the spinor indices for the sigma matrices, we will not distinguish between sub- or superscript spinor indices for gamma matrices and write them all as subscripts. Usually, spinor indices are suppressed altogether. Gamma matrices are defined through the Clifford algebra 
\begin{align}\label{eq:SClifford}
\{ \gamma^\mu, \gamma^\nu\} = 2 \eta^{\mu\nu} \, .
\end{align}
with the mostly minus Minkowski metric $\eta^{\mu\nu} = (+, -, \ldots, -)$. For the gamma matrices, we are mostly interested in the trace relations
\begin{gather}
\begin{gathered}
\tr(\gamma^\mu) = 0 \, , \\
\tr( \gamma^{\mu_1} \ldots \gamma^{\mu_n}) = \sum_{i=2}^n (-1)^i \ \eta^{\mu_1 \mu_i} \ \tr (\gamma^{\mu_2} \ldots \hat{\gamma}^{\mu_i} \ldots \gamma^{\mu_n})\, ,
\end{gathered}
\end{gather}
where the hat indicates that $\hat{\gamma}^{\mu_i}$ is excluded from the product. In particular, the trace over an odd number of gamma matrices is zero. The anti-symmetrized product of two gamma matrices is given by
\begin{align}
\gamma^{\mu\nu} \coloneqq \frac{1}{2} \left( \gamma^\mu \gamma^\nu - \gamma^\nu \gamma^\mu \right) \,.  
\end{align}
This implies
\begin{align}
\begin{aligned}
\gamma^\mu \gamma^{\nu \rho} = \gamma^{\mu\nu\rho} + \eta^{\mu[\nu} \gamma^{\rho]} \, , \\
\gamma^{\mu\nu} \gamma^\rho =  \gamma^{\mu\nu\rho} + \gamma^{[\mu} \eta^{\nu]\rho} \, .
\end{aligned}
\end{align}
Additionally, one may introduce the charge conjugation matrix $\mathcal{C}$. It is defined via
\begin{align}
\mathcal{C} \gamma^\mu \mathcal{C}^{-1} = - (\gamma^\mu)^T \, ,
\end{align}
where $( \,  \cdot \,  )^T$ denotes the matrix transpose. The charge conjugation matrix satisfies $\mathcal{C}^{-1} = \mathcal{C}^\dagger = - \mathcal{C}^T = \mathcal{C}$. There are two special dimensions demanding some more attention.

\subsection{Gamma Matrices in Four Dimensions}
In four dimensions we have the additional $\gamma^5$-matrix $\gamma^5 \coloneqq  i \gamma^0 \gamma^1 \gamma^2 \gamma^3$. It anti-commutes with all other gamma matrices, \emph{i.e.}
\begin{align}
\{ \gamma^5, \gamma^\mu \} = 0 
\end{align}
and it obeys $\gamma^5 \gamma^5 = 1$. Moreover,
\begin{gather}
\tr(\gamma^5) = \tr(\gamma^5 \gamma^\mu \gamma^\nu) = 0 
\end{gather}
and also the trace of an odd number of gamma matrices times $\gamma^5$ is zero. For the dimensional reduction of 10-dimensional $\mathcal{N}=1$ super Yang-Mills to 4-dimensional $\mathcal{N}=4$ super Yang-Mills, we are required to choose a gamma matrix representation. Let
\begin{align}\label{eq:S4DGamma}
\gamma^\mu = \begin{pmatrix} 0 & \sigma^\mu \\ \bar{\sigma}^\mu & 0 \end{pmatrix} \, , \quad \gamma^5 = \begin{pmatrix} - \mathbbm{1}_2 & 0 \\ 0 & \mathbbm{1}_2 \end{pmatrix} \, .
\end{align}
In this representation, the charge conjugation matrix is $\mathcal{C} = i \gamma_2 \gamma_0$. It is straightforward to check that these 4-dimensional gamma matrices satisfy the Clifford algebra \eqref{eq:SClifford}.

\subsection{Gamma Matrices in Ten Dimensions}
In ten dimensions, the spacetime indices are $M,N = 0, \ldots, 9$. The 10-dimensional gamma matrices are defined via the $\alpha$- and $\beta$-matrices \eqref{eq:SAlphaBeta1} and the 4-dimensional gamma matrices \eqref{eq:S4DGamma}
\begin{align}
\begin{aligned}\label{eq:S10DGamma}
&\Gamma^\mu \coloneqq \gamma^\mu \otimes \mathbbm{1}_8 \, , &&\mu = 0,1,2,3 \, , \\
&\Gamma^{3+i} \coloneqq \gamma_5 \otimes \begin{pmatrix} 0 & i \alpha^i \\ -i \alpha^i & 0 \end{pmatrix} \, , \quad \quad &&i = 1,2,3 \, , \\
&\Gamma^{6+i} \coloneqq \gamma_5 \otimes \begin{pmatrix} 0 & \beta^i \\ \beta^i & 0 \end{pmatrix} \, , &&i = 1,2,3 \, . 
\end{aligned}
\end{align}
Furthermore, we introduce the 11th gamma matrix $\Gamma_{11} \coloneqq \Gamma_0 \cdots \Gamma_9$. The 10-dimensional charge conjugation matrix is 
\begin{align}
\mathcal{C}_{10} = \mathcal{C}_4 \otimes \mathcal{C}_6 \quad \text{with} \quad \mathcal{C}_4 = i \gamma_2 \gamma_0 \quad \text{and} \quad \mathcal{C}_6 = \begin{pmatrix} 0 & \mathbbm{1}_4 \\ \mathbbm{1}_4 & 0 \end{pmatrix} \, .
\end{align}
Also, these 10-dimensional gamma matrices satisfy the Clifford algebra \eqref{eq:SClifford}.

\section{Weyl Spinors in Four Dimensions}\label{sec:SWeylSPinors}
Weyl spinors exist in every even spacetime dimension. Here we state their properties in four spacetime dimensions. They are anti-commuting objects $\psi_\alpha$ (and $\bar{\psi}_{\dot{\alpha}}$) with one spinor index $\alpha = 1,2$ (respectively $\dot{\alpha}=1,2$), that can be raised and lowered by
\begin{align}
\psi^\alpha = \epsilon^{\alpha\beta} \psi_\beta \, , \quad \psi_\alpha = \epsilon_{\alpha\beta} \psi^\beta 
\end{align}
and respectively for $\bar{\psi}_{\dot{\alpha}}$. Usually, we suppress Weyl spinor indices via $(\psi \chi) \equiv \psi^\alpha \chi_\alpha$. The most important relations of two-component Weyl spinors are
\begin{align}
\begin{aligned}
(\psi \chi) &= (\chi \psi) \, , \\
(\bar{\psi} \bar{\chi}) &= (\bar{\chi} \bar{\psi}) \, ,  \\
(\psi \chi)^\dagger &= (\bar{\psi} \bar{\chi}) \, ,  \\
(\psi \sigma^\mu \bar{\chi}) &= - (\bar{\chi} \bar{\sigma}^\mu \psi) \, , \\
(\psi \sigma^\mu \bar{\chi})^\dagger &= (\psi \sigma^\mu \bar{\chi}) \, ,  \\
(\psi \sigma^\mu \bar{\sigma}^\nu \chi)  &= (\chi \sigma^\nu \bar{\sigma}^\mu \psi) \, ,  \\
(\psi \sigma^\mu \bar{\sigma}^\nu \chi)^\dagger  &= (\bar{\chi} \bar{\sigma}^\nu \sigma^\mu \bar{\psi}) \, .
\end{aligned}
\end{align}
Here $(\, \cdot \,)^\dagger$ denotes the conjugate transpose.

\section{Dirac Spinors}\label{sec:SDiracSPinors}
Dirac spinors $\lambda_\alpha$ are labeled by a single spinor index $\alpha = 1, \ldots, r$, where $r$ is the dimension of the corresponding Clifford algebra representation. Dirac spinors are complex and they exist in any spacetime dimensions $d \ge 1$. In even spacetime dimensions, they have $2^{d/2}$ complex degrees of freedom. In odd spacetime dimensions, they have $2^{(d-1)/2}$ complex degrees of freedom. In either case, the real degrees of freedom are twice that. 

In any even number of spacetime dimensions, we can impose the Weyl condition
\begin{align}\label{eq:SWeylCond}
\lambda = \frac{1}{2} ( \mathbbm{1} - \gamma^{d+1}) \lambda
\end{align}
on a Dirac spinor. This reduces their degrees of freedom by a factor of 2. In any spacetime dimension $d \equiv 1,2,3,4 \mod 8$, we may impose the Majorana condition
\begin{align}\label{eq:SMajoranaCond}
\bar{\lambda} = (\lambda^T \mathcal{C}) \, , 
\end{align}
where $\mathcal{C}$ is the charge conjugation matrix. This also halves the degrees of freedom of the Dirac spinor. Finally, if the dimension is $d = 2 \mod 8$, we may impose both the Majorana and Weyl condition. These results are summarized in table \ref{tab:SSpinorTypes}. 
\begin{table}
\begin{tabular}{|l|ccccccccc|}
\hline
\textbf{Dimension} & 2 & 3 & 4 & 5 & 6 & 7 & 8 & 9 & 10 \TBspace \\
\hline
Weyl Spinors & $\times$ & &  $\times$ & &  $\times$ & &  $\times$ & &  $\times$ \Tspace \\
Majorana Spinors & $\times$ & $\times$ &  $\times$ & &  & & & $\times$ &  $\times$ \Tspace \\
Majorana-Weyl Spinors & $\times$ & & & & & & & & $\times$  \TBspace \\
\hline
\end{tabular}
\caption{Spacetime dimensions and their possible spinor representations.}
\label{tab:SSpinorTypes}
\end{table} 
In the following, we list further properties of Majorana spinors.

\section{Majorana Spinors}\label{sec:SMajoranaSPinors}
A Majorana spinor $\lambda_\alpha$ is a Dirac spinor that satisfies the Majorana condition \eqref{eq:SMajoranaCond}. Majorana spinors are real. They have $2^{d/2}$ (respectively $2^{(d-1)/2}$ in odd spacetime dimensions) real degrees of freedom. 

\subsection{Majorana Spinors in Two Dimensions}
In two spacetime dimensions, the irreducible Clifford algebra representations are 2-dimensional. The four matrices $1$, $\gamma^\mu$ and $i\gamma^{\mu\nu}$ form a basis of the $2 \times 2$ matrices. On this basis, we have the following identities for Majorana spinors
\begin{align}
(\bar{\lambda} M \chi) = \begin{cases}
+ (\bar{\chi} M \lambda) & M = 1 \, ,  \\
- (\bar{\chi} M \lambda) & M = \gamma^\mu \, , \ i \gamma^{\mu\nu} \, . \end{cases} 
\end{align}
Furthermore, we can obtain the 2-dimensional Fierz identity
\begin{align}
\bar{\lambda}_{\alpha} \chi_{\beta} = \frac{1}{2} \sum_A \, \mathcal{O}_{\beta\alpha}^A \,  (\bar{\lambda} \mathcal{O}_A \chi) \quad \text{with} \quad \mathcal{O}^A = \left(1 \, , \ \gamma^\mu \, , \  i \gamma^{\mu\nu} \right) 
\end{align}
from these relations. 
\subsection{Majorana Spinors in Four Dimensions}
In four spacetime dimensions, the irreducible Clifford algebra representations are 4-dimensional. A basis of the $4 \times 4$ matrices is given by the 16 matrices $1$, $\gamma_5$, $\gamma_\mu$, $i\gamma_\mu \gamma_5$ and $\frac{i}{\sqrt{2}}\gamma^{\mu\nu}$. On this basis, we have the following identities for Majorana spinors
\begin{align}
(\bar{\lambda} M \chi) = \begin{cases}
+ (\bar{\chi} M \lambda) & M = 1 \, , \ \gamma^5 \, , \ i \gamma^\mu \gamma^5 \, , \\
- (\bar{\chi} M \lambda) & M = \gamma^\mu \, , \ \frac{i}{\sqrt{2}} \gamma^{\mu\nu} \, . \end{cases} 
\end{align}
The Fierz identity in four dimensions is
\begin{align}
\bar{\lambda}_{\alpha} \chi_{\beta} = \frac{1}{4} \sum_A \mathcal{O}_{\beta\alpha}^A (\bar{\lambda} \mathcal{O}_A \chi) \quad \text{with} \quad \mathcal{O}^A = \left(1 \, , \ \gamma^5 \, , \ \gamma^\mu \, , \ i \gamma^\mu \gamma^5 \, , \ \frac{i}{\sqrt{2}} \gamma^{\mu\nu} \right) \, .
\end{align}
Given the 4-dimensional gamma matrix representation \eqref{eq:S4DGamma} the Majorana spinors are written in the Weyl basis
\begin{align}
\lambda_\alpha = \begin{pmatrix} \psi_\alpha \\ \bar{\psi}^{\dot{\alpha}} \end{pmatrix} \, ,
\end{align}
where $\psi_\alpha$ is a Weyl spinor.

\section{Majorana-Weyl Spinors in Ten Dimensions}\label{sec:SMajoranaWeylSPinors}
Dirac spinors in 10 spacetime dimensions have $2^5 = 32$ complex components. Given the 10-dimensional gamma matrix representation \eqref{eq:S10DGamma} and imposing the Majorana condition \eqref{eq:SMajoranaCond}
\begin{align}
\bar{\Lambda} = (\Lambda^T \mathcal{C}_{10}) \, , 
\end{align}
as well as the Weyl condition \eqref{eq:SWeylCond}
\begin{align}
\Lambda = \frac{1}{2} \left( \mathbbm{1}_{32} - \Gamma_{11} \right) \Lambda
\end{align}
the Majorana-Weyl spinors take the form
\begin{align}
\Lambda = \begin{pmatrix} \chi \\ \bar{\chi} \end{pmatrix} \quad \text{with} \quad
\chi = \begin{pmatrix} \mathbf{0} \\ \psi_1 \\ \mathbf{0} \\ \psi_2 \end{pmatrix}  \quad \text{and} \quad 
\psi_i = \begin{pmatrix} \omega_{1i} \\ \omega_{2i} \\ \omega_{3i} \\ \omega_{4i} \end{pmatrix} \, \quad 
\mathbf{0} = \begin{pmatrix} 0 \\ 0 \\ 0 \\ 0 \end{pmatrix} \, .
\end{align}

\chapter{Functional Determinants}\label{app:FD}
In this appendix, we give an introduction to the calculation of bosonic and fermionic functional determinants. 
\section{Bosonic Functional Determinants}
Our starting point for bosonic functional determinants is the Gaussian integral
\begin{align}
\int_{-\infty}^\infty \mathrm{d}x \ e^{-a(x+b)^2} = \sqrt{\frac{\pi}{a}} \, .
\end{align}
This can be generalized by considering the expression
\begin{align}
\left( \prod_{k=1}^n \int \mathrm{d}\xi_k \right) \exp\left[ - \sum_{i,j=1}^n \xi_i A_{ij} \xi_j \right] \, , 
\end{align}
where $A$ is a symmetric $n \times n$ matrix with non-zero eigenvalues $a_i$. Then there exists an orthogonal matrix $O$, \emph{i.e.} $O^{-1} = O^T$ and $\det(O) = 1$, such that $O^{-1} A O$ is diagonal. Thus, we can perform a change of variables to coordinates $x_i$ with $\xi_i = \sum_{j=1}^n O_{ij} x_j$ such that
\begin{align}
\begin{aligned}
\left( \prod_{k=1}^n  \int_{-\infty}^\infty \mathrm{d}\xi_k \right) \exp\left[ - \sum_{i,j=1}^n \xi_i A_{ij} \xi_j \right]  &= \left(  \prod_{k=1}^n\int_{-\infty}^\infty \mathrm{d}x_k \right) \exp[ - \sum_{i=1}^n a_i x_i^2]  \\
&= \prod_{i=1}^n \left( \int_{-\infty}^\infty \mathrm{d}x_i \ \exp[ -  a_i x_i^2] \right) \\
&=  \prod_{i=1}^n \sqrt{\frac{\pi}{a_i}} = \pi^{n/2} \ [\det(A)]^{-1/2} \, .
\end{aligned}
\end{align}
With this calculation in mind, it is easy to evaluate the bosonic path integral for a gauge theory
\begin{align}
\int \mathcal{D}A \ e^{- \frac{i}{2} \int \mathrm{d}^dx \  A_\mu^a (\mathcal{M}^{\mu\nu} \, A_\nu)^a} = [ \det(\mathcal{M})]^{-1/2} \, ,
\end{align}
where we have appropriately normalized the measure $\mathcal{D}A$. $\det(\mathcal{M})$ is called the functional determinant. The definitions for scalar theories and Euclidean space follow accordingly.

\section{Fermionic Functional Determinants}
For anti-commuting variables, the functional integration behaves a little differently. It requires the technique of the Berezin integral \cite{Berezin:1966nc}. We start with the 1-dimensional case. Let $\theta$ be a Grassmann variable, \emph{i.e.} $\theta^2 = 0$. Then the Berezin integral over $\theta$ is
\begin{align}
\int \mathrm{d}\theta \ \theta = 1 \, , \quad \int \mathrm{d}\theta = 0 \, .
\end{align}
Any function $f(\theta)$ can be expanded in a Taylor series, which terminates after two terms since $\theta^2 = 0$. Thus, we have $f(\theta) = a + b \theta$ and 
\begin{align}
\int \mathrm{d}\theta \ f(\theta) = \int \mathrm{d}\theta \ (a + b \theta) = b \, .
\end{align}
The generalization to multidimensional Grassman numbers $\theta_1 \ldots \theta_n$ with $\theta_i \theta_j = - \theta_j \theta_i$ and thus $\theta_i \theta_i = 0$ (no summation over $i$) is immediate. For example, we have
\begin{align}
\int \mathrm{d}\theta_1 \int \mathrm{d}\theta_2 \ \theta_1 \theta_2 =  - \int \mathrm{d}\theta_1 \int \mathrm{d}\theta_2 \ \theta_2 \theta_1 =   - \int \mathrm{d}\theta_1 \  \theta_1 = -  1 \, .
\end{align}
Let $B$ be a anti-symmetric $n \times n$ matrix with matrix elements $B_{ij}$ and consider the integral
\begin{align}\label{eq:FDIn}
I_n(B) = \int \mathrm{d}\theta_1 \ldots \mathrm{d}\theta_n \ e^{- \sum_{i,j=1}^n \theta_i B_{ij} \theta_j} \, . 
\end{align}
It is not hard to see that $I_n(B) = 0$ for odd $n$ since all terms in the Taylor series expansion of the exponential come with an even number of $\theta$s. So there are either more than $n$ $\theta_i$s which gives zero since $\theta_i \theta_i = 0$ (no summation over $i$), or there are less than $n$ $\theta_i$s in which case the integral gives zero. For $n$ even there exist a unitary matrix $U$ such that
\begin{align}
C = U^T B U = \begin{pmatrix}
0 & \lambda_1 & \cdots & 0 & 0 \\
- \lambda_1 & 0 & \cdots & 0 & 0 \\
\vdots & \vdots & \ddots & \vdots & \vdots \\
0 & 0 & \cdots & 0 & \lambda_n \\
0 & 0 & \cdots & -\lambda_n & 0 \\ 
\end{pmatrix} \, .
\end{align}
Then we introduce new Grassmann variables $\tau_i$ in \eqref{eq:FDIn} via $\theta_i = \sum_{j=1}^n U_{ij} \tau_j$
\begin{align}
\begin{aligned}
I_n(B) &= \int \mathrm{d}\theta_1 \ldots \mathrm{d}\theta_n \ e^{- \sum_{i,j=1}^n \theta_i B_{ij} \theta_j} \\
&= \frac{1}{\det(U)}  \int \mathrm{d}\tau_1 \ldots \mathrm{d}\tau_n \ e^{- \sum_{i,j=1}^n \tau_i C_{ij} \tau_j} \\
&= \frac{1}{\det(U)} \int \mathrm{d}\tau_1 \ldots \mathrm{d}\tau_n \ e^{- 2 ( \lambda_1 \tau_1 \tau_2 + \ldots + \lambda_n \tau_{n-1} \tau_n)} \\
&= \frac{(-1)^{n/2}}{\det(U)} \int \mathrm{d}\tau_1 \ldots \mathrm{d}\tau_n \ \tau_1 \ldots \tau_n \ \prod_{i=1}^n \lambda_i \\
&= \frac{1}{\det(U)} \prod_{i=1}^n \lambda_i  = \frac{\sqrt{\det(C)}}{\det(U)}  = \sqrt{\det(B)} \, .
\end{aligned}
\end{align}
One can use this Berezin integral as the definition of the Pfaffian of a complex $n \times n$ matrix $B$
\begin{align}
\Pf(B) \coloneqq \begin{cases}
0 & n \text{ odd} \, ,  \\
\sqrt{\det(B)} & n \text{ even} \, .
\end{cases}
\end{align}
Thus, we have $I_n(B) = \Pf(B)$. These relations give us the functional determinant of a path integral over Majorana spinors
\begin{align}
\int \mathcal{D}\lambda \ e^{- \frac{1}{2} \int \mathrm{d}^dx \    \bar{\lambda}^a (\mathcal{M} \, \lambda)^a} =  \sqrt{\det(\mathcal{M})} \, .
\end{align}
Again the normalization is hidden in the path integral measure. 

Finally, we want to discuss the complex Berezin integral. To this end, we introduce the complex Grassmann variables
\begin{align}
\eta = \frac{1}{\sqrt{2}} (\theta_1 + i \theta_2) \, , \quad \bar{\eta} = \frac{1}{\sqrt{2}} (\theta_1 - i \theta_2) \, ,
\end{align}
such that
\begin{align}
\mathrm{d}\theta_1 \ \mathrm{d}\theta_2 =  \mathrm{d}\bar{\eta} \ \mathrm{d}\eta \, . 
\end{align}
This definition easily extends to the $n$-dimensional case. Then let $D$ be any $n \times n$ matrix with entries $D_{ij}$ and consider the integral
\begin{align}
J_n(D) = \int \mathrm{d}\bar{\eta}_1 \mathrm{d}\eta_1 \ldots  \mathrm{d}\bar{\eta}_n \mathrm{d}\eta_n \ e^{- \sum_{i,j=1}^n \bar{\eta}_i D_{ij} \eta_j} \, .
\end{align}
In the Taylor expansion of the exponential function, only the terms proportional to $\bar{\eta}_1 \eta_1 \ldots \bar{\eta}_n \eta_n$ will survive. We find that
\begin{align}
\begin{aligned}
J_n(D) &=  (-1)^n\int \mathrm{d}\bar{\eta}_1 \mathrm{d}\eta_1 \ldots \ \mathrm{d}\bar{\eta}_n \mathrm{d}\eta_n \  \bar{\eta}_1 \eta_1 \ldots \bar{\eta}_n \eta_n \, \sum_{\sigma \in S_n} \sgn(\sigma) \prod_{i=1}^n D_{i\sigma_i} \\
&=  (-1)^n \int \mathrm{d}\bar{\eta}_1 \mathrm{d}\eta_1 \ldots \ \mathrm{d}\bar{\eta}_n \mathrm{d}\eta_n \  \bar{\eta}_1 \eta_1 \ldots \bar{\eta}_n \eta_n \,  \det(D) \\
&= \det(D) \, .
\end{aligned}
\end{align}
Thus, for the path integral over anti-commuting ghost fields $C$ and $\bar{C}$ we obtain
\begin{align}
\int \mathcal{D}\bar{C} \ \mathcal{D}C \ e^{- \frac{i}{2} \int \mathrm{d}^dx \  \bar{C}^a (\mathcal{M} \, C)^a} = \det(\mathcal{M}) \, .
\end{align}
Here, too, the path integral measure has been appropriately normalized.

\chapter{The Fourth-Order Nicolai Map}\label{app:FourthOrder}
We give the `on-shell' $\mathcal{N}=1$ super Yang-Mills Nicolai map $(\mathcal{T}_g \, A)_\mu^a$ in Landau gauge up to and including $\mathcal{O}(g^4)$. The result is valid in all critical dimensions $d=3$, 4, 6, 10. It was first obtained in \cite{Malcha:2021ess} and extends \eqref{eq:YM2Result} by one order. We have checked that it satisfies all three tests from section \ref{sec:YM2Tests}. Moreover, upon dimensional reduction, it agrees with the 2-dimensional result in \cite{Lechtenfeld:2021zgd}. The inverse of the map given here has been used in \cite{Malcha:2022fuc} to compute the vacuum expectation value of the infinite straight line Maldacena-Wilson loop to the sixth order. 

In the expression below, anti-symmetrizations over five spacetime indices occur. These terms vanish if the result is taken to be in $d=3$ or $d=4$ dimensions. In that sense, the Nicolai map depends on the number of dimensions from the fourth order onwards. The fourth-order `on-shell' $\mathcal{N}=1$ super Yang-Mills Nicolai map in Landau gauge reads \cite{Malcha:2021ess}

\begin{footnotesize}
\begin{align}
\begin{aligned}\label{eq:FourthOrderResult}
(\mathcal{T}_g \, A)_\mu^a(x) &= A_\mu^a(x) + g f^{abc} \int \mathrm{d}^d y \ \partial^\rho C(x-y) A_\mu^b(y) A_\rho^c(y) \\
&\quad +  \frac{3g^2}{2} f^{abc} f^{bde} \int \mathrm{d}^d y \ \mathrm{d}^d z \ \partial^\rho C(x-y) A^{c \, \lambda}(y) \partial_{[\rho} C(y-z) A_\mu^d(z) A_{\lambda]}^e(z) \\
&\quad + \frac{g^3}{2} f^{a b c} f^{b d e} f^{c f g} \int \mathrm{d}^d y \ \mathrm{d}^d z \ \mathrm{d}^d w \ \partial^\rho C(x-y) \\
&\quad \quad \quad \times \partial^\lambda C(y-z) A_\lambda^d(z) A^{e \,  \sigma}(z) \partial_{[\rho} C(y-w) A_\mu^f(w) A_{\sigma]}^g(w) \\
&\quad +  g^3 f^{a b c} f^{b d e} f^{d f g} \int \mathrm{d}^d y \ \mathrm{d}^d z \ \mathrm{d}^d w \ \partial^\rho C(x-y) A^{c \, \lambda}(y) \ \Big\{ \\
&\quad \quad \quad - \partial^{\sigma} C(y-z)A_{\sigma}^e(z) \partial_{[\rho} C(z-w) A_\mu^f(w) A_{\lambda]}^g(w) \\
&\quad \quad \quad + \partial_{[\rho} C(y-z) A_\mu^e(z) \partial^\sigma C(z-w) A_{\lambda ]}^f(w) A_\sigma^g(w) \Big\} \\
&\quad +  \frac{g^3}{3} f^{abc} f^{bde} f^{dfg} \int \mathrm{d}^d y \ \mathrm{d}^d z \ \mathrm{d}^d w \ \Big\{ \\
&\quad \quad \quad + 6 \, \partial_\rho C(x-y) A^{c \, \lambda}(y) \partial^{[\rho} C(y-z) A^{e \, \sigma]}(z) \partial_{[\lambda} C(z-w) A_\mu^f(w) A_{\sigma]}^g(w) \\
&\quad \quad \quad - 6 \, \partial^\rho C(x-y) A_\lambda^c(y) \partial^{[\lambda} C(y-z) A^{\sigma] \, e}(z) \partial_{[\rho} C(z-w) A_\mu^f(w) A_{\sigma]}^g(w) \\
&\quad \quad \quad - 6 \, \partial_\rho C(x-y) A_\lambda^c(y) \partial_{[\sigma} C(y-z) A_{\mu]}^e(z) \partial^{[\rho} C(z-w) A^{f \, \lambda}(w) A^{g \, \sigma]}(w) \\
&\quad \quad \quad + 2 \, \partial^\rho C(x-y) A_{[\rho}^c(y) \partial_{\mu]} C(y-z) A^{e \,  \lambda}(z) \partial^\sigma C(z-w) A_\lambda^f(w) A_\sigma^g(w) \\
&\quad \quad \quad - \partial_\mu C(x-y) \, \partial^\rho \left( A_\rho^c(y) C(y-z) \right) A^{e \,  \lambda}(z) \partial^\sigma C(z-w) A_\lambda^f(w) A_\sigma^g(w) \Big\} \\
&\quad -  \frac{g^3}{3} f^{abc} f^{bde} f^{dfg} \int \mathrm{d}^d y \ \mathrm{d}^d z  \ A_\mu^c(x) C(x-y) A^{e \, \rho}(y) \partial^\lambda C(y-z) A_\rho^f(z) A_\lambda^g(z) 
\end{aligned}
\end{align}
\end{footnotesize}%
\newpage
\begin{footnotesize}
\begin{align*}
\hphantom{(\mathcal{T}_g \, A)_\mu^a(x)} 
&\quad + \frac{g^4}{12} f^{a b c} f^{b d e} f^{d f g} f^{c h i} \int \mathrm{d}^d y \ \mathrm{d}^d z \ \mathrm{d}^d w \\
&\quad \quad \quad \times C(x-y) A^{e \,  \lambda}(y) \partial^\rho C(y-z) A_\lambda^f(z) A_\rho^g(z) \partial^\sigma C(x-w) A_\sigma^h(w) A_\mu^i(w) \\
&\quad + \frac{g^4}{8} f^{a b c} f^{b d e} f^{d f g} f^{c h i} \int \mathrm{d}^d y \ \mathrm{d}^d z \ \mathrm{d}^d w \ \mathrm{d}^d v \ \partial^\lambda C(x-y) \partial_\rho C(y-z) \ \Big\{ \\
&\quad \quad \quad - 9 A_\sigma^e(z) \partial^{[\rho} C(z-w) A^{f \,  \sigma}(w) A^{g \, \nu]}(w) \partial_{[\mu} C(y-v) A_\lambda^h(v) A_{\nu]}^i(v) \\ 
&\quad \quad \quad + 4 A^{e\, [\rho}(z) \partial^{|\sigma|} C(z-w) A_\sigma^f(w) A^{g \ ,\nu]}(w) \partial_{[\mu} C(y-v) A_\lambda^h(v) A_{\nu]}^i(v) \\
&\quad \quad \quad - 2 A^{e \, \rho}(z) \partial_{[\mu} C(z-w) A_\lambda^f(w) A_{\nu]}^g(w) \partial^\sigma C(y-v) A_\sigma^h(v) A^{i \, \nu}(v) \Big\} \\
&\quad - \frac{g^4}{12} f^{a b c} f^{b d e} f^{d f g} f^{c h i} \int \mathrm{d}^d y \ \mathrm{d}^d z \ \mathrm{d}^d w \ \mathrm{d}^d v \ \partial_\mu C(x-y) \partial^\lambda C(y-z) \\
&\quad \quad \quad \times A^{e \,  \rho}(z) \partial^\sigma C(z-w) A_\sigma^f(w) A_\rho^g(w) \partial^\tau C(y-v) A_\tau^h(v) A_\lambda^i(v) \\ 
&\quad + \frac{g^4}{2} f^{a b c} f^{b d e} f^{d f g} f^{c h i} \int \mathrm{d}^d y \ \mathrm{d}^d z \ \mathrm{d}^d w \ \mathrm{d}^d v \ \partial_\lambda C(x-y) \ \Big\{ \\
&\quad \quad \quad + \partial_{[\mu} C(y-z) A_{\rho]}^e(z) \partial^{[\lambda} C(z-w) A^{f \, \rho}(w) A^{g \, \nu]}(w) \partial^\sigma C(y-v) A_\sigma^h(v) A_\nu^i(v) \\ 
&\quad \quad \quad - \partial^{[\lambda} C(y-z) A^{e \, \rho}(z) \partial_{[\mu} C(z-w) A_\rho^f(w) A_{\nu]}^g(w) \partial^\sigma C(y-v) A_\sigma^h(v) A_\nu^i(v) \Big\} \\
&\quad + \frac{g^4}{6} f^{a b c} f^{b d e} f^{d f g} f^{c h i} \int \mathrm{d}^d y \ \mathrm{d}^d z \ \mathrm{d}^d w \ \mathrm{d}^d v \ \partial^\lambda C(x-y) \ \Big\{ \\
&\quad \quad \quad + 3 \partial^{[\rho} C(y-z) A^{e \, \nu]}(z) \partial_{[\mu} C(z-w) A_\lambda^f(w) A_{\rho]}^g(w) \partial^\sigma C(y-v) A_\sigma^h(v) A_\nu^i(v) \\ 
&\quad \quad \quad + \partial_{[\lambda} C(y-z) A^{e \, \nu}(z) \partial^\sigma C(z-w) A_{|\sigma}^f(w) A_\nu^g(w) \partial^\rho C(y-v) A_{\rho|}^h(v) A_{\mu]}^i(v) \Big\}  \\
&\quad - \frac{g^4}{3} f^{a b c} f^{b d e} f^{d f g} f^{e h i} \int \mathrm{d}^d x \ \mathrm{d}^d y \ \mathrm{d}^d z \ \mathrm{d}^d w \\
&\quad \quad \quad \times A_\mu^c(x) C(x-y) \partial^\lambda C(y-z) A_\lambda^f(z) A^{g \, \rho}(z) \partial^\sigma C(y-w) A_\sigma^h(w) A_\rho^i(w) \\
&\quad - \frac{g^4}{3} f^{a b c} f^{b d e} f^{d f g} f^{e h i} \int \mathrm{d}^d y \ \mathrm{d}^d z \ \mathrm{d}^d w \ \mathrm{d}^d v \ \partial_\mu C(x-y) \\
&\quad \quad \quad \times \partial^\lambda \left( A_\lambda^c(y) C(y-z) \right) \partial^\rho C(z-w) A_\rho^f(w) A^{g \, \nu}(w) \partial^\sigma C(z-v) A_\sigma^h(v) A_\nu^i(v) \\
&\quad + \frac{g^4}{12} f^{a b c} f^{b d e} f^{d f g} f^{e h i} \int \mathrm{d}^d y \ \mathrm{d}^d z \ \mathrm{d}^d w \ \mathrm{d}^d v \ \partial^\lambda C(x-y) A^{\rho \, c}(y) \ \Big\{ \\
&\quad \quad \quad - 3 \, \partial_\rho C(y-z) \partial_{[\mu} C(z-w) A_\lambda^f(w) A_{\nu]}^g(w) \partial^\sigma C(z-v) A_\sigma^h(v) A_\nu^i(v) \\ 
&\quad \quad \quad - 3 \, \partial^\nu C(y-z) \partial_{[\mu} C(z-w) A_\rho^f(w) A_{\nu]}^g(w) \partial^\sigma C(z-v) A_\sigma^h(v) A_\lambda^i(v) \\ 
&\quad \quad \quad + 3 \, \partial^\nu C(y-z) \partial_{[\lambda} C(z-w) A_\rho^f(w) A_{\nu]}^g(w) \partial^{\sigma} C(z-v) A_{\sigma}^h(v) A_\mu^i(v) \\ 
&\quad \quad \quad - 3  \, \partial_{\mu} C(y-z) \partial_{[\lambda} C(z-w) A_\rho^f(w) A_{\nu]}^g(w) \partial^{\sigma} C(z-v) A_{\sigma}^h(v) A_\nu^i(v) \\ 
&\quad \quad \quad + 3 \, \partial_{\lambda} C(y-z) \partial_{[\mu} C(z-w) A_\rho^f(w) A_{\nu]}^g(w) \partial^{\sigma} C(z-v) A_{\sigma}^h(v) A^{i \, \nu}(v) \\ 
&\quad \quad \quad - 2 \, \partial_{[\lambda} C(y-z) \partial^\nu C(z-w) A_{|\nu|}^f(w) A_{\mu]}^g(w) \partial^{\sigma} C(z-v) A_\sigma^h(v) A_\rho^i(v) \\ 
&\quad \quad \quad + \partial_{\rho} C(y-z) \partial^\nu C(z-w) A_{\nu}^f(w) A_{\mu}^g(w) \partial^\sigma C(z-v) A_\sigma^h(v) A_\lambda^i(v) \Big\} \\ 
&\quad + \frac{g^4}{6} f^{a b c} f^{b d e} f^{d f g} f^{e h i} \int \mathrm{d}^d y \ \mathrm{d}^d z \ \mathrm{d}^d w \ \mathrm{d}^d v \ \partial^\lambda C(x-y) \ \Big\{ \\
&\quad \quad \quad - 7 A_{[\mu}^c(y) \partial_{\lambda]} C(y-z) \partial^\rho C(z-w) A_\rho^f(w) A^{g \, \nu}(w) \partial^\sigma C(z-v) A_\sigma^h(v) A_\nu^i(v) \\ 
&\quad \quad \quad + 3 A^{c \, [\nu}(y) \partial^{\rho]} C(y-z) \partial_{[\mu} C(z-w) A_\lambda^f(w) A_{\rho]}^g(w) \partial^\sigma C(z-v) A_\sigma^h(v) A_\nu^i(v) \Big\} 
\end{align*}
\newpage
\begin{align*}
\hphantom{(\mathcal{T}_g \, A)_\mu^a(x)} 
&\quad - \frac{g^4}{2} f^{a b c} f^{b d e} f^{d f g} f^{f h i} \int \mathrm{d}^d y \ \mathrm{d}^d z \ \mathrm{d}^d w \ \\
&\quad \quad \quad \times \partial^\lambda C(x-y) A_{[\mu}^c(y) A_{\lambda]}^e(y) C(y-z) A^{g \, \rho}(z) \partial^\sigma C(z-w) A_\sigma^h(w) A_\rho^i(w) \\
&\quad + \frac{g^4}{12} f^{a b c} f^{b d e} f^{d f g} f^{f h i} \int \mathrm{d}^d x \ \mathrm{d}^d y \ \mathrm{d}^d z \ \mathrm{d}^d w \ A_\mu^c(x) C(x-y) \ \Big\{ \\
&\quad \quad \quad + 9 A^{e \,  \lambda}(y) \partial^\rho C(y-z) A^{g \, \sigma}(z) \partial_{[\lambda} C(z-w) A_\rho^h(w) A_{\sigma]}^i(w) \\ 
&\quad \quad \quad + 4 A^{e \, [\lambda}(y) \partial^{\rho]} C(y-z) A_\lambda^g(z) \partial^\sigma C(z-w) A_\sigma^h(w) A_\rho^i(w) \\
&\quad \quad \quad - 3 A_\lambda^e(y) \partial^\lambda C(y-z) A^{\rho \, g}(z) \partial^\sigma C(z-w) A_\sigma^h(w) A_\rho^i(w) \\ 
&\quad \quad \quad - 3 \, \partial^\lambda \left( A_\lambda^e(y) C(y-z) \right) A^{g \, \rho}(z) \partial^\sigma C(z-w) A_\sigma^h(w) A_\rho^i(w) \Big\} \\ 
&\quad + \frac{g^4}{12} f^{a b c} f^{b d e} f^{d f g} f^{f h i} \int \mathrm{d}^d y \ \mathrm{d}^d z \ \mathrm{d}^d w \ \mathrm{d}^d v \ \partial_\mu C(x-y) \partial^\lambda \left( A_\lambda^c(y) C(y-z) \right) \ \Big\{ \\
&\quad \quad \quad + 9 A^{e \,  \rho}(z) \partial^\sigma C(z-w) A^{g \, \nu}(w) \partial_{[\rho} C(w-v) A_\sigma^h(v) A_{\nu]}^i(v) \\ 
&\quad \quad \quad + 4 A^{e \, [\rho}(z) \partial^{\nu]} C(z-w) A_\rho^g(w) \partial^\sigma C(w-v) A_\sigma^h(v) A_\nu^i(v) \\
&\quad \quad \quad - 3 \,  \partial^\rho \left( A_\rho^e(z) C(z-w) \right) A^{g \, \nu}(w) \partial^\sigma C(w-v) A_\sigma^h(v) A_\nu^i(v) \\ 
&\quad \quad \quad - 3 A_\rho^e(z) \partial^\rho C(z-w) A^{g \, \nu}(w) \partial^\sigma C(w-v) A_\sigma^h(v) A_\nu^i(v) \Big\} \\ 
&\quad + \frac{g^4}{2} f^{a b c} f^{b d e} f^{d f g} f^{f h i} \int \mathrm{d}^d y \ \mathrm{d}^d z \ \mathrm{d}^d w \ \mathrm{d}^d v \ \partial^\lambda C(x-y) A_{[\mu}^c(y) \partial_{\lambda]} C(y-z) \ \Big\{ \\
&\quad \quad \quad - \partial^\rho \left( A_\rho^e(z) C(z-w) \right) A^{g \, \nu}(w) \partial^\sigma C(w-v) A_\sigma^h(v) A_\nu^i(v) \\ 
&\quad \quad \quad - A_\rho^e(z) \partial^\rho C(z-w) A^{g \, \nu}(w) \partial^\sigma C(w-v) A_\sigma^h(v) A_\nu^i(v) \Big\}  \\
&\quad + \frac{2g^4}{3} f^{a b c} f^{b d e} f^{d f g} f^{f h i} \int \mathrm{d}^d y \ \mathrm{d}^d z \ \mathrm{d}^d w \ \mathrm{d}^d v \ \partial^\lambda C(x-y) \\
&\quad \quad \quad \times A_{[\mu}^c(y) \partial_{\lambda]} C(y-z) A^{e \,  \rho}(z) \partial^\nu C(z-w) A_{[\rho}^g(w) \partial^\sigma C(w-v) A_{|\sigma|}^h(v) A_{\nu]}^i(v)  \\
&\quad + \frac{3g^4}{2} f^{a b c} f^{b d e} f^{d f g} f^{f h i} \int \mathrm{d}^d y \ \mathrm{d}^d z \ \mathrm{d}^d w \ \mathrm{d}^d v \ \partial_\lambda C(x-y)  \ \Big\{ \\
&\quad \quad \quad + 4 A^{c \, \rho}(y) \partial^{[\lambda} C(y-z) A^{e \, \nu]}(z) \partial^\sigma C(z-w) A_{[\mu}^g(w) \partial_\rho C(w-v) A_\sigma^h(v) A_{\nu]}^i(v) \\
&\quad \quad \quad - 4 A_{\rho}^c(y) \partial_{[\mu} C(y-z) A_{\nu]}^e(z) \partial_\sigma C(z-w) A^{g\, [\lambda}(w) \partial^\rho C(w-v) A^{h \, \sigma}(v) A^{i \, \nu]}(v) \\
&\quad \quad \quad - A^{c \, \rho}(y) \partial_{[\rho} C(y-z) A_{\sigma]}^e(z) \partial_\mu C(z-w) A_\nu^g(w) \partial^{[\lambda} C(w-v) A^{h \, \sigma}(v) A^{i \, \nu]}(v) \\
&\quad \quad \quad + A_\rho^c(y) \partial^{[\rho} C(y-z) A^{e \, \sigma]}(z) \partial^\lambda C(z-w) A^{g \, \nu}(w) \partial_{[\mu} C(w-v) A_\sigma^h(v) A_{\nu]}^i(v) \Big\} \\
&\quad + \frac{3g^4}{2} f^{a b c} f^{b d e} f^{d f g} f^{f h i} \int \mathrm{d}^d y \ \mathrm{d}^d z \ \mathrm{d}^d w \ \mathrm{d}^d v \ \partial^\lambda C(x-y) \ \Big\{ \\
&\quad \quad \quad - A_{[\lambda}^c(y) \partial_{\mu]} C(y-z) A^{e \,  \rho}(z) \partial^\sigma C(z-w) A^{g \, \nu}(w) \partial_{[\rho} C(w-v) A_\sigma^h(v) A_{\nu]}^i(v) \\
&\quad \quad \quad - A^{c \, [\rho}(y) \partial_\lambda C(y-z) A^{e \, \sigma]}(z) \partial_\rho C(z-w) A^{g \, \nu}(w) \partial_{[\mu} C(w-v) A_\sigma^h(v) A_{\nu]}^i(v) \\
&\quad \quad \quad + A^{c \, [\rho}(y) \partial_\mu C(y-z) A^{e \, \sigma]}(z) \partial_\rho C(z-w) A^{g \, \nu}(w) \partial_{[\lambda} C(w-v) A_\sigma^h(v) A_{\nu]}^i(v) \\
&\quad \quad \quad + A^{c \, [\rho}(y) \partial^{\sigma]} C(y-z) A_\lambda^e(z) \partial_\rho C(z-w) A^{g \, \nu}(w) \partial_{[\mu} C(w-v) A_\sigma^h(v) A_{\nu]}^i(v) \\
&\quad \quad \quad - A^{c \, [\rho}(y) \partial^{\sigma]} C(y-z) A_\mu^e(z) \partial_\rho C(z-w) A^{g \, \nu}(w) \partial_{[\lambda} C(w-v) A_\sigma^h(v) A_{\nu]}^i(v) \Big\} 
\end{align*}
\newpage
\begin{align*}
\hphantom{(\mathcal{T}_g \, A)_\mu^a(x)}
&\quad + \frac{g^4}{2} f^{a b c} f^{b d e} f^{d f g} f^{f h i} \int \mathrm{d}^d y \ \mathrm{d}^d z \ \mathrm{d}^d w \ \mathrm{d}^d v \ \partial^\lambda C(x-y) A^{\rho \, c}(y) \ \Big\{ \\
&\quad \quad \quad + 8 \, \partial^\nu C(y-z) A_{[\mu}^e(z) \partial_\lambda C(z-w) A_\rho^g(w) \partial^\sigma C(w-v) A_{|\sigma|}^h(v) A_{\nu]}^i(v) \\
&\quad \quad \quad  + 2 \,  \partial_\lambda C(y-z) A^{e \, \nu}(z) \partial_{[\mu} C(z-w) A_\rho^g(w) \partial^\sigma C(w-v) A_{|\sigma|}^h(v) A_{\nu]}^i(v) \\ 
&\quad \quad \quad - 2 \, \partial_\rho C(y-z) A^{e \, \nu}(z) \partial_{[\mu} C(z-w) A_\lambda^g(w) \partial^\sigma C(w-v) A_{|\sigma|}^h(v) A_{\nu]}^i(v) \\ 
&\quad \quad \quad  - 2 \,  \partial_\mu C(y-z) A^{e \, \nu}(z) \partial_{[\lambda} C(z-w) A_\rho^g(w) \partial^\sigma C(w-v) A_{|\sigma|}^h(v) A_{\nu]}^i(v) \\ 
&\quad \quad \quad - \frac{3}{2} \partial_{[\mu} C(y-z) A_{\rho]}^e(z) \partial^\sigma C(z-w) A^{g \, \nu}(w) \partial_{[\lambda} C(w-v) A_\sigma^h(v) A_{\nu]}^i(v) \\ 
&\quad \quad \quad + \frac{3}{2} \partial_{[\lambda} C(y-z) A_{\rho]}^e(z) \partial^\sigma C(z-w) A^{g \, \nu}(w) \partial_{[\mu} C(w-v) A_\sigma^h(v) A_{\nu]}^i(v) \\ 
&\quad \quad \quad + \frac{3}{2} \partial_{[\mu} C(y-z) A_{\lambda]}^e(z) \partial^\sigma C(z-w) A^{g \, \nu}(w) \partial_{[\rho} C(w-v) A_\sigma^h(v) A_{\nu]}^i(v) \\ 
&\quad \quad \quad + \partial_{[\mu} C(y-z) A_\lambda^e(z)  \partial^\nu C(z-w) A_{\rho]}^g(w) \partial^\sigma C(w-v) A_\sigma^h(v) A_\nu^i(v) \\ 
&\quad \quad \quad - \partial_{[\mu} C(y-z) A_\lambda^e(z) \partial^\nu C(z-w) A_{|\nu}^g(w) \partial^\sigma C(w-v) A_{\sigma|}^h(v) A_{\rho]}^i(v) \Big\} \\
&\quad + \frac{3g^4}{4} f^{a b c} f^{b d e} f^{d f g} f^{f h i} \int \mathrm{d}^d y \ \mathrm{d}^d z \ \mathrm{d}^d w \ \mathrm{d}^d v \ \partial^\lambda C(x-y) A^{\rho\, c}(y) \ \Big\{ \\
&\quad \quad \quad - 20 \,  \partial^\nu C(y-z) A^{e \,  \sigma}(z) \partial_{[\nu} C(z-w) A_\mu^g(w) \partial_\sigma C(w-v) A_\lambda^h(v) A_{\rho]}^i(v) \\
&\quad \quad \quad - 4 \, \partial_\rho C(y-z) A^{e \, \nu}(z) \partial^\sigma C(z-w) A_{[\mu}^g(w) \partial_\lambda C(w-v) A_\sigma^h(v) A_{\nu]}^i(v) \\
&\quad \quad \quad + 4 \, \partial^\sigma C(y-z) A_\sigma^e(z) \partial^\nu C(z-w) A_{[\mu}^g(w) \partial_\lambda C(w-v) A_\rho^h(v) A_{\nu]}^i(v) \\
&\quad \quad \quad - 2 \, \partial_{[\mu} C(y-z) A^{e \,  \sigma}(z) \partial_{\lambda]} C(z-w) A^{g \, \nu}(w) \partial_{[\rho} C(w-v) A_\sigma^h(v) A_{\nu]}^i(v) \\
&\quad \quad \quad + 2 \,  \partial^\sigma C(y-z) A_{[\mu}^e(z) \partial_{\lambda]} C(z-w) A^{g \, \nu}(w) \partial_{[\rho} C(w-v) A_\sigma^h(v) A_{\nu]}^i(v) \\
&\quad \quad \quad - 2 \, \partial^\sigma C(y-z) A_{[\sigma}^e(z) \partial_{\rho]} C(z-w) A^{g \, \nu}(w) \partial_{[\mu} C(w-v) A_\lambda^h(v) A_{\nu]}^i(v) \\
&\quad \quad \quad - 2  \, \partial_{[\mu} C(y-z) A_\lambda^e(z) \partial_{\rho]} C(z-w) A^{g \, \nu}(w) \partial^\sigma C(w-v) A_\sigma^h(v) A_\nu^i(v) \\ 
&\quad \quad \quad - 2 \, \partial^\sigma C(y-z) A_\rho^e(z) \partial^\nu C(z-w) A_{[\mu}^g(w) \partial_\lambda C(w-v) A_{\sigma]}^h(v) A_\nu^i(v) \\
&\quad \quad \quad - \partial_\rho C(y-z) A_\sigma^e(z) \partial^\sigma C(z-w) A^{g \, \nu}(w) \partial_{[\mu} C(w-v) A_\lambda^h(v) A_{\nu]}^i(v) \\
&\quad \quad \quad - \partial^\sigma C(y-z) A_\sigma^e(z) \partial_\mu C(z-w) A^{g \, \nu}(w) \partial_{[\lambda} C(w-v) A_\rho^h(v) A_{\nu]}^i(v) \\
&\quad \quad \quad - \partial^\sigma C(y-z) A_\rho^e(z) \partial^\nu C(z-w) A_{[\mu}^g(w) \partial_{|\nu|} C(w-v) A_\lambda^h(v) A_{\sigma]}^i(v) \\ 
&\quad \quad \quad + \partial^\sigma C(y-z) A_\rho^e(z) \partial^\nu C(z-w) A_\nu^g(w) \partial_{[\mu} C(w-v) A_\lambda^h(v) A_{\sigma]}^i(v) \\
&\quad \quad \quad + \partial^\sigma C(y-z) A_\sigma^e(z) \partial_\lambda C(z-w) A^{g \, \nu}(w) \partial_{[\mu} C(w-v) A_\rho^h(v) A_{\nu]}^i(v) \Big\} \\
&\quad + \mathcal{O}(g^5) \, .
\end{align*}
\end{footnotesize}%

\bibliography{PhD_Thesis_Main}

\end{document}